\PassOptionsToPackage{pdfpagelabels=false}{hyperref} 
\documentclass[fleqn]{mnras}

\usepackage{natbib}
\usepackage{graphicx}	
\usepackage{amsmath}	
\usepackage{amssymb}	
\usepackage{xcolor}
\usepackage{float}
\usepackage{mathptmx}
\usepackage{ulem}
\usepackage[utf8]{inputenc}
\usepackage{rotating}
\usepackage{booktabs}
\usepackage{tabularx}
\usepackage{longtable}
\usepackage{mathtools}
\usepackage{tipa}

\def\eq{\begin{equation}}
\def\en{\end{equation}}

\def\etal{{\it et al}\thinspace}

\def\apj{{\it Ap.J.}\thinspace}
\def\apjl{{\it Ap.J. Lett.}\thinspace}
\def\apjs{{\it Ap.J. Suppl.}\thinspace}
\def\aj{{\it A.J.}\thinspace}

\def\aap{{\it A\&A}\thinspace}
\def\aaps{{\it A\&A Suppl.}\thinspace}
\def\mnras{{\it MNRAS}\thinspace}
\def\P3hat{{\mathaccent 94 P}_3}

\def\nat{{\it Nature}\thinspace}

\def\etal{{\it et al.}\thinspace}

\def\eg{{\it e.g.,}\thinspace}

\title[Pulsar Beam Geometry at Lower Frequency:  Sources Outside the Arecibo Sky]{Radio Pulsar Beam Geometry at Lower Frequencies:  Bright Sources Outside the Arecibo Sky\thanks{This paper is dedicated to our colleagues at the Institute for Astronomy, Kharkiv, Ukraine}}

\author[Joanna Rankin]
{Joanna Rankin\thanks{E-mail: Joanna.Rankin@uvm.edu} \\
Physics Department, University of Vermont, Burlington, VT 05405, USA} 

\date{Accepted XXX. Received XXX; in original form XXX}

\pubyear{XX}

\begin{document}
\label{firstpage}
\pagerange{\pageref{firstpage}--\pageref{lastpage}}
\maketitle

\renewcommand{\thetable}{A\arabic{table}}
\setcounter{table}{0}

\begin{abstract}
We present pulsar emission beam analyses and models in an effort to examine pulsar geometry and physics at the lowest frequencies scattering permits.  We consider two populations of well-studied pulsars that lie outside the Arecibo sky, the first drawing on the Jodrell Bank Gould \& Lyne survey down to --35\degr\ declination and a second using Parkes surveys in the far south.  These assemble the full sky population of 487 pulsars known before the late 1990s which conveniently all have ``B'' names.  We make full use of the core/double-cone emission beam model to assess its efficacy at lower frequencies, and we outline how different pair plasma sources probably underlie its validity.  The analysis shows that with a very few exceptions pulsar radio emission beams can be modeled quantitatively with two concentric conal beams and a core beam of regular angular dimensions at 1 GHz. Further, the beamforms at lower frequencies change progressively in size but not in configuration.  Pulsar emission-beam properties divide strongly depending on whether the plasma excitation is central within the polar fluxtube producing a core beam or peripheral along the edges generating conal beams, and this seems largely determined by whether their spindown energy is greater or less than about 10$^{32.5}$ ergs/s.  Core emission dominated pulsars tend concentrate closely along the Galactic plane and in the direction of the Galactic center; whereas conal pulsars are somewhat more uniformly distributed both in Galactic longitude and latitude.  Core dominated pulsars also tend to be more distant and particularly so in the inner Galaxy region.  
\end{abstract}




\section{Introduction}
Pulsars emit mostly in the radio frequency regime, with rotation periods typically around 1 s down to a few ms. Radiation is generated primarily by outflowing plasma along the polar flux tube that connects a pulsar's magnetic polar cap to the external environment \citep{ruderman}. The physics and dynamics of these processes are still an open and ongoing field of study \citep{Harding_2017}, so observations and analyses pertaining to the emission region are essential in constraining physical theories of the magnetosphere. Key such issues are the angular structures of pulsar emission beams and their spectral variations, the subjects of our analyses below.

As the pulsar rotates, so too will its radiation pattern. As the majority of the radiation generation is confined to a narrow polar fluxtube, this leads to a high directionality in the subsequent radiation pattern, which is called the pulsar beam.  While in the far field, an observer will see intensity along a narrow line of sight through the radiation pattern. Integrating over a large number of rotations accrues to a stable average radiation pattern called the average profile\footnote{Average profiles are largely time-stable and characteristic of each individual pulsar, which make them useful for examining emission properties, pulse structure, and lastly the emission geometry.  However, a few pulsars have ``modes'' with different associated profiles.  Most pulsars have only a main pulse, but a few also have an interpulse roughly half a rotation away.}, while substructure within the profile are called components.  

Multi-component pulse profiles together with their frequency evolution provide the strongest evidence for an organized emission geometry.  This in turn suggests \citep[\eg][]{Radhakrishnan} their magnetic fields are approximately dipolar in the radio emission region. It should be stressed that this same assumption does not hold for other subpopulations, such as millisecond pulsars (MSPs)\footnote{There is evidence that a few MSPs have component structures similar to those seen in the canonical pulsar population and thus most likely possess approximately dipolar fields. See \cite{Rankin_2017} for more details}, or magnetars.    
 
Observational efforts to delineate the topology of pulsar emission beams are complicated by the accidental traverses our sightline makes through each pulsar's beam.  As we have no general means of mapping a pulsar's full emission beam pattern, investigations have proceeded by proposing a reasonable beam topology and then attempting to verify it using a large ensemble of pulsars. The first such beam model consisted of a single hollow cone \citep{Radhakrishnan, Komesaroff}.  Subsequent efforts added a central pencil beam and the possibility of two concentric hollow conical beams \citep{backer}, called the core/double-cone model.  Building on the this work, \citep{rankin1983a} identified two distinct types of frequency evolution that single profiles exhibit: one often assumed a triple form at high frequencies and the other showed an increasing bifurcating width at lower frequencies.  She attributed the one to a central traverse through the core (pencil) beam (often with a pair of ``outriding'' conal components appearing due to a flatter spectrum) and the other to an oblique traverse through a conal beam. And that a few pulsars had five components suggested a double cone configuration with a central core beam.

The spectral evolution of pulsar profiles is then crucial to interpreting their underlying emission beamforms.  However, most pulsars have been observed only within about an octave of 800 MHz where their flux relative to the background Galactic noise is maximal. Despite the fact that the first pulsars were discovered in the 80-MHz band \citep{hewish}, many pulsars are difficult to observe at lower frequencies both because of spectral turnovers and the difficulty of correcting for the dispersing of their signals in traversing the ionized Galactic interstellar medium (ISM) \citep{2004hpa..book.....L}.  In addition, broadening by scattering in the ISM can so distort profiles as to obliterate their intrinsic structure.  Pulsar beaming studies then have necessarily focused on determining configurations around 1 GHz (as do those we reference below in \cite[together hereafter ET VI]{rankin1993a,rankin1993b} while knowing that there is much to learn by extending their analyses to the lowest possible frequencies. 

The Pushchino Radio Astronomy Observatory (PRAO) in Russia has long pioneered 100 MHz studies of pulsar emission using their Large Phased Array (LPA). Recent surveys by \cite{kuzmin1999} and \citet{MM10} provide a foundation for our work here.  More recently, the Low Frequency Array (LOFAR) in the Netherlands has produced an abundance of high-quality profiles with their High Band Surveys by \cite{bilous2016, pilia2016} in the 100-200 MHz band and supplemented with their Low Band Surveys \citep{bilous2019, Bondonneau} below 100 MHz. Further, the Institute for Radio Astronomy (IRA) in Ukraine has long led in making decametric observations using their UTR-2 instrument, most recently by \citet{Zakharenko2013,Kravtsov22}.  

Our purpose here is to assess and extend the efficacy of the core/double-cone beam model at frequencies down to 100 MHz or below, and to compare this geometry with new and existing 1-GHz models from ET VI. Our goal in this work is to identify the physical implications of pulsar beamform variations with radio frequency, within the context and limitations imposed by interstellar scattering. 

Here we consider the emission-beam geometries of two large groups of bright pulsars lying outside the Arecibo sky.  The first group parallels the remarkable multiband polarimetric, Jodrell Bank survey \citet[GL98]{GL98} of the entire sky down to about --35\degr\ declination, as supplemented by numerous more recent efforts \citep[\eg][]{JKMG2008}.  The GL98 survey usually provides profiles at 1.6 and 1.4 GHz as well as others at 950, 600, 400 and 240 MHz.  The second group of Far South pulsars assembles and interprets the polarimetric observations carried out using the Parkes telescope in Australia.  This is comprised of two subgroups, one of multiband observations \citep[\eg][]{HMAK,MHM,MHMA}, as supplemented by various later polarimetry programs \citep[\eg][hereafter JK18]{WMLQ, QMLG95, MHQ, kj06, JohnstonI, JohnstonII, jk18} and a second with only the JK18 1.4-GHz observations. Most of this group lies roughly in the Galactic Center direction and focuses on higher frequencies because scattering is often so severe.  Together these include most of the pulsars studied (and often discovered) by the two great 76-m Lovell and 70-m Parkes telescopes---and conveniently this joint population consists of the 487 objects known before 1993 and thus have ``B'' discovery names.  

The first group of 195 includes most of the pulsars that have been or can be observed down into the 100-MHz band or below as northern instruments were used to conduct them:  the PRAO surveys \citep[KL99] {kuzmin1999}\footnote{Some of these profiles reflect processing aimed at deconvolving the effects of scattering; in some cases this seems to have worked well and in others not.} and \citep[MM10]{MM10}; the LOFAR High Band \citep[BKK+, PHS+]{bilous2016, pilia} and Low Band Surveys \citep[BKK++, BGT+]{bilous2019, Bondonneau}; and the Kharkov survey \citep[ZVK+]{Zakharenko2013}.  The second group of 148 includes only a few pulsars that have been observed down to 170 MHz \citep{MHMb}. Readers can readily access most of these profiles on the European Pulsar Network (hereafter EPN) Database.\footnote{http://www.epta.eu.org/epndb/ and in some cases there are minor differences between the published and the database profiles.}  

In what follows, \S\ref{sec:models} reviews the geometry and theory of core and conal beams and describes how our beaming models are computed and displayed,  \S\ref{sec:scattering} discusses scattering and its effects at low frequency, \S\ref{sec:discussion} assesses our results, and \S\ref{sec:summary} gives a short summary.  In the Appendices, we discuss the interpretation and beam geometry of each pulsar and show the results of analyses clarifying the beam configurations. 

Appendix A Tables~A1--A3 show the observational sources, physical parameters and beam geometries of the Jodrell GL98 pulsar population (see the samples below), beam models for which are in turn plotted in Figs.~A1--A15.  Appendix B Tables~B1--B3 together with Figs.~B1--B6 and Tables~B4--B5 give parallel information for the Parkes Far South multifrequency and JK18 1.4-GHz groups, respectively. Machine readable ascii versions of each table are provided in the supplementary material.

\section{Core and Conal Beams}
\label{sec:models}
\subsection{Radio Pulsar Speciation by Beam Traverse}
Canonical pulsar average profiles are observed to have up to five components \citet{rankin1983a}. This places an important constraint on the emission-beam topology and underlies the conception of the core/double-cone beam model as originally proposed by \citet{backer}. 

Following this model, pulsar profiles divide into two major categories depending on whether core or conal emission components are dominant at about 1 GHz.  Prominent core components occur in single ({\textbf S$_{t}$}) profiles consisting of an isolated core component, in core-cone triple ({\textbf T}) profiles with a core component flanked by a pair of outriding conal components, or in five-component ({\textbf M}) profiles where the central core component is flanked by both an inner and outer pair of conal components. 

By contrast, entirely conal profiles include those with a single conal ({\textbf S$_{d}$}) component, double profiles ({\textbf D}) consisting of a pair of conal components (occasionally with a weak core component in-between), or conal triple (c{\textbf T}) or quadruple (c{\textbf Q}) profiles where the sightline encounters both conal beams. Outer conal component pairs tend to have an increasing separation with wavelength, whereas inner conal pairs tend to have more constant separations.  Also important to these profile classes is single-pulse phenomenology. Subpulse drift has long been considered a defining feature of conal emission, and provides an important role in interpreting a profile \citep[\eg][]{et3}. 

Each profile class tends to evolve with frequency in a characteristic manner:  core single ({\textbf S$_{t}$}) profiles often ``grow'' a pair of conal outriders at high frequency, whereas conal single ({\textbf S$_{d}$}) profiles tend to broaden and bifurcate at low frequency.  Triple ({\textbf T}) profiles usually show all three components over a very broad band, but the relative component intensities can change greatly.  Five-component ({\textbf M}) profiles tend to exhibit their individual components most clearly at meter wavelengths; at high frequency the components often become conflated into a ``boxy'' form and at low frequency the inner cone often weakens relative to the outer one.

A large proportion of pulsars are well associated with a particular profile class and their beam geometry modeled (see \S\ref{sec:quantgeom}) with the core/double-cone emission geometry as described in the \textit{Empirical Theory of Pulsar Emission} series. Clearly, however, these profile species can only give partial guidance, because the three emission beams with their different dimensions and intensities accrue to the profile in widely different ways due to their accidental encounters of our sightline.

\subsection{Quantitative Beam Geometry}
\label{sec:quantgeom}
Once a pulsar's emission-beam configuration is identified, then application of spherical geometry opens the possibility of measuring the angular beam dimensions---resulting in a quantitative emission-beam model for a given pulsar.  Then, 
application to larger groups of pulsars show how the beam dimensions scale with pulsar rotational period $P$ and perhaps other factors.

Two key angles describing the geometry are the magnetic colatitude (angle between the rotation and magnetic axes) $\alpha$ and the sightline-circle radius (the angle between the rotation axis and the observer’s sightline) $\zeta$, where the sightline impact angle $\beta$ = $\zeta-\alpha$.  The three beams are found to have specific angular dimensions at 1 GHz in terms of a pulsar's polar cap angular diameter, {$\Delta_{PC}$} = $2.45\degr P^{-1/2}$ \citep{rankin1990}.  The outside half-power radii of the inner and outer cones, {$\rho_{i}$} and {$\rho_{o}$} are $4.33\degr P^{-1/2}$ and $5.75\degr P^{-1/2}$ \citep{rankin1993a}.  Other studies such as \citet{gil}, \citet{kramer}, \citep{Bhattacharya}, and \citep{mitra1999} have come to very similar conclusions 

In practice, the magnetic colatitude $\alpha$ can be estimated from the width of the core component when present, as its expected half-power width at 1 GHz, $W_{\rm core}$ has empirically been shown to scale as {$\Delta_{\rm PC}/\sin\alpha$} (ET IV).  The sightline impact angle $\beta$ can then in turn be estimated from the steepest gradient of the polarization angle (PPA) $\chi_0$ traverse (at the inflection point in longitude $\varphi$). $R$=$|d\chi/d\varphi|$ measures the ratio $\sin\alpha/\sin\beta$.  Conal beam radii can similarly be estimated from the outside half-power width of a conal component or conal component pair at 1 GHz $W_{\rm cone}$ together with $\alpha$ and $\beta$ using eq.(4) in ET VIa.  The characteristic height of the emission can then be computed assuming dipolarity using eq.(6).

These 1-GHz inner and outer conal emission heights have typically been seen to concentrate around 130 and 220 km, respectively.  However, it is important to note that these are {\it characteristic} emission heights, not physical ones, estimated using the convenient but perhaps problematic assumption that the emission occurs adjacent to the ``last open'' field lines at the polar fluxtube edge.  More physical emission heights can be estimated using aberration/retardation (see \citet{Blaskiewic} as corrected by \citet{dyks}), and these are typically somewhat larger than the characteristic emission heights.

A number of studies have followed expanding the population of pulsars with classifications.  Most have looked at the  frequency evolution between $\mathtt{\sim}${300} MHz and $\mathtt{\sim}$1500 MHz in order to model a pulsar's emission geometry at 1 GHz \citep{Weisberg1999, Weisberg2004, mitra2011, brinkman_freire_rankin_stovall}. 

However successful the core/double-cone model has been in assessing and quantitatively modeling pulsar beam topology, 
it is only a model and a means for identifying potentially significant regularities and/or anomalies.  Probably its validity stems from the magnetic field structure being nearly dipolar in the emission region.  So far a few pulsars are found to exhibit ``pre-/postcursor'' features \citep{Basu2015} and several others are known to have weak emission in regions of their rotation cycles far from their main pulse or interpulse \citep[\eg][]{rankin1995}.  The core/double-cone model currently has no ability to comprehend these aspects of pulsar radiation.

\begin{figure}
\begin{center}
\includegraphics[width=80mm,angle=0.]{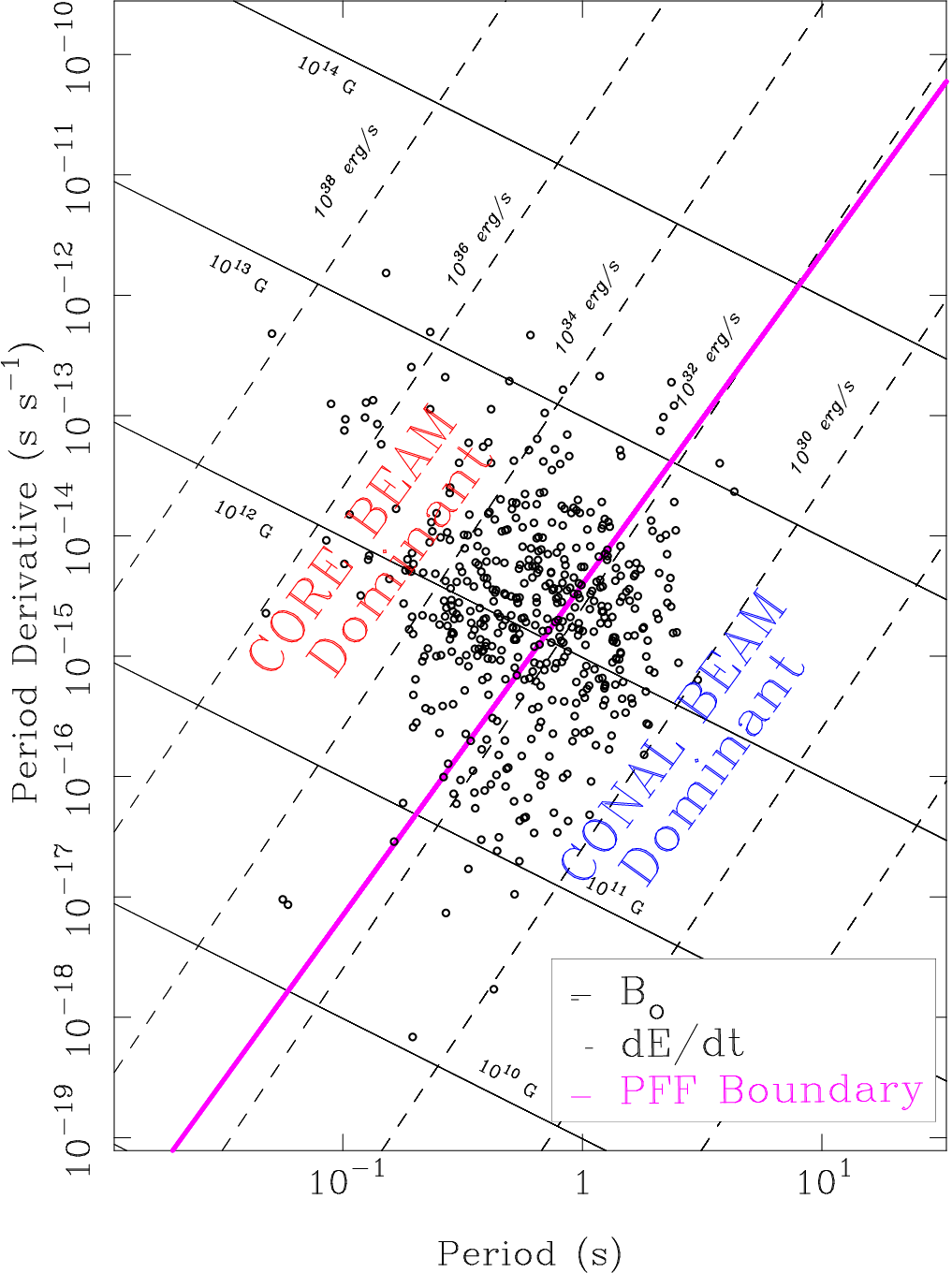}
\caption{$P$-$\dot P$ diagram showing the position of ``B'' population pulsars along with the PFF boundary line. Core-emission dominated pulsars tend to lie to the upper left of the boundary line while those mainly emitting conal radiation fall to the lower right (see text).}
\label{fig1}
\end{center}
\end{figure}

\subsection{Radio Pulsar Speciation by Plasma Source}
\label{sec:plasma}
The outflowing electron-positron plasma that gives rise to a pulsar's emission is partly and or fully generated by a polar "gap" \citep{ruderman}, just above the stellar surface.  \citet{Timokhin} find that this plasma is generated in one of two pair-formation-front (PFF) configurations:  for the younger, energetic part of the pulsar population, pairs are created at some 100 m above the polar cap in a central, uniform (1-D) gap potential that produces copious backflow heating and thus thermal X-rays---thus a 2-D PFF; whereas for older pulsars the pair-formation front has a lower, annular shape and extends up along the conducting walls of the polar flux tube, thus becoming three-dimensional (cup shaped, 3-D) with a 2-D gap potential and greatly reduced backflow heating. 

Curvature radiation generates the pair plasma in both cases, dominating the inverse-Compton process. An approximate boundary line between the flat and cup-shaped pair-formation geometries---and thus pulsar populations---is plotted on the $P$-$\dot P$ diagram of Fig~\ref{fig1}, so that the more energetic pulsars are to the top left and those less so at the bottom right.  Its dependence is $\dot P=$4.29$\times10^{-29} \rho P^{9/4}$, where the fieldline curvature $\rho$ = 9.2$\times$10$^7 P^{1/2}$ cm, overall giving $\dot P=$3.95$\times$$10^{-15}P^{11/4}$.  

We emphasize in this context that pulsars with dominant core emission tend to lie to the upper left of the line, whereas those with dominant conal emission tend to lie to the lower right of the line.  This boundary line seemingly divides the pulsar population between the younger, more energetic stars whose radiation is core dominated and those older pulsars whose emission is mainly conal.

This division is fundamental to the core/double-cone beam model of ET VI, discussed just above, where the radio pulsar profile classes were first defined.  Pulsars with conal single ({\textbf S$_d$}), double ({\textbf D}) and five-component ({\textbf M}) profiles tend to fall below the boundary lines to the right, whereas those with core-single ({\textbf S$_t$}) profiles are found to the upper left of the boundary.  Those with triple ({\textbf T}) profiles are found on both sides of the boundary, but divide roughly into core-dominated and conal-dominated groups, which are delineated by the boundary.  In the parlance of ET VI, the division corresponds to an acceleration potential parameter $B_{12}/P^2$ of about 2.5, which in turn represents an energy loss $\dot E$ of 10$^{32.5}$ ergs/s.  This delineation also squares well with \citet{Weltevrede2008}'s observation that high energy pulsars have distinct properties and \citet{basu2016}'s demonstration that conal drifting occurs only for pulsars with $\dot E$ less than about $10^{32}$ ergs/s.

\subsection{Computation and Presentation of Geometric Models}
Two key observational values underlie the computation of conal radii at each frequency, and thus the model overall: the profile width and the polarization position-angle (PPA) sweep rate $R$. The former gives the angular scale of the emission beam and the latter the impact angle $\beta$ showing how the sightline crosses the beam.  Figure 2 of ET VI depicts this configuration and the spherical geometry underlying the emission.

Empirically, core radiation is found to have a bivariate Gaussian (von Mises) beamform such that its 1-GHz (and often invariant) width measures $\alpha$ but provides no $\beta$ information. If a pulsar has a core component, we attempt to use its width at around 1-GHz to estimate the magnetic colatitude $\alpha$, and when this is possible the $\alpha$ value is bolded in Appendix Tables~A3, B3 and C3 (see the sample below of Table A3) below.  $\beta$ is then estimated from $\alpha$ and the polarization position angle sweep rate $R$. The outside half-power (3 db) widths of conal components or pairs are measured, and the spherical geometry above then used to estimate the outside half-power conal beam radii.  Where $\alpha$ can be measured, the value is used, when not an $\alpha$ value is estimated by using the established conal radius or characteristic emission height for an inner or outer cone.  These conal radii and core widths are then computed for different frequencies wherever possible.  

\begin{table*}
\begin{center}
\caption{Sample: Observational Information; see Appendices}
\begin{tabular}{lccc|l|l|}
 \hline
 Pulsar & P & DM & RM  & References &  References \\
    (B1950) & (s) & ($pc/cm^{3}$) & ($rad$-$m^{2}$) &   & \\
    \hline
    & & & &  \multicolumn{1}{c|}{100 MHz} &   \multicolumn{1}{c|}{$<$100 MHz} \\
    \hline
    \hline
B0011+47 & 1.24 & 30.4 & -15.6 & GL98; Han+09; FDR; BKK+; PHS+; KL99 & BKK++; BGT+ \\
B0031--07 & 0.94 & 10.9 & 9.9 & HMAK;W93;JKMG;JK18;PHS+;BMM+;KL99 & MHMb;BKK++;BGT+;ZVK+ \\
B0037+56 & 1.12 & 92.5 & 15.3 & GL98; BKK+; PHS+; MM10 &  \\
B0052+51 & 2.12 & 44.0 & -64.1 & GL98; BKK+; PHS+; MM10 & BKK++ \\
B0053+47 & 0.47 & 18.1 & -34.2 & GL98; BKK+; PHS+; Han+09; MM10 & BKK++; BGT+; ZVK+ \\
\\
B0059+65 & 1.68 & 65.9 & -94.0 & GL98; PHS+; MM10 &  \\
B0105+65 & 1.28 & 30.5 & -27.1 & GL98; BKK+; PHS+; MM10 & BGT+ \\
B0105+68 & 1.07 & 61.1 & -33.0 & GL98; BKK+;MM10 & BKK++ \\
B0114+58 & 0.10 & 49.4 & -8.1 & GL98;Han+09; BKK+; PHS+; KL99 & BKK++ \\
B0136+57 & 0.27 & 73.8 & -94.1 & GL98; BKK+; FDR; PHS+; KL99 & BKK++ \\
\hline
\end{tabular}
\end{center}
Legend: For each pulsar the period, dispersion and rotation measures are given along with references to profiles in the bands both above and below 100 MHz.
\end{table*}

\begin{table*}
\caption{Sample: Pulsar Parameters; see Appendices}
\begin{center}
\begin{tabular}{lccc|ccccccc}
\hline
 Pulsar & L & B & Dist. & P & $\dot{P}$ & $\dot{E}$ & $\tau$ & $B_{surf}$ & $B_{12}/P^2$ & 1/Q \\ 
 (B1950) & (\degr) & (\degr) & (kpc) & (s) & ($10^{-15}$ s/s) & ($10^{32}$ ergs/s)  & (Myr) & ($10^{12}$ G) &   &   \\
\hline
\hline
B0011+47 & 116.50 & -14.63 & 1.78 & 1.241 & 0.56 & 0.12 & 34.8 & 0.63 & 0.4 & 0.3 \\
B0031--07 & 110.42 & -69.82 & 1.03 & 0.943 & 0.41 & 0.19 & 36.6 & 0.63 & 0.7 & 0.4 \\
B0037+56 & 121.45 & -5.57 & 2.42 & 1.118 & 2.88 & 0.81 & 6.2 & 1.82 & 1.5 & 0.7 \\
B0052+51 & 123.62 & -11.58 & 2.86 & 2.115 & 9.54 & 0.40 & 3.5 & 4.55 & 1.0 & 0.5 \\
B0053+47 & 123.80 & -14.93 & 1.12 & 0.472 & 3.33 & 12.0 & 2.3 & 1.27 & 5.7 & 1.8 \\
\\
B0059+65 & 124.08 & 2.77 & 2.50 & 1.679 & 5.95 & 0.50 & 4.5 & 3.20 & 1.1 & 0.6 \\
B0105+65 & 124.65 & 3.33 & 2.13 & 1.284 & 13.05 & 2.40 & 1.6 & 4.14 & 2.5 & 1.1 \\
B0105+68 & 124.46 & 6.28 & 1.98 & 1.071 & 0.05 & 0.02 & 353 & 0.23 & 0.2 & 0.1 \\
B0114+58 & 126.28 & -3.46 & 1.77 & 0.101 & 5.85 & 2200 & 0.3 & 0.78 & 75.8 & 12.6 \\
B0136+57 & 129.22 & -4.04 & 2.60 & 0.272 & 10.71 & 210 & 0.4 & 1.73 & 23.3 & 5.4 \\
\hline
\end{tabular}
\end{center}
Legend: For each pulsar the Galactic longitude, latitude and distance are given, along with the period ($P$), spindown ($\dot P$) and quantities computed from it: the energy loss rate ($\dot E$), spindown age ($\tau$), magnetic field ($B_{surf}$), $B_{12}/P^2$ and $1/Q$.  Values are based on the ATNF Pulsar Catalogue \citep{ATNF}.  
\end{table*}

\begin{table*}
\caption{Sample: Emission Beam Model Geometry; see Appendices}
\begin{center}
\begin{tabular}{lc|cccc|cccc|ccccc|ccc}
\hline
      Pulsar &  Class & $W_{c}$ & $\alpha$ & $R$ & $\beta$ &  $W_i$ & $\rho_i$ & $W_o$  & $\rho_o$ & $W_{c}$ & $W_i$ & $\rho_i$    & $W_o$  & $\rho_o$ & $W_{c}$ & $W_{i,o}$  & $\rho_{i,o}$ \\
  &   & (\degr) & (\degr) & (\degr/\degr) & (\degr) & (\degr) & (\degr) & (\degr) & (\degr) & (\degr) & (\degr) & (\degr) & (\degr) & (\degr) & (\degr) & (\degr) & (\degr) \\
  \hline
  & & \multicolumn{4}{c|}{1-GHz Geometry} & \multicolumn{4}{c|}{1-GHz Cone Sizes} & \multicolumn{5}{c|}{100-MHz Cone Sizes} & \multicolumn{3}{c}{$<$100 MHz} \\
  \hline
  \hline
B0011+47 & cT? &  --- & 6.7 & -1.8 & +3.7 & 19.6 & 4.0 & 48 & 5.1 &  --- & 27 & 4.2 &  --- &  --- &  --- & 23 & 4.1 \\
B0031--07 & Sd &  --- & 6 & -1.0 & +6.0 &  --- &  --- & 17.0 & 6.1 &  --- &  --- &  --- & 34 & 6.5 &  --- & 39 & 6.6 \\
B0037+56 & Sd? &  --- & 90 & +18 & +3.2 & 3.7 & 3.7 &  --- &  --- &  --- & 19 & 10.0 &  --- &  --- &  --- &  --- &  --- \\
B0052+51 & D/T? &  --- & 50 & $\infty$ & 0.0 &  --- &  --- & 10.1 & 3.9 &  --- &  --- &  --- & 15.5 & 5.9 &  --- &  --- &  --- \\
B0053+47 & St? & 8.0 & {\bf 26} &  --- &  --- &  --- &  --- &  --- &  --- & 14.9 &  --- &  --- &  --- &  --- & 88 &  --- &  --- \\
\\
B0059+65 & T & 4.0 & {\bf 28} & -16 & +1.7 &  --- &  --- & 16.9 & 4.4 &  --- &  --- &  --- & 21.8 & 5.5 &  --- &  --- &  --- \\
B0105+65 & Sd &  --- & 18 & -5 & +3.5 & 7.5 & 3.8 &  --- &  --- &  --- & 7.7 & 3.8 &  --- &  --- &  --- &  --- &  --- \\
B0105+68 & T & 5.7 & {\bf 25} & $\infty$ & 0.0 &  --- &  --- & 26.7 & 5.5 & $\sim$6 &  --- &  --- & $\sim$33 & 6.8 &  --- &  --- &  --- \\
B0114+58 & St & 11.7 & {\bf 41} & +1.2 &  --- &  --- &  --- &  --- &  --- & 29 &  --- &  --- &  --- &  --- &  --- &  --- &  --- \\
B0136+57 & St & 7.0 & {\bf 42} & +5.3 & -7.3 & $\sim$10 & 7.9 &  --- &  --- &  --- & $\sim$18 & 9.2 &  --- &  --- &  --- &  --- &  --- \\
\hline
\end{tabular}
\end{center}
Legend: For each pulsar the profile class is given along with profile measurements and beam-geometry-model values.  Core widths ($W_c$) and PPA sweep rates ($R$) contributing to the magnetic colatitude ($\alpha$) and sightline impact angle ($\beta$) comprise the first section; 1-GHz inner/outer conal dimensions ($W_{i,o}$) and beam radii ($\rho_{i,o}$) the second section; 100-MHz core and conal dimensions and beam radii the third; and lower frequency core and conal values the fourth section.
\end{table*}

For Group I our 1.6 GHz to 240 MHz profile measurements here are based on the GL98 compendium; however, some have also benefited from more recent multifrequency observations \citep[\eg][]{JKMG2008} or from particular studies \citep[\eg][]{Wu1998} as summarized in Table A1 (see the sample below).  For profiles with a single component, we have usually used the GL98 width measurements directly, modifying them only when asymmetries, multiple components, or judgment about noise levels required them.  We then extend the analysis using PRAO 102-MHz profiles, LOFAR High Band 100-200-MHz, and in some cases below 100 MHz using LOFAR Low Band, PRAO or Kharkov profiles.  For Group II the profiles at the available frequencies were measured individually.  Appendix Table~B1 gives the sources for these multifrequency profiles in the principal bands as well as each pulsar's observational parameters.   

{\textbf Appendix Tables~A2, B2 and B4 give the Galactic positions, distances and physical parameters that can be computed from the period $P$ and spindown $\dot P$ (see the sample below of Table A2)}---that is, the spindown energy $\dot E$, spindown age $\tau$, surface magnetic field $B_{\rm surf}$, the acceleration parameter $B_{12}/P^2$ and the reciprocal of \citet{pulsar_magnetosphere_book}'s similar $Q$ (=$0.5\ 10^{15} \dot P^{0.4} P^{-1.1}$) parameter.  

Following the analysis procedures of ET VI, we have measured outside conal half-power (3 db) widths and half-power core widths wherever possible.  However, we do not plot these directly.  Rather we use the beam dimensions to model the core and conal geometry as above, but here emphasizing as low a frequency range as possible. The model results are given in Appendix Tables~A3, B3 and B5 for the 1-GHz regime, and where possible the 100-200-MHz band and $<$100-MHz band as well (see the sample below of Table A3).  $W_{c}$,  $\alpha$, $R$ and $\beta$ are the 1-GHz core width, the magnetic colatitude, the PPA sweep rate and the sightline impact angle;  $W_i$/$W_o$ and $\rho_i$/$\rho_o$ are the inner and outer conal component widths and the respective beam radii at 1 GHz, the lowest frequency values in the 100-MHz band and the decametric band.

\begin{figure}
\begin{center}
\includegraphics[width=80mm,height=80mm,angle=-90.]{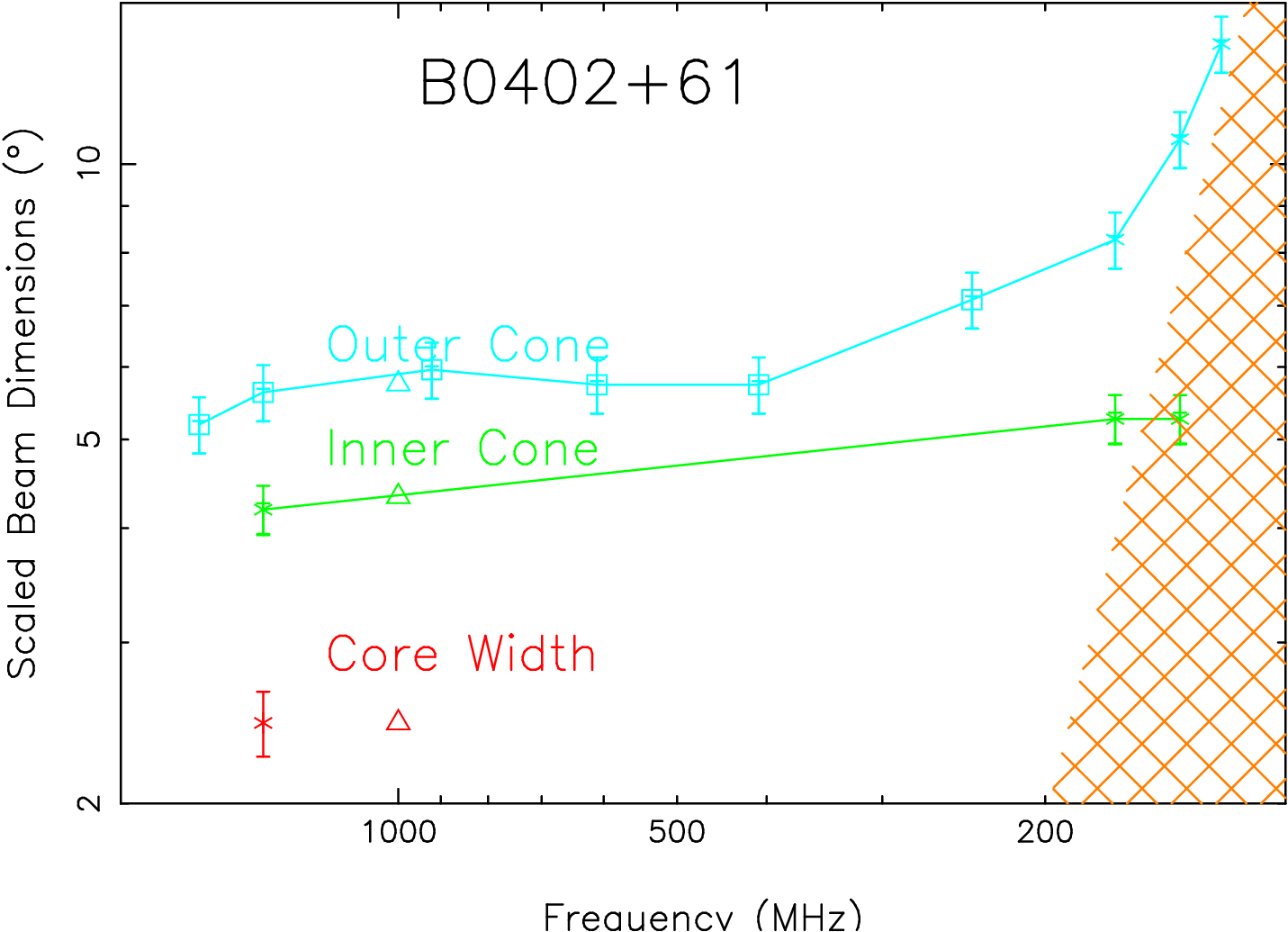}
\includegraphics[width=80mm,height=80mm,angle=-90.]{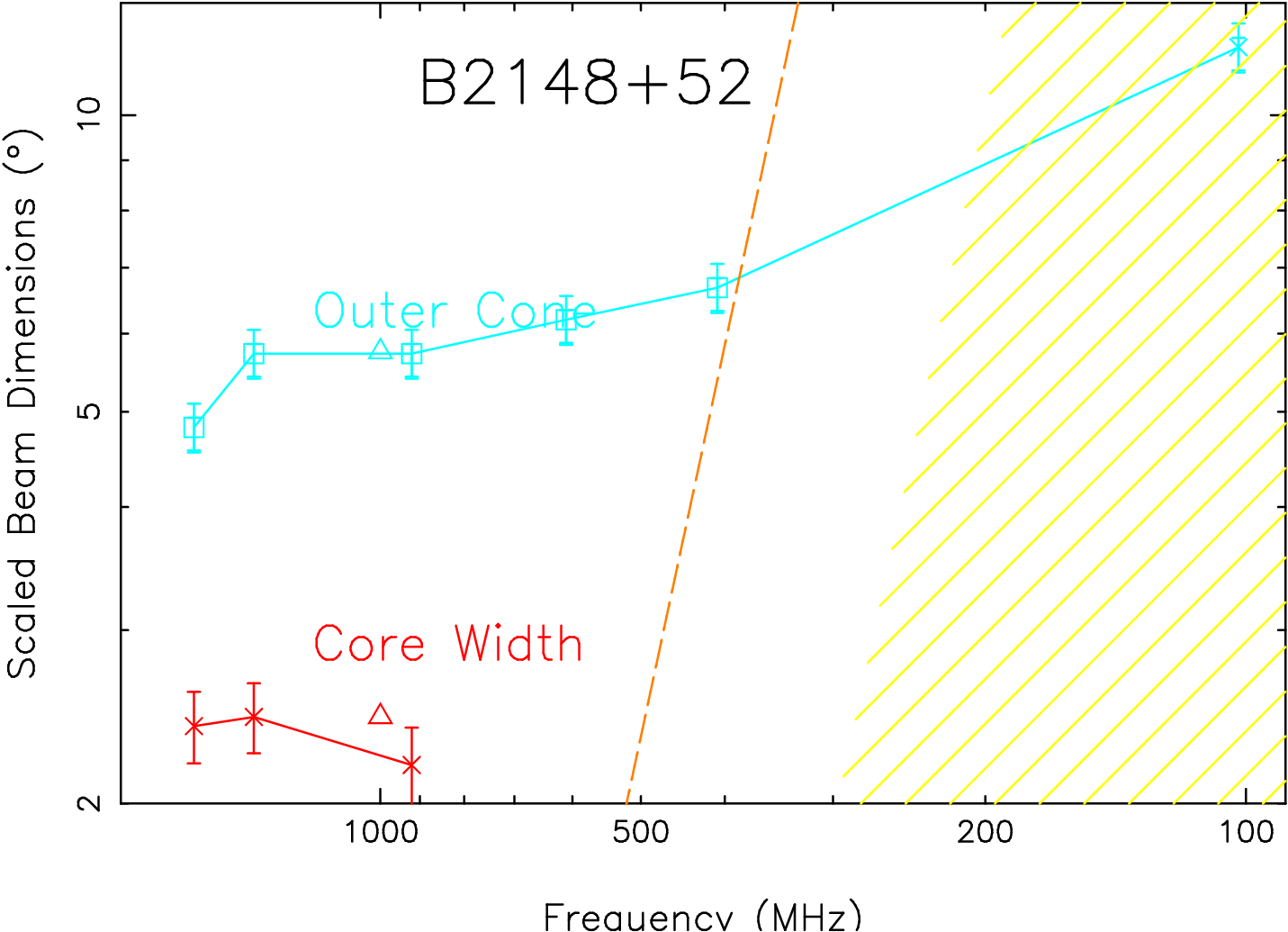}
\caption{Sample core/double-cone beam model display for pulsars B0402+61 and B1451--68.  Curves for the scaled outer and inner conal radii and core width are shown---the former by $\sqrt{P}$ and the latter by $\sqrt{P}\sin\alpha$.  Conal error bars reflect the rms of 10\% uncertainties in both the profile widths and PPA rate (see text), whereas the core errors reflect a uniform 10\%.  The triangles at 1 GHz indicate the established nominal dimensions for the beams.  The upper display gives an example of known low frequency scattering as indicated by the double hatching.}
\label{fig20}
\end{center}
\end{figure}

We depart from past practice by plotting our results in terms of core and conal beam dimensions as a function of frequency, not profile widths as has been the almost universal practice as described in Figure \ref{fig20}. The results of the model for each pulsar are plotted in Figs~A1 to A15 for Group I and Figs~B1 to B6 for Group II.  Model plots are omitted for pulsars when only a single frequency is available.  The plots are logarithmic on both axes, and labels are given only for exponents of base 10 in orders of 1, 2 and 5.  For each pulsar the plotted values represent the {\textit scaled} inner and outer conal beam radii and the core angular width, respectively.  The scaling plots each pulsar's beam dimensions as if it were an orthogonal rotator with a 1-sec period---thus the conal beam radii are scaled by a factor of $\sqrt{P}$ and the core width (diameter) by $\sqrt{P}\sin{\alpha}$.  This facilitates easy comparison of the beaming models for different objects as well as showing how each evolves with frequency relative to expected 1-GHz dimensions. The outer and inner conal radii are plotted with blue and green lines and the core diameter in red.  The nominal values of the three beam dimensions at 1 GHz are shown in each plot by a small triangle.  

Estimating and propagating the observational errors in the width values is very difficult; they come from different sources at different frequencies and sometimes entail unknown instrumental errors and issues in addition to errors we have made in interpreting the profiles and measuring them. Therefore, we have chosen rather to provide error bars reflecting the (scaled) beam-radii {$\rho_{\rm scaled}$} errors for a 10\% uncertainty in the width values and a 10\% uncertainty in the polarization position-angle (PPA) sweep rate---the former $0.1(1-\beta/\rho)\rho_{\rm scaled}$ and the latter $0.1(\beta/\rho)\rho_{\rm scaled}$.  The error bars shown reflect the {\it rms} of the two sources with the former indicated in the lower bar and the latter in the upper one.  For many pulsars only one of the errors is dominant so the bars corresponding to the two individual error sources are hard to see; however, B1727--18 provides a case where both can be seen clearly.  The errors shown for the core-beam angular diameters are also 10\% in the scaled width.

\section{Low Frequency Scattering Effects}
\label{sec:scattering}
Scattering in the local interstellar medium distorts and broadens profiles by delaying a portion of the pulsar's signal as it traverses through the interstellar medium.  For many pulsars the effect is as if the intrinsic profile is convolved with a truncated exponential function.  This results in an exponential ``tail'' that can go from being imperceptible to dominant within an octave or so due to its steep ($f^{\sim-4}$) frequency dependence. Scattering also reduces the detectability of pulsars at low frequency to the end that the lowest frequency profiles available often entail significant scattering.  Because our beam modeling efforts require us to distinguish between intrinsic profile dimensions and those distorted by scattering, we must also estimate the level of measured or estimated scattering for each pulsar and show this in relation to our beam models at low frequency.  

Fortunately, many of the pulsars in both groups have published scattering or scintillation studies that can be used to accurately estimate the scattering time at a given frequency. We are indebted to \citet{kuzmin_LL2007} for their extensive compendia of 100-MHz scattering times as well as other studies by \citet{abs86}, \citet{geyer} and \citet{Zakharenko2013}.  When these are available, they are shown on the model plots as double-hatched orange regions where the boundary reflects the scattering timescale at that frequency in scaled rotational degrees (\eg see the model plot for B0402+61 in Fig~\ref{fig20}).  

For pulsars where no scattering study is available, we use the mean scattering level determined in the foregoing study for a large ensemble of pulsars in the 100-MHz band as a function of dispersion measure (DM), where $t_{\rm scatt}$ is some $240\ DM^{2.2}/f(MHz)^{4.1}$ secs \citep{Kuzmin2001}.  While this mean scattering level is well-determined, the authors found that actual levels can depart from the average by up to an order of magnitude.  Therefore, our model plots show the average scattering level (where applicable) as yellow single hatching and with an orange line indicating 10 times this value as a rough upper limit (\eg see the model plot for pulsar B2148+52 in Fig~\ref{fig20}).

\section{Analysis and Discussion}
\label{sec:discussion}
\noindent\textit{\textbf{``B'' Populations}}: The ATNF Catalog\footnote{https://www.atnf.csiro.au/research/pulsar/psrcat/} lists about 487 normal pulsars with ``B'' discovery names---that is, sources that were discovered before the mid-1990s or so.  Of these, some 325 lie north of about --35\degr\ declination, the southern declination limit for the Jodrell Bank Lovell telescope. Remarkably, the 1998 \citeauthor{GL98} compendium includes fully 300 of these pulsars in their six frequency polarimetric survey.  Given that a few of the 325 were awkwardly positioned or too weak to provide useful polarized profiles, the GL98 survey provides the most complete large scale survey of pulsar characteristics ever conducted.  

In the south there are about 160 pulsars with ``B'' discovery names at declinations less than --35\degr.  Some 148 of these have been studied with the 70-m Parkes telescope, an instrument of comparable sensitivity to the Lovell instrument.  Together then, the polarimetric surveys of the two instruments encompass all the known ``B'' pulsars within the entire sky down to a similar sensitivity.  

Here, we focus on the population of pulsars outside the Arecibo sky---that is, declinations above 37\degr\ and below the equator.  Of the 130 such Arecibo ``B'' objects, 100 are included in the GL98 survey\footnote{Some 123 of the 130 have been studied in various publications \citep[\eg][]{rankin1993b,Olszanski2019}} so we are left with some 195 of these at declinations greater than --35\degr---the Group I population, and 148 of the 160 in the remaining far south sky---the Group II population.   The two groups are the main focus of our analyses here, and as we will see they have significant differences in the character of the available observations and in the characteristics of the areas of sky they represent.  

The \cite{GL98} population of 300 pulsars represents a fairly complete and coherent group that are bright enough to provide good quality profiles at some or most of the survey frequencies, 240, 400, 600, 920, 1400 and 1600 MHz. The objects were known to be bright enough to qualify for this survey from the extensive, on-going Jodrell timing programs that remain the foundation of such efforts to this day.  This is also the pulsar population that has been observed at frequencies down to 100 MHz or below, in part due its accessibility to other northern instruments.  The population of pulsars now observed using the PRAO LPA (KL99 and MM10) together with others observed with the LOFAR High Band (BKK+ and PHS+) provides a rich environment for investigating pulsar low frequency emission, and in a few cases observations are available at decametric wavelengths as well (BKK++, BGT+, ZVK+).  More than half (107) of the 195 Group I pulsars include observations down to the 100-200-MHz band and some 20 into the decameter regime; see Appendix Table~A1.  This is the more significant because the weakest part of the GL survey was its 240-MHz observations, often because of the difficulty at that time of achieving adequate dedispersion.  

The 148 pulsars of the Group II population based on Parkes observations have very different properties.  For most, only 2-3 polarized profiles are available, usually at 1.4 GHz and 600 MHz---and a third with only a 1.4-GHz observation.  The recent 1.4-GHz \citet{jk18} profiles are generally of very high quality and have clarified older less well resolved and measured ones.  Many of the profiles show significant scattering up toward 1 GHz, so for emission-beam studies these lower frequency observations are not useful.  Only in a few cases do observations extend down into the 100-MHz regime, and none at all into the decameter band; see Appendix Table~B1.  Fortunately, a number of fine observations are available at 3.1-GHz \citep[\eg][]{kj06}, and these have been useful in cases when scattering distorts the 1.4-GHz profile.  

\noindent\textit{\textbf{Status of 1-GHz Core/ouble-cone Modeling Results}}:  The quality and extent of the available polarized profiles have permitted us to identify the beam structure and usually construct quantitative geometric beam models of 324 pulsars; see Appendix Tables~A3, B3 and B5.  In only seventeen cases were we unable to do so for differing reasons:  For the Group I population where the observations are generally better with more redundancy:  the profiles of B1809--173 and B1822--14 are inconsistent with any known path of frequency evolution, possibly due to moding.  For B1834--04, no interpretation can be made due to the poor quality of the observations, and for B1842--02 only 1.4/1.6 GHz profiles are available and neither permits any estimate of the PPA rate.   In Group II, the single B0529--66 profile could not be interpreted due to the ostensibly flat PPA traverse. As a 100-ms energetic interpulsar B0906--49 requires further detailed study.  B1054--62 provides the most interesting case of a well studied pulsar for which no core/double cone model seems appropriate.  For B1436--62 the existing observations are inadequate.  Finally, 9 of the objects with only a 1.4-GHz profile could not be interpreted, usually because of scattering but also for lack of a PPA rate.

The models for most pulsars are strongly motivated by the profile characteristics:  When $\alpha$ is well determined by the core width, the PPA rate well defined and the conal width well determined, then the spherical geometry computes an inner or outer conal beam radius close to that expected for a pulsar of its rotation period.  We encountered no good example to the contrary.  These are the models in Appendix Tables~A3, B3 and B5 with $\alpha$ shown bolded and no qualifications.  

For many pulsars, of course, one or another of the three measurements is more difficult for a variety of reasons:  Very commonly the core component is conflated with other components, and its width can only be  estimated---and thus denoted by a $\sim$; in other cases where a core is so conflated that no direct measurement is possible, its width can still be estimated using the conal geometry---and then marked with a $\approx$.\footnote{Component fitting would doubtlessly improve many of these estimates, but here it is beyond the scope of this paper.}  Estimating the PPA rate can also be challenging, because it is poorly defined, (too) flat, or its form hard to interpret.  In a few cases where a central sightline traverse is probable, $R$ is taken as infinite to build the model.  Further, one or the other of the conal ``outriders'' are often weak and conflated in \textbf{S$_t$} profiles at 1.4 GHz.  Sometimes they are clearer in a higher frequency profile, but in other cases their dimensions had to be gleaned from inflections or the form of the linearly polarized ($L$) profile.  Also, a number of our models reflect only high frequency profiles (due to scattering or other reasons), so we were not able to distinguish between the \textbf{S$_t$} and \textbf{T} classes---and so are marked \textbf{S$_t$/T?} in the tables.  

Conal profiles have the immediate issue that no $\alpha$ value can be determined independently, so models reflect the established 1-GHz model width of an inner or outer cone.  And distinguishing between the two geometries is often challenging or impossible when lower frequency profiles are unavailable or in the presence of scattering.  So we have modeled the conal profiles with outer geometries only when a reason was clear---low frequency width escalation or additional inner components in c\textbf{T} or c\textbf{Q} situations.  Thus the $\alpha$ values of some conal pulsars may be underestimated by some 10\degr\ or so.  Further, in conal triple or quadruple profiles the inner conal dimensions can usually only be estimated, and sometimes not well or at all, so outer conal dimension can here be measured in weaker or lower frequency profiles.   

\noindent\textit{\textbf{Frequency of Profile Types}}:  In this large sample pulsars with bright core beams (\textbf{S$_t$} and \textbf{T} profiles) were most numerous (178).  There were 116 pulsars with entirely conal profiles (\textbf{S$_d$}, \textbf{D}, c\textbf{T} and c\textbf{Q}) as well as some 25 with various hybrid profiles (\textbf{M} or possibly so).  Seventeen pulsars could not be modeled, and five had single profiles that could be either conal or core beams.  More interesting is that core-beam dominated pulsars represent 60\% of the Group II population but only 50\% of Group I, a matter than we will return to below.  

\noindent\textit{\textbf{Beamform Evolution at Low Frequency}}:  Just over half (110) of the Group I pulsars have been observed down into the 100-200-MHz band, and many suffer from significant scattering in this regime.  Only a handful (21) have been detected in the decameter band, and the great majority of these lie in the Galactic anticenter direction.  Only six of the Group II have 100-MHz band observations and none at lower frequencies.  Nonetheless, general patterns of core and conal beam spectral evolution are emerging---
\begin{itemize}
\item Core beams tend to have a dimension similar to the angular size of a pulsar's polar cap.  This may be more true at around 1 GHz, in that other factors may contribute to core widths at both higher and lower frequencies:  above 1 GHz conal power often seems to emerge on the wings of a core component; and at lower frequencies core widths are most susceptible to broadening by scattering.
\item Remarkably a few core beams narrow from their 1-GHz values in certain frequency intervals.  A possibility is that the differently polarized leading and trailing parts of the core have different spectra.  We see instances of these parts being displaced and partially resolved (\eg B1409--62), and in several cases (\eg B1046--58) the leading and trailing parts of the core have different amplitudes.  
\item Inner and outer conal beams can only be positively distinguished when $\alpha$ is fixed by a core width.  In \textbf{T} profiles, the inner conal beam size tends to be vary little, unlike that of outer cones which tend to increase with wavelength.   
\item This said, it is difficult to distinguish between inner and outer cones in single conal profiles.  Some inner cones do show increases with wavelength and some outer cones seem to not.  There are several good examples of c\textbf{T} or c\textbf{Q} profiles where neither the inner or outer beam increase very much to very low frequency (\eg B1039--19). 
There are instances where an outer cone is ruled out because it would require too large an $\alpha$ value.  
\end{itemize}

\begin{figure}
\begin{center}
\includegraphics[width=80mm,height=80mm,angle=0.]{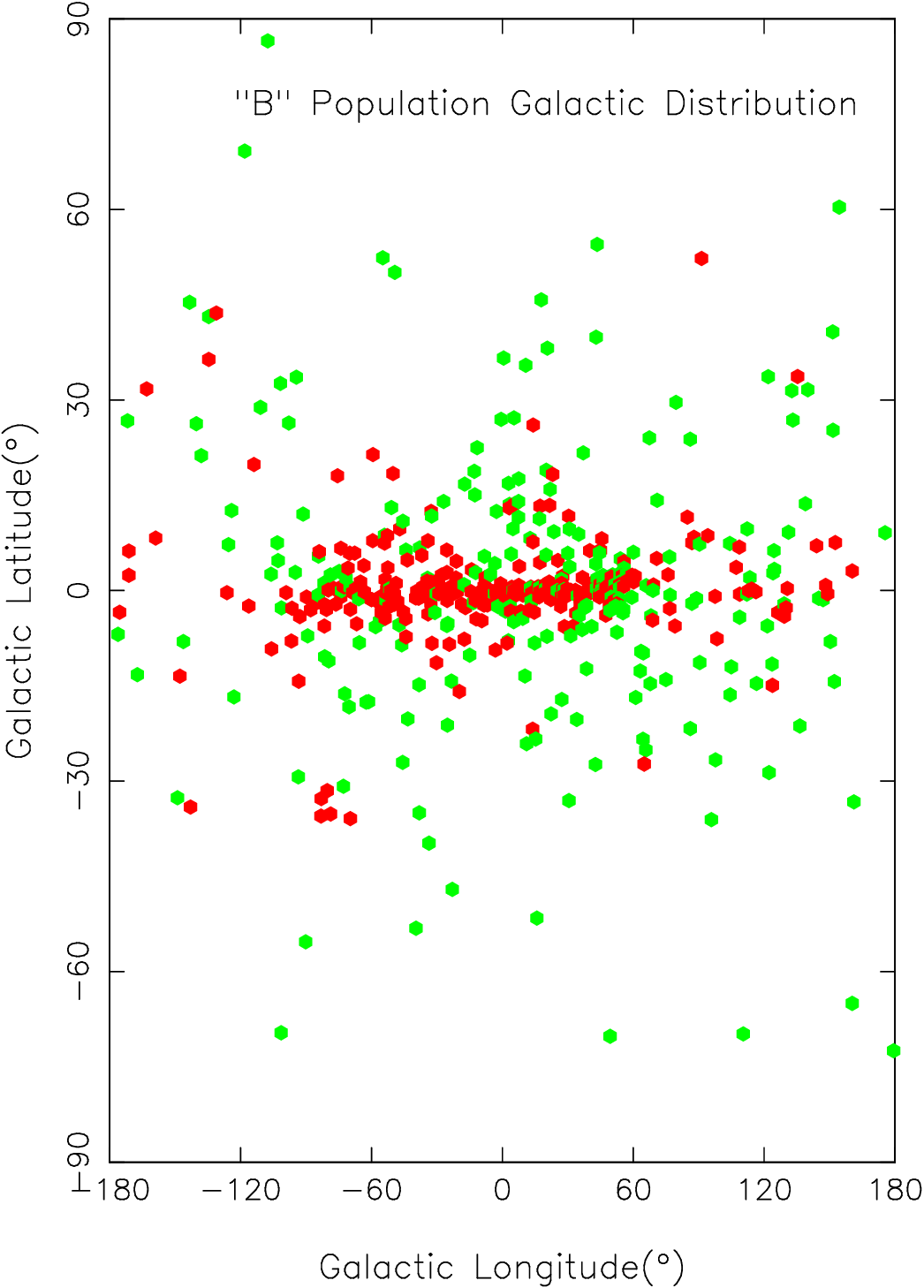}
\caption{Plot showing the distribution of the ``B'' pulsar population on the sky in Galactic coordinates.  Pulsars with {$\dot E$} greater than or less than 10$^{32.5}$ ergs/s are shown with red and green dots, respectively.  Clearly the core-emission dominated energetic objects lie both closer to the Galactic plane and Galactic Center than their less energetic conal cousins.}  
\label{fig30}
\end{center}
\end{figure}

\begin{figure}
\begin{center}
\includegraphics[width=80mm,angle=0.]{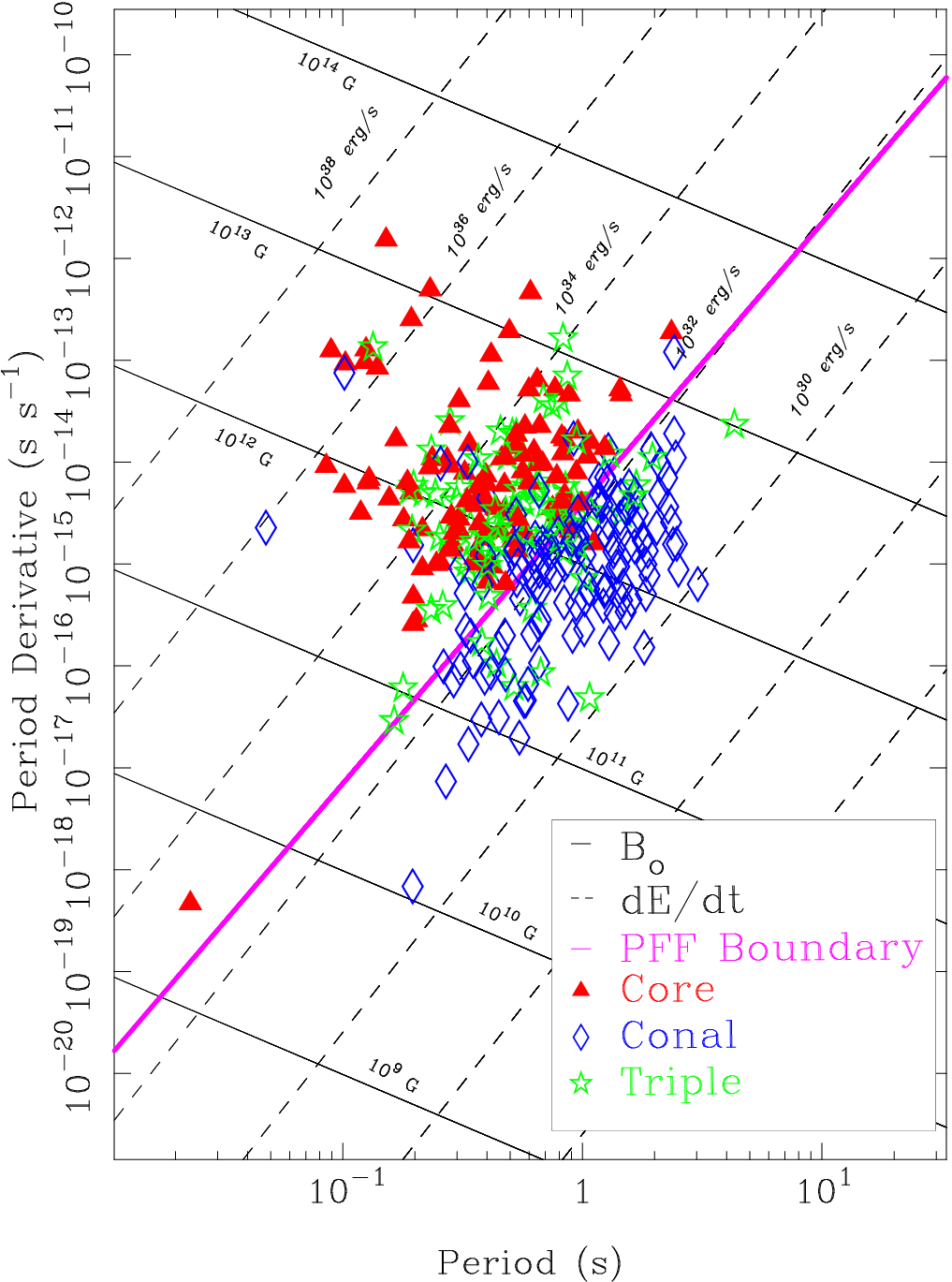}
\caption{$P$-$\dot P$ Diagram showing the position of the ``B''-pulsar populations considered here along with the PFF boundary line. Core- and conal-dominated pulsars as well as core-cone triples are indicated by symbols in the legend to the bottom right.}  
\label{fig31}
\end{center}
\end{figure}

\noindent\textit{\textbf{Physical Characteristics and Distributions}}:  Remarkably Groups I and II both divide strongly on whether $\dot E$ is greater or less than $10^{32.5}$ ergs/s---or what is the same from ET VI, an acceleration parameter $B_{12}/P^2$ of 2.5:  Fig~\ref{fig31} provides a $P$-$\dot P$ diagram showing that conal single, double, triple and quadruple profiles (blue diamond symbols) occupy the region at lower energies, whereas core single beams (red triangles) are all associated with higher energies.  Triple {T} profiles (green stars) fall on both sides of the boundary but tend to themselves divide on the basis of whether core or conal energy is dominant.  Only a couple of core singles fall below the boundary or conal dominated profiles above it, and this is also the region in which a few single profiles are difficult to distinguish as core or conal perhaps because of some hybrid properties.  

There is the interesting and unusual case of B1259--63 which seems to have an outer conal double profile, but a huge {$\dot E$} of 8.3x10$^{35}$ ergs/s.  It is a 50-ms MSP with a binary companion and both X-ray and $\gamma$-ray emission are detected during periastron passage.  Nonetheless, core emission would normally be expected from such an energetic pulsar, and there is a case to be made that our sightline through an outer cone would miss the core beam in this case.  

\noindent\textit{\textbf{Emission Beamform Evolution}}:  Given that pulsars spin down over time, we are able to estimate their ages in the conventional manner as the ratio of their rotation period $P$ to their spindown rate {$\dot P$}.  This obviously implies---following \S \ref{sec:plasma}---that young pulsars with large {$\dot E$} will generate pair plasma mainly in an axial manner within their polar flux tubes and thus emit core-beam radiation.  

Observationally, we also know that most pulsars emit a mixture of core and conal radiation---as evidenced for instance by the conal ``outriders'' in many energetic pulsars.  Why this is so is unclear, but a possibility is that the gap potential undergoes variations that permit peripheral plasma generation that results in conal emission.  This might be a temporal effect [\eg pulsar B0823+26's strongly varying core intensity---\citet{Rankin_2017}] or perhaps due to spacial variations between a central and peripheral PFF across the polar cap.  

As pulsars spin down, peripheral plasma generation is increasingly favored and the angular size of their emission beams decreases in accordance with their shrinking polar caps.  Evolving radio pulsars will inevitably cross the {$\dot E$} boundary of 10$^{32.5}$ ergs/s, but it is a ``soft'' boundary wherein plasma generation in central and peripheral PFFs is equally probable and thus comparably admixed for a variety of conditions in particular pulsars.

\noindent\textit{\textbf{Galactic Distribution of ``B'' Pulsars}}:  The 487 ``B'' population pulsars can be located on the sky in Galactic coordinates, and a plot showing their distribution appears as Figure~\ref{fig30}.  The energetic---we now know mostly core-dominated---pulsars are plotted using red symbols, and the lower energy mostly conal objects in green.  That the core- and conal dominated populations have different sky distributions is obvious.  The core emitters are concentrated more closely along the Galactic plane and primarily in inner Galaxy directions.  The conal objects by contrast are found at higher Galactic latitudes and are distributed more evenly in Galactic longitude.  This ``B'' core population has an rms latitude of only 9\degr, whereas the conal portion is much larger at 22\degr.  

\noindent\textit{\textbf{Conjuring With the Luminosity Distribution of ``B'' Pulsars}}:  Distances have been computed for nearly all of the ``B'' pulsars, and unsurprisingly the distances of both the core and conal populations peak strongly in the Galactic center direction.  Overall, the average distances to a core or conal emission dominated pulsar are not too different, 4.0 and 3.1 kpc, respectively.  Nor are the planar distances very different, 4.0 and 3.0 kps, respectively.  What is markedly different is their Z-direction, out-of-plane average distances, some 230 and 510 parsecs, respectively.  This squares with the different average log$_{10}$ ages of the two populations,  6.0 for the core emitters and 7.2 for the conal population.  There is also a small difference in their average log$_{10}$ B-field values, 12.18 for the former and 11.95 for the latter.  

Some pulsars are known to be bright and others faint, but more than half a century after their discovery, no systematics of this property are well established.  Unlike normal stars, pulsars radiate in beams of different configurations and dimensions, and our accidental sightlines through their emission samples their extent very poorly.  Moreover, pulsar radio spectra vary significantly, and pulsar distances may or may not be accurate.  Nonetheless, average flux densities are tabulated for most pulsars at 1.4 GHz and 400 MHz and often measured at many other frequencies, so in principle a total luminosity could be integrated.  

Such an effort is far beyond the scope of this analysis.  However, the ATNF Catalog tabulates luminosity estimates for most pulsars at both 400 MHz and 1.4 GHz, and one might expect these values to show some correlation with the core and conal properties of the pulsars above.  Crude estimates of a pulsar's luminosity can be computed from these values, and remarkably they range over nearly six orders of magnitude.  Puzzlingly, there is no obvious increase in in the luminosity with {$\dot E$} for either the core or conal populations, although \citet{KJ17} model pulsar luminosity as proportional to {$\dot E^{1/2}$}.  These luminosity estimates are poor for a number of reasons, but it is difficult to understand how any of the uncertainties, or their aggregate effects, could be responsible for the enormous variance.  Equally puzzling is that the variance is roughly the same for both the core-dominated energetic population as well as the conal one.  Apparently, a wide dispersion of radio emission efficiencies are characteristic of the emission process \citep{Szary+14}.

\begin{figure}
\begin{center}
\includegraphics[width=80mm,height=80mm,angle=0.]{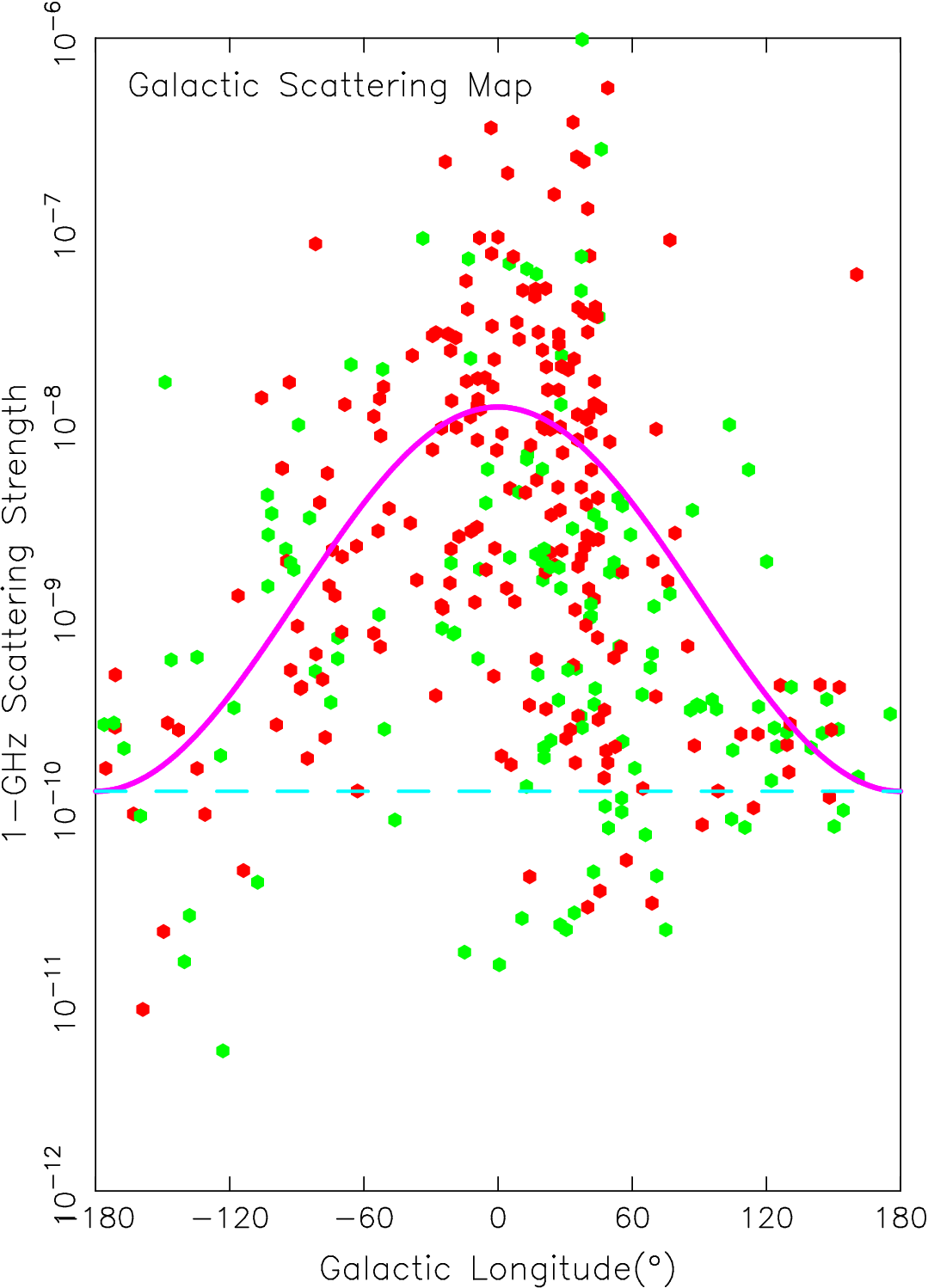}
\caption{Scattering strengths of the two ``B'' pulsar populations as a function of Galactic longitude.  Here we have normalized the scattering-time values by the $DM^{2.2}$ factor in the \citet{kuzmin_LL2007} relationship.  Further the horizontal stripe shows the level of mean scattering from the above work. }  
\label{fig33}
\end{center}
\end{figure}

\noindent\textit{\textbf{Scattering Levels of ``B'' Population Pulsars}}: The above analyses have required attentiveness to the effects of scattering in the ISM, and are particularly apropos here as most of the measured scattering times in the ATNF Catalog pertain to ``B'' pulsars (280 our of 364).  In most of our work we have been concerned with the scattering effects for individual pulsars, but we are now able here to consider the overall levels of Galactic scattering.   Figure~\ref{fig33} plots normalized 1-GHz scattering times against Galactic longitude.  We have chosen to normalize the scattering by the $DM^{2.2}$ factor in the mean scattering level discussed in \S\ref{sec:scattering}.  The map shows immediately that while the scattering level varies greatly over different paths, its effect is concentrated in directions toward the inner Galaxy.  Notably, the putative mean level that is shown as a cyan stripe woefully underestimates the scattering level here, because the PRAO telescopes had access to pulsars at 100 MHz only in the outer Galaxy region.  We therefore averaged the log of the scattering time in segments and used this in the model plots of Figs~A1--A15 and Figs~B1--B6.  The mean levels of scattering here for the energetic core-emission population is 2.4x10$^{-9}$, some 20 times greater than the ``mean'' and that for the conal population 6x10$^{-10}$, 5 times the ``mean''.  The variance around the mean at different Galactic longitudes does seem roughly compatible with a level of $\pm$10 the local mean in most directions; however, it breaks down entirely in the inner Galaxy region.  Also puzzling is that the scattering levels are not symmetrical about 0\degr\ longitude, perhaps due to our position relative to local spiral arms.

\section{Summary}
\label{sec:summary}
We have begun a process of examining how pulsar beams evolve at lower frequencies in an attempt to interpret observed changes in terms of pulsar emission geometry and physics as possible.  The 343 pulsars considered here show beam configurations across all of the core/double-cone model classes (ET VI).  Some half of this population have $\dot E$ values $\ge$ $10^{32.5}$ ergs/s and either core-cone triple \textbf{T} or core-single {\textbf S$_t$} profiles.  The remainder tend to have profiles dominated by conal emission---that is, conal single \textbf{S$_d$}, double \textbf{D}, triple c\textbf{T}, or quadruple c\textbf{Q} geometries. We were able to construct quantitative beam geometry models for all but a handful of these pulsars, though some are better established than others on the basis of the available information.  Lack of reliable PPA rate estimates was a limiting factor in a number of cases, either due to low fractional linear polarization or difficulty interpreting it.  Usually it was possible to trace a fixed number of profile components across the observed bands, sometimes despite very different spectral behavior.  We thus conclude that the emission-beam configuration encountered by a given pulsar is largely fixed, and that profile variations over the observable spectrum mainly reflect spectral variations in the one or more beams that comprise the profile.  

Together, the \citet{GL98} and several Parkes surveys provide a coherent all-sky population of the pulsars known in the mid-1990s, with roughly comparable sampling in in both Galactic longitude and latitude.  This ``B'' pulsar population has been accessible for study by one or the other of the 70--80m-class Lovell and Parkes telescopes, making it a pulsar near-equivalent to a visual magnitude star catalog.  Given that we so far lack for pulsars any systematic understanding of their radio luminosity, a population of comparably detectable pulsars with well studied radio beaming characteristics provides a useful foundation.

\section{Observational Data availability}
The paper draws on published observations, and so includes no original observational data.

\section*{Acknowledgements}
Many thanks to Prof. Patrick Weltevrede who kindly reviewed the manuscript in several forms and helped to improve it substantially.  Much of the work was made possible by support from the US National Science Foundation grant 18-14397. We especially thank our colleagues who maintain the ATNF Pulsar Catalog and the European Pulsar Network Database as this work drew heavily on them both.  This work made use of the NASA ADS astronomical data system.

\bibliography{biblio.bib}



\label{lastpage}

\appendix
\setcounter{figure}{0}
\renewcommand{\thefigure}{A\arabic{figure}}
\renewcommand{\thetable}{A\arabic{table}}
\setcounter{table}{0}
\renewcommand{\thefootnote}{A\arabic{footnote}}
\setcounter{footnote}{0}
\newpage

\section{Gould \& Lyne Population Tables, Notes and Model Plots}
\noindent\textit{\textbf{B0011+47}}: \cite{mitra2011} identified triplicity in this profile.  However, it seems to have an interesting conal triple (c\textbf{T}) structure, wherein both cones can be distinguish at higher frequencies, but only the inner one survives to LOFAR frequencies.  \cite{Weltevrede2006,Weltevrede2007} find a low frequency excess in its pulse sequences, and \citet[KLL07]{kuzmin_LL2007} provide a scattering timescale.  
\vskip 0.06in
\noindent\textit{\textbf{B0031--07}} is a well-studied pulsar with three prominent drift modes and a conal single \textbf{S$_d$} beam configuration \citep[\eg][]{Weltevrede2006,Weltevrede2007}.  We reiterate the ET VI classification, and the recent studies by \cite{McSweeney2019}, \cite{Ilie2020} and \citet{BMM2020} provide links to the chain of earlier studies. \cite{Zakharenko2013} include profiles at both 25 and 20 MHz, and \citet{kuzmin_LL2007} provide a $t_{\rm scatt}$.
\vskip 0.06in
\noindent\textit{\textbf{B0037+56}}: This pulsar is scattered at LOFAR frequencies, and the GL98 234-MHz profile may not show an intrinsic width.  \citet{Weltevrede2006} find evidence of drifting at 1.4 GHz, but do not confirm it at the lower frequency \citep{Weltevrede2007}.  We model it as having an inner conal single \textbf{S$_d$} beam structure.  The $t_{\rm scatt}$ is from \citet{geyer}.
\vskip 0.06in
\noindent\textit{\textbf{B0052+51}}: The pulsar has a conal double profile with a hint of additional structure. \citet{Weltevrede2006} find evidence of an odd-even modulation.  We model its outer conal double structure but with further study it may have a conal triple c{\textbf T} one.  \citet{kuzmin_LL2007} provide a $t_{\rm scatt}$.
\vskip 0.06in
\noindent\textit{\textbf{B0053+47}}: We have little information to go off of for this pulsar; it has a possible three-part profile at 606 MHz, but we do not have much information in terms of PPA sweep rate.  Probably a core single; thus so modeled.  \cite{Zakharenko2013} detect the pulsar at both 25 and 20 MHz but the profiles are too scattered to be useful.  No $t_{\rm scatt}$ is available.
\vskip 0.06in
\noindent\textit{\textbf{B0059+65}}:  This pulsar has nice outer cone {\textbf T} triple; however the central core component's width is difficult to measure accurately.  No $t_{\rm scatt}$ is available.
\vskip 0.06in
\noindent\textit{\textbf{B0105+65}}: The pulsar was earlier thought to have a core-single geometry, but \citet{Weltevrede2006} find an odd-even modulation at 1.4 GHz that, however, was not confirmed at 327 MHz \citep{Weltevrede2007}.  More study is needed, but we model it here as probably having an inner single conal \textbf{S$_d$} configuration.  \citet{kuzmin_LL2007} provide a $t_{\rm scatt}$.
\vskip 0.06in
\noindent\textit{\textbf{B0105+68}}: The pulsar seems to have a {\textbf T} geometry with a clear core component at 149 MHz. The core widths cannot be measured accurately, but a value just under 6\degr\ seems plausible and together with the assumption of a central sightline traverse (as the PPA rate cannot be measured) indicates an outer cone.  No $t_{\rm scatt}$ is available.
\vskip 0.06in
\noindent\textit{\textbf{B0114+58}}: The pulsar has a very large spindown energy loss rate and seems to have an {\textbf S$_t$} structure with a possible preceding ``pedestal'' feature.  It shows the effects of substantial scattering at LOFAR frequencies \citep{kuzmin_LL2007}.
\vskip 0.06in
\noindent\textit{\textbf{B0136+57}}: This pulsar is either scattered or poorly resolved) in the GL98 234-MHz profile and down into LOFAR band.  The three highest frequency profiles show structure that we interpret at the core and conal outriders of a core-single {\textbf S$_t$} configuration.   \cite{Weltevrede2006,Weltevrede2007} find a diffuse modulation that seems unlikely to indicate an organized drift.  We include the 103-MHz profile \citep{kuzmin1999} but the significance of its width is uncertain because the authors have attempted to correct it for scattering \citep[see][]{kuzmin_LL2007}.
\vskip 0.06in
\noindent\textit{\textbf{B0138+59}}:  ET VI classified the pulsar as a possible {\textbf M} and \cite{mitra2011} considered a conal quadruple {c\textbf Q} structure.  We agree that the evidence favors the latter, and we model it as such here.  \cite{Weltevrede2006,Weltevrede2007} found a broad drift feature at 21 cms. but were unable to confirm it at their lower frequency.  A 1-GHz double-cone geometry requires an $\alpha$ of about 20\degr\ which then implies a $\beta$ of some 2.2\degr, such that the core would be broad and skirted by the sightline if present.  \citet{kuzmin_LL2007} provide a $t_{\rm scatt}$.
\vskip 0.06in
\noindent\textit{\textbf{B0144+59}}:  This fast poorly studied pulsar seems to have a core single or triple profile, but little more can be said.  GL98's three highest frequency profiles show a central feature that appears to be a core component, but its width is narrower than the polar cap width; it is accompanied by only negative Stokes $V$, so might be an incomplete core.  Moreover, the \cite{Weltevrede2006} 21 cms. profile has a very different form, and they find a weak modulation, but it does not seem to be conal. The 610 and 408-MHz profiles show structure that may be conal, but not in a way that can be related to the higher frequencies.  The PPA rate is only a guess at --5\degr/\degr.  \citet{kuzmin_LL2007} provide a $t_{\rm scatt}$.
\vskip 0.06in
\noindent\textit{\textbf{B0148--06}}: We support the ET VI conal double {\textbf D} configuration for this pulsar \citep[see also][]{JKMG2008}, and any doubt is resolved by the strong drift fluctuation feature identified in both bands by \citet{BHMM1985}, \cite{Weltevrede2006, Weltevrede2007} and \citet{basu2016}. The two components seen at 103 MHz \citep{kuzmin1999} may connect to the outer conal evolution.  \citet{Zakharenko2013} seem to detect the pulsar in their decameter band, but it is far too scattered to be useful.  No $t_{\rm scatt}$ is available.
\vskip 0.06in
\noindent\textit{\textbf{B0149--16}}:  ET VI raised the possibility of a triple structure in this double resolved profile.  We do see three features in the linear polarization of some profiles as well as antisymmetric Stokes $V$ \citep[see also][]{JKMG2008}.  Modeling the profile as an inner conal double requires a nearly orthogonal geometry—and were there a core component, its width would then be some 3\degr---indeed, the rough width of the above features. \cite{Weltevrede2006,Weltevrede2007} and \citet{basu2016} find a weak drift feature.  The KL99 profile cannot be meaningfully interpreted. No $t_{\rm scatt}$ is available.
\vskip 0.06in
\noindent\textit{\textbf{B0153+39}}: The only available GL98 606-MHz profile has an {\textbf S$_d$} or {\textbf D} profile.  We do not have adequate polarimetry to estimate the PPA sweep rate, so the model is a guestimate.  No $t_{\rm scatt}$ is available.
\vskip 0.06in
\noindent\textit{\textbf{B0154+61}}: Following ET VI this pulsar has an {\textbf S$_t$} beam configuration where the core component is detected at 103 MHz \citep{kuzmin1999}.  \cite{Weltevrede2006} find a flat fluctuation spectrum, supporting this model.  Moreover, a hint of conal outriders is seen in GL98's 1.6-GHz $L$ profile that permits a rough model (that we provisionally trace down to 1.4 GHz).  \citet{kuzmin_LL2007} provide a $t_{\rm scatt}$.
\vskip 0.06in
\noindent\textit{\textbf{B0226+70}}: This little studied pulsar has a double profile form in the three GL98 profiles, but a surprisingly clear {\textbf T} profile in the BKK+ 149-MHz one, where both core and conal widths are measurable.  At 1 GHz a core component would have a 3\degr\ width for an outer cone ($\alpha$ $\sim$ 39\degr) and 4\degr\ ($\alpha$ $\sim$ 30\degr) for an inner one.  The core width at 149 MHz cannot be measured accurately, but it is hardly the smaller value, so we model the configuration with an outer cone.  \citet{kuzmin_LL2007} provide a $t_{\rm scatt}$.

\onecolumn

\begin{center}
\begin{longtable}{lccc|l|l|}
\caption{Gould \& Lyne Population Observation Information.} \label{tabA1}  \\
   \hline
 Pulsar & P & DM & RM  & References &  References \\
    (B1950) & (s) & ($pc/cm^{3}$) & ($rad$-$m^{2}$) &   & \\
    \hline
    & & & &  \multicolumn{1}{c|}{100 MHz} &   \multicolumn{1}{c|}{$<$100 MHz} \\
    \hline
    \hline
\endfirsthead
    \hline
 Pulsar & P & DM & RM  & References &  References \\
    (B1950) & (s) & ($pc/cm^{3}$) & ($rad$-$m^{2}$) &   & \\
    \hline
    & & & &  \multicolumn{1}{c|}{100 MHz} &   \multicolumn{1}{c|}{$<$100 MHz} \\
    \hline
    \hline
\endhead
B0011+47 & 1.24 & 30.4 & -15.6 & GL98; Han+09; FDR; BKK+; PHS+; KL99 & BKK++; BGT+ \\
B0031--07 & 0.94 & 10.9 & 9.9 & HMAK;W93;JKMG;JK18;PHS+;BMM+;KL99 & MHMb;BKK++;BGT+;ZVK+ \\
B0037+56 & 1.12 & 92.5 & 15.3 & GL98; BKK+; PHS+; MM10 &  \\
B0052+51 & 2.12 & 44.0 & -64.1 & GL98; BKK+; PHS+; MM10 & BKK++ \\
B0053+47 & 0.47 & 18.1 & -34.2 & GL98; BKK+; PHS+; Han+09; MM10 & BKK++; BGT+; ZVK+ \\
\\[-3pt]
B0059+65 & 1.68 & 65.9 & -94.0 & GL98; PHS+; MM10 &  \\
B0105+65 & 1.28 & 30.5 & -27.1 & GL98; BKK+; PHS+; MM10 & BGT+ \\
B0105+68 & 1.07 & 61.1 & -33.0 & GL98; BKK+;MM10 & BKK++ \\
B0114+58 & 0.10 & 49.4 & -8.1 & GL98;Han+09; BKK+; PHS+; KL99 & BKK++ \\
B0136+57 & 0.27 & 73.8 & -94.1 & GL98; BKK+; FDR; PHS+; KL99 & BKK++ \\
\\[-3pt] 
B0138+59 & 1.22 & 34.9 & -48.0 & GL98; ETIX; PHS+; KL99 & BGT+ \\
B0144+59 & 0.20 & 40.1 & -19.0 & GL98; MM10  &  \\
B0148--06 & 1.46 & 25.7 & 2.0 & W93; GL98; JKMG; JK18; MM10; BMM+ & ZVK+ \\
B0149--16 & 0.83 & 11.9 & 6.6 & Q95; GL98; JKMG; FDR; JK18; BMM+  & KL99; BGT+ \\
B0153+39 & 1.81 & 59.8 & -68.6 & GL98; Han+09; BKK+ & BKK++ \\
\\[-3pt] 
B0154+61 & 2.35 & 30.2 & -29.0 & GL98; KL99; MM10 &  \\
B0226+70 & 2.35 & 30.2 & -43.8 & GL98; BKK+; MM10 & BKK++; BGT+ \\
B0320+39 & 3.03 & 26.2 & 60.0 & GL98;  KL99; FDR; BKK+; PHS+ & PHS+;BKK++;BGT+;ZVK+ \\
B0329+54 & 0.71 & 26.8 & -64.3 & GL98; Mitra \etal\ (2006); PHS+; KL99 & PHS+; BGT+; ZVK+ \\
B0331+45 & 0.27 & 47.1 & 5.6 & GL98; KL99; MM10; BKK+;  PHS+ &  \\
\\[-3pt] 
B0339+53 & 1.93 & 67.3 & -84.0 & GL98; KL99; PHS+ &  \\
B0353+52 & 0.20 & 103.7 & 261.0 & GL98; KL99 &  \\
B0355+54 & 0.16 & 57.1 & 79.0 & GL98; KL99; ETIX; PHS+ & BGT+ \\
B0402+61 & 0.59 & 65.4 & 8.2 & GL98; KL99; BKK+; PHS+; MM10 &  \\
B0410+69 & 0.39 & 27.4 & -21.4 & GL98; KL99; Han+09; BKK+; PHS+; MM10 & BKK++ \\
\\[-3pt] 
B0447--12 & 0.44 & 37.0 & 13.0 & GL98; KL99; MM10; JK18; PHS+  &  \\
B0450--18 & 0.55 & 39.9 & 11.1 & GL98; MHM; JK18; KL99 &  \\
B0450+55 & 0.34 & 14.6 & 5.8 & GL98; KL99; ETIX; BKK+; PHS+  & PHS+;BKK++;BGT+;ZVK+ \\
B0458+46 & 0.64 & 41.8 & -175.2 & GL98; KL99; Han09+; FDR; BKK+; KL99 &  \\
B0559--05 & 0.40 & 80.5 & 64.0 & GL98;Q95; JKMG; MM10 &  \\
\\[-3pt] 
B0621--04 & 1.04 & 70.8 & 49.0 & GL98; MM10; JK18 &  \\
B0628--28 & 1.24 & 34.4 & 46.5 & GL98; HMAK; MHMA; JKMG; JI; BMM+ & MHMb; BGT+ \\
B0643+80 & 1.21 & 33.3 & -31.8 & GL98; BKK+; ETIX; PHS+; MM10 & BKK++ \\
B0655+64 & 0.20 & 8.8 & -18.1 & GL98; KL99; BKK+  & BKK+; BGT+; BKK++ \\
B0727--18 & 0.51 & 61.3 & 51.0 & GL98; Q95; JKMG; JK18 &  \\
\\[-3pt] 
B0740--28 & 0.17 & 73.7 & 150.0 & MHM; MHMA; MHQ; W93; JKMG; JK18; JI  & MHMb; ETIX \\
B0756--15 & 0.68 & 63.3 & 55.0 & GL98; MM10; JK18; BMM+  &  \\
B0809+74 & 1.29 & 5.8 & -14.0 & GL98; KL99; ETIX; BKK+; MM10 & PHS+;BKK++;BGT+;ZVK+ \\
B0818--13 & 1.24 & 40.9 & -1.2 & GL98; HMAK; Q95; PHS+; JI; JK18; BMM+ & MHMb; KL99; BGT+ \\
B0826--34 & 1.85 & 52.2 & 59.0 & GL98; \cite{BiggsMHML1985}; & MHMb  \\
\\[-3pt] 
B0841+80 & 1.60 & 34.8 & -21.8 & GL98; KL99; Han09+; BKK+; MM10 & BKK++ \\
B0844--35 & 1.12 & 94.2 & 136.5 & GL98; Q95; JK18; BMM+ &  \\
B0853--33 & 1.27 & 86.6 & 165.0 & GL98 &  \\
B0906--17 & 0.40 & 15.9 & -32.1 & GL98; MM10; ETIX; JI; FDR; JK18; PHS+ & BGT+ \\
B0917+63 & 1.57 & 13.2 & -14.9 & GL98; KL99; Han09+; MM10; BKK+; PHS+ & BKK++; PHS+; BGT+; KRU+ \\
\\[-3pt] 
B0942--13 & 0.57 & 12.5 & -7.0 & GL98; W93; MM10; BMM+ & KL99 \\
B1010--23 & 2.52 & 22.5 & 52.0 & GL98 &  \\
B1016--16 & 1.80 & 48.8 & * & GL98; MM10; JK18  &  \\
B1039--19 & 1.39 & 33.8 & -22.8 & GL98; JK18; \citet{vH_thesis}; BMM+ &  \\
B1112+50 & 1.66 & 9.2 & 2.4 &  GL98; KL99; ETIX; BKK+; MM10; FDR & PHS+;BKK++;BGT+;ZVK+ \\
\\[-10pt]
B1254--10 & 0.62 & 29.6 & 8.0 & GL98; MM10  &  \\
B1309--12 & 0.45 & 36.2 & -13.0 & GL98; MM10  &  \\
B1322+83 & 0.67 & 13.3 & -23.7 & GL98; MM10; ETIX; FDR; BKK+; PHS+ & KL99; BKK++; BGT+; ZVK+ \\
B1508+55 & 0.74 & 19.6 & 1.5 & GL98; KL99; BKK+; PHS+; MM10; FDR & PHS+;BKK++;BGT+;ZVK+ \\
B1540--06 & 0.71 & 18.4 & -1.8 & GL98; W93; KL99; ETIX; FDR; JK18 & MHMb; PHS+; BGT+; ZVK+ \\
\\[-1pt] 
B1552--31 & 0.52 & 73.0 & -49.0 & GL98; BMM+ &  \\
B1552--23 & 0.53 & 51.9 & -22.0 & GL98 &  \\
B1600--27 & 0.78 & 46.2 & -5.0 & GL98; JK18 &  \\
B1607--13 & 1.02 & 49.1 & -61.0 & GL98; MM10 &  \\
B1612--29 & 2.48 & 44.8 & -30.0 & GL98 &  \\
\\[-1pt]
B1620--09 & 1.28 & 68.2 & -85.0 & GL98; MM10  &  \\
B1620--26 & 0.01 & 62.9 & -8.0 & GL98 &  \\
B1642--03 & 0.39 & 35.8 & 15.8 & HMAK; MHMA; JKMG; JI; JK18; BMM+ & MHMb; KL99; PHS+; BGT+ \\
B1648--17 & 0.97 & 33.5 & 4.0 & GL98 &  \\
B1649--23 & 1.70 & 68.4 & -24.0 & GL98; JK18 &  \\
\\[-1pt] 
B1657--13 & 0.64 & 60.4 & 35.0 & GL98; MM10  &  \\
B1700--32 & 1.21 & 110.3 & -21.7 & GL98; MHM; MHMA; JKMG; JK18; BMM+ &  \\
B1700--18 & 0.80 & 49.6 & -29.0 & GL98;  ETIX; MM10  & BGT+ \\
B1702--19 & 0.30 & 22.9 & -19.2 & GL98; JKMG; MM10; JK18  &  \\
B1706--16 & 0.65 & 24.9 & -1.3 & GL98; HMAK; MHMA; JKMG; JI; JK18 & MHMb; KL99 \\
\\[-1pt] 
B1709--15 & 0.87 & 59.9 & 8.0 & GL98; MM10  &  \\
B1714--34 & 0.66 & 587.7 & -191.0 & GL98; JK18 &  \\
B1717--16 & 1.57 & 44.8 & -12.0 & GL98; MM10  &  \\
B1717--29 & 0.62 & 42.6 & 21.0 & GL98; JK18; BMM+ &  \\
B1718--02 & 0.48 & 67.0 & 6.0 & GL98; MM10; FDR  &  \\
\\[-1pt] 
B1718--19 & 1.00 & 75.7 & * & GL98 &  \\
B1718--35 & 0.28 & 496.0 & 159.0 & GL98; Q95; JKW06; JK18 &  \\
B1718--32 & 0.48 & 126.1 & 70.4 & GL98; MHMA; W93; FDR; JK18; BMM+ &  \\
B1727--33 & 0.14 & 261.3 & -142.0 & GL98; \citep{JKW06}; JK18 &  \\
B1730--22 & 0.87 & 41.1 & -12.0 & GL98; JKMG; JK18; BMM+ &  \\
\\[-1pt] 
B1732--02 & 0.84 & 65.1 & 48.0 & GL98; MM10  &  \\
B1732--07 & 0.42 & 73.5 & 34.5 & GL98; JKMG; ETIX; JII; FDR; JK18; BMM+ &  \\
B1734--35 & 0.40 & 89.4 & 50.0 & GL98; JK18 &  \\
B1735--32 & 0.77 & 49.6 & 7.0 & GL98; JK18 &  \\
B1736--31 & 0.53 & 600.1 & 32.0 & GL98; JK18 &  \\
\\[-1pt] 
B1736--29m & 0.32 & 138.6 & -236.0 & GL98; W93; MHQ; JK18 &  \\
B1737--30 & 0.61 & 152.0 & -168.0 & GL98; W93; Q95; JKMG; JK18 &  \\
B1738--08 & 2.04 & 74.9 & 124.0 & GL98; W93; JK18; BMM+ &  \\
B1740--13 & 0.44 & 30.3 & * & GL98; MM10  &  \\
B1740--31 & 2.41 & 193.1 & -215.0 & GL98; Q95; JK18 &  \\
\\[-1pt] 
B1740--03 & 0.41 & 116.3 & * & GL98; MM10  &  \\
B1742--30 & 0.37 & 88.4 & 101.0 & GL98; MHMA; W93; JKMG;  ETIX; JK18 &  \\
B1745--12 & 0.39 & 99.4 & 67.0 & GL98;  ETIX; MM10; BMM+  &  \\
B1746--30 & 0.61 & 509.4 & -290.0 & GL98; Q95; JK18 &  \\
B1747--31 & 0.91 & 206.3 & 111.0 & GL98; Q95; JK18 &  \\
\\[-1pt] 
B1749--28 & 0.56 & 50.4 & 96.0 & HMAK; MHMA; MHM; GL98; PHS+; JK18 & MHMb; KL99 \\
B1750--24 & 0.53 & 672.0 & 21.0 & GL98 &  \\
B1753+52 & 2.39 & 35.0 & 27.9 & GL98; BKK+; MM10  &  \\
B1753--24 & 0.67 & 367.1 & -130.0 & GL98 &  \\
B1754--24 & 0.23 & 179.5 & 16.0 & GL98; JK18 &  \\
\\[-10pt] 
B1756--22 & 0.46 & 177.2 & 6.0 & GL98; ETIX &  \\
B1757--24 & 0.12 & 291.6 & 605.7 & GL98; FDR; JK18 &  \\
B1758--23 & 0.42 & 1073.9 & -1156 & GL98; KKWJ; JK18 &  \\
B1758--03 & 0.92 & 120.4 & 32.0 & GL98; MHQ; MM10; MHMb  &  \\
B1800--21 & 0.13 & 234.0 & -36.1 & GL98; W93; FDR; JK18 &  \\
\\[-1pt] 
B1802--07 & 0.02 & 186.3 & * & GL98 &  \\
B1804--27 & 0.83 & 313.0 & -47.0 & GL98; JK18 &  \\
B1804--08 & 0.16 & 112.4 & 166.0 & GL98; MM10  &  \\
B1805--20 & 0.92 & 606.8 & 93.0 & GL98 &  \\
B1806--21 & 0.70 & 381.9 & 256.0 & GL98 &  \\
\\[-1pt] 
B1809--173 & 1.21 & 255.1 & 70.6 & GL98; JK18 &  \\
B1809--176 & 0.54 & 518.0 & 342.6 & GL98; JK18 &  \\
B1811+40 & 0.93 & 41.6 & 49.4 & GL98; KL99: BKK+ & BKK++; BGT+ \\
B1813--17 & 0.78 & 525.5 & 82.9 & GL98; JK18 &  \\
B1813--26 & 0.59 & 128.1 & 90.0 & GL98; JK18; BMM+ &  \\
\\[-1pt] 
B1813--36 & 0.39 & 94.3 & 66.0 & W93; Q95; GL98; JK18 &  \\
B1815--14 & 0.29 & 622.0 & 1174 & GL98; JK18 &  \\
B1817--13 & 0.92 & 776.7 & 893.0 & GL98 &  \\
B1817--18 & 0.31 & 436.0 & -70.0 & GL98; JK18 &  \\
B1818--04m & 0.60 & 84.4 & 69.2 & HMAK; MHMA; JKMG; JII; PHS+; JK18 & MHMb; KL99 \\
\\[-1pt] 
B1819--22 & 1.87 & 121.2 & 124.0 & W93; Q95; GL98; JKMG; JK18; BMM+ &  \\
B1820--14 & 0.21 & 651.1 & 897.0 & GL98 &  \\
B1820--11 & 0.28 & 428.6 & -354.0 & GL98 &  \\
B1820--30B & 0.38 & 87.0 & * & GL98 &  \\
B1820--31 & 0.28 & 50.2 & 95.0 & W93; GL98; JK18 &  \\
\\[-1pt] 
B1821--19 & 0.19 & 224.4 & -302.2 & W93; GL98; FDR; JK18 &  \\
B1821--11 & 0.44 & 603.0 & 213.0 & GL98 &  \\
B1822--14 & 0.28 & 352.2 & -899.0 & W93; GL98; JK18 &  \\
B1822--09m & 0.77 & 19.4 & 65.2 & MHMA; MHM; JKMG; ETIX; JII; JK18 & MHMb; KL99; PHS+; BGT+ \\
B1823--11 & 2.09 & 320.6 & 229.0 & GL98; MM10  &  \\
\\[-1pt] 
B1823--13 & 0.10 & 231.0 & 10.0 & GL98; JK18 &  \\
B1824--10 & 0.25 & 430.0 & -42.0 & GL98; JK18 &  \\
B1826--17 & 0.31 & 217.1 & 304.7 & MHM; MHQ; GL98; FDR; JK18 &  \\
B1828--11 & 0.41 & 159.7 & 47.0 & GL98; JK18 &  \\
B1829--08 & 0.65 & 300.9 & 39.0 & GL98; JK18 &  \\
\\[-1pt] 
B1829--10 & 0.33 & 475.7 & 103.8 & GL98; JK18 &  \\
B1830--08 & 0.09 & 411.0 & -470.0 & GL98; JK18 &  \\
B1831--03 & 0.69 & 234.5 & -41.0 & GL98; MM10; JK18 &  \\
B1831--04 & 0.29 & 79.3 & 100.0 & W93; GL98; KL99; PHS+ &  \\
B1832--06 & 0.31 & 467.9 & 44.0 & GL98; JK18 &  \\
\\[-1pt] 
B1834--04 & 0.35 & 231.5 & 14.9 & GL98; JK18 &  \\
B1834--10 & 0.56 & 317.0 & 826.6 & GL98; JK18 &  \\
B1834--06 & 1.91 & 316.1 & -268.4 & GL98; JK18 &  \\
B1839+56 & 1.65 & 26.8 & -3.9 & GL98;KL99; BKK+; PHS+; MM10; FDR & BKK++; BGT+; ZVK+ \\
B1838--04 & 0.19 & 325.5 & 416.0 & GL98; JII; JK18; BMM+ &  \\
\\[-1pt] 
B1839--04 & 1.84 & 196.0 & 326.0 & W93; GL98; KL99; MM10; JK18 &  \\
B1841--05 & 0.26 & 411.7 & 16.0 & GL98; JK18 &  \\
B1841--04 & 0.99 & 123.2 & 7.0 & GL98 &  \\
B1842--02 & 0.51 & 429.0 & 123.0 & GL98; JK18 &  \\
B1842--04 & 0.49 & 230.8 & -248.0 & GL98; JK18 &  \\
\\[-10pt] 
B1844--04 & 0.60 & 142.0 & 117.0 & GL98; MM10; JK18  &  \\
B1845--19 & 4.31 & 18.2 & 7.0 & GL98 &  \\
B1845--01 & 0.66 & 159.5 & 580.0 & GL98; MHMA; MHM; JKMG; MM10; JK18 &  \\
B1846--06 & 1.45 & 148.2 & -35.0 & MHQ; GL98; MM10  &  \\
B1851--14 & 1.15 & 130.4 & 103.0 & GL98; ETIX &  \\
\\[-5pt] 
B1857--26 & 0.61 & 38.0 & -9.3 & MHMA; MHM; MHQ; JKMG; JK18; BMM+ & MHMb; PHS+  \\
B1900--06 & 0.43 & 195.6 & 203.0 & GL98; MM10  &  \\
B1905+39 & 1.24 & 31.0 & 5.4 & GL98; FDR; BKK+; PHS+ & BKK++ \\
B1907--03 & 0.50 & 205.5 & 152.0 & GL98; MM10; ETIX  &  \\
B1911--04 & 0.83 & 89.4 & 4.0 & HMAK; GL98; JI; FDR; JK18; PHS+ & MHMb; KL99 \\
\\[-5pt] 
B1937--26 & 0.40 & 50.0 & -33.5 & W93; Q95; GL98; JKMG; ETIX; JI; JK18  &  \\
B1940--12 & 0.97 & 28.9 & -75.4 & GL98; KL99: JK18 &  \\
B1941--17 & 0.84 & 56.3 & -40.0 & GL98; JK18 &  \\
B1943--29 & 0.96 & 44.3 & -28.0 & MHQ; GL98; JK18  &  \\
B1946--25 & 0.96 & 23.1 & -13.0 & MHQ; GL98 &  \\
\\[-5pt] 
B1953+50 & 0.87 & 59.9 & -23.8 & GL98;  KL99; MM10; BKK+; PHS+; & BKK++; BGT+ \\
B2000+40 & 0.66 & 587.7 & 145.0 & GL98;KL99; BKK+ &  \\
B2003--08 & 0.58 & 32.4 & -62.0 & GL98; JKMG; MM10; JK18; BMM+ &  \\
B2011+38 & 0.23 & 238.2 & 78.0 & GL98; MM10 &  \\
B2021+51 & 1.57 & 44.8 & -6.7 & GL98; KL99; MM10; ETIX; BKK+; PHS+ & BGT+ \\
\\[-5pt] 
B2022+50m & 0.62 & 42.6 & 44.8 & GL98; KL99; Han09+; MM10; BKK+; PHS+ & BKK++; BGT+ \\
B2036+53 & 0.48 & 67.0 & -103.1 & GL98;KL99; Han09+; MM10; BKK+; &  \\
B2043--04 & 1.55 & 35.8 & -1.0 & GL98; KL99; MM10; ETIX; PHS+; BMM+  &  \\
B2045+56 & 0.48 & 101.8 & 1.3 & GL98; KL99; Han09+; BKK+ &  \\
B2045--16 & 1.96 & 11.5 & -10.0 & HMAK; MHMA; MHM; JII; JK18; BMM+ & MHMb \\
\\[-5pt] 
B2106+44 & 0.41 & 139.8 & -433.0 & GL98; MM10; FDR  &  \\
B2111+46 & 1.01 & 141.3 & -218.7 & GL98; KL99; FDR &  \\
B2148+63 & 0.38 & 129.7 & -157.6 & GL98; KL99; FDR; BKK+; PHS+ &  \\
B2148+52 & 0.33 & 148.9 & -44.0 & GL98; MM10  &  \\
B2152--31 & 1.03 & 14.9 & 21.0 & GL98 &  \\
\\[-5pt] 
B2154+40 & 1.53 & 71.1 & -42.0 & GL98; KL99; FDR; BKK+; PHS+ & BKK++ \\
B2217+47 & 0.54 & 43.5 & -35.9 & GL98; KL99;  ETIX; FDR; BKK+; PHS+ & BKK++; PHS+; BGT+ \\
B2224+65 & 0.68 & 36.4 & -23.0 & GL98; KL99;  ETIX; FDR; BKK+; PHS+ & BKK++; BGT+ \\
B2227+61 & 0.44 & 124.6 & -105.9 & GL98; KL99; MM10; BKK+; PHS+  & BKK++ \\
B2241+69 & 1.66 & 40.9 & -16.8 & GL98; KL99; MM10; BKK+; PHS+  &  \\
\\[-5pt] 
B2255+58 & 0.37 & 151.1 & -323.5 & GL98;KL99; FDR; PHS+  &  \\
B2303+46 & 1.07 & 62.1 & -22.1 & GL98; KL99; BKK+; PHS+ & BKK++ \\
B2306+55 & 0.48 & 46.5 & -29.4 & GL98; KL99; BKK+; PHS+  & BKK++; BGT+ \\
B2310+42 & 0.35 & 17.3 & 5.0 & GL98; KL99; MM10; FDR; BKK+; PHS+ & BKK++; BGT+; ZVK+ \\
B2319+60 & 2.26 & 94.6 & -232.6 & GL98; KL99; MM10; FDR &  \\
\\[-5pt] 
B2323+63 & 1.44 & 197.4 & -102.0 & GL98 &  \\
B2324+60m & 0.23 & 122.6 & -220.7 & GL98; FDR &  \\
B2327--20 & 1.64 & 8.5 & 9.2 & MHMA;JKMG;ETIX;JII;FDR;JK18;BMM+ & MHMb; BGT+ \\
B2334+61 & 0.50 & 58.4 & -100.0 & GL98; KL99; MM10 &  \\
B2351+61 & 0.94 & 94.7 & -75.9 & GL98; KL99; FDR &  \\
\hline
\end{longtable}
Notes: BGT+: \citet{Bondonneau}; BKK+: \citet{bilous2016}; BKK++: \citet{bilous2019}; BMM+: \citet{basu2016}; CMH: \citet{CMH}; ETIX: \citet{ETIX}; ETV: \citet{rankin1993b}; FDR: \citet{Force2015}; GL98: \citet{GL98}; Han09+: \citet{han2009}; HMAK: \citet{HMAK}; JKMG: \citet{JKMG2008}; JK18: \citet{jk18}; JI: \citet{JohnstonI}; JII: \citet{JohnstonII}; KKWJ: \citet{kkwj98}; KL99: \citet{kuzmin1999}; KRU+: \citet{Kravtsov22}; MHM: \citet{MHM}; MHMA: \citet{MHMA}; MHMb: \citet{MHMb}; MHQ: \citet{MHQ}; MM10: \citet{MM10}; Q95: \citet{QMLG95}; PHS+: \citet{pilia}; WW09: \citet{WeWr09}; ZVK+: \citet{Zakharenko2013}
\end{center}

\begin{center}
\begin{longtable}{lccc|ccccccc}
\caption{Gould \& Lyne Pulsar Population Parameters} \label{tabA2}  \\

\hline
 Pulsar & L & B & Dist. & P & $\dot{P}$ & $\dot{E}$ & $\tau$ & $B_{surf}$ & $B_{12}/P^2$ & 1/Q \\ 
 (B1950) & (\degr) & (\degr) & (kpc) & (s) & ($10^{-15}$ s/s) & ($10^{32}$ ergs/s)  & (Myr) & ($10^{12}$ G) &   &   \\
\hline
\hline
\endfirsthead
    \hline
  Pulsar & L & B & Dist. & P & $\dot{P}$ & $\dot{E}$ & $\tau$ & $B_{surf}$ & $B_{12}/P^2$ & 1/Q \\ 
 (B1950) & (\degr) & (\degr) & (kpc) & (s) & ($10^{-15}$ s/s) & ($10^{32}$ ergs/s)  & (Myr) & ($10^{12}$ G) &   &   \\
    \hline
    \hline
\endhead B0011+47 & 116.50 & -14.63 & 1.78 & 1.241 & 0.56 & 0.12 & 34.8 & 0.63 & 0.4 & 0.3 \\
B0031--07 & 110.42 & -69.82 & 1.03 & 0.943 & 0.41 & 0.19 & 36.6 & 0.63 & 0.7 & 0.4 \\
B0037+56 & 121.45 & -5.57 & 2.42 & 1.118 & 2.88 & 0.81 & 6.2 & 1.82 & 1.5 & 0.7 \\
B0052+51 & 123.62 & -11.58 & 2.86 & 2.115 & 9.54 & 0.40 & 3.5 & 4.55 & 1.0 & 0.5 \\
B0053+47 & 123.80 & -14.93 & 1.12 & 0.472 & 3.33 & 12.0 & 2.3 & 1.27 & 5.7 & 1.8 \\
\\[-2pt] 
B0059+65 & 124.08 & 2.77 & 2.50 & 1.679 & 5.95 & 0.50 & 4.5 & 3.20 & 1.1 & 0.6 \\
B0105+65 & 124.65 & 3.33 & 2.13 & 1.284 & 13.05 & 2.40 & 1.6 & 4.14 & 2.5 & 1.1 \\
B0105+68 & 124.46 & 6.28 & 1.98 & 1.071 & 0.05 & 0.02 & 353 & 0.23 & 0.2 & 0.1 \\
B0114+58 & 126.28 & -3.46 & 1.77 & 0.101 & 5.85 & 2200 & 0.3 & 0.78 & 75.8 & 12.6 \\
B0136+57 & 129.22 & -4.04 & 2.60 & 0.272 & 10.71 & 210 & 0.4 & 1.73 & 23.3 & 5.4 \\
\\[-2pt] 
B0138+59 & 129.15 & -2.11 & 2.30 & 1.223 & 0.39 & 0.08 & 49.5 & 0.70 & 0.5 & 0.3 \\
B0144+59 & 130.06 & -2.72 & 2.13 & 0.196 & 0.26 & 13.0 & 12.1 & 0.23 & 5.9 & 1.7 \\
B0148--06 & 160.37 & -65.00 & 25.00 & 1.465 & 0.44 & 0.06 & 52.4 & 0.82 & 0.4 & 0.2 \\
B0149--16 & 179.31 & -72.46 & 0.92 & 0.833 & 1.30 & 0.89 & 10.2 & 1.05 & 1.5 & 0.7 \\
B0153+39 & 136.37 & -21.33 & 4.92 & 1.812 & 0.15 & 0.01 & 189 & 0.53 & 0.2 & 0.1 \\
\\[-2pt] 
B0154+61 & 130.59 & 0.33 & 1.79 & 2.352 & 188.9 & 5.70 & 0.2 & 21.3 & 3.9 & 1.6 \\
B0226+70 & 131.16 & 9.18 & 1.76 & 1.467 & 3.11 & 0.39 & 7.5 & 2.16 & 1.0 & 0.5 \\
B0320+39 & 152.18 & -14.34 & 0.95 & 3.032 & 0.64 & 0.01 & 75.6 & 1.40 & 0.2 & 0.1 \\
B0329+54 & 145.00 & -1.22 & 1.69 & 0.715 & 2.05 & 2.20 & 5.5 & 1.22 & 2.4 & 1.0 \\
B0331+45 & 150.35 & -8.04 & 2.44 & 0.269 & 0.01 & 0.15 & 580 & 0.05 & 0.6 & 0.3 \\
\\[-2pt] 
B0339+53 & 147.02 & -1.43 & 1.71 & 1.934 & 13.42 & 0.73 & 2.3 & 5.16 & 1.4 & 0.7 \\
B0353+52 & 149.10 & -0.52 & 3.57 & 0.197 & 0.48 & 25.0 & 6.6 & 0.31 & 8.0 & 2.2 \\
B0355+54 & 148.19 & 0.81 & 1.00 & 0.156 & 4.39 & 450 & 0.6 & 0.84 & 34.3 & 7.0 \\
B0402+61 & 144.03 & 7.05 & 4.55 & 0.595 & 5.57 & 10.0 & 1.7 & 1.84 & 5.2 & 1.8 \\
B0410+69 & 138.91 & 13.67 & 1.37 & 0.391 & 0.08 & 0.51 & 80.8 & 0.18 & 1.1 & 0.5 \\
\\[-2pt] 
B0447--12 & 211.08 & -32.63 & 1.77 & 0.438 & 0.10 & 0.48 & 67.6 & 0.22 & 1.1 & 0.5 \\
B0450--18 & 217.08 & -34.09 & 0.40 & 0.549 & 5.75 & 14.0 & 1.5 & 1.80 & 6.0 & 1.9 \\
B0450+55 & 152.62 & 7.55 & 1.18 & 0.341 & 2.37 & 24.0 & 2.3 & 0.91 & 7.8 & 2.3 \\
B0458+46 & 160.36 & 3.08 & 1.32 & 0.639 & 5.58 & 8.50 & 1.8 & 1.91 & 4.7 & 1.6 \\
B0559--05 & 212.20 & -13.48 & 2.08 & 0.396 & 1.30 & 8.30 & 4.8 & 0.73 & 4.6 & 1.5 \\
\\[-2pt] 
B0621--04 & 213.79 & -8.04 & 1.94 & 1.039 & 0.83 & 0.29 & 19.8 & 0.94 & 0.9 & 0.4 \\
B0628--28 & 236.95 & -16.76 & 0.32 & 1.244 & 7.12 & 1.50 & 2.8 & 3.01 & 1.9 & 0.9 \\
B0643+80 & 133.18 & 26.83 & 2.35 & 1.214 & 3.80 & 0.84 & 5.1 & 2.17 & 1.5 & 0.7 \\
B0655+64 & 151.55 & 25.24 & 0.41 & 0.196 & 0.00 & 0.04 & 4520 & 0.01 & 0.3 & 0.2 \\
B0727--18 & 233.76 & -0.34 & 2.00 & 0.510 & 18.96 & 56.0 & 0.4 & 3.15 & 12.1 & 3.4 \\
\\[-2pt] 
B0740--28 & 243.77 & -2.44 & 2.00 & 0.167 & 16.82 & 1400 & 0.2 & 1.69 & 60.8 & 11.1 \\
B0756--15 & 234.46 & 7.22 & 2.71 & 0.682 & 1.62 & 2.00 & 6.7 & 1.06 & 2.3 & 0.9 \\
B0809+74 & 140.00 & 31.62 & 0.43 & 1.292 & 0.17 & 0.03 & 122 & 0.47 & 0.3 & 0.2 \\
B0818--13 & 258.75 & -2.73 & 1.90 & 1.238 & 2.11 & 0.44 & 9.3 & 0.47 & 0.3 & 0.5 \\
B0826--34 & 253.97 & 2.56 & 0.35 & 1.849 & 1.00 & 0.06 & 29.4 & 1.37 & 0.4 & 0.3 \\
\\[-2pt] 
B0841+80 & 132.65 & 31.46 & 3.30 & 1.602 & 0.45 & 0.04 & 56.9 & 0.86 & 0.3 & 0.2 \\
B0844--35 & 257.19 & 4.71 & 0.54 & 1.116 & 1.60 & 0.46 & 11.0 & 1.35 & 1.1 & 0.5 \\
B0853--33 & 256.85 & 7.52 & 0.50 & 1.268 & 6.32 & 1.20 & 3.2 & 2.86 & 1.8 & 0.8 \\
B0906--17 & 246.12 & 19.85 & 0.80 & 0.402 & 0.67 & 4.10 & 9.5 & 0.53 & 3.3 & 1.2 \\
B0917+63 & 151.43 & 40.73 & 1.00 & 1.568 & 3.61 & 0.37 & 6.9 & 2.41 & 1.0 & 0.5 \\
\\[-2pt] 
B0942--13 & 249.13 & 28.84 & 0.69 & 0.570 & 0.05 & 0.10 & 200 & 0.16 & 0.5 & 0.3 \\
B1010--23 & 262.13 & 26.38 & 0.98 & 2.518 & 0.88 & 0.02 & 45.3 & 1.51 & 0.2 & 0.2 \\
B1016--16 & 258.26 & 32.61 & 25.00 & 1.805 & 1.74 & 0.12 & 16.4 & 1.79 & 0.5 & 0.3 \\
B1039--19 & 265.59 & 33.59 & 2.53 & 1.386 & 0.94 & 0.14 & 23.2 & 1.16 & 0.6 & 0.3 \\
B1112+50 & 154.41 & 60.37 & 0.92 & 1.656 & 2.49 & 0.22 & 10.5 & 2.06 & 0.8 & 0.4 \\
\\[-10pt] 
B1254--10 & 305.21 & 52.40 & 25.00 & 0.617 & 0.36 & 0.61 & 27.0 & 0.48 & 1.3 & 0.6 \\
B1309--12 & 310.72 & 50.10 & 25.00 & 0.448 & 0.15 & 0.66 & 47.0 & 0.26 & 1.3 & 0.6 \\
B1322+83 & 121.89 & 33.67 & 1.09 & 0.670 & 0.57 & 0.74 & 18.7 & 0.62 & 1.4 & 0.6 \\
B1508+55 & 91.33 & 52.29 & 2.10 & 0.740 & 5.00 & 4.90 & 2.3 & 1.95 & 3.6 & 1.3 \\
B1540--06 & 0.57 & 36.61 & 3.23 & 0.709 & 0.88 & 0.97 & 12.8 & 0.80 & 1.6 & 0.7 \\
\\ 
B1552--31 & 348.44 & 22.50 & 5.30 & 0.518 & 0.06 & 0.18 & 132 & 0.18 & 0.7 & 0.3 \\
B1552--23 & 342.70 & 16.76 & 3.75 & 0.533 & 0.69 & 1.80 & 12.2 & 0.62 & 2.2 & 0.9 \\
B1600--27 & 347.13 & 18.77 & 2.31 & 0.778 & 3.01 & 2.50 & 4.1 & 1.55 & 2.6 & 1.0 \\
B1607--13 & 359.43 & 26.95 & 3.25 & 1.018 & 0.23 & 0.09 & 70.2 & 0.49 & 0.5 & 0.3 \\
B1612--29 & 347.39 & 15.06 & 1.72 & 2.478 & 1.58 & 0.04 & 24.8 & 2.00 & 0.3 & 0.2 \\
\\ 
B1620--09 & 5.30 & 27.18 & 1.67 & 1.276 & 2.58 & 0.49 & 7.8 & 1.84 & 1.1 & 0.6 \\
B1620--26 & 350.98 & 15.96 & 1.80 & 0.011 & 0.00 & 190 & 262 & 0.00 & 22.5 & 3.8 \\
B1642--03 & 14.11 & 26.06 & 3.85 & 0.388 & 1.78 & 12.0 & 3.5 & 0.84 & 5.6 & 1.8 \\
B1648--17 & 2.81 & 16.88 & 0.84 & 0.973 & 3.04 & 1.30 & 5.1 & 1.74 & 1.8 & 0.8 \\
B1649--23 & 357.32 & 12.45 & 3.37 & 1.704 & 3.16 & 0.25 & 8.6 & 2.35 & 0.8 & 0.4 \\
\\[-2pt] 
B1650--13 & 7.51 & 17.59 & 0.48 & 0.641 & 0.62 & 0.93 & 16.4 & 0.64 & 1.6 & 0.7 \\
B1700--32 & 3.23 & 13.56 & 3.17 & 1.212 & 0.66 & 0.15 & 29.1 & 0.91 & 0.6 & 0.3 \\
B1700--18 & 351.79 & 5.39 & 2.86 & 0.804 & 1.73 & 1.30 & 7.4 & 1.19 & 1.8 & 0.8 \\
B1702--19 & 3.19 & 13.03 & 0.75 & 0.299 & 4.14 & 61.0 & 1.1 & 1.13 & 12.6 & 3.3 \\
B1706--16 & 5.78 & 13.66 & 0.56 & 0.653 & 6.31 & 8.90 & 1.6 & 2.05 & 4.8 & 1.7 \\
\\ 
B1709--15 & 7.42 & 14.01 & 1.08 & 0.869 & 1.10 & 0.66 & 12.5 & 0.99 & 1.3 & 0.6 \\
B1714--34 & 352.12 & 2.03 & 25.00 & 0.656 & 9.80 & 14.0 & 1.1 & 2.57 & 6.0 & 2.0 \\
B1717--16 & 20.13 & 18.94 & 0.99 & 1.566 & 5.80 & 0.60 & 4.3 & 3.05 & 1.2 & 0.6 \\
B1717--29 & 7.37 & 11.54 & 1.10 & 0.620 & 0.75 & 1.20 & 13.2 & 0.69 & 1.8 & 0.8 \\
B1718--02 & 356.51 & 4.25 & 2.36 & 0.478 & 0.08 & 0.30 & 91.4 & 0.20 & 0.9 & 0.4 \\
\\ 
B1718--19 & 4.86 & 9.74 & 8.60 & 1.004 & 1.62 & 0.63 & 9.8 & 1.29 & 1.3 & 0.6 \\
B1718--35 & 351.69 & 0.67 & 4.60 & 0.280 & 25.19 & 450 & 0.2 & 2.69 & 34.2 & 7.4 \\
B1718--32 & 354.56 & 2.53 & 2.93 & 0.477 & 0.65 & 2.30 & 11.7 & 0.56 & 2.5 & 0.9 \\
B1727--33 & 354.13 & 0.09 & 3.49 & 0.139 & 84.83 & 12000 & 0.0 & 3.48 & 179 & 25.8 \\
B1730--22 & 4.03 & 5.75 & 1.11 & 0.872 & 0.04 & 0.03 & 323 & 0.20 & 0.3 & 0.2 \\
\\ 
B1732--02 & 21.90 & 15.93 & 1.80 & 0.839 & 0.42 & 0.28 & 31.6 & 0.60 & 0.9 & 0.4 \\
B1732--07 & 17.27 & 13.28 & 6.67 & 0.419 & 1.21 & 6.50 & 5.5 & 0.72 & 4.1 & 1.4 \\
B1734--35 & 353.18 & -2.27 & 2.49 & 0.398 & 6.12 & 38.0 & 1.0 & 1.58 & 10.0 & 2.8 \\
B1735--32 & 356.47 & -0.49 & 1.27 & 0.768 & 0.79 & 0.69 & 15.3 & 0.79 & 1.3 & 0.6 \\
B1736--31 & 359.21 & 1.06 & 4.41 & 0.529 & 18.58 & 49.0 & 0.5 & 3.17 & 11.3 & 3.2 \\
\\ 
B1736--29 & 357.10 & -0.22 & 2.91 & 0.323 & 7.87 & 92.0 & 0.7 & 1.61 & 15.4 & 4.0 \\
B1737--30 & 358.29 & 0.24 & 0.40 & 0.607 & 466.1 & 820 & 0.0 & 17.0 & 46.2 & 10.1 \\
B1738--08 & 16.96 & 11.30 & 3.57 & 2.043 & 2.27 & 0.11 & 14.2 & 2.18 & 0.5 & 0.3 \\
B1740--13 & 21.65 & 13.40 & 0.20 & 0.445 & 1.56 & 7.00 & 4.5 & 0.84 & 4.3 & 1.5 \\
B1740--31 & 12.70 & 8.21 & 3.33 & 2.415 & 120.8 & 3.40 & 0.3 & 17.3 & 3.0 & 1.3 \\
\\ 
B1740--03 & 357.30 & -1.15 & 3.50 & 0.405 & 0.48 & 2.80 & 13.4 & 0.45 & 2.7 & 1.0 \\
B1742--30 & 358.55 & -0.96 & 0.20 & 0.367 & 10.67 & 85.0 & 0.5 & 2.00 & 14.8 & 3.9 \\
B1745--12 & 14.02 & 7.66 & 2.40 & 0.394 & 1.21 & 7.80 & 5.2 & 0.70 & 4.5 & 1.5 \\
B1746--30 & 359.46 & -1.24 & 12.73 & 0.610 & 7.87 & 14.0 & 1.2 & 2.22 & 6.0 & 2.0 \\
B1747--31 & 357.98 & -2.52 & 4.34 & 0.910 & 0.20 & 0.10 & 73.4 & 0.43 & 0.5 & 0.3 \\
\\ 
B1749--28 & 1.54 & -0.96 & 0.20 & 0.563 & 8.13 & 18.0 & 1.1 & 2.16 & 6.8 & 2.2 \\
B1750--24 & 4.27 & 0.51 & 5.31 & 0.528 & 14.12 & 38.0 & 0.6 & 2.76 & 9.9 & 2.9 \\
B1753+52 & 79.61 & 29.63 & 6.25 & 2.391 & 1.56 & 0.05 & 24.2 & 1.96 & 0.3 & 0.2 \\
B1753--24 & 5.03 & 0.04 & 3.83 & 0.670 & 0.28 & 0.37 & 37.3 & 0.44 & 1.0 & 0.5 \\
B1754--24 & 5.28 & 0.05 & 3.12 & 0.234 & 13.00 & 400 & 0.3 & 1.77 & 32.3 & 6.9 \\
\\ 
B1756--22 & 7.47 & 0.81 & 3.26 & 0.461 & 10.85 & 44.0 & 0.7 & 2.26 & 10.6 & 3.0 \\
B1757--24 & 23.60 & 9.26 & 3.80 & 0.125 & 127.9 & 26000 & 0.0 & 4.04 & 259 & 34.3 \\
B1758--23 & 6.84 & -0.07 & 4.00 & 0.416 & 112.9 & 620 & 0.1 & 6.93 & 40.1 & 8.7 \\
B1758--03 & 5.25 & -0.88 & 5.74 & 0.921 & 3.31 & 1.70 & 4.4 & 1.77 & 2.1 & 0.9 \\
B1800--21 & 8.40 & 0.15 & 4.40 & 0.134 & 134.4 & 22000 & 0.0 & 4.29 & 240 & 32.5 \\
\\ 
B1802--07 & 20.79 & 6.77 & 7.80 & 0.023 & 0.00 & 15.0 & 784 & 0.00 & 6.2 & 1.5 \\
B1804--27 & 20.06 & 5.59 & 15.36 & 0.828 & 12.17 & 8.50 & 1.1 & 3.21 & 4.7 & 1.7 \\
B1804--08 & 3.84 & -3.26 & 1.50 & 0.164 & 0.03 & 2.60 & 90.1 & 0.07 & 2.6 & 0.9 \\
B1805--20 & 9.45 & -0.40 & 4.58 & 0.918 & 17.08 & 8.70 & 0.9 & 4.01 & 4.8 & 1.7 \\
B1806--21 & 9.42 & -0.72 & 4.12 & 0.702 & 3.82 & 4.40 & 2.9 & 1.66 & 3.4 & 1.3 \\
\\ 
B1809--173 & 13.11 & 0.54 & 3.68 & 1.205 & 19.08 & 4.30 & 1.0 & 4.85 & 3.3 & 1.3 \\
B1809--176 & 12.90 & 0.39 & 4.52 & 0.538 & 0.98 & 2.50 & 8.7 & 0.74 & 2.5 & 1.0 \\
B1811+40 & 67.41 & 24.03 & 4.92 & 0.931 & 2.55 & 1.20 & 5.8 & 1.56 & 1.8 & 0.8 \\
B1813--17 & 13.43 & -0.42 & 4.45 & 0.782 & 7.26 & 6.00 & 1.7 & 2.41 & 3.9 & 1.4 \\
B1813--26 & 5.22 & -4.91 & 3.59 & 0.593 & 0.07 & 0.13 & 141 & 0.20 & 0.6 & 0.3 \\
\\ 
B1813--36 & 356.80 & -9.37 & 4.40 & 0.387 & 2.02 & 14.0 & 3.0 & 0.90 & 6.0 & 1.9 \\
B1815--14 & 16.41 & 0.61 & 5.46 & 0.291 & 2.04 & 32.0 & 2.3 & 0.78 & 9.2 & 2.6 \\
B1817--13 & 25.46 & 4.73 & 5.90 & 0.921 & 4.50 & 2.30 & 3.3 & 2.06 & 2.4 & 1.0 \\
B1817--18 & 17.16 & 0.48 & 14.04 & 0.310 & 0.09 & 1.20 & 52.5 & 0.17 & 1.8 & 0.7 \\
B1818--04m & 13.20 & -1.72 & 2.86 & 0.598 & 6.33 & 12.0 & 1.5 & 1.97 & 5.5 & 1.8 \\
\\ 
B1819--22 & 9.35 & -4.37 & 3.26 & 1.874 & 1.35 & 0.08 & 21.9 & 1.61 & 0.5 & 0.3 \\
B1820--14 & 17.25 & -0.18 & 4.76 & 0.215 & 0.91 & 36.0 & 3.8 & 0.45 & 9.7 & 2.6 \\
B1820--11 & 19.77 & 0.95 & 5.37 & 0.280 & 1.38 & 25.0 & 3.2 & 0.63 & 8.0 & 2.3 \\
B1820--30B & 2.79 & -7.92 & 12.10 & 0.379 & 0.03 & 0.22 & 201 & 0.11 & 0.8 & 0.4 \\
B1820--31 & 2.12 & -8.27 & 1.59 & 0.284 & 2.93 & 50.0 & 1.5 & 0.92 & 11.4 & 3.1 \\
\\ 
B1821--19 & 19.81 & 0.74 & 3.70 & 0.189 & 5.24 & 300 & 0.6 & 1.01 & 28.2 & 6.0 \\
B1821--11 & 12.28 & -3.11 & 6.02 & 0.436 & 3.55 & 17.0 & 1.9 & 1.26 & 6.6 & 2.1 \\
B1822--14 & 21.45 & 1.32 & 4.44 & 0.279 & 22.67 & 410 & 0.2 & 2.55 & 32.7 & 7.1 \\
B1822--09m & 16.81 & -1.00 & 0.30 & 0.769 & 52.36 & 45.0 & 0.2 & 6.42 & 10.9 & 3.3 \\
B1823--11 & 19.80 & 0.29 & 3.98 & 2.093 & 4.91 & 0.21 & 6.8 & 3.24 & 0.7 & 0.4 \\
\\ 
B1823--13 & 18.00 & -0.69 & 3.61 & 0.101 & 75.25 & 28000 & 0.0 & 2.80 & 272 & 34.9 \\
B1824--10 & 21.29 & 0.80 & 5.03 & 0.246 & 1.00 & 27.0 & 3.9 & 0.50 & 8.3 & 2.3 \\
B1826--17 & 14.60 & -3.42 & 5.94 & 0.307 & 5.55 & 76.0 & 0.9 & 1.32 & 14.0 & 3.6 \\
B1828--11 & 20.81 & -0.48 & 3.15 & 0.405 & 59.92 & 360 & 0.1 & 4.99 & 30.4 & 6.9 \\
B1829--08 & 23.27 & 0.30 & 5.20 & 0.647 & 63.90 & 93.0 & 0.2 & 6.51 & 15.5 & 4.3 \\
\\ 
B1829--10 & 21.59 & -0.60 & 4.69 & 0.330 & 4.20 & 46.0 & 1.3 & 1.19 & 10.9 & 3.0 \\
B1830--08 & 27.66 & 2.27 & 4.50 & 0.085 & 9.18 & 5800 & 0.1 & 0.90 & 123 & 18.2 \\
B1831--03 & 23.39 & 0.06 & 2.50 & 0.687 & 41.56 & 51.0 & 0.3 & 5.41 & 11.5 & 3.4 \\
B1831--04 & 27.04 & 1.75 & 2.49 & 0.290 & 0.07 & 1.20 & 63.9 & 0.15 & 1.7 & 0.7 \\
B1832--06 & 25.09 & 0.55 & 5.04 & 0.306 & 40.46 & 560 & 0.1 & 3.56 & 38.1 & 8.1 \\
\\ 
B1834--04 & 27.17 & 1.13 & 4.36 & 0.354 & 1.66 & 15.0 & 3.4 & 0.78 & 6.2 & 1.9 \\
B1834--10 & 22.26 & -1.42 & 5.34 & 0.563 & 11.80 & 26.0 & 0.8 & 2.61 & 8.2 & 2.5 \\
B1834--06 & 25.19 & 0.00 & 4.14 & 1.906 & 0.77 & 0.04 & 39.1 & 1.23 & 0.3 & 0.2 \\
B1839+56 & 86.08 & 23.82 & 1.45 & 1.653 & 1.49 & 0.13 & 17.5 & 1.59 & 0.6 & 0.3 \\
B1838--04 & 27.82 & 0.28 & 4.40 & 0.186 & 6.39 & 390 & 0.5 & 1.10 & 31.7 & 6.7 \\
\\ 
B1839--04 & 28.35 & 0.17 & 3.71 & 1.840 & 0.51 & 0.03 & 57.3 & 0.98 & 0.3 & 0.2 \\
B1841--05 & 29.73 & 0.24 & 5.40 & 0.256 & 9.71 & 230 & 0.4 & 1.59 & 24.3 & 5.6 \\
B1841--04 & 28.10 & -0.55 & 3.07 & 0.991 & 3.91 & 1.60 & 4.0 & 1.99 & 2.0 & 0.9 \\
B1842--02 & 27.07 & -0.94 & 4.92 & 0.508 & 16.74 & 50.0 & 0.5 & 2.95 & 11.4 & 3.3 \\
B1842--04 & 28.19 & -0.79 & 4.09 & 0.487 & 11.33 & 39.0 & 0.7 & 2.38 & 10.0 & 2.9 \\
\\[-8pt] 
B1844--04 & 28.88 & -0.94 & 3.42 & 0.598 & 51.69 & 96.0 & 0.2 & 5.63 & 15.8 & 4.3 \\
B1845--19 & 31.34 & 0.04 & 0.75 & 4.308 & 23.28 & 0.11 & 2.9 & 10.1 & 0.5 & 0.4 \\
B1845--01 & 14.77 & -8.25 & 4.40 & 0.659 & 5.25 & 7.20 & 2.0 & 1.88 & 4.3 & 1.5 \\
B1846--06 & 26.77 & -2.50 & 3.85 & 1.451 & 46.24 & 6.00 & 0.5 & 8.29 & 3.9 & 1.5 \\
B1851--14 & 20.46 & -7.21 & 6.91 & 1.147 & 4.16 & 1.10 & 4.4 & 2.21 & 1.7 & 0.8 \\
\\ 
B1857--26 & 10.34 & -13.45 & 0.70 & 0.612 & 0.20 & 0.35 & 47.4 & 0.36 & 1.0 & 0.5 \\
B1900--06 & 28.48 & -5.68 & 12.45 & 0.432 & 3.40 & 17.0 & 2.0 & 1.23 & 6.6 & 2.1 \\
B1905+39 & 70.95 & 14.20 & 2.46 & 1.236 & 0.54 & 0.11 & 36.2 & 0.83 & 0.5 & 0.3 \\
B1907--03 & 32.28 & -5.68 & 14.44 & 0.505 & 2.19 & 6.70 & 3.7 & 1.06 & 4.2 & 1.5 \\
B1911--04 & 31.31 & -7.12 & 4.04 & 0.826 & 4.07 & 2.90 & 3.2 & 1.85 & 2.7 & 1.1 \\
\\ 
B1937--26 & 13.90 & -21.82 & 3.56 & 0.403 & 0.96 & 5.80 & 6.7 & 0.63 & 3.9 & 1.3 \\
B1940--12 & 27.26 & -17.16 & 1.20 & 0.972 & 1.66 & 0.71 & 9.3 & 1.28 & 1.4 & 0.6 \\
B1941--17 & 22.31 & -19.43 & 4.12 & 0.841 & 0.99 & 0.65 & 13.5 & 0.92 & 1.3 & 0.6 \\
B1943--29 & 11.11 & -24.12 & 3.13 & 0.959 & 1.49 & 0.67 & 10.2 & 1.21 & 1.3 & 0.6 \\
B1946--25 & 15.26 & -23.38 & 1.05 & 0.958 & 3.27 & 1.50 & 4.6 & 1.79 & 2.0 & 0.8 \\
\\ 
B1953+50 & 84.79 & 11.55 & 2.19 & 0.519 & 1.37 & 3.90 & 6.0 & 0.85 & 3.2 & 1.2 \\
B2000+40 & 76.61 & 5.29 & 6.39 & 0.905 & 1.74 & 0.93 & 8.3 & 1.27 & 1.6 & 0.7 \\
B2003--08 & 34.10 & -20.30 & 2.63 & 0.581 & 0.05 & 0.09 & 200 & 0.17 & 0.5 & 0.3 \\
B2011+38 & 75.93 & 2.48 & 7.12 & 0.230 & 8.85 & 290 & 0.4 & 1.44 & 27.2 & 6.0 \\
B2021+51 & 87.86 & 8.38 & 1.80 & 0.529 & 3.06 & 8.20 & 2.7 & 1.29 & 4.6 & 1.6 \\
\\ 
B2022+50m & 86.86 & 7.54 & 2.06 & 0.373 & 2.51 & 19.0 & 2.4 & 0.98 & 7.1 & 2.1 \\
B2036+53 & 90.37 & 7.31 & 8.19 & 1.425 & 0.94 & 0.13 & 23.9 & 1.17 & 0.6 & 0.3 \\
B2043--04 & 42.68 & -27.39 & 6.25 & 1.547 & 1.47 & 0.16 & 16.7 & 1.53 & 0.6 & 0.4 \\
B2045+56 & 94.20 & 8.64 & 4.34 & 0.477 & 11.12 & 41.0 & 0.7 & 2.33 & 10.3 & 3.0 \\
B2045--16 & 30.51 & -33.08 & 0.95 & 1.962 & 10.96 & 0.57 & 2.8 & 4.69 & 1.2 & 0.6 \\
\\ 
B2106+44 & 86.91 & -2.01 & 4.35 & 0.415 & 0.09 & 0.48 & 76.2 & 0.19 & 1.1 & 0.5 \\
B2111+46 & 89.00 & -1.27 & 2.17 & 1.015 & 0.71 & 0.27 & 22.5 & 0.86 & 0.8 & 0.4 \\
B2148+63 & 104.26 & 7.41 & 2.78 & 0.380 & 0.17 & 1.20 & 35.8 & 0.26 & 1.8 & 0.7 \\
B2148+52 & 97.52 & -0.92 & 3.61 & 0.332 & 10.11 & 110 & 0.5 & 1.85 & 16.8 & 4.2 \\
B2152--31 & 15.85 & -51.58 & 1.30 & 1.030 & 1.24 & 0.45 & 13.2 & 1.14 & 1.1 & 0.5 \\
\\ 
B2154+40 & 90.49 & -11.34 & 2.90 & 1.525 & 3.43 & 0.38 & 7.0 & 2.32 & 1.0 & 0.5 \\
B2217+47 & 98.39 & -7.60 & 2.39 & 0.538 & 2.77 & 7.00 & 3.1 & 1.23 & 4.2 & 1.5 \\
B2224+65 & 108.64 & 6.85 & 0.90 & 0.683 & 9.66 & 12.0 & 1.1 & 2.60 & 5.6 & 1.9 \\
B2227+61 & 107.15 & 3.65 & 3.01 & 0.443 & 2.26 & 10.0 & 3.1 & 1.01 & 5.1 & 1.7 \\
B2241+69 & 112.22 & 9.70 & 2.01 & 1.665 & 4.82 & 0.41 & 5.5 & 2.87 & 1.0 & 0.5 \\
\\ 
B2255+58 & 108.83 & -0.58 & 3.00 & 0.368 & 5.75 & 45.0 & 1.0 & 1.47 & 10.8 & 3.0 \\
B2303+46 & 108.73 & -4.21 & 3.16 & 1.066 & 0.57 & 0.19 & 29.7 & 0.79 & 0.7 & 0.4 \\
B2306+55 & 104.41 & -16.42 & 2.07 & 0.475 & 0.20 & 0.73 & 37.7 & 0.31 & 1.4 & 0.6 \\
B2310+42 & 112.10 & -0.57 & 1.06 & 0.349 & 0.11 & 1.00 & 49.3 & 0.20 & 1.6 & 0.7 \\
B2319+60 & 113.42 & 2.01 & 2.70 & 2.256 & 7.04 & 0.24 & 5.1 & 4.03 & 0.8 & 0.4 \\
\\ 
B2323+63 & 112.95 & 0.00 & 4.86 & 1.436 & 2.83 & 0.38 & 8.1 & 2.04 & 1.0 & 0.5 \\
B2324+60m & 49.39 & -70.19 & 2.73 & 0.234 & 0.35 & 11.0 & 10.5 & 0.29 & 5.3 & 1.6 \\
B2327--20 & 114.28 & 0.23 & 0.86 & 1.644 & 4.63 & 0.41 & 5.6 & 2.79 & 1.0 & 0.5 \\
B2334+61 & 114.28 & 0.23 & 0.70 & 0.495 & 193.4 & 6300 & 0.0 & 9.91 & 40.4 & 8.9 \\
B2351+61 & 116.24 & -0.19 & 2.44 & 0.945 & 16.26 & 7.60 & 0.9 & 3.97 & 4.4 & 1.6 \\

\hline
\end{longtable}
\end{center}

\twocolumn  

\noindent\textit{\textbf{B0320+39}}:  This pulsar is well known for its regular 6.4-$P$ modulation \citep{Weltevrede2006, Weltevrede2007}, and its profile consists of two components at all frequencies.  We then model it using an outer conal {\textbf S$_d$} configuration.  However, the profile narrows at low frequency, perhaps because one component fades out in a manner similar to the ``absorption'' \citep[\eg][]{Rankin2006} in B0809+74 or B0943+10. If scattering is responsible for the larger width at 38 MHz, then perhaps an inner conal model would be more appropriate.  \cite{Zakharenko2013} detect the pulsar at 25 MHz, but the profile is too scattered to be useful here \citep{kuzmin_LL2007}.
\vskip 0.09in
\noindent\textit{\textbf{B0329+54}}: This pulsar has a classic and very well studied {\textbf T} profile (\eg \citet{mitra2007}) with inner conal features that are seen in single pulses.  Its core component is known to be ``notched'' and thus often does not reflect the full width of the polar cap.  In addition, the pulsar exhibits prominent moding  \citep[\eg][]{brinkman2019} that complicates interpreting average profiles.  \cite{Zakharenko2013} detect the pulsar at 25 MHz, but the profile is too scattered to be useful here \citep{kuzmin_LL2007}.
\vskip 0.09in
\noindent\textit{\textbf{B0331+45}}: There is every indication that this pulsar has an inner cone {\textbf S$_d$} geometry, as the profile widths vary little down to 100 MHz.  However, there is no adequate polarimetry to estimate a PPA sweep rate; the $R$ value is a guess, so the geometry is therefore conjectural.  \citet{kuzmin_LL2007} provide a $t_{\rm scatt}$.
\vskip 0.09in
\noindent\textit{\textbf{B0339+53}}:  This pulsar has two unresolved components down to the LOFAR band and seems to have an inner conal {\textbf S$_d$} beam system.  The PPA sweep rate can be roughly estimated. \citet{kuzmin_LL2007} provide a $t_{\rm scatt}$.
\vskip 0.09in
\noindent\textit{\textbf{B0353+52}}: The profiles have an asymmetric scattered shape even at the highest frequencies.  It may have a core-single {\textbf S$_t$} geometry, as the increased width above 1 GHz may indicate conal outriders.  \citet{Weltevrede2006} find a flat fluctuation spectrum. The KL99 103-MHz profile has undergone a deconvolution to correct for the substantial scattering distortion \citep[see][]{kuzmin_LL2007}.
\vskip 0.09in
\noindent\textit{\textbf{B0355+54}}: This bright pulsar has a well-studied {\textbf S$_t$} (or {\textbf T} in that the conal outriders are visible at 1 GHz) profile. \citep[e.g.][]{MSFB1980}.  Its PPA traverse is well defined but steepens under the trailing side of the profile as if there is aberration/retardation.  \cite{Weltevrede2006,Weltevrede2007} find a low frequency excess at both frequencies consistent with sporadic modulation.  We reiterate the ET VI and ET IX outer conal model, but it can be discerned down only to about 600 MHz.  Note that the core width increases at longer wavelengths; however, the \cite{kuzmin1999} scattering-deconvolved width is more consistent with high frequency values \citep[see][]{kuzmin_LL2007}.  
\vskip 0.09in
\noindent\textit{\textbf{B0402+61}}: ET VI discussed this pulsar as possibly having a five-component profile; however, neither the core nor inner cone was readily seen.  Rather, suggestions for this structure came from the detailed $L$ and total power profiles then available, and some of these suggestions are seen in both the GL98 and LOFAR profiles.  Here we see clear evidence for triplicity as well as in the \cite{Weltevrede2007} profile. The outer cone is nicely traced to low frequency with some scatter broadening.  A core width of about 3\degr\ is implied by this geometry---and the interior components, in the few profiles they are seen, would have about this width.  Single-pulse analyses are needed to fully discern this pulsar's structure.  MM10 detect the pulsar at 103 MHz in a manner that may be scattered per the \citet{kuzmin_LL2007} value.
\vskip 0.09in
\noindent\textit{\textbf{B0410+69}}: This pulsar has two barely resolved components, the leading one much weaker.  It probably has an inner cone {\textbf S$_d$} geometry though it is too weak for a meaningful fluctuation spectral analysis \citep{Weltevrede2007}. Its width is fairly consistent down to the LOFAR band. The PPA rate is poorly determined, so the beam model is poorly determined.  MM10 seem to detect the pulsar at 103 MHz, but the form and width is entirely different that the other profiles.  No $t_{tscatt}$ value is available.
\vskip 0.09in
\noindent\textit{\textbf{B0447--12}}: The GL98 profiles suggests a tripartite structure that could be entirely conal, but \citep{Weltevrede2007} find no modulation feature, so we model its geometry as a core-conal {\textbf T} triple.  Scattering at 149 MHz is predicted to be negligible and the KL99 profile cannot be reliably interpreted.  The widths are measured to include the weak trailing component, and an estimate of the core width can be made from the 149-MHz profile. \citet{kmn+15} provide a scattering measurement.
\vskip 0.09in
\noindent\textit{\textbf{B0450--18}}:  We follow ET VI in modeling the pulsar using a core-inner cone triple {\textbf T} beam system.  \cite{Weltevrede2006,Weltevrede2007} find a flat fluctuation spectrum at both frequencies.  Core widths can be estimated at both 408 and 149 MHz which agree with that stemming from the 1-GHz geometry. Scattering is negligible down to the LOFAR band (KLL07).  
\vskip 0.09in
\noindent\textit{\textbf{B0450+55}}: We again follow the ET VI core-inner cone triple model, and the ET IX observation of its delayed PPA traverse.  Conal drift is not observed at either frequency \citep{Weltevrede2006,Weltevrede2007}.  The core and leading conal component are nearly conflated at low frequency, but the width of the bright core can be traced down to 65 MHz using the PHS+ observation.  Narrower widths in the LOFAR High Band suggest that the core is incomplete here or that the profile evolves in an uncharacteristic manner.  \cite{Zakharenko2013} detect the pulsar at 25 MHz, but the profile is too scattered to be useful here (see KLL07).
\vskip 0.09in
\noindent\textit{\textbf{B0458+46}}:  ET VI first suggested that the profiles might have a triple configuration as the leading feature shows structure and \cite{Force2015} seconded the question.  Most of the GL98 profiles show this complexity as well as three parts to the $L$ profile.  \citet{Weltevrede2006,Weltevrede2007} further find flat fluctuation spectra at both frequencies.  We model the geometry using a core/inner conal beam system.  This requires an $\alpha$ value of about 31\degr, and while the core width cannot be measured at any frequency, the roughly 6\degr\ it implies is quite plausible.  Scattering is negligible down to 400 MHz; however, both LOFAR profiles show long scattering tails \citep[see][]{kmn+15}.  MM10's narrower 111-MHz profile is difficult to interpret.
\vskip 0.09in
\noindent\textit{\textbf{B0559--05}}: The pulsar may have a core-cone triple profile wherein the leading conal component is either very weak or missing.  A conal width of some 22\degr\ may follow from the \cite{XRSS} 1720-MHz profile and several of the GL ones.  The frequency evolution nicely depicted by \cite{vH_thesis} seems to support this, and fluctuation spectra show no modulation features \citep{Weltevrede2006,Weltevrede2007}. \cite{JKMG2008} report scattering in their 327 and 243-MHz profiles, and the \citet{MM10} shows a very substantial distortion. \cite{kuzmin_LL2007} give a 100-MHz scattering time.
\vskip 0.09in
\noindent\textit{\textbf{B0621--04}}:  ET VI proposed that this pulsar had a five-component core/double-cone profile; however, we see no evidence for a core, nor can the inner cone be measured accurately at any frequency.  \citet{Weltevrede2006} find an odd-even drift modulation, so we here model the pulsar using a double cone c{\textbf Q} geometry.  KLL07 give a $t_{tscatt}$ value.  
\vskip 0.09in
\noindent\textit{\textbf{B0628--28}}: We follow ET VI in modeling the pulsar as having an inner conal single {\textbf S$_d$} profile, and \citet{Weltevrede2006,Weltevrede2007} and \citet{basu2016} do find evidence of sporadic drift modulation.  The profile broadens some below 100 MHz, but the reason is unclear: \citet{cor86} reports an unusually small scattering time from scintillation analyses, and no other study as yet confirms it. 
\vskip 0.09in
\noindent\textit{\textbf{B0643+80}}: The GL98 profiles show what appears to be a conal single or double profile with a stronger leading component---but no hint of two cones.  However, the LOFAR profiles seem to show both cones, so we use a conal quadruple {c\textbf Q} beam system to model its geometry.  Although the profile seems entirely conal, \citet{Weltevrede2007} find only a flat fluctuation spectrum. No $t_{tscatt}$ value is available. 
\vskip 0.09in
\noindent\textit{\textbf{B0655+64}}:  PSR B0655+64 is a fast pulsar with an unusually small spindown, and it seems to be entirely conal with two barely resolved components at high frequency.  However, there are hints of weaker outer conal components on the profile edges as can be seen in GL98's 408 and 610-MHz profiles---but these cannot be measured with enough accuracy to include them.  We then use an inner conal {\textbf D} geometry, which interestingly seems to require an $\alpha$ value of about 90\degr.  In this context we cannot see how to interpret the KL99 103-MHz observation.No $t_{tscatt}$ value is available. 
\vskip 0.09in
\noindent\textit{\textbf{B0727--18}}: Some of the GL98 profiles suggest a triple form, and the \cite{JKMG2008} profile shows what seems to be a core feature as well as perhaps inner conal features.  The various profiles differ enough between them as to suggest moding.  We model the pulsar with an outer conal geometry.  No core width can be measured, but the computed with of some 3.7\degr\ seems plausible within the profiles that best show it.  \citet{Weltevrede2007} find that the pulsar has a flat fluctuation spectrum.  No $t_{tscatt}$ value is available. 
\vskip 0.09in
\noindent\textit{\textbf{B0740--28}}: This is a difficult pulsar to model.  The GL98 profiles at 1 GHz and above are hardly compatible with more recent ones by \citet{vH_thesis} and \citet{kj06} that show much more structure.  \citet{Weltevrede2006,Weltevrede2007} report low frequency modulation, perhaps due to moding.  In addition, the PPA rate is inconsistent among the various observations; there is a steep rotation on the trailing edge of some profiles, but most show a shallower rate.  While there is little doubt that the pulsar's profiles are core-dominated---and there seems to be a triparite structure at high frequency---no consistent measurements are possible at lower frequencies, and our model reflects only the core width at best.  Scattering becomes prominent \citep{JKMG2008} by 100 MHz as measured by \citet{kuzmin_LL2007}.
\vskip 0.09in
\noindent\textit{\textbf{B0756--15}}: This pulsar has a single profile up to 5 GHz \citep{kkwj98}, so we model it as having a conal single {\textbf S$_d$} geometry.  The PPA sweep rate is poorly determined by GL98's observations and thus limits the accuracy of the model. \citet{Weltevrede2006,Weltevrede2007} find no clear signature of a conal modulation; whereas \citet{basu2016} find a 50-60 $P$ amplitude modulation.  No $t_{tscatt}$ value is available. 
\vskip 0.09in
\noindent\textit{\textbf{B0809+74}}:  The pulsar is well known for its prominent 11-$P$ drifting subpulses \citep{Weltevrede2006,Weltevrede2007}, memory across nulls, and thus conal single {\textbf S$_d$} profile \citep[and the references there cited] {Lyne1983}.  The pulsar's width behavior has been much studied, being a first example of a partial decrease with wavelength dubbed ``absorption'' \citep{Bartel1981}.  We follow ET VI and ET IX as well as the geometry model in \cite{Rankin2006}.   Scattering is negligible to 25 MHz or so \citep{Zakharenko2013,kuzmin_LL2007}.
\vskip 0.09in
\noindent\textit{\textbf{B0818--13}}: \citet{Lyne1983} studied this conal single {\textbf S$_d$} drifter \citep{Weltevrede2006,Weltevrede2007,basu2016}, and we follow ET VI in modeling its geometrical behavior. We used an outer cone because the LOFAR width was greater; however, the scattering-corrected KL99 103-MHz profile seems in line with those at high frequency.  Scattering is expected to be small at 103 MHz \citep{kuzmin_LL2007}.
\vskip 0.09in
\noindent\textit{\textbf{B0826--34}}:  The {\textbf M} configuration of ET VI is used here with some corrections; however, only the 606-MHz GL98 profile and the \cite{BiggsMHML1985} 610- and 408-MHz profiles are interpretable.  The PPA rate is shallower than that used above.  The pulsar seems to be a single-pole interpulsar such that the leading component in the GL 606-MHz profile can be interpreted as a core component lying half a period away from the center of the following filled conal double structure.  Complex drifting is seen in several regions within this latter region leaving little question about its conal character.  No scattering value is available.  
\vskip 0.09in
\noindent\textit{\textbf{B0841+80}}: A conal quadruple structure can clearly be seen from the 149-MHz LOFAR observation.   The GL 606-MHz detection is poor and cannot be interpreted reliably.  As no polarimetry is available, a geometric model is attempted using a central sightline traverse.  No scattering measurement is reported. 
\vskip 0.09in
\noindent\textit{\textbf{B0844--35}}: The \cite{WangMJ2007} study shows that this pulsar exhibits ``swooshes'' similar to those of B0919+06 and B1859+07 \citep[see][and its citations]{wahl} that distort a probable core-cone triple {\textbf S$_t$} profile.  The three GL98 profiles together with the PPA rate confirmation from \citet{JKMG2008} then provide a rough geometry.  \citet{basu2016} find a nearly odd-even modulation in the trailing component.  A core component seems indicated by the antisymmetric $V$ seen in most profiles \citep[\eg][]{TvO,QMLG95}; though its width cannot be well estimated in any profile, a 7\degr\ value for an inner cone seems more plausible than the somewhat narrower width for the alternative.  \citet{kmn+15} find significant scattering at 327 MHz. 
\vskip 0.09in
\noindent\textit{\textbf{B0853--33}}: The GL98 and \citep{TvO} profiles all have an unresolved double form with about the same widths, and the latter indicates a PPA rate of about +12\degr/\degr.  We model its beam with an inner conal double {\textbf D} geometry, and \citet{kmn+15} show that the profile becomes ``scattered out'' at lower frequencies.
\vskip 0.09in
\noindent\textit{\textbf{B0906--17}}: This pulsar has a core component and apparent conal outriders making its profile either core-single {\textbf S$_t$} or triple {\textbf T} \citep[see also][]{JKMG2008}.  \citet{Weltevrede2006,Weltevrede2007} find a flat fluctuation spectrum at both frequencies.  The conal outriders are only discernible at the highest frequencies.  \citet{cor86} report a scattering time. 
\vskip 0.09in
\noindent\textit{\textbf{B0917+63}}: PSR B0917+63 has a classic conal double {\textbf D} profile that can be traced down to 65 MHz.  We model it with an outer conal geometry, but little width increase is observed.  
\vskip 0.09in
\noindent\textit{\textbf{B0942--13}}: ET VI regarded this pulsar at having a triple structure, but we now believe this was incorrect.  \citet{Weltevrede2007} find a 3-$P$ drift feature, and so an inner conal single {\textbf S$_d$} geometry appears much more appropriate.  The PRAO 103-MHz point may signal a width increase or may simply reflect scattering.  No $t_{\rm scatt}$ is available. 
\vskip 0.09in
\noindent\textit{\textbf{B1010--23}}: The GL 610- and 410-MHz profiles show a similar structure and width as does their 234-MHz plot along with what many be a scattering tail.  The PPA rate is hardly guessible, but we take it as +6\degr/\degr.  No other evidence exists to confirm this being an conal single {\textbf S$_d$} structure.  No $t_{\rm scatt}$ is available. 
\vskip 0.09in
\noindent\textit{\textbf{B1016--16}}: A single profile with a leading ``bump'' indicating an unresolved component is seen in this pulsar's profile.  Many conal single profiles have this sort of form, so we model it with an inner conal {\textbf S$_d$} geometry.  No $t_{\rm scatt}$ is available. 
\vskip 0.09in
\noindent\textit{\textbf{B1039--19}}: We support that this pulsar has a five-component {\textbf M} geometry.  The central core component is barely discernible only at 4.85 GHz \citep{vH_thesis} where the inner conal components can also be discerned in the linear polarization; see also \citet{vH_thesis}.   \citet{Weltevrede2006,Weltevrede2007} find clear drift bands under the outer components at both frequencies.  No obvious scattering effects down to 234 MHz and no measurement reported.  
\vskip 0.09in
\noindent\textit{\textbf{B1112+50}}: This pulsar exhibits moding with a bright and weak leading component in the two modes \citep{wright} and clear drifting \citep{Weltevrede2006,Weltevrede2007} in one of them.  We have thus modeled the profiles with an inner conal single {\textbf S$_d$} geometry.  We include the decametric profiles of \cite{Zakharenko2013}, and \citet{cor86} report a scattering timescale.
\vskip 0.09in 
\noindent\textit{\textbf{B1254--10}}: The pulsar seems to have a triple {\textbf T} profile with a weak trailing conal component that is conflated with the core.  \citet{Weltevrede2006,Weltevrede2007} find no evidence of a conal modulation feature.  GL98's 1.7-GHz profile is poor, and the MM10 111-MHz width value is a guess.  No $t_{\rm scatt}$ reported. 
\vskip 0.09in
\noindent\textit{\textbf{B1309--12}}: The five GL98 profiles give all the available information.  All are unimodal and show little width increase with wavelength.  The $L/I$ is low and the PPA rate is shallow and inaccurately measured.  We thus model the pulsar using an inner conal single beam.  No scattering measurement has been reported.
\vskip 0.09in
\noindent\textit{\textbf{B1322+83}}: This pulsar has an unusually broad profile with a very weak leading feature.  A flat fluctuation spectrum was reported by \citet{Weltevrede2007} at 327 MHz pertaining only the trailing component.  Interestingly, the broad weak preceding feature can be discerned at LOFAR frequencies, shows RFI, and may also be lightly linearly polarized.  We model the pulsar using an outer conal double {\textbf D} geometry.  The pulsar is detected at both 60 \citep{bilous2019} and 25 MHz  \citep{Zakharenko2013}, but only the trailing bright component, so is not useful for our purposes here.  No $t_{\rm scatt}$ is reported. 
\vskip 0.09in
\noindent\textit{\textbf{B1508+55}}:  As discussed in ET VI and \citet{Force2015}, the pulsar has a closely spaced triple profile that we model using a core/inner conal {\textbf T} beam geometry.  \citet{Weltevrede2006,Weltevrede2007} find a complex drifting subpulse modulation in the leading and trailing conal components.  The pulsar is well detected at 25 and 20 MHz by \cite{Zakharenko2013}, but substantial scattering \citep{kuzmin_LL2007} makes its interpretation impossible for our purposes.
\vskip 0.09in
\noindent\textit{\textbf{B1540--06}}: As with many {\textbf S$_d$} pulsars, this pulsar's profile seems to have two unresolved components, the trailing one weaker than the leading one, giving it an asymmetric shape that resembles scattering but is broadband.  We measure the halfwidths because we have no other recourse.  \citet{Weltevrede2006,Weltevrede2007} find a 3-$P$ drift modulation that provides most of the profile power.  A scattered profile is detected at 25 MHz by \citet{Zakharenko2013} \citep[see][]{cor86}.  
\vskip 0.09in
\noindent\textit{\textbf{B1552--31}}: The five GL98 and the JK18 profiles show a conal double structure and a filled ``boxy'' profile at 1.6 GHz as seen in many {\textbf M} profiles.  Moreover, both the 1408- and 925-MHz profiles suggest five components.  We take a central traverse per JK18 and thus model the profiles with a conal quadruple {c\textbf Q} beam structure (the inner conal component dimensions are rough estimates).  \citet{basu2016} find a 10-$P$ drift modulation.  A weak core of the needed 4.3\degr\ width would fit comfortably between the inner conal components.  No scattering time value is available.  
\vskip 0.09in
\noindent\textit{\textbf{B1552--23}}: The five GL98 profiles seem to show a triple structure throughout.  The PPA rate can only be guessed at, and measurements of the poorer profiles must be guided by the 1408- and 610-MHz forms.  We thus provide a rough core-inner cone triple {\textbf T} model.  No scattering is visible in the profiles nor measurement available. 
\vskip 0.09in
\noindent\textit{\textbf{B1600--27}}: Again the \cite{jk18} profile provides the high quality key to interpreting the five GL98 observations.  It shows a strong leading and weak trailing conal component around a bright central core feature---whereas, none of the latter show this clearly.  \citet{Weltevrede2007} find a strong low frequency excess but no clear drifting in the fluctuation spectra.  We then model the geometry using a core-inner conal triple {\textbf T}  configuration, and we measure the GL98 profiles guided by the foregoing structure.  No scattering measurement seems to be available.  
\vskip 0.09in
\noindent\textit{\textbf{B1607--13}}:  \citet{Weltevrede2007} detect drifting subpulses clearly.  The GL98 610- and \citet{SGG+95} 1420-MHz profiles seem to have three parts, although the GL98 1408- and 925-MHz profiles are too poor to confirm this.  We therefore model the geometry using a conal triple {c\textbf T} configuration.  Only the 610- and 410-MHz profiles can be measured with any accuracy, and the latter seems to provide a reliable PPA rate.  No scattering time measurement is available.  
\vskip 0.09in
\noindent\textit{\textbf{B1612--29}}: The 606- and 410-MHz GL profiles show a partially resolved double form and poor total power 1.4 GHz profiles \citep{D'Amico+98} may have a similar form.  All have widths of about 8\degr, and the PPA rate can only be guessed at, perhaps +6\degr/\degr.  We used an outer conal single model, though an inner one is also clearly possible.  No $t_{\rm scatt}$ is reported. 
\vskip 0.09in
\noindent\textit{\textbf{B1620--09}}: This pulsar shows pretty strong evidence of having a conal single {\textbf S$_d$} geometry.  \citet{Weltevrede2007} find a low frequency fluctuation feature, and the single form persists to 5 GHz \citep{kkwj98}.  The poor PPA rate value limits the accuracy of the model.  The nearly constant width of the profile at 100 MHz shows that scattering here is weak; no $t_{\rm scatt}$ is reported.  
\vskip 0.09in
\noindent\textit{\textbf{B1620--26}}: This 11-ms MSP seems to have a usual core-cone triple {\textbf T} profile.  The PPA rate is very shallow, and the --2/\degr/\degr\ value is only a gross estimate compatible with an inner cone.  The 408-MHz profile may be broadened by scattering, but no timescale measurement as been reported.
\vskip 0.09in 
\noindent\textit{\textbf{B1642--03}}: The pulsar has a well-studied core-single {\textbf S$_t$} profile as discussed in ET VI. Its emission comes in bursts per \citet{Weltevrede2006, Weltevrede2007} with some subpulse organization, but differently at the two frequencies where conal ``outriders'' are present around and above 1 GHz and not at 300 MHz \citep[see also][]{JKMG2008}.  More study is needed to understand their significance.  As with a number of GL98's 234-MHz profiles, this one is poorly resolved. KLL07 report a $t_{\rm scatt}$ value.
\vskip 0.09in
\noindent\textit{\textbf{B1648--17}}: The pulsar shows a conal double {\textbf D} profile, and we so model it.  \citet{Weltevrede2007} find no fluctuation feature, no scattering timescale is reported.  
\vskip 0.09in
\noindent\textit{\textbf{B1649--23}}: The GL98 profiles show two clear components along with hints of a weak leading one.  However, we see no clear confirmation of this in the high quality \citet{jk18} profile, so we model the beam structure with a conal double {\textbf D} model.  \citet{Weltevrede2006, Weltevrede2007} find fluctuations that may indicate conal emission.   No scattering timescale measurement.  
\vskip 0.09in
\noindent\textit{\textbf{B1657--13}}: This pulsar seems to have a steep spectrum given GL98's poor quality detection at 925 MHz, and MM10's 100-MHz detection seems reliable.  We model it as a conal {c\textbf T} triple.  At risk of overinterpretation, we use both widths in MM10's profile, which square well with the higher frequency values, suggesting little scattering on this pulsar's path. The poor fluctation spectrum of \citet{Weltevrede2007} found no modulation feature.  No $t_{\rm scatt}$ is reported.
\vskip 0.09in
\noindent\textit{\textbf{B1700--32}}: We follow ET VI; the pulsar has a discernible triple form up to about 1 GHz and a squarish profile at higher frequencies \citep{JKMG2008}.  \citet{basu2016} find a 5-$P$ modulation in the outer components and a different one in the central feature.  The width increases at lower frequencies, so we take the conal beam as an outer one.  The central component seems to be a core, given that it evolves similarly to other {\textbf T} objects, but we find no published fluctuation-spectral analysis to support this.  The core width cannot be estimated accurately from any profile, but some 3\degr\ seems a plausible guess at 610 MHz.  Evidence of a scattering ``tail'' is seen at 243 MHz, which seems compatible with the \cite{kmn+15} $t_{\rm scatt}$ value. 
\vskip 0.09in
\noindent\textit{\textbf{B1700--18}}: This pulsar was classified as a conal single {\textbf S$_d$} pulsar in ET IX, and so we model it here. Drifting subpulses were detected by \citet{Weltevrede2007} both by eye and in their fluctuation spectra with a $P_2$ of 3.7 $P$.  No $t_{\rm scatt}$ is reported.
\vskip 0.09in
\noindent\textit{\textbf{B1702--19}}: PSR B1702--19 is a well known interpulsar with cores in both its main and interpulses as discussed in ET VI \citep[see also][]{JKMG2008}.  It also shows a prominent 11-$P$ non-drift modulation \cite[see][and the references there cited]{Weltevrede2006, Weltevrede2007}.  The main pulse changes little with frequency, and the leading conal component shows only as an inflection on the leading edge of the core.  The width of MM10's 100-MHz profile suggests that scattering is minimal in this direction. No $t_{\rm scatt}$ is reported.
\vskip 0.09in
\noindent\textit{\textbf{B1706--16}}: This pulsar was classified as as a core-cone triple {\textbf T} in ET VI, and we so regard it here as well.  The conal ``outriders'' are only seen as inflections across the spectrum, so are not readily or accurately measured \citep[see also][]{JKMG2008}.  Its fluctuation spectra show a strong low frequency excess, probably indicating sporadic emission \citep{Weltevrede2006, Weltevrede2007}.  The core width shows an unusual narrowing behavior, perhaps due to incompleteness at some frequencies. KLL07 report a $t_{\rm scatt}$ value.
\vskip 0.09in
\noindent\textit{\textbf{B1709--15}}: This pulsar has a conal single {\textbf S$_t$} profile configuration, but there is so little polarization that no reliable PPA sweep rate can be determined.  We model it as an inner cone and guestimate a small negative PPA rate.  The well resolved MM10 100-MHz profiles suggests that scattering is unimportant in this direction.  \cite{Weltevrede2007} find no significant modulation, but the sensitivity was poor.  No $t_{\rm scatt}$ is reported.
\vskip 0.09in
\noindent\textit{\textbf{B1714--34}}: The \citet{jk18} profile shows a clear scattering ``tail'' at 1.4 GHz, whereas this is clear only in the GL98 606-MHz profile---so no beam model is possible on this basis.  Intrinsically, this is probably another core-single profile.  \citet{okp+21}measure a very large scattering timescale.  
\vskip 0.09in
\noindent\textit{\textbf{B1717--16}}: Two conflated conal components over the entire frequency range with negligible scattering in this pulsar.  We thus model it with a conal single {\textbf S$_d$} geometry.  \cite{Weltevrede2007} find a flat fluctuation spectrum.No $t_{\rm scatt}$ is reported.
\vskip 0.09in
\noindent\textit{\textbf{B1717--29}}: Both the GL98 and \citet{jk18} profiles show a clear four-component structure representing a conal quadruple {c\textbf Q} beam geometry.  This is confirmed by \cite{Weltevrede2006, Weltevrede2007} who find a coherent 2.5-$P$ modulation at both frequencies as well as by \citet{basu2016} who also see what may be a harmonic.  No scattering measurement available. 
\vskip 0.09in
\noindent\textit{\textbf{B1718--02}}: As suggested by \cite{Force2015} this pulsar has a broad profile with a conal single evolution at high frequency, bifurcating by 400 MHz as is very clear in the 327-MHz \cite{Weltevrede2007} observation---which also shows a strong 5.4-$P$ stationary modulation.  A surprise is MM10's wide triple {c\textbf T} profile at 100 MHz; its central component cannot be measured accurately, but were it some 13\degr\ as seems possible, it would reflect the inner cone width.  No $t_{\rm scatt}$ is reported.
\vskip 0.09in
\noindent\textit{\textbf{B1718--19}}: In addition to the published GL98 profiles at 606 and 408 MHz, another at 1.4 GHz is found on the EPN Database; all are poor but the latter gives a PPA rate.  We model it with an inner conal single beam.  No scattering timescale been measured.  No $t_{\rm scatt}$ is reported.
\vskip 0.09in
\noindent\textit{\textbf{B1718--35}}: GL98, \citet{QMLG95} and especially \citet{joh90} show a beautifully scattered profile at 1.4 GHz.  However, the \citet{JKW06} 8.4-GHz observation shows what may be a core feature with a negative PPA traverse. It thus seems very probable that this is another core-single profile intrinsically.  \citet{okp+21} measure a very large scattering timescale.  
\vskip 0.09in
\noindent\textit{\textbf{B1718--32}}: Several of the five GL98 profiles suggest triplicity, and we attempted to model them with a conal triple configuration.  However, the well defined PPA rate \citep{jk18} is incompatible with this geometry, so we have modeled the profile using a core-cone triple {\textbf T} geometry successfully.  \citet{basu2016} find a rough 50-$P$ modulation in the first component.  Scattering with the \citet{kmn+15} value may be compatible with the asymmetry of the 410-MHz profile.  
\vskip 0.09in
\noindent\textit{\textbf{B1727-33}}:  Both the well measured \citet{jk18} and the single 1.4-GHz GL98 profiles show a clear scattering ``tail'' probably compatible with the measurement of \citet{joh90}.  The 8.4-GHz profile \citep{JKW06} shows a narrow single feature which may represent a core beam.  \citet{okp+21} measure a very large scattering timescale.  
\vskip 0.09in
\noindent\textit{\textbf{B1730-22}}:  \cite{Weltevrede2007} find a drift feature in this pulsar, and \citet{basu2016} a 50-$P$ mostly amplitude modulation in the first component.  Its profile suggests a conal triple or quadruple form---more the latter at higher frequencies and the former in the meter wavelength profiles.  We model the profile as an outer conal triple, but no reliable dimension for the inner features is possible given the overall quality of the profiles.  Nor is the PPA rate well determined, and appears to change sign in both the GL98 and \cite{JKMG2008} profiles.  No obvious evidence of scattering is seen down to 243 MHz, and no published value is reported.  
\vskip 0.09in
\noindent\textit{\textbf{B1732--02}}: This pulsar seems to have an asymmetric inner conal single {\textbf S$_d$} profile, but no modulation feature is identified in the \citet{Weltevrede2007} weak fluctuation spectra.  MM10's asymmetric profile is dominated by scattering, but no measurement is reported.
\vskip 0.09in
\noindent\textit{\textbf{B1732-07}}:  Earlier work strongly suggested a core-cone triple {\textbf T} geometry for this pulsar \citep{mitra2011,Force2015}, now strengthened by a flat fluctuation spectrum at 21 cms \citep{Weltevrede2006} and improved observations over a broader band.  \citet{basu2016} see an interesting 50-$P$ mostly amplitude modulation in the core feature at meter wavelengths.  Especially in the \cite{JKMG2008} profiles we see evidence of a very steep PPA traverse that we model as infinite---and a scattering ``tail is seen at 243 MHz---but no measurement is available.  
\vskip 0.09in
\noindent\textit{\textbf{B1734--35}}:  Probably a core-single profile; however, scattering seems to be present in both of GL98's 1408- and 606-MHz profile.  No higher frequency measurement nor scattering value seems to be available.  
\vskip 0.09in
\noindent\textit{\textbf{B1735--32}}: The five GL98 profiles seem to show a conal structure.  There are hints that it might be a conal triple one, but the quantitative analysis does not support it.  So we model the profile with an inner conal single {\textbf S$_d$} configuration.  We see no scattering ``tail'' at 410 MHz, and no measurement is available.  
\vskip 0.09in
\noindent\textit{\textbf{B1736--31}}: Probably a core-single beam structure.  However, no model is possible because a scattering ``tail'' may even be present at 1.6 GHz.  The \cite{okp+21} measurement indicated that this is one of the most scattered objects in our population.  No scattering measurement is available.
\vskip 0.09in
\noindent\textit{\textbf{B1736--29}}: Only the \cite{jk18} observation at 1.4 GHz adds to the discussion in ET VI.  While the GL98 profiles show little or no structure in either the main pulse (MP) or interpulse (IP), the newer observation clearly shows a profile of parts in both the total power and PPA---and a central PPA rate that may be about +8\degr/\degr.  If the bright 1.4-GHz MP component is a core as seems likely, its width can be estimated by doubling that from its peak to trailing 3db point---and this gives a value very close to that of the polar cap size, strengthening this interpretation.  Thus $\alpha$ is very close to 90\degr, indicating a two-pole interpulsar.  We then interpret the full widths of the GL 98 profiles as if they were conal component pairs, and they model an inner cone-core triple  {\textbf T} configuration.  Less can be said about the IP apart from its width being close to that of the MP.  \citet{Weltevrede2006} find a flat fluctuation spectrum as expected.  No $t_{\rm scatt}$ is available.
\vskip 0.09in
\noindent\textit{\textbf{B1737--30}}:  Apparently a well identified core-single {\textbf S$_t$} profile with no obvious conal ``outriders'' at 1.6 GHz.  The PPA traverse, however, is well defined suggesting that some conal radiation is admixed with that of the core.  A scattering ``tail'' is seen in both the 610- and 408-MHz profiles that seems compatible with the \cite{kmn+15} value.  Additionally, \citet{JKMG2008} failed to see the pulsar at both 243 and 327 MHz, probably due to scattering \citep{kmn+15}.  
\vskip 0.09in
\noindent\textit{\textbf{B1738--08}}:  The GL 98 profiles together with the Lyne \& Manchester 409-MHz and \citet{jk18} 1.4-GHz observations show what is clearly a double cone {c\textbf Q} structure, and this interpretation is confirmed by the 5-$P$ drifting subpulses found by \citet{Weltevrede2006,Weltevrede2007} and \citet{basu2016}.  As usual the inner conal dimensions are difficult to measure accurately so are estimated.  \citet{kmn+15} measure a scattering timescale.  
\vskip 0.09in
\noindent\textit{\textbf{B1740--13}} A triple profile seems to be present in this pulsar, which is clearest in GL98's 1408-MHz profile, though others have a triangular form with a bright central putative core component. None of the conal widths can be determined accurately, but the rough model with a {\textbf T} geometry seems adequate.  The MM10 profile shows no obvious scattering \citep{kmn+15} , and its width is difficult to interpret in a manner compatible with the higher frequency profiles---so its smaller width is probably a ``measurement error''.
\vskip 0.09in
\noindent\textit{\textbf{B1740--31}}: Three GL98 profiles are supplemented by a well measured 1.4-GHz \citet{jk18} one, where the PPA rate is very well defined.  The emission could be either core or conal, but we tilt toward conal given the prominent edge depolarization in the latter profile---and either an inner or outer conal geometry is possible, but given the small width increase that might be due to scattering, our model uses an inner one.  There is also a possible weak trailing component in several of the high frequency profiles, but apparently the highest quality \citet{jk18} profile did not find it.  The strange 410-MHz profile could then reflect a combination of interference and scattering \citep{kmn+15}.  
\vskip 0.09in
\noindent\textit{\textbf{B1740--03}}: PSR B1740--03 shows a compatible triple form in the three GL98 profiles.  We model it with a triple {\textbf T} geometry, and the MM10 profile seems to have a compatible form and width.  No scattering value has been reported.
\vskip 0.09in
\noindent\textit{\textbf{B1742-30}}: We follow the analysis of \citet{mitra2011} who corrected the errant interpretation in ET VI.  They clearly identified the weak trailing emission that suggests a core/double conical geometry.  Despite the several 90\degr\ ``jumps'' the pulsar seems to have a peripheral PPA traverse of some --3.6\degr/\degr.  Unusually, the core appears to be the entire central unresolved double structure with a width of about 10\degr, their relative amplitudes varying over the observing bands but maintaining about the same overall width.  We model the pulsar here as a core/cone triple given that the inner conal features can only be discerned via a single-pulse analysis.  A scattering ``tail'' is seen at 243 MHz \citep{JKMG2008}, and $t_{\rm scatt}$ has been determined by \cite{kmn+15}.  
\vskip 0.09in
\noindent\textit{\textbf{B1745--12}}: Both ET VI and ET IX discussed this pulsar as having multiple parts to its profile, however no core component is seen. \citet{basu2016} find evidence of weak conal modulation though \citep{Weltevrede2006,Weltevrede2007} did not. We model it as having a four-part conal quadruple c{\textbf Q} configuration, and its profile forms seem compatible with other such pulsars that often seem both to show stronger leading components and modal activity wherein both halves of the profile are illuminated comparably.  As usual the inner conal dimensions are estimates.  KLL07 report a $t_{\rm scatt}$ value.
\vskip 0.09in
\noindent\textit{\textbf{B1746--30}}:  GL98 provide 1.408- and 606-MHz profiles, and \citet{jk18} another high quality 1.4-GHz observation.  The former show a triple or probably five component structure with a well defined PPA rate.  We model it with an {\textbf M} core-double cone configuration very satisfactorily.  The 606-MHz profile appears to be dominated by scattering, perhaps compatible with the \cite{joh90} measurement.  We use its width as an outer conal value, but it could well be the scattered core as well; we have no basis for deciding.  
\vskip 0.09in
\noindent\textit{\textbf{B1747--31}}:  We are dependent on the GL98 profiles at 1408 and 606 MHz as well as the more recent \citet{jk18} 1.4-GHz observation that suggests a steep PPA traverse.  The 1408-MHz profile clearly has three components, and we so model it as a core-cone {\textbf T} configuration, though a conal triple structure is also possible.  In any case the width of the middle component can only be estimated, and if a core could accommodate either an inner or outer conal geometry, though we have tilted toward an outer one. No $t_{\rm scatt}$ is available.
\vskip 0.09in
\noindent\textit{\textbf{B1749--28}}:  ET VI identified this pulsar as having an {\textbf S$_t$} beam geometry.  Conal outriders are barely perceptible at 1.7 GHz but are clearly present at higher frequencies---\eg 3.1 GHz \citep{kj06}, so we can be sure about the core/inner conal structure.  The core is traceable down to the LOFAR 149-MHz observation, and its width shows an increase perhaps due to scattering, though \cite{abs86} measure a very small timescale.
\vskip 0.09in
\noindent\textit{\textbf{B1750--24}}:  Only two high frequency profiles are available from GL98, and it is difficult to discern from their form whether they represent a very broad profile with weak emission following an early peak or rather a long scattering ``tail''.  A beam model is impossible from this information, but given the pulsar's large {$\dot E$} it is likely a core-dominated beam system. Interestingly, the \citet{kkwj98} 4.85-GHz profile is double, and could represent the flatter spectrum of conal ``outriders'' relative to the core.  \citet{ldk+13} measured a large $t_{\rm scatt}$ value that may be compatible.  
\vskip 0.09in
\noindent\textit{\textbf{B1753+52}}: The pulsar shows evidence of multiple highly conflated components at high frequency with mostly the outer conal components surviving at 149 MHz. \citet{Weltevrede2006, Weltevrede2007} find a 7-$P$ largely stationary modulation, mostly in the trailing components that can be seen by eye as well as in the fluctuation spectra.  We model the profile as having a conal quadruple c{\textbf Q} beam system, but in practice only the outer conal components can be measured well enough to include.  No $t_{\rm scatt}$ is available.
\vskip 0.09in
\noindent\textit{\textbf{B1753--24}}: Three components are seen in the GL98 1.4-GHz profile and perhaps compatible structures in the 1.6 GHz and 925-MHz profiles.  No fluctuation spectrum is available for possible confirmation, but this seems to be a conal triple c{\textbf T} beam configuration.  The inner conal dimensions can only be estimated above 1 GHz, and the much broader 408-MHz profile seems to have a scattering ``tail'' compatible with the large scattering timescale measured by \citet{wpc+90}.  
\vskip 0.09in
\noindent\textit{\textbf{B1754--24}}: A clear triple structure is seen in the GL98, JK18 and \citet{Weltevrede2006} 1.4-GHz profiles.  The latter paper finds no fluctuation feature but a low frequency excess.  The PPAs suggest a central traverse rather than a slight rate, and we take $\beta$ as zero in the triple {\textbf T} beam model.  A core width of about 7\degr\ can only be estimated at 1.4 GHz.  Scattering appears to set in at 408 MHz in a manner that is probably compatible with the \citet{rmd+97} measurement.  
\vskip 0.09in 
\noindent\textit{\textbf{B1756--22}}: The GL98 profiles show single forms across the band of observation with strong hints of a conal ``outrider'' pair in their wings.  The PPA rate is poorly defined, but some profiles show a steepening on the trailing edge to perhaps --15\degr/\degr.  \citet{Weltevrede2006,Weltevrede2007} find flat fluctuation spectra, so we follow ET IX in modeling the profile using an inner cone triple \textbf {S$_t$/T} model.  We use an ``outrider'' pair width of about 10\degr\ as in no profile can it be determined accurately.  The 408-MHz profile may be broadened by scattering, and \cite{kmn+15} measure a $t_{\rm scatt}$.
\vskip 0.09in
\noindent\textit{\textbf{B1757--24 (B1758--24 in GL98)}}:  \cite{jk18} and \citet{GL98} show 1.4-GHz profiles with minimal scattering that probably represent a core single structure.  The large $L/I$ gives a well determined PPA rate, but we see no indication of conal outriders.  The GL98 606-MHz profile though is highly scattered with no measured value seemingly available.  
\vskip 0.09in
\noindent\textit{\textbf{B1758--23}} is one the most highly dispersed and scattered objects in our population.  Both the GL98 1.6- and 1-4-GHz profiles have long scattering ``tails''.  No beam model is therefore possible from them; however, the \citet{kkwj98} 4.85-GHz profile appears to show an unscattered core component (as well as a possible leading conal ``outrider''), but the profile is not of sufficient quality to fully confirm the classification.  \cite{okp+21} measure a very large scattering timescale.    
\vskip 0.09in
\noindent\textit{\textbf{B1758--03}}: The pulsar seems to have a fairly usual core-single {\textbf S$_t$} or triple {\textbf T} geometry---the issue being that the conal outriders are apparent down to perhaps 400 MHz, though the core seems to dominate at very low frequency.  \citet{Weltevrede2007} find no drift modulation, but they do find burst-like emission that probably includes nulling. \citet{basu2016} see a strong slow modulation at perhaps 50-$P$.  \citet{kuzmin_LL2007} measure a $t_{\rm scatt}$ value.
\vskip 0.09in
\noindent\textit{\textbf{B1800--21}}: This fast energetic pulsar seems to have a triple {\textbf T} profile where the sightline misses most of the core emission.  A weak conflated core is seen in only some of the profiles \citep{GL98,WMLQ} with a plausible width of about 30\degr.  The profile is very broad, and the PPA rate is about 2\degr/\degr, but difficult to determine accurately.  For an inner/outer sightline traverse, an inner/outer cone is indicated.  We model the profile using the latter as the profile broadens appreciably below 1 GHz.  See also \citet{vH_thesis}. \citet{Weltevrede2006} find only ``red'' noise in the fluctuation spectra, and \cite{ldk+13} measure a scattering time.
\vskip 0.09in
\noindent\textit{\textbf{B1802--07}}:  Of the three GL98 profiles of this MSP, only the 1.4 GHz is well resolved, and no reliable PPA rate can be estimated.  Maybe we are seeing a core component at 1.4 GHz, and maybe the much larger width at 1.6 GHz is the effect of conal ``outriders''.  No beam model is possible with this limited information. No $t_{\rm scatt}$ is available.
\vskip 0.09in
\noindent\textit{\textbf{B1804--27}}:  The GL98 profiles show no reliable structure nor PPA rate.  The profile may represent a core-single configuration, and we model it as such.  \cite{kmn+15} give a $t_{\rm scatt}$ value.
\vskip 0.09in
\noindent\textit{\textbf{B1804--08}}: Following the ET VI discussion, this pulsar shows a very clear core-cone triple profile as well as a weak set of outer conal components on the flanks of the high frequency profiles.  \citet{Weltevrede2006, Weltevrede2007} find no evidence of a drift modulation.  Here, we model the inner conal geometry and add the outer conal features as possible though they are difficult to distinguish at low frequency.  Scattering greatly broadens the 102-MHz profile \citep{kmn+15}.
\vskip 0.09in
\noindent\textit{\textbf{B1805--20}}:  Only the 1.6- and 1.4-GHz GL98 profiles are little affected by scattering.  They show a double structure that may be conal.  \citet{Weltevrede2006} find a flat fluctuation spectrum, but perhaps without very much sensitivity.  Further, the PPA rate is difficult to determine, and the tracks probably incur a 90\degr\ ``jump'' on the leading edge of the profile, often a conal feature.  Visible scattering sets in below 1 GHz, and \citet{lkm+01} have measured a timescale.  
\vskip 0.09in
\noindent\textit{\textbf{B1806--21}}:  The four GL98 profiles, all single, are all that are available---and the PPA rate could only be guessed at.  Given their similar widths and small $L$/$I$, we tilt toward seeing the pulsar as having a core-single geometry, but a conal single configuration is also possible, perhaps answerable with fluctuation spectra. No $t_{\rm scatt}$ is available.
\vskip 0.09in
\noindent\textit{\textbf{B1809--173}}:  The two GL98 profiles at 1.6 and 1.4 GHz along with that of JK18 are single and seem incompatible with the conflated double description in \cite{WangMJ2007}, so there is insufficient information to even hazard a guess at the classification or beamform. \cite{okp+21} measure a $t_{\rm scatt}$ value. 
\vskip 0.09in
\noindent\textit{\textbf{B1809--176}}:  The two GL98 profiles at 1.6 and 1.4 GHz as well as that of JK18 have forms similar to many conal structures with a conflated brighter leading and weaker trailing component.  They are very broad and have very gradual PPA rates.  No reliable model is possible for such a pulsar where the sense of $\beta$ is very significant.  \citet{okp+21} do measure a scattering time.  
\vskip 0.09in
\noindent\textit{\textbf{B1811+40}}: The pulsar seems to have a c{\textbf T} or c{\textbf Q} geometry as three components are clearly seen at LOFAR frequencies and the higher quality GL98 profiles may also have a triple or maybe 4-component form.  The outer pair show little width increase down to the LOFAR band, and we interpret the trailing ``tail'' as incipient scattering. \citet{Weltevrede2007} identify a 2.3-$P$ modulation that is strong enough to be seen in the single-pulse sequence, and their 326-MHz profile provides something of a bridge between the GL98 and LOFAR regimes.  THen, the 102-MHz \citet{MIS89} observation bridges to the 60-MHz LOFAR ones (BKK++, BGT+). 
\vskip 0.09in
\noindent\textit{\textbf{B1813--17}}:  The 1.6 and 1.4-GHz GL98 profiles as well as that of JK18 are single, and together with the flat fluctuation spectrum \citep{Weltevrede2006} we can guess this pulsar has a core-single geometry.  The slightly large width at 1.6 GHz may suggest developing conal outriders, but no reliable estimate is possible either of the conal width or PPA rate.  \cite{okp+21} measure a $t_{\rm scatt}$ value. 
\vskip 0.09in
\noindent\textit{\textbf{B1813--26}}:  Both the GL98 and JK18 profiles show a conal double structure with the suggestion of inner conal emission, and \citet{basu2016} find a clear 4-$P$ ``drift'' modulation.  Thus the pulsar seems to have a conal triple or quadruple geometry, but the inner cone width can only be estimated at 1.6 GHz.  A $t_{\rm scatt}$ value is available from \cite{kmn+15}.
\vskip 0.09in
\noindent\textit{\textbf{B1813--36}}:  Both the GL98 610- and 925-MHz profiles show a conflated triple structure that is particularly clear in the JK18 1.4-GHz profile.  The published width at 925 MHz seems to provide a good core value, while the JK18 gives a PPA rate estimate.  We then model the beam geometry with a core-cone triple {\textbf T} model.  Otherwise, little to go on, and no scattering measurement.  
\vskip 0.09in
\noindent\textit{\textbf{B1815--14}}:  All the GL98 and JK18 profiles show scattering ``tails'', and \citet{Weltevrede2006} find a flat fluctuation spectrum.  Only the 4.9-GHz profile \citep{kkwj98} shows an intrinsic profile.  We model it with a core-single {\textbf S$_t$} configuration, and \cite{okp+21} measure a $t_{\rm scatt}$ value. 
\vskip 0.09in
\noindent\textit{\textbf{B1817--13}}:  All four GL98 profiles show scattering, so only the poor 4.9-GHz observation is largely free of it.  \citet{Weltevrede2006} find a flat fluctuation spectrum, so we tilt toward interpreting the profile as core emission, but a conal one is also possible.  The lower frequency detections are poor but may indicate rapidly increasing scattering that would be compatible with the huge \citet{okp+21} $t_{\rm scatt}$ value.
\vskip 0.09in
\noindent\textit{\textbf{B1817--18}}:  Profiles exist only at 1.4 and 1.6 GHz \citep{GL98,jk18}, and these profiles give only a guess at the PPA rate.  Nor do we have any fluctuation-spectral information.  However, it seems likely that this is a conal single {\textbf S$_d$} configuration and that the 1.6-GHz profile would be very similar to that at 1.4 GHz if better resolved.  \citet{joh90} measure a $t_{\rm scatt}$ value. 
\vskip 0.09in  
\noindent\textit{\textbf{B1818--04m}}:  ET VI identified this pulsar as having a core/cone triple profile, but the evidence now suggests that a core-single {\textbf S$_t$} beam geometry is more likely.  Three components are seen as structure in profiles at and above 1 GHz, but are not seen at meter wavelengths.  The 1.4- JK18 and 4.9-GHz \citep{kkwj98} show the usual evolution of the steeper core spectrum clearly.  However, the pulsar shows a clear scattering tail at 327 MHz \citep{Weltevrede2007}, so any structure could be smoothed out.  The above paper finds a 3-$P$ stationary modulation.  The LOFAR 149-MHz profile shows dominant scattering \citep{kuzmin_LL2007}, and the PRAO 103-MHz has a narrower width because of attempted scattering correction.  
\vskip 0.09in
\noindent\textit{\textbf{B1819--22}}: \citet{Weltevrede2006,Weltevrede2007} and \citet{basu2016} find a strong 17-$P$ drift modulation complete with clear drift bands at the higher frequency, and \citet{SSW09} study its complexity.  The JK18 profile shows a clear and linear PPA rate.  We thus model the pulsar using a conal single {\textbf S$_d$} beam geometry.  \citet{kkwj98} report a 4.9-GHz detection; \citet{JKMG2008} detected the pulsar at 327 MHz but not at 243 MHz, which may be compatible with the scattering measured by \citet{kmn+15}.  
\vskip 0.09in
\noindent\textit{\textbf{B1820--14}}:  We only have GL98's four profiles to go on, and none gives any reliable estimate of the PPA rate.  We thus model the pulsar using a core-single geometry, where the larger width at 1.6 GHz is probably due to the growth of conal ``outriders'' but this conal width cannot be estimated.  Scattering sets in below 1 GHz as measured by \cite{okp+21}.  
\vskip 0.09in
\noindent\textit{\textbf{B1820--11}}:  The broad profile shows hints of structure at 1.4 GHz, but little can be discerned clearly.  The profile might be entirely core or it might have a highly conflated core-cone triple form.  \citet{Weltevrede2006} find a flat fluctuation spectrum, but this does not resolve the possibilities here.  A well defined linear PPA traverse is seen, however, which may indicate conal contributions to the emission.  We thus model the emission beam with a core-cone triple structure, guessing that the core width could be as large as about 25\degr.  The width increases below 1 GHz are largely due to scattering as evidenced by the very large $t_{\rm scatt}$ value measured by \citet{lkm+01}.  
\vskip 0.09in
\noindent\textit{\textbf{B1820--30B}}:  Little can be done with the two 606- and 408-MHz GL98 profiles that only permit a guessimate of the PPA rate.  Nonetheless, the profile is probably conal, and we model it with an inner cone {\textbf S$_d$}  geometry.  No $t_{\rm scatt}$ is available.
\vskip 0.09in
\noindent\textit{\textbf{B1820--31}}:  All the available profiles \citep{GL98,WMLQ,jk18} are single, but with strong $L$ only in the leading part.   Some of the better profiles suggest a steepening of the PPA rate on the trailing edge as seen in pulsar B0540+23 and a number of others.  We model the profile with a core-single geometry, but with all the doubts that the other pulsars with similar profiles raise.  No fluctuation spectra nor $t_{\rm scatt}$ measurement is available.
\vskip 0.09in
\noindent\textit{\textbf{B1821--19}}:  The broad orderly PPA traverse at 1.4 GHz \citep[\eg][]{jk18} suggests weak conal ``outriders'' and the much wider (but poorly resolved) 1.6-GHz profile may support this conjecture.  \citet{Weltevrede2006,Weltevrede2007} find no conal modulation feature.  We thus model the profiles provisionally with a core-single geometry.  Scattering sets in rapidly below 1 GHz as is also clear from the large $t_{\rm scatt}$ value \citep{kmn+15}.
\vskip 0.09in
\noindent\textit{\textbf{B1821--11}}:  The GL98 profiles show narrow single profiles at 1.4 and 1.6 GHz with scattering setting in dramatically at lower frequencies. \citet{Weltevrede2006} find no fluctuation feature, so we model it with a core single geometry and no reliable PPA rate.  A very large $t_{\rm scatt}$ value was measured by \cite{okp+21}.
\vskip 0.09in
\noindent\textit{\textbf{B1822--14}}: JK18's well measured 1.4-GHz profile shows two features, a bright trailing and weak leading, separated by about 30\degr.  Little definite can be made of this profile in that none of the other \citep{GL98,JKW06} profiles show the leading feature.  Moreover, the 4.9-GHz profile is much narrower, arguing that all the lower frequencies are distorted by scattering.  One possibility is that the trailing feature has a core-single {\textbf S$_t$} beam geometry and that the weak leading one is a precursor.  \cite{Weltevrede2006} find a flat fluctuation spectrum.  \citet{okp+21} provide a scattering time measurement.
\vskip 0.09in
\noindent\textit{\textbf{B1822--09}}: This well studied pulsar has a confused history of interpretation because of its prominent precursor component (\eg ET VI) and interpulse.  We follow \cite{Backus2010} in showing that its main pulse has a core-single {\textbf S$_t$} beam configuration.  However, the conal parts of the profile are subtle and difficult to measure independently from the core, so we include only a 1.4- (see JK18) and 0.4 GHz value---although there are suggestions that the profiles may show three parts both at 4.9 GHz \citep{vonHoensbroech1997} and in the LOFAR 149-MHz profile \citep[see also][] {JKMG2008}.  \cite{Weltevrede2006, Weltevrede2007} detect the well known 11-$P$ modulation in the higher band but not the lower.  The KL99 102-MHz profile includes a correction for the  scattering that is seen clearly in both the 65- and 25-MHz \citep{Zakharenko2013} profiles as measured by \citet{cor86}.
\vskip 0.09in
\noindent\textit{\textbf{B1823--11}}: PSR B1823--11 seems to have an inner cone double {\textbf D} profile with a weaker trailing component conflated with the leading one---a very usual configuration.  The GL98 profiles seem to show a positive PPA traverse where the two leading and trailing parts of the profile have different OPMs, giving a roughly +7\degr/\degr rate.  The 408-MHz profile may have a scattering tail and the MM10 profile show a long scattering tail as probably compatible with the \citet{kmn+15} timescale.
\vskip 0.09in
\noindent\textit{\textbf{B1823--13}}:  This fast energetic pulsar has two well resolved components with substantial $L$, and the JK18 1.4-GHz PPA traverse is well defined but does not show whether it becomes steeper in the region between them.  For $\alpha$ values of 14-19\degr---and guessing at the PPA rate---a conal double configuration can be computed with a $\beta$ value of only several degrees.  Core emission would be expected for such an energetic pulsar, and if the foregoing interpretation is correct, the sightline would not miss it.  Therefore, we seem unable to understand how to interpret the geometry that this profile implies.  See also \citet{vH_thesis}.  \cite{ldk+13} measure a very large scattering time. 
\vskip 0.09in
\noindent\textit{\textbf{B1824--10}}:  Little can be done with the pulsar in terms of a reliable beam model, however, the 1.4-GHz profiles (GL98, JK18) do appear single and minimally scattered, so we take it as a core component.  The 1.6-GHz profile may be more complex but not in a manner we can interpret, and in any case the nearly complete depolarization leaves little indication of the PPA rate.  Both of GL98's lower frequency profiles are poor, and it unclear how their widths were measured.  The scattering is very large as measured by \cite{wpc+90}.  
\vskip 0.09in
\noindent\textit{\textbf{B1826--17}}:  The GL98 and JK18 profiles show a very clear core-cone triple {\textbf T} beam configuration with the usual softer core relative spectrum---dramatically shown in the \citet{Seiradakis1995} 4.9-GHz profile where the core is absent!  The sightline traverse appears to imply a negligible $\beta$.  \cite{Weltevrede2006, Weltevrede2007} detect no modultation in either band.  \citet{kmn+15} measure a large $t_{\rm scatt}$ value that seems compatible with the visible scattering ``tails'' below 1 GHz.  
\vskip 0.09in
\noindent\textit{\textbf{B1828--11 (née B1828--10)}}:  A very clear core single {\textbf S$_t$} profile with no apparent conal ``outriders'' at 1.6 GHz.  \cite{Weltevrede2006} find no fluctuation features, and \cite{ldk+13} measure significant scattering.  
\vskip 0.09in
\noindent\textit{\textbf{B1829--08}}:  The profiles seem to have three features, the first well resolved and the others conflated---best shown in JK18's 1.4-GHz profile.  Apparently the bright feature is a core component with a weak conal ``outrider'' on its trailing flank.  The PPA rate is difficult to discern but may be estimated from several profiles accounting for 90\degr\ ``jumps''.   \cite{Weltevrede2006} finds no fluctuation feature, so we model the geometry using a core-cone triple {\textbf T} beam configuration.  \citet{kmn+15} measure a large $t_{\rm scatt}$ value. 
\vskip 0.09in
\noindent\textit{\textbf{B1829--10}}:  Both the JK18 and GL98 1.4-GHz profiles probably show a core component, and the poorly resolved 1.6 GHz profile may be broadened by conal ``outriders''.  The lower frequency profiles are all scattered in line with the very large  $t_{\rm scatt}$ value determined by \citet{okp+21}.
\vskip 0.09in
\noindent\textit{\textbf{B1830--08}}:  GL98's 1408- and 925-MHz---and particularly the JK18 1.4-GHz---profiles show three components with the characteristic evolution of a core-single {\textbf S$_t$} beam geometry.  The 1.6-GHz profile is poorly resolved but much wider, probably due to more prominent conal ``outriders''.  \cite{Weltevrede2006} find no fluctuation feature.  Overall, the profile is very broad, and the modeling is very sensitive to the PPA rate value---which can only be estimated---as well as the sign of $\beta$.  Solutions can be found for either sense as well as both cone types, but we model it using an inner cone and --ve $\beta$.  The 610-MHz profile form is obliterated by scattering, and a large $t_{\rm scatt}$ value has been measured by \citet{ldk+13}.
\vskip 0.09in
\noindent\textit{\textbf{B1831--03}}: While the GL98 profiles only hint at a triple structure, the 1.4-GHz profiles of \cite{Seiradakis1995}, \cite{JKMG2008} and \citet{jk18} exhibit it clearly---and the latter shows a very steep PPA rate that is apparently unresolved in the GL98 profiles.  \cite{Weltevrede2006,Weltevrede2007} detect no modulation in either band. Therefore, we model the profile with a cone/cone {\textbf T} triple structure.  The GL98 408-MHz profile shows substantial scattering, and the narrow peak in MM10's profile seems to be the residual of a ``scattered out'' response \citep[see][]{kmn+15}.
\vskip 0.09in
\noindent\textit{\textbf{B1831--04}}: ET VI identified this pulsar as having a five-component {\textbf M} profile.  The profile is very broad and the core width can be tracked down to the LOFAR band and the PRAO 103-MHz profiles---and in the 4.9-GHz band \citep{HKK98} the core is weak or absent as usual.  \citet{Weltevrede2006, Weltevrede2007} detect no modultation in either band. The inner conal width is difficult to measure, but the $L$ profile gives good indications---and about 85\degr\ is a good estimate for all the GL98 profiles. \citet{kuzmin_LL2007} provide a scattering measurement.
\vskip 0.09in
\noindent\textit{\textbf{B1832--06}}: This is one of the very most scattered pulsars in this population, where a scattering ``tail'' is seen even at 1.4 GHz \citep[\eg][]{jk18}.  Fortunately, \cite{vH_thesis} has provided a profile at 4.85 GHz that may have three components, a weak leading conal feature before the core and a conflated trailing ``outrider''.  This interpretation is seriously undermined by the stated resolution, but the profile structure appears better resolved than were it so smoothed.  Only at 1.6 GHz do we see any indication of the PPA rate, and this will be flattened by the scattering.  So, stretching very far, we propose this possibly very incorrect 4.85-GHz core-single beam model for the pulsar.  \citet{okp+21} provide a measurement of the scattering time.  
\vskip 0.09in
\noindent\textit{\textbf{B1834--04}}: The poor quality of the three GL98 profiles does not support modeling, but the excellent \citet{jk18} 1.4-GHz profile provides a PPA rate and suggests a core-cone triple structure in which the core is conflated with the trailing conal ``outrider''.  \cite{Weltevrede2006} find a flat fluctuation spectrum that supports a core beam structure perhaps with conal ``outriders''.  We model this above configuration as a best guess of the beam configuration.  The 606-MHz profile shows substantial scattering as also indicated by the \cite{ldk+13} measurement.  
\vskip 0.09in
\noindent\textit{\textbf{B1834--10}}: The pulsar's high frequency profiles seem to show a core-single evolution---that of JK18 very clearly.  GL98's poorly resolved 1.6-GHz profile is broader than the 1.4-GHz one, and the \citet{XRSS} 1720-MHz observation seems to show some structure.  \citet{Weltevrede2006} find no fluctuation feature.  Scattering sets in rapidly below 1 GHz as also seen in the \cite{kmn+15} value. 
\vskip 0.09in
\noindent\textit{\textbf{B1834--06}}: Only the GL98 and JK18 1.4-GHz profiles are well measured, and they show an inconsistent PPA traverse.  We model the profile as having a conal double or perhaps conal quadruple geometry, and only then by guessing that $\beta$ may be close to 0; the quality of the profiles permit no further interpretation.  No fluctuation analysis or scattering time measurement is available.
\vskip 0.09in
\noindent\textit{\textbf{B1839+56}}: PSR B1839+56 has an unusual {\textbf M} profile that shows its full core/double cone structure only in the LOFAR profiles.  Some 3\degr\ is a good value for the core width, suggesting that the profile is a core-cone composite even at 4.9 GHz \citep{HKK98}, and perhaps incomplete at low frequency.  \cite{Force2015} saw it as a {\textbf T} profile, but the LOFAR profiles pretty clearly show a strong central core component flanked by two sets of conal ``outriders'', and scattering does not fully destroy its structure even at 65 MHz.  \cite{Weltevrede2006, Weltevrede2007} find a low frequency modulation that may signal sporadicity.  \cite{Zakharenko2013} detect the pulsar at both 25 and 20 MHz, but the profile is too scattered \citep{kuzmin_LL2007} to be of use here.
\vskip 0.09in
\noindent\textit{\textbf{B1838--04}}:  Again, the LM98 and JK18 1.4-GHz profiles seem to be mostly core emission, and the less well resolved 1.6-GHz one is broader, apparently due to conal ``outriders''.  Here we can see this clearly because the \citet{vH_thesis} 4.9-GHz profile shows both conal features with a central notch for a weakening core.  We thus model the emission using a core-single {\textbf S$_t$} beam configuration and use the higher frequency conal width in place of the poorly resolved 1.6-GHz value.  The PPA rate is clear from the JK18 profile.  Scattering is seen in all the bands below 1 GHz, and a measurement has been published by \citet{okp+21}.
\vskip 0.09in
\noindent\textit{\textbf{B1839--04}}: This pulsar has a classic outer cone double {\textbf D} profile that is nicely seen in JK18's profile, and we model it as such.  \cite{Weltevrede2006} and \citet{basu2016} find a 12-$P$ drift modulation in both components.  The MM10 profile is far too narrow to interpret in any compatible manner and may represent a noise response.  \citet{kuzmin_LL2007} measure a scattering time.
\vskip 0.09in
\noindent\textit{\textbf{B1841--05}}: Again we seem to see a core component at 1.4 GHz in the JK18 profile.  GL98's 1.6 GHz profile seems broader due to incipient conal ``outriders'' and \cite{Weltevrede2006} find a flat fluctuation spectrum.  All the profiles below I GHz show scattering ``tails'', and \citet{okp+21} have provided a measurement.
\vskip 0.09in
\noindent\textit{\textbf{B1841--04}}: The evidence is strong that this is a conal profile, including the strong, coherent drift feature identified by \cite{Weltevrede2006}.  Like in many conal single {\textbf S$_d$} profiles we seem to be seeing a strong leading component and a much weaker and conflated trailing one.  The PPA rate seems to be about --4\degr/\degr\ under the first component.  The total width of the profile including the weak trailing component is difficult or impossible to estimate, so we model the width of the first component only (an error, when corrected that would lead to a better value of $\alpha$).  We modeled the geometry using an outer cone, but it remains unclear how much of the low frequency width increase is due to scattering \citep{kmn+15}.
\vskip 0.09in
\noindent\textit{\textbf{B1842--02}}: The two GL98 1.4- and 1.6-GHz and even the \citet{jk18} profiles tell us little, especially as none permits even an estimate of the PPA rate.  \citet{okp+21} provide a scattering value.  
\vskip 0.09in
\noindent\textit{\textbf{B1842--04}}: Little can be said about this pulsar on the basis of the GL98 1.4- and 1.6-GHz profiles; moreover, their widths seem to be about twice that of the higher quality \cite{jk18} profile.  The latter suggests a possible triple structure with a small $\beta$ and so we model it guided by the PPAs on the trailing edge signaling a weak conal ``outrider''.  \cite{Weltevrede2006} finds a flat fluctuation spectrum.  We use the halved GL widths with caution.  No scattering value is available.  No $t_{\rm scatt}$ value has been reported.  
\vskip 0.09in
\noindent\textit{\textbf{B1844--04}}: The the JK18 observation shows a core-cone triple {\textbf T} profile with the core and trailing ``outrider'' barely resolved.  The GL98 profiles conflate these features as the \cite{JKMG2008} 1.4-GHz profile also shows with its very steep PPA slope and bifurcated trailing component. The 5-GHz \citep{Seiradakis1995} seems to show the core on its own with a width of some 4\degr\ which agrees with the earlier feature of the bifurcated trailing component above.  On this basis we model the geometry as an outer cone/core triple, though it is difficult to reconcile the 610-MHz profile form with that above 1 GHz.  Scattering seems a factor in the shape of GL98's 408-MHz profile, and the small width of MM10's 102-MHz profile seems in conflict with the largish scattering time measured by \citet{kmn+15}.  \cite{Weltevrede2006, Weltevrede2007} find a weak 12-$P$ stationary modulation at 1.4 GHz that was not confirmed at the lower frequency.
\vskip 0.09in
\noindent\textit{\textbf{B1845--19}}:  The GL98 profiles of this slow pulsar show what seems to be an inner cone double {\textbf D} profile where the PPA rate is poorly defined, but may be about --9\degr/\degr.  A surprise is that the \cite{Weltevrede2007} profile is triple (and no modulation periodicity is identified), and GL98's 410-MHz profile may be incipiently so.  Therefore, it seems to behave as a triple, and the prominent $V$ suggests core radiation.  We then model it as a {\textbf T} wherein the core seems to be conflated with the conal features in most of the profiles.  No $t_{\rm scatt}$ value is available.
\vskip 0.09in
\noindent\textit{\textbf{B1845--01}}:  ET VI regarded this pulsar as a conal triple {c\textbf T}, and we support this geometry.  \citet{deich} and \citet{hankins1987} studied the organization of its subpulses, and \cite{Weltevrede2006} and \citet{basu2016} reiterate these results.  Polarized profiles [MHM, MHMA, GL98, \citet{rankin1989}] show the profile structure above 1 GHz and the onset of scattering below.  Given the largish $t_{\rm scatt}$ value measured by \citet{kmn+15}, it is hard to understand the significance of the apparent 100-MHz detection by \cite{MM10}.
\vskip 0.09in
\noindent\textit{\textbf{B1846--06}}: This pulsar seems to have a core-single {\textbf S$_t$} geometry with the peculiarity that the 1.4-GHz cone seems to be the inner whereas the 1.6-GHz appears an outer.  \cite{Weltevrede2006, Weltevrede2007} find a weak longitude-stationary modulation.  KLL07 give a $t_{\rm scatt}$ value.
\vskip 0.09in
\noindent\textit{\textbf{B1851--14}}: Following ET IX the pulsar seems to have a conal single {\textbf S$_d$} geometry.  We thus model it using an outer cone, although an inner cone is also possible, and \cite{Weltevrede2007} find no clear fluctuation feature.  Scattering is not apparent in the 325-MHz profile, which is perhaps compatible with the small scattering time measured by \cite{kmn+15}.
\vskip 0.09in
\noindent\textit{\textbf{B1857--26}}: The pulsar is well known for its five-component {\textbf M} profile \citep[see also][]{JKMG2008}, wherein a ``boxy'' conflated profile is seen at high frequency \citep[\eg][]{jk18} and the components are more separated at low frequencies. \cite{Weltevrede2006,Weltevrede2007} and \citet{basu2016} find a strong and consistent 7.5-$P$ stationary modulation; see also \citet{MR1857}.  The core survives to 149 MHz, shows a clear scattering tail.  No $t_{\rm scatt}$ value has been reported.
\vskip 0.09in
\noindent\textit{\textbf{B1900--06}}: This pulsar has a core/cone triple {\textbf T} profile, where its core is most clearly exhibited in GL98's 925-MHz profile and its triplicity in the 610-MHz one,  The PPA slope is readily estimated and the geometry well so modeled. GL98's 410-MHz profile is poorly resolved and MM10's 102-MHz may show a scattering tail \citep[see][]{kmn+15}. \cite{Weltevrede2006} find no modulation feature.  
\vskip 0.09in
\noindent\textit{\textbf{B1905+39}}: Following ET VI and the references there cited, the pulsar has a well studied five-component {\textbf M} profile; however, as usual the individual components are conflated at the higher frequencies, and the core is not clearly discernible at any frequency.  \cite{Weltevrede2006, Weltevrede2007} find a 4.1-$P$ stationary modulation feature at the lower frequency.  \citet{cor86} give a scattering value.
\vskip 0.09in
\noindent\textit{\textbf{B1907--03}}: Though it was  regarded earlier as a core single pulsar (ET VI), its conal outriders are clear at frequencies below 1 GHz (ET IX), so we model it here with a core/inner cone {\textbf T} configuration.  The pulsar has a flat fluctuation spectrum at 327 MHz \citep{Weltevrede2007}.  \citet{kmn+15} measure a scattering timescate.
\vskip 0.09in
\noindent\textit{\textbf{B1911--04}}: This pulsar has long been classified as having a core-single {\textbf S$_t$} profile [\eg ET VI and \citet{Force2015}; see also \citet{JKMG2008}].  \cite{Weltevrede2006, Weltevrede2007} find a 15-$P$ stationary modulation at 1.4 GHz.  Scattering ``tails'' are seen on the LOFAR profiles, and a small scattering time scale has been measured by \citet{kuzmin_LL2007}, but the 102-MHz observation has had the scattering deconvoled \citep{kuzmin1999}.
\vskip 0.09in
\noindent\textit{\textbf{B1937--26}}: This putative ``partial cone'' pulsar (Lyne \& Manchester 1988) was investigated in ET IX, and we follow this analysis in treating the pulsar as having a core-cone triple configuration.  The GL98 profiles show only a core component with a trailing conal feature; however both the \citet{JohnstonI} and \citet{jk18} 1.4-GHz profiles show a leading edge feature, so a weak ``outrider'' is present.  Further, most of the profiles show a relatively flat PPA traverse, but the well resolved latter profile above shows the steeper negative rate.  No scattering timescale is available.
\vskip 0.09in
\noindent\textit{\textbf{B1940--12}}: PSR B1940--12 seems to have a conal single {\textbf S$_d$} profile, although its fluctuation spectrum is flat at 327 MHz.  KL99 detect the pulsar at 103 MHz, but the surviving core is narrower than the expected width.  However, this is a scattering-corrected profile, so it is difficult to compare to those at higher frequencies.  \citet{abs86} measure a small scattering time.  
\vskip 0.09in
\noindent\textit{\textbf{B1941--17}}:  We have only the 606- and 410-MHz GL98 profiles to go on as those at higher frequencies are very poor.  Fortunately, the JK18 1.4-GHz profile is better.  The PPA rate may be about +9\degr/\degr.  No fluctuation spectra nor scattering time is available.  Possibly a conal single beam structure.  No scattering timescale is reported.
\vskip 0.09in
\noindent\textit{\textbf{B1943--29}}:  The triple structure of the pulsar's profile is clear in the recent 658- and 434-MHz profiles of \citet{MHQ}, more so than in the GL98 profiles that may not be as well resolved.  Our modeling use both, and in particular a rough estimate of the PPA rate of --8\degr/\degr.  \cite{Weltevrede2006, Weltevrede2007} find flat fluctuation spectra at both frequencies, so we tilt to a core-cone triple {\textbf T} model rather than a conal triple. No scattering time has been published.
\vskip 0.09in
\noindent\textit{\textbf{B1946--25}}:  The profile remains single in all the GL98 observations as well as in \cite{Weltevrede2007} and \citet{MHQ}.  We can only guess at the PPA rate, but it may be about +9\degr/\degr.  We then use an inner cone {\textbf S$_d$} model.  No fluctuation feature has been detected nor scattering time measured. 
\vskip 0.09in
\noindent\textit{\textbf{B1953+50}}: ET VI suggested that the pulsar had a core single profile; however, we believe that an inner conal single {\textbf S$_d$} beam geometry is more likely.  \cite{Weltevrede2006, Weltevrede2007} find a roughly 20-$P$ modulation in both bands.  The dimensions of the four LOFAR profiles (BKK+: 129 and 168 MHz; PHS+: 143 MHz; NSK+: 151 MHz) are identical, and neither the 103- nor 65-MHz profiles can be measured accurately.  
\vskip 0.09in
\noindent\textit{\textbf{B2000+40}}:  Both LOFAR profiles have long scattering ``tails'', so are worthless for our purposes. (The timescale $t_{\rm scatt}$ at 168 MHz is much larger than average.)  \cite{Weltevrede2006} find a drift feature at 21 cms, and the profile has the asymmetric form and breaks that suggest a c{\textbf T} or c{\textbf Q} configuration.
\vskip 0.09in
\noindent\textit{\textbf{B2003--08}}:  Following ET VI the pulsar is a fine example of the core/double-cone {\textbf M} structure.  The inner and outer conal components are conflated at high frequency as is so for most such pulsars, but they can be seen clearly around 400 MHz.   \cite{Weltevrede2006, Weltevrede2007} find no clear modulation feature at either frequency apart from a low frequency excess; however. the \citet{basu2016} analysis shows an interesting coherent 50-$P$ cycle that needs more detailed study.   The small \cite{cor86} measurement shows that scattering has minimal effect down to 100 MHz.  
\vskip 0.09in
\noindent\textit{\textbf{B2011+38}}:  The pulsar's wide single profile is unimodal across the bands of observation.  The only hint of structure is that the $L$ profile is delayed within the total power profile.  We thus can only interpret the profiles as core emission features, and their widths escalate little down to 408 MHz.  The PPA traverse is very shallow with a hint of steepening on the trailing edge.  \citep{Weltevrede2006, Weltevrede2007} find a slow periodicity at their higher frequency but are not able to confirm it at 327 MHz.  The \citet{MM10} profile does not square with the GL98 series and may be spurious.  A scattering time has been measured by \citet{kmn+15}.  
\vskip 0.09in
\noindent\textit{\textbf{B2021+51}}: This pulsar's strange conal profile, thought to be double in ET VI and single in ET IX, has either a basically conal single {\textbf S$_d$} or double {\textbf D} profile.  Fluctuation spectra show somewhat different modulation periods in the bands \citep{Weltevrede2006, Weltevrede2007}, but both suggest conal emission.  We model it as an inner conal single, and \citet{kuzmin_LL2007} measure a very small scattering time.  
\vskip 0.09in
\noindent\textit{\textbf{B2022+50m}}:  PSR B2022+50m is an {\textbf S$_t$} but has no indication of conal ``outriders''.   \citet{Weltevrede2006, Weltevrede2007} find no signature of periodic modulation.  The decametric detections \citep{bilous2019,Bondonneau} are too scattered to be of use here.  No scattering measurement is available.  
\vskip 0.09in
\noindent\textit{\textbf{B2036+53}}: A conal single beam traverse is the most likely possibility for this pulsar.  \citet{Weltevrede2006} find an 11-$P$ modulation. However, the polarimetry gives no reliable PPA sweep estimate. GL98's EPN 606-MHz profile suggests a sweep rate of perhaps +15, and this is what is modeled.  Both the 168- the 149-MHz LOFAR observations are included, but scattering is setting in (though no measurment has been published), and both the 129- and MM10 103-MHz observations are useless for our purposes.    
\vskip 0.09in
\noindent\textit{\textbf{B2043--04}}: ET VI and IX found the profiles to be conal, and \citet{Weltevrede2006, Weltevrede2007} and \citet{basu2016} measured a strong and stable 2.75-$P$ drift modulation. The PPA rate can be estimated reliably, so we model the  pulsar with an inner conal single {\textbf S$_d$} geometry.  The PHS+ 135-MHz LOFAR profile is noisy but seems compatible, whereas the MM10 102-MHz one seems to have a scattering ``tail'' despite the very small \citep{cor86} measurement.  
\vskip 0.09in
\noindent\textit{\textbf{B2045+56}}: This pulsar apparently has a core-single {\textbf S$_t$} geometry, but the 1.4-GHz profiles are too weak to identify conal outriders.  $t_{\rm scatt}$ is estimated at 13\degr\ at 129 MHz, but the 10.5\degr\ core there shows no obvious scattering distortion.  
\vskip 0.09in
\noindent\textit{\textbf{B2045--16}}:  \citet{Weltevrede2007} and  \citet{basu2016} and find a clear drift feature in the pulsar's outer components and a different modulation in the central one.  We considered whether the pulsar might rather be a conal triple, but no way was found to square the geometrical model with this conjecture, so we accept the ET VI model with a somewhat less steep PPA rate.  \citet{cor86} reports an unusually small scattering time from scintillation measurements, and no other study as yet confirms it. 
\vskip 0.09in
\noindent\textit{\textbf{B2106+44}}:  Most of the GL98 profiles clearly show two conflated components and several seem to indicate additional structure on the leading and trailing edges.  \citet{Weltevrede2006, Weltevrede2007} find no drift features, but do see a low frequency excess that may be produced by a population of null pulses.  Given the limited quality of the profiles, we model the beams using an inner conal single {\textbf S$_d$} geometry.  The $L/I$ is low, but a PPA rate can be determined at 1.4/1.6 GHz.  Scattering is apparent in the GL98 234-MHz profile, and the BKK+ 159-MHz profile shows nothing more than a long scattering ``tail'' presumably compatible with the \cite{kmn+15} $t_{\rm scatt}$ value.  The 102-MHz profile \citep{MM10} shows too narrow a width given the clear scattering at higher frequencies and may be ``scattered out''.
\vskip 0.09in
\noindent\textit{\textbf{B2111+46}}:  See ET VI, \cite{Force2015} and the references therein for a discussion of this classic well studied core/cone triple pulsar.  \citet{Weltevrede2006, Weltevrede2007} find modulation features, especially in the leading conal component.  The core widths vary and decrease substantially at meter wavelengths.  The \cite{Noutsos} 151-MHz profile shows substantial scattering that obliterates the trailing conal component.  The KL99 103-MHz profile has had the scattering \citep{kuzmin_LL2007} deconvolved, and shows a width roughly comparable with that seen at high frequency.
\vskip 0.09in
\noindent\textit{\textbf{B2148+63}}:  Both ET VI and \cite{Force2015} understood this pulsar to have an inner conal single {\textbf S$_d$} geometry, and we so model it here.  \citet{Weltevrede2006, Weltevrede2007} find three distinct coherent features which would probably repay further detailed study.  Scattering clearly broadens the broad profile in the LOFAR band roughly as expected, but no measurement is available.
\vskip 0.09in
\noindent\textit{\textbf{B2148+52}}: PSR B2148+52 seems to have three components the trailing one much weaker than the others, and a core width can be estimated from several of the higher frequency profiles.  A 4.9-GHz profile \citep{kkwj98} supports the evolution, and a possible PPA rate is suggested poorly by the 610-MHz profile.  We then model it as an outer conal {\textbf T} triple.  \citet{Weltevrede2006, Weltevrede2007} find a flat fluctuation spectrum and no features.  The MM10 102-MHz profile is broader and may reflect scattering though no measurement is available.  
\vskip 0.09in
\noindent\textit{\textbf{B2152--31}}:  The GL98 profiles consist of two conflated components---the leading stronger than the trailing---that show a similar structure over the available band.  The PPA rate seems to be about --9\degr/\degr, and all the widths seems to be about 10\degr.  No fluctuation spectra nor scattering measurements are available.  We model the emission using an inner conal double {\textbf D} geometry.  No scattering value is available.
\vskip 0.09in
\noindent\textit{\textbf{B2154+40}}: Both ET VI and \cite{Force2015} regarded this pulsar as having a conal triple {c\textbf T} configuration.  At all frequencies there is a bright leading, middle and weak trailing component---a situation that is common in this type of profile---and the trailing feature is often difficult to discern. At high frequency the components are also conflated, so a single-pulse analysis is needed to fully decipher their actions.  \citet{Weltevrede2006, Weltevrede2007} find a 3-4-$P$ stationary modulation feature in both bands.  The inner conal widths are all the same because they are little more than plausible guesses.  The MM10 102-MHz seems compatible with little scattering \citep{kuzmin_LL2007}, and the KL99 profile also seems so if somehow the core is absent.  
\vskip 0.09in
\noindent\textit{\textbf{B2217+47}}:  A difficult pulsar with a component reported to change position with time \citep{Suleymanova1994}, however overall it seems to have a core-single or triple configuration as uniformly specified in ET VI, IX and \cite{Force2015}. The core width remains constant down to 100 MHz, apart from the 240-MHz GL98 profile that seems to be poorly resolved---and the 38-MHz profile shows a substantial scattering tail as expected.  \citet{Weltevrede2006, Weltevrede2007} find a flat fluctuation spectrum at 1.4 GHz but some evidence for a weak 4-$P$ stationary modulation in the lower band.  The profiles at all frequencies are inconsistent in their structure: the two highest frequencies show breaks that may signal a conflated inner conal pair, and only the 925-MHz GL98 profile seems to show a weak outer conal pair.  Then in the 100-MHz band, some show weak structure and others do not, but much of what is seen seems to be consistently outer conal.  The 38-MHz detection is too scattered as expected to be useful.  Similarly, the decametric profiles of \cite{bilous2019} and \citet{Bondonneau} show the progression of increasing scattering presumably compatible with the \citet{kuzmin_LL2007} time scale.  
\vskip 0.09in
\noindent\textit{\textbf{B2224+65}}: This pulsar is a puzzle.  At high frequency it superficially looks like a well resolved double profile, but on closer inspection the two components are so dissimilar in polarization, spectrum and a lack of separation with wavelength that no such model can be sustained.  One other possibility is a core-single component with a postcursor \citep{basu2018}, and the flat fluctuation spectra at both frequencies \citep{Weltevrede2006, Weltevrede2007} perhaps strengthens this case.  We thus model it with a core single {\textbf S$_t$} beam configuration, and we do nothing with the trailing (putative ``postcursor'') feature that becomes ever weaker in the LOFAR band. The decametric profiles \citep{bilous2019,Bondonneau} are too scattered \citep{kuzmin_LL2007} to be of use here.
\vskip 0.09in
\noindent\textit{\textbf{B2227+61}}: GL98 provide a full set of profiles from which a PPA rate can only be estimated at 1.4 GHz.  The core is never clearly seen; however, if the middle component is a core with a plausible width of some 5.5\degr, it would be compatible with the outer conal {\textbf S$_t$} geometry modeled. The LOFAR profiles of PSR B2227+61 are corrupted by scattering, roughly at a level of 2-3 times the average model estimate of 25\degr\ at 129 MHz.  Even the 240-MHz profile is scattered or poorly observed.  \citep{Weltevrede2007} find a flat fluctuation spectrum.  
\vskip 0.09in
\noindent\textit{\textbf{B2241+69}}: This pulsar seems to have an inner conal single {\textbf S$_d$} geometry.  However, there is little else to go on:  the PPA rate can only be guessed at and no fluctuation spectra are available.  The two LOFAR profiles, however, are very well measured, and the MM10 103-MHz profile may be compatible.  No scattering value has been published.  
\vskip 0.09in
\noindent\textit{\textbf{B2255+58}}: Both ET VI and \cite{Force2015} regarded this pulsar as having a core feature, and the highest frequency EPN profiles show a structure with weak ``skirts'' that have about the right dimensions to be conflated inner conal components. \citet{Weltevrede2006, Weltevrede2007} find a 10-$P$ stationary modulation feature in both bands.  We thus model it with a {\textbf S$_t$} configuration.  The 149-MHz LOFAR profile is too scattered \citep{geyer} to be useful for our purposes.  \
\vskip 0.09in
\noindent\textit{\textbf{B2303+46}}: This pulsar shows a clear triple form at 149 MHz, but the components are conflated in the mostly low quality GL98 higher frequency profiles.  No fluctuation spectra are available, and the PPA rate is only guessable, but we suggest this pulsar has a conal triple {c\textbf T} configuration.  We estimate the inner cone dimensions from the $L$ profile at 606 MHz.  Scattering is minimal in this pulsar \citep{kuzmin_LL2007}.
\vskip 0.09in
\noindent\textit{\textbf{B2306+55}}: ET VI established that this pulsar had a classic outer conal double {\textbf D} profile.  \citet{Weltevrede2006, Weltevrede2007}, however, find no signature of periodic modulation to support this.  Profiles down to 100 MHz seem little affected by scattering; however, the BKK++ 60 MHz profile is single, probably with its components conflated by the thus far unmeasured level of scattering.
\vskip 0.09in
\noindent\textit{\textbf{B2310+42}}: Both ET VI and \cite{Force2015} suggested that this pulsar may have an {\textbf M} profile, but the components are always conflated, and the core component is never clearly distinguishable.  A strong, coherent 2.1-$P$ modulation is seen at both frequencies \citep{Weltevrede2006, Weltevrede2007}.  The 37-MHz detection---if a detection at all---is very poor, but the BKK++ detections are better and seem to reflect the outer conal dimensions.  Similarly, \citet{Zakharenko2013} detect the pulsar at 25 MHz, but with very substantial scattering \citep{kuzmin_LL2007}.
\vskip 0.09in
\noindent\textit{\textbf{B2319+60}}: The pulsar has a well-studied conal quadruple c{\textbf Q} profile; see ET VI and \cite{Force2015} and the references there cited.  A strong, coherent 7.7-$P$ modulation is seen at 1.4 GHz but seems weaker at the lower frequency.  No core component is seen in any of the GL98 profiles; however, the descattered PRAO 103-MHz profile \citep[see the EPN Database]{kuzmin1999} has a triple form and the central component has a 5.5\degr\ width---exactly that expected for a core component.  \citet{kmn+15} measure substantial scattering that would be more noticeable in a faster pulsar.  
\vskip 0.09in
\noindent\textit{\textbf{B2323+63}}: The GL98 profiles show a filled conal profile that suggests either a conal triple or conal quadruple form.  We model it as the latter, although the inner cone dimension can only be estimated.   \citep{Weltevrede2006} find a low frequency excess but no modulation feature.  The 234-MHz profile seems to have a scattering ``tail'' that may be compatible with the mean level of scattering.  No measurement seems to exist.  
\vskip 0.09in
\noindent\textit{\textbf{B2324+60m}}:  \cite{Force2015} were probably incorrect in regarding this as a core-single profile.  The orderly PPA traverse seen in most of the profiles argues for conal emission, and close examination of the $L$ profiles in the forgoing work and in GL98 suggest a strong leading and weak trailing ``outrider'' such that the latter is conflated with the trailing part of a central core component---that is itself seen only as an inflection in the total power and a dip in $L$.  The \citep{Weltevrede2006} analysis tends to support this structure of the leading conal and core component.  The 234-MHz profile is poorly resolved and probably also scattered---as no measurement seems to be available.  Clearly, all the ``measurements'' in the model here are rough estimates, but exemplify what seems to be the overall case.  
\vskip 0.09in
\noindent\textit{\textbf{B2327--20}}:  The pulsar's profile shows three components over the entire band, more and more clearly at low frequency where the central one is brighter, as clearly seen in \citet{JKMG2008}.  Overall, the leading one is strongest with the other two at decreasing intensities.  \citep{Weltevrede2006, Weltevrede2007} find no drift but modulation that could be due to sporadic emission, and \citet{basu2016} find a highly interesting 50-$P$ amplitude modulation that surely begs for more study.  If the central component is a core, its width can best be estimated at low frequency, where it is little more than 2\degr.  We follow ET VI in modeling the geometry as a core-cone triple {\textbf T}, but it could equally well be a conal triple c{\textbf T}.  The 270- and 170-MHz profiles are thanks to MHMb.  \citet{brg99} measure a very small scattering value.  The BGT+ 53-MHz profile is poorly plotted but seems hardly more than about 10\degr---squaring with the very small level of scattering reported by \citet{brg99}.
\vskip 0.09in
\noindent\textit{\textbf{B2334+61}}: A weak pulsar with little study, it seems to have a core-single {\textbf S$_t$} beam configuration.  No fluctuation features were found in either band \citep{Weltevrede2006, Weltevrede2007}. The core in this pulsar seems to have a ``pedestal'' as in several other cases.  KL07 provide a scattering value. 
\vskip 0.09in
\noindent\textit{\textbf{B2351+61}}: This pulsar is difficult to classify, but it may have a core/cone {\textbf T} beam system.  If so, the main component is a core component, the trailing one a conal outrider and the leading outrider difficult to discern at high frequency.  A 17-$P$ modulation is seen at both frequencies \citep{Weltevrede2006, Weltevrede2007}.  The one decametric profile at 102.5 MHz \citep{kuzmin1999} has the scattering \citep{kuzmin_LL2007} deconvolved and  may show both the cone and cone.

\onecolumn
\begin{figure*}
\begin{center}
\includegraphics[width=180mm,height=218mm,angle=0.]{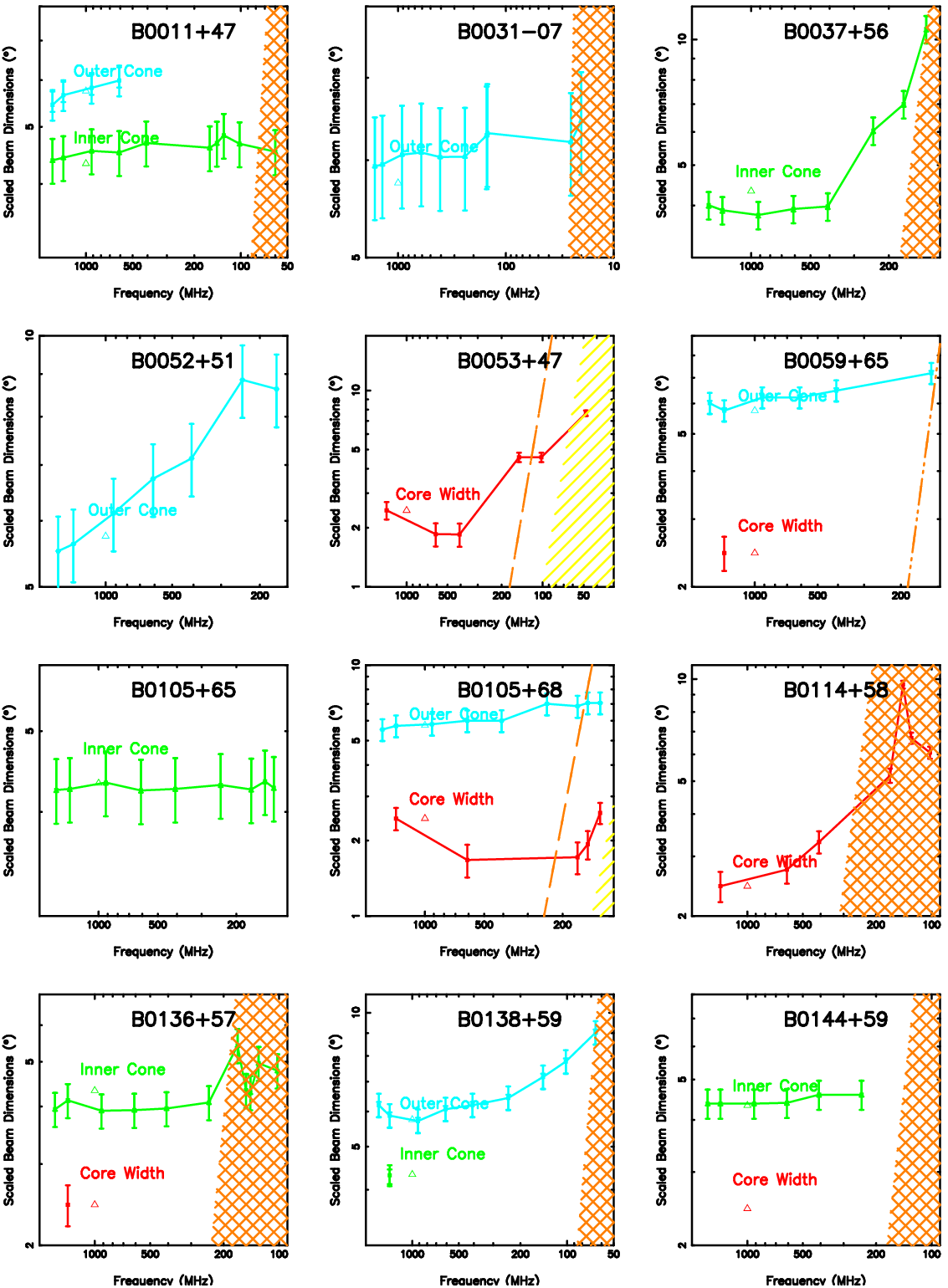}
\caption{Emission-beam geometry models for the Gould \& Lyne population.  The inner and outer conal beam radii and core widths are plotted as a function of radio frequency, scaled to a 1-second orthogonal rotator configuration (see text).  The error bars reflect 10\% uncertainties in measuring both the widths and the PPA sweep rate $R$.  The triangles at 1 GHz show the nominal beam dimensions.}
\label{figA1}
\end{center}
\end{figure*}

\begin{figure*}
\begin{center}
\includegraphics[width=180mm,height=225mm,angle=0.]{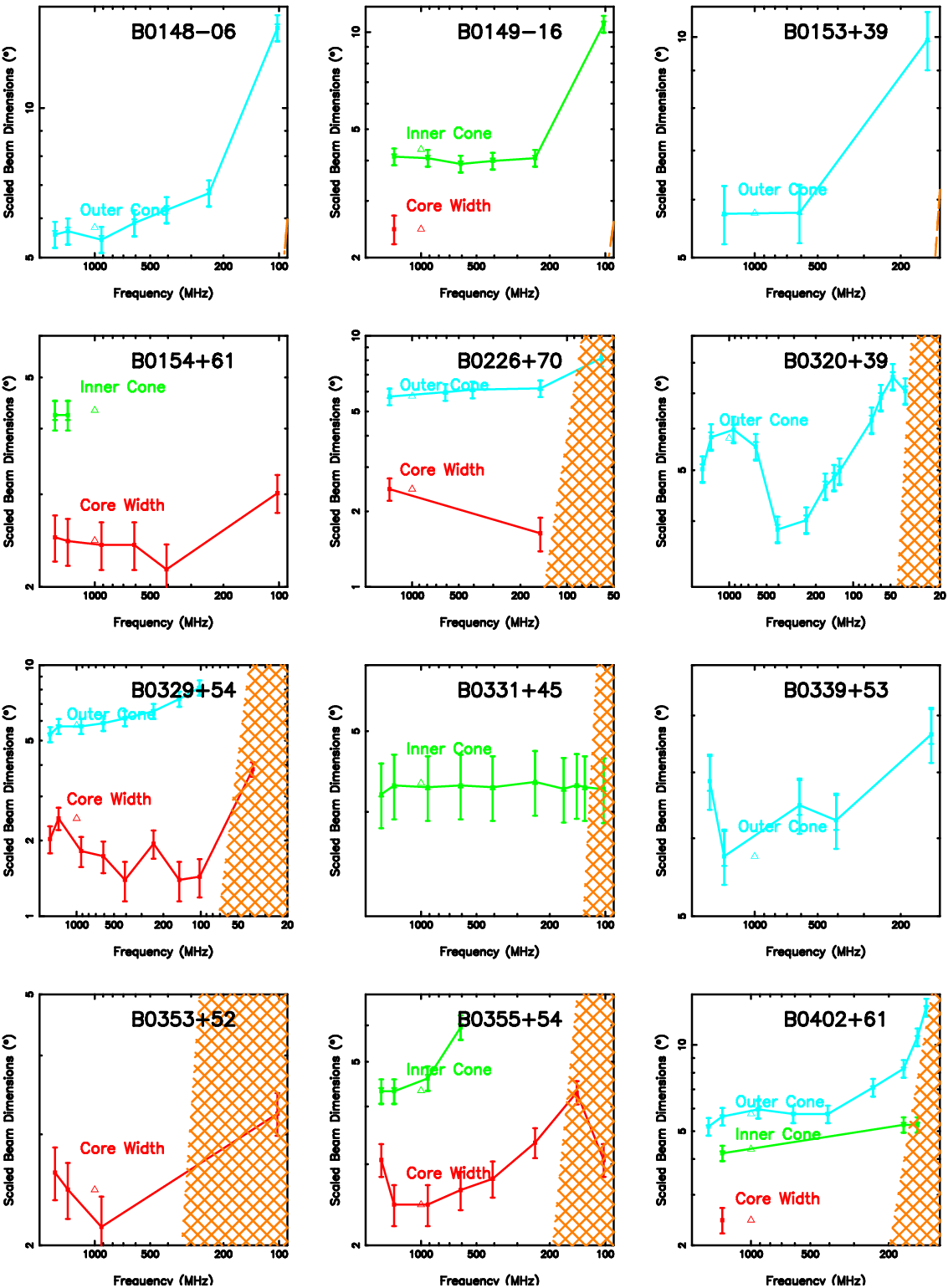}
\caption{Emission-beam geometry models as in Fig~\ref{figA1}.}
\label{figA2}
\end{center}
\end{figure*}

\begin{figure*}
\begin{center}
\includegraphics[width=180mm,height=225mm,angle=0.]{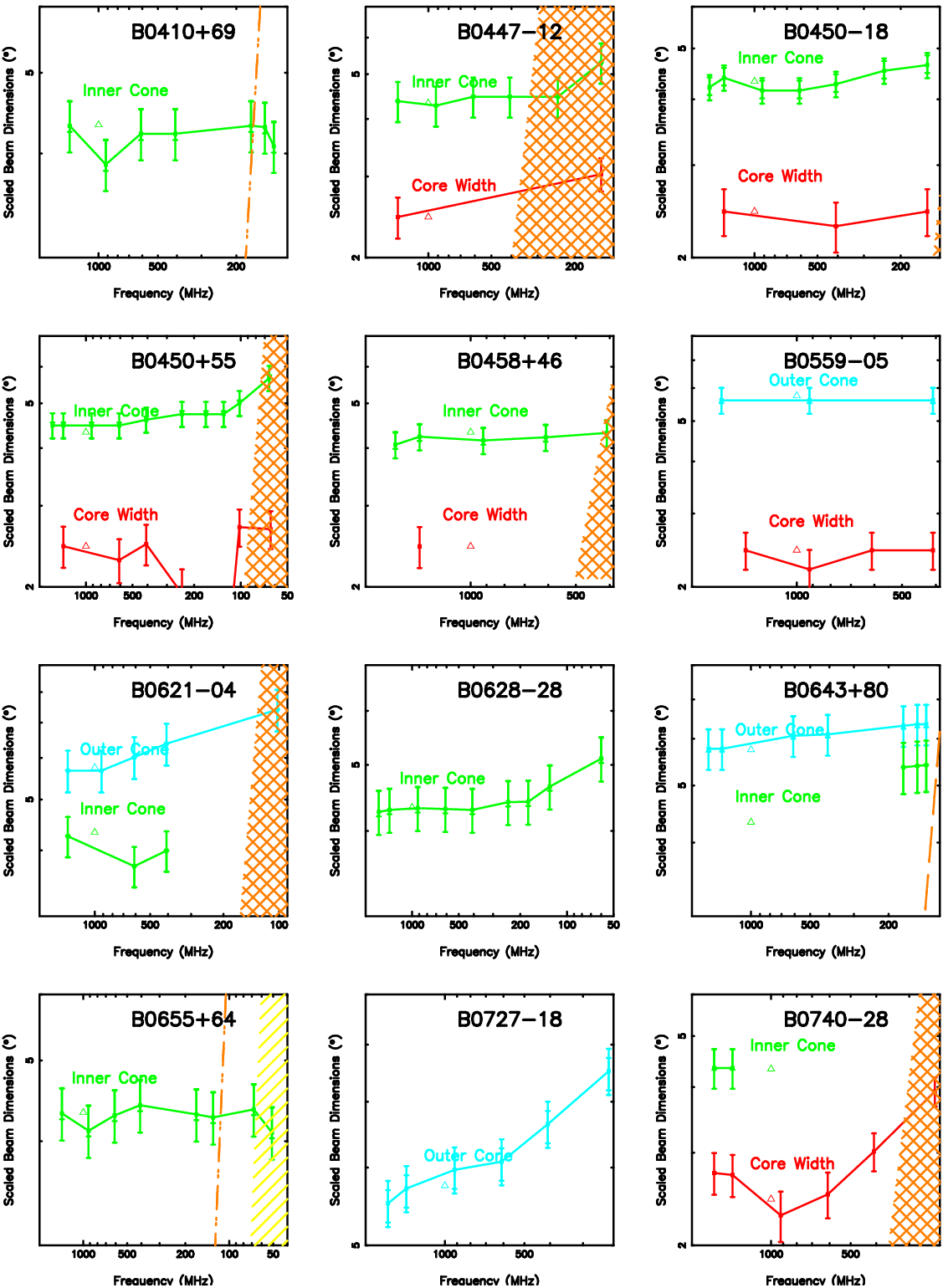}
\caption{Emission-beam geometry models as in Fig~\ref{figA1}.}
\label{figA3}
\end{center}
\end{figure*}

\begin{figure*}
\begin{center}
\includegraphics[width=180mm,height=225mm,angle=0.]{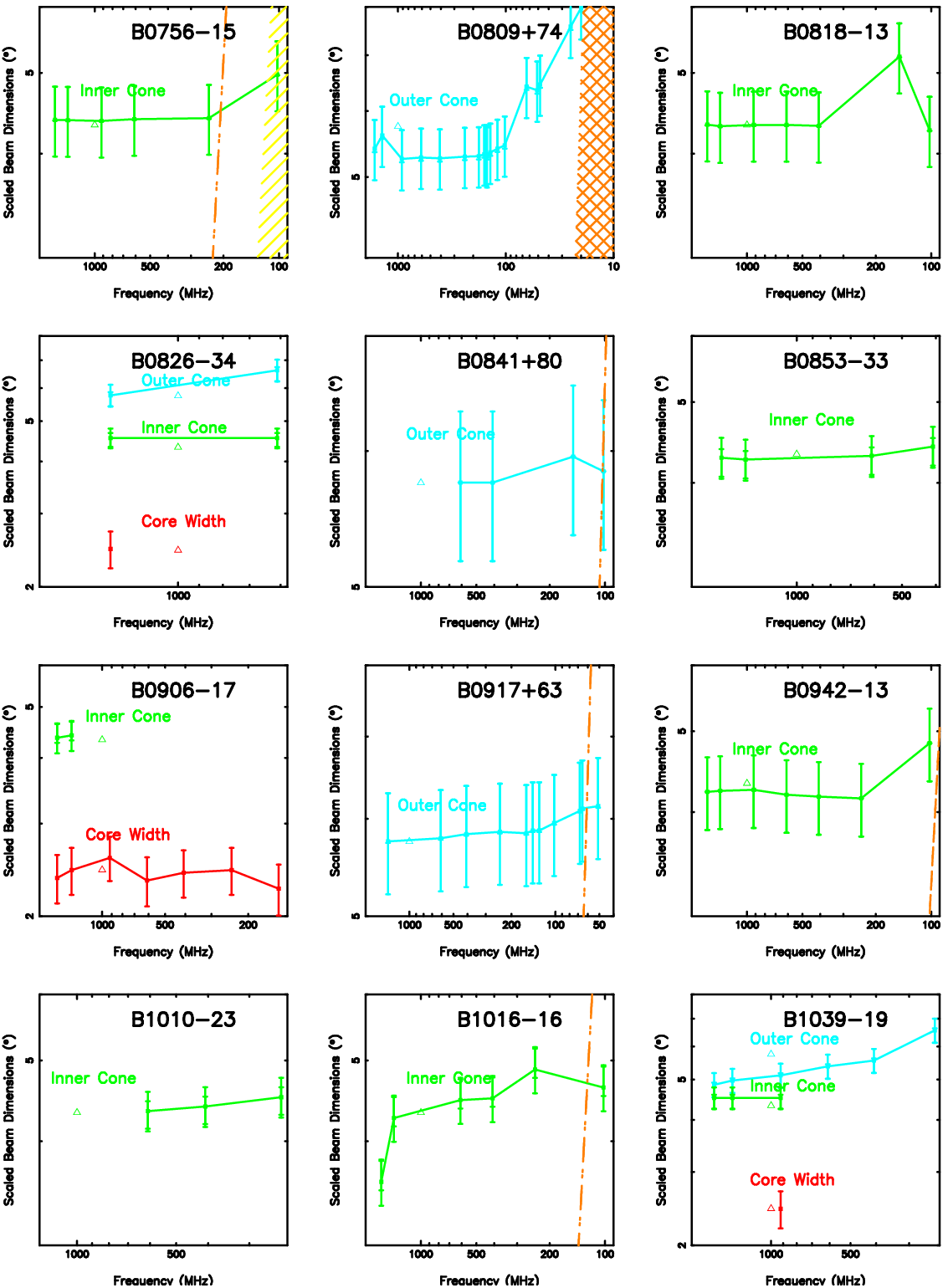}
\caption{Emission-beam geometry models as in Fig~\ref{figA1}.}
\label{figA4}
\end{center}
\end{figure*}

\begin{figure*}
\begin{center}
\includegraphics[width=180mm,height=225mm,angle=0.]{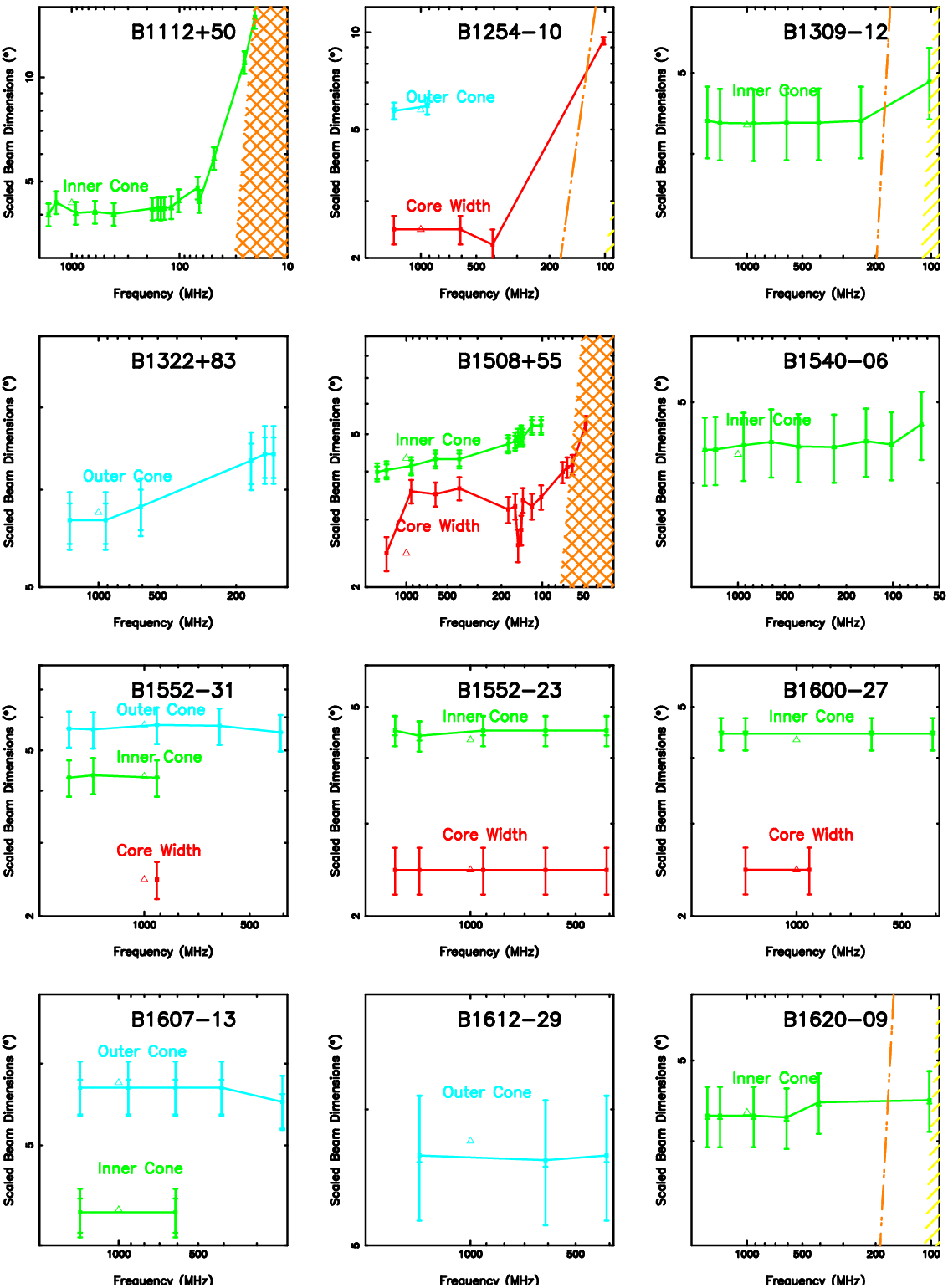}
\caption{Emission-beam geometry models as in Fig~\ref{figA1}.}
\label{figA5}
\end{center}
\end{figure*}

\begin{figure*}
\begin{center}
\includegraphics[width=180mm,height=225mm,angle=0.]{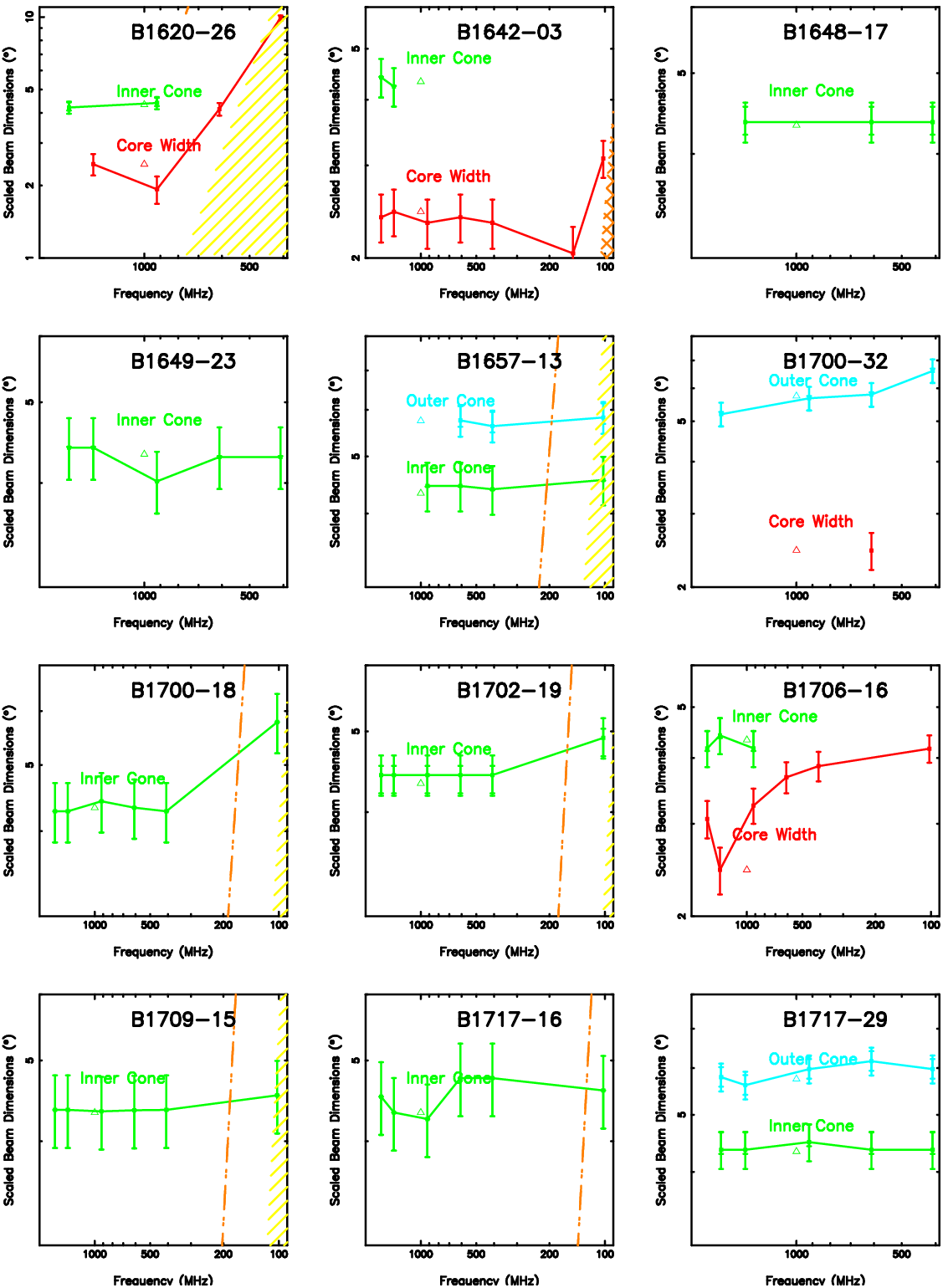}
\caption{Emission-beam geometry models as in Fig~\ref{figA1}.}
\label{figA6}
\end{center}
\end{figure*}

\begin{figure*}
\begin{center}
\includegraphics[width=180mm,height=225mm,angle=0.]{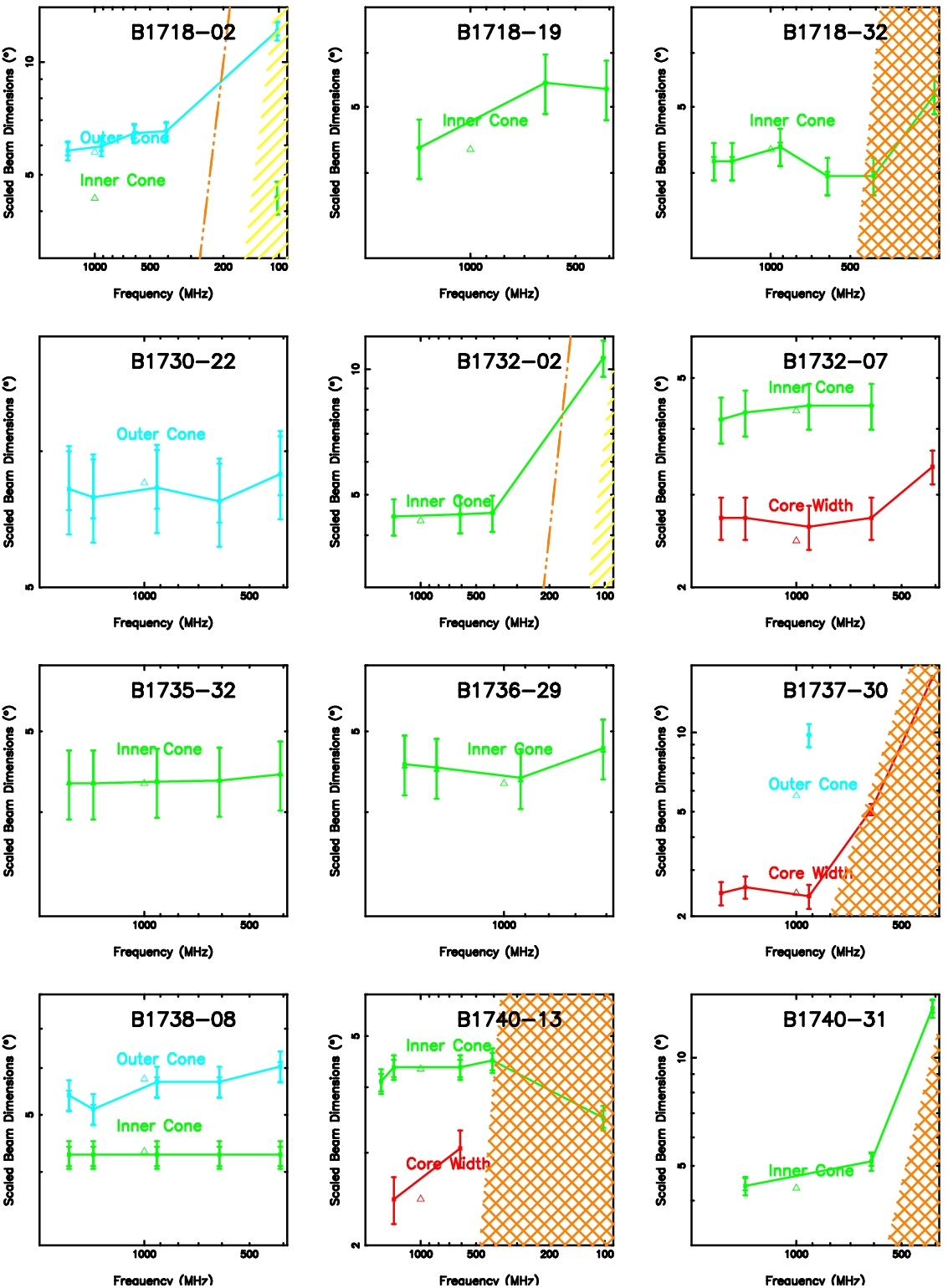}
\caption{Emission-beam geometry models as in Fig~\ref{figA1}.}
\label{figA7}
\end{center}
\end{figure*}

\begin{figure*}
\begin{center}
\includegraphics[width=180mm,height=225mm,angle=0.]{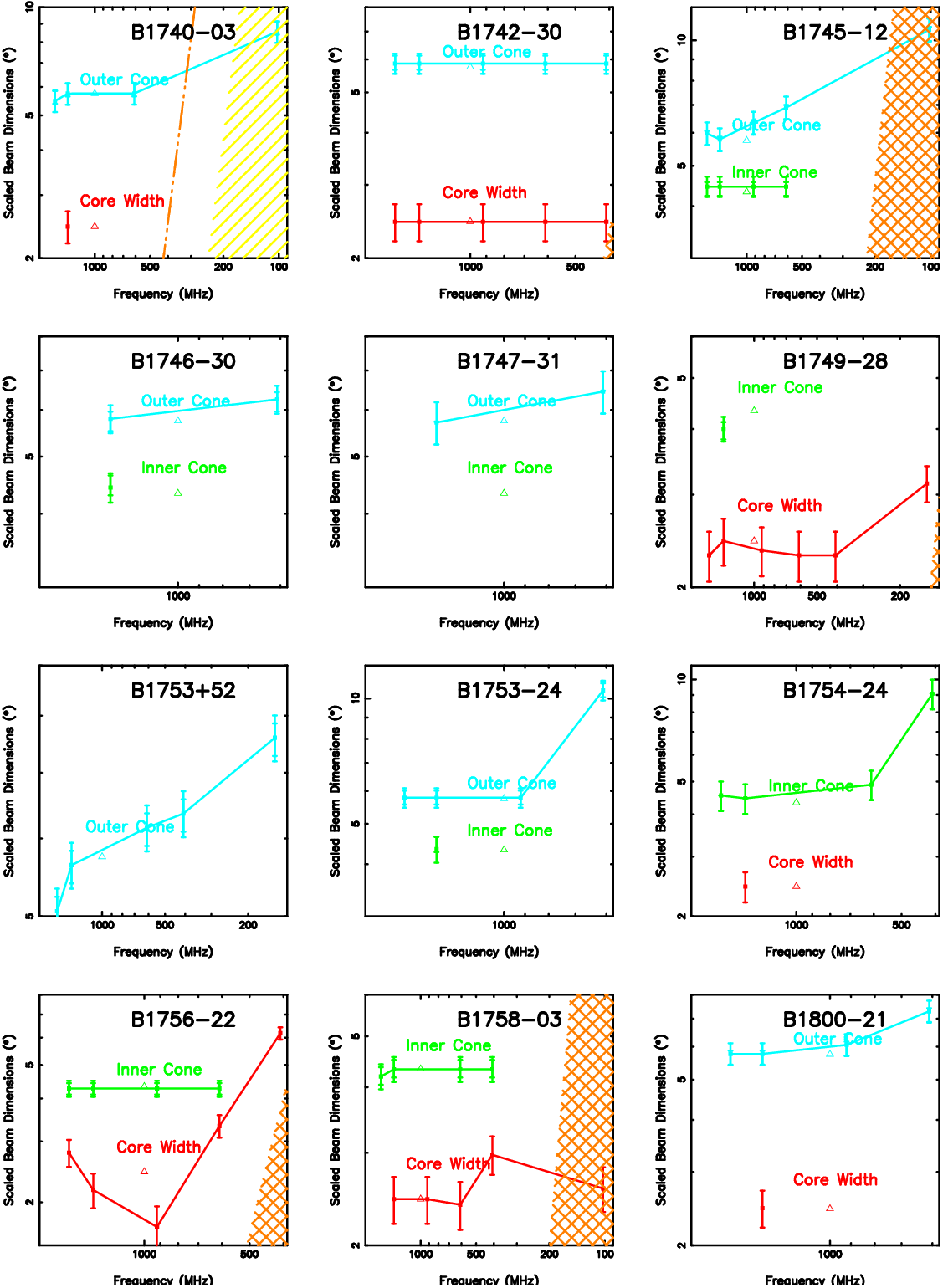}
\caption{Emission-beam geometry models as in Fig~\ref{figA1}.}
\label{figA8}
\end{center}
\end{figure*}

\begin{figure*}
\begin{center}
\includegraphics[width=180mm,height=225mm,angle=0.]{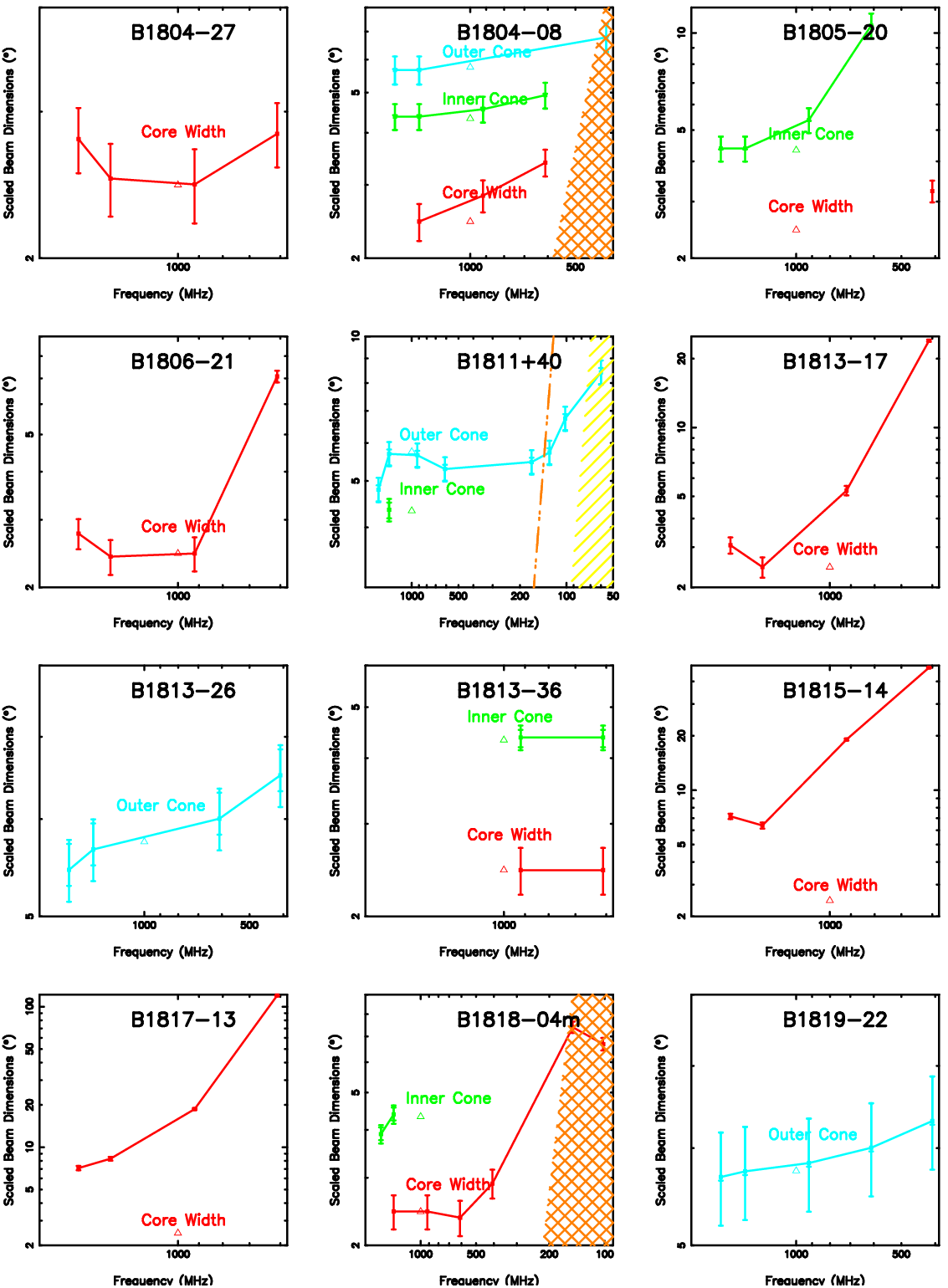}
\caption{Emission-beam geometry models as in Fig~\ref{figA1}.}
\label{figA9}
\end{center}
\end{figure*}

\begin{figure*}
\begin{center}
\includegraphics[width=180mm,height=225mm,angle=0.]{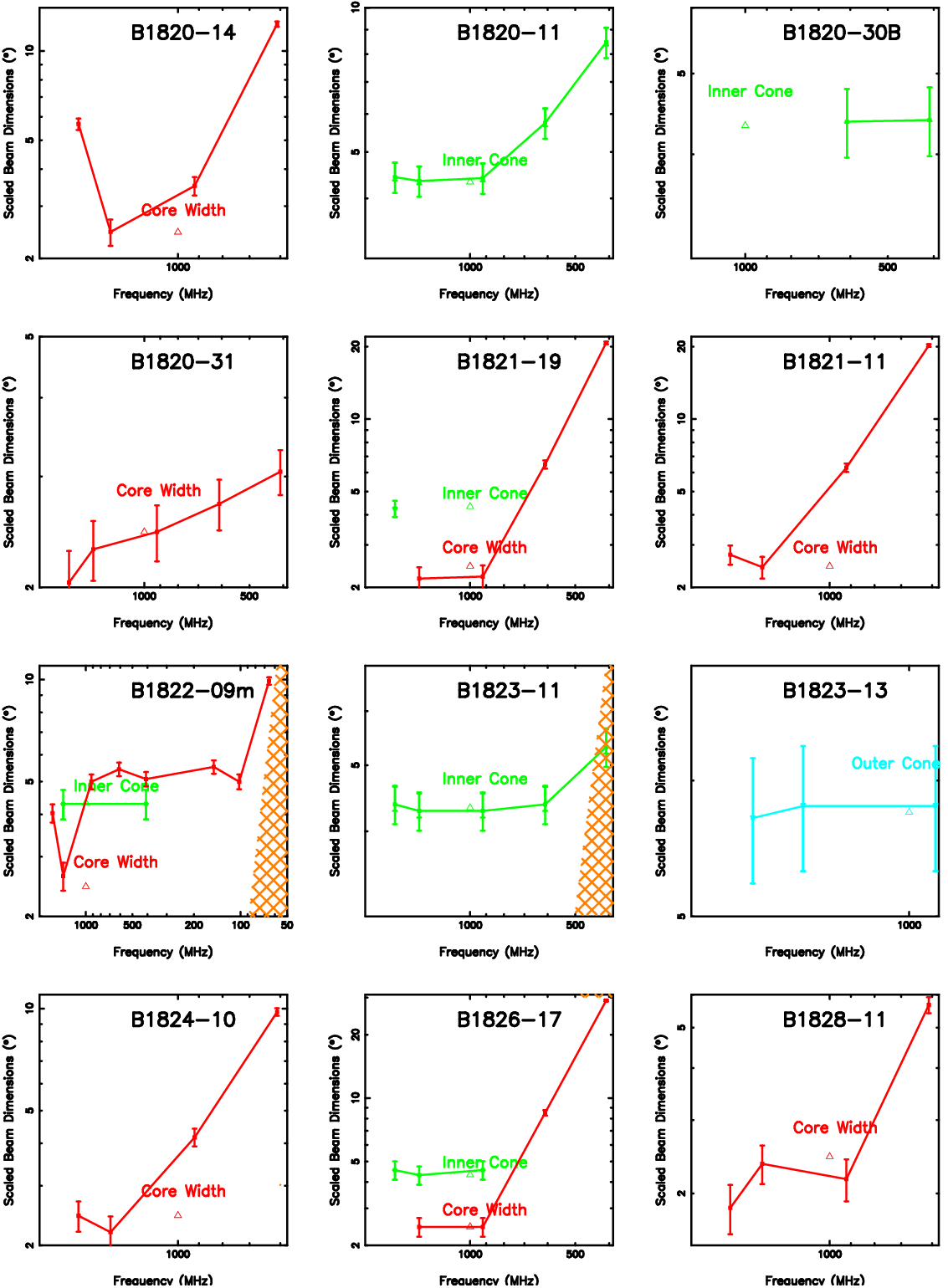}
\caption{Emission-beam geometry models as in Fig~\ref{figA1}.}
\label{figA10}
\end{center}
\end{figure*}

\begin{figure*}
\begin{center}
\includegraphics[width=180mm,height=225mm,angle=0.]{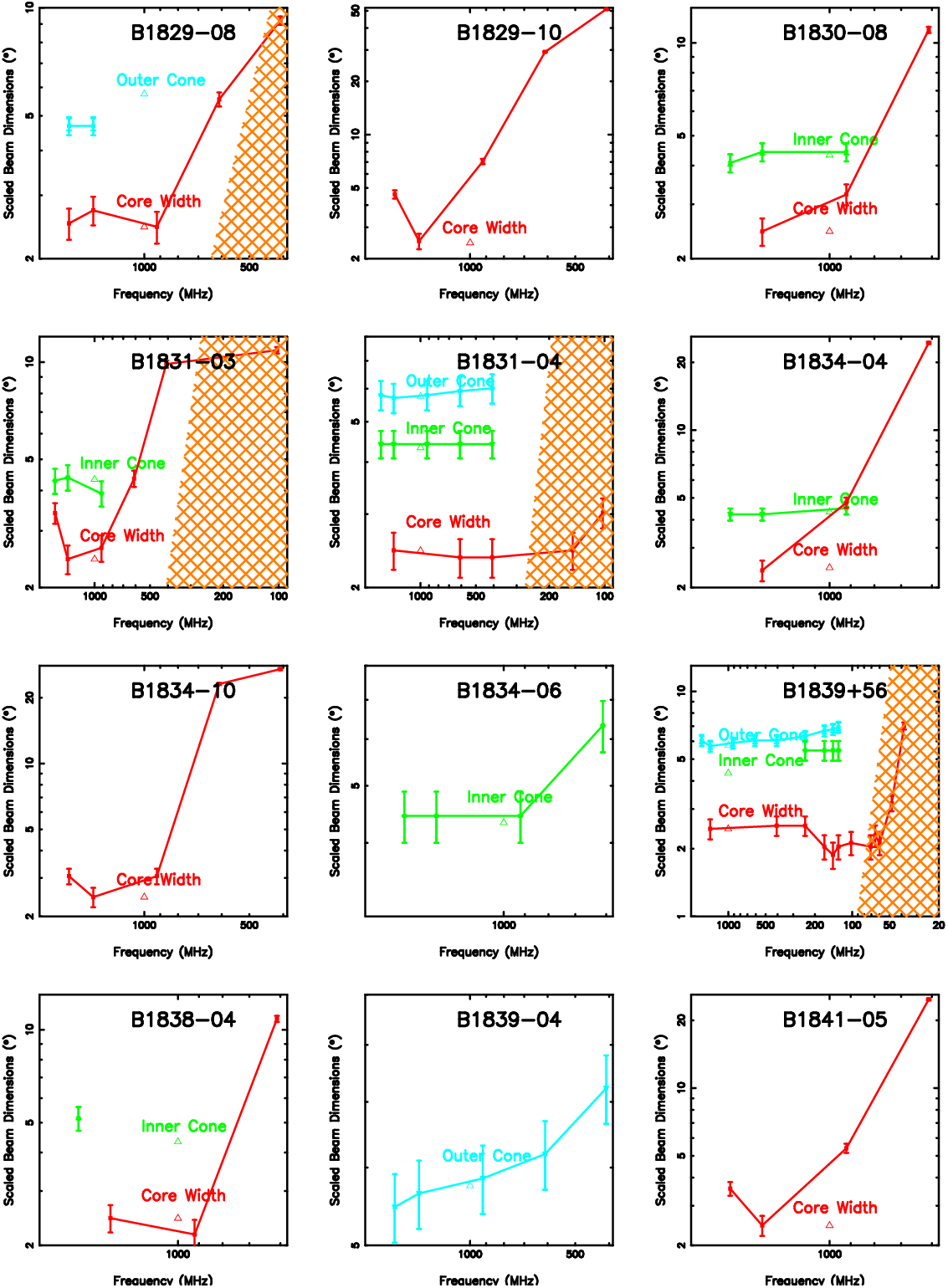}
\caption{Emission-beam geometry models as in Fig~\ref{figA1}.}
\label{figA11}
\end{center}
\end{figure*}

\begin{figure*}
\begin{center}
\includegraphics[width=180mm,height=225mm,angle=0.]{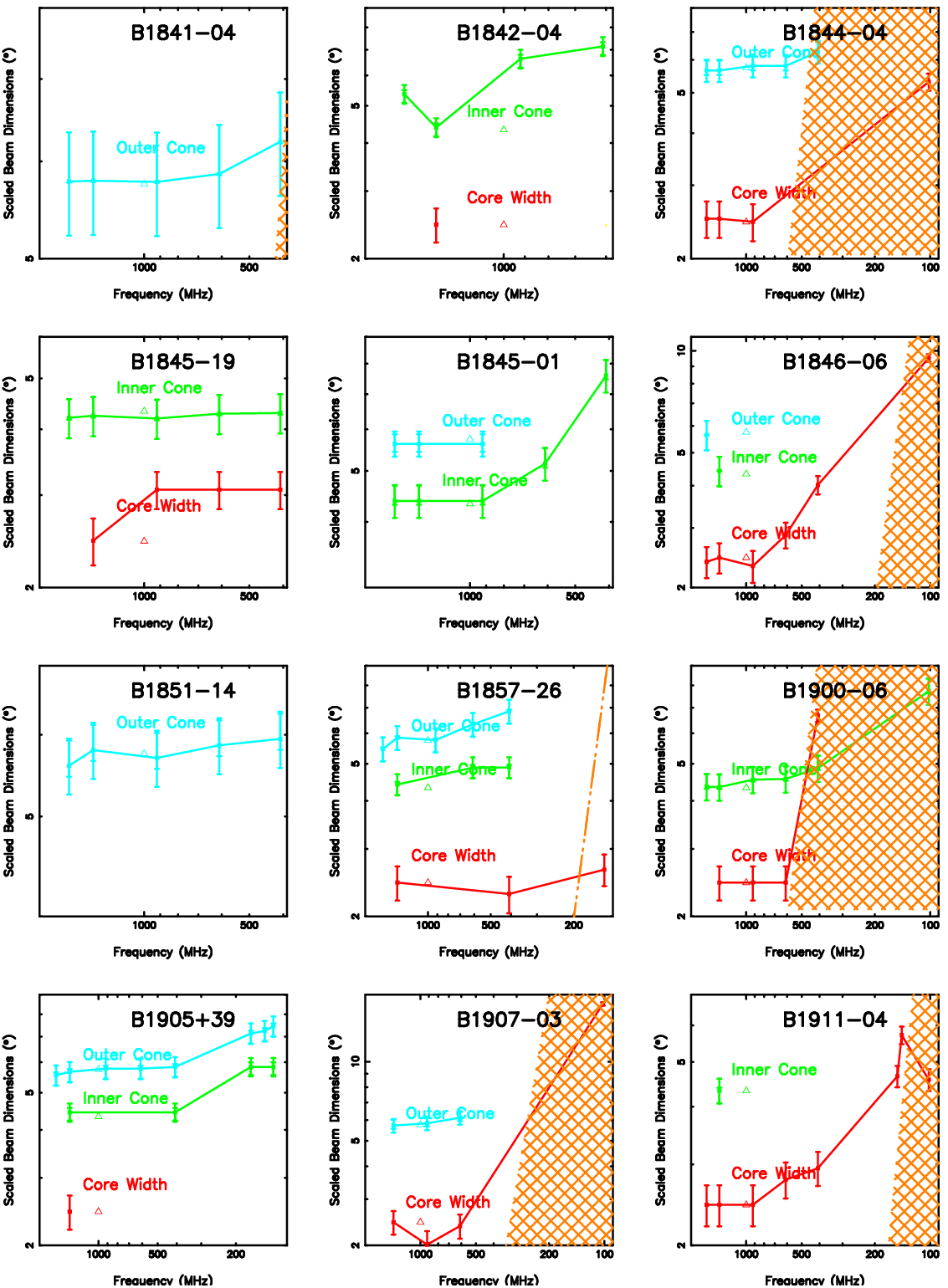}
\caption{Emission-beam geometry models as in Fig~\ref{figA1}.}
\label{figA12}
\end{center}
\end{figure*}

\begin{figure*}
\begin{center}
\includegraphics[width=180mm,height=225mm,angle=0.]{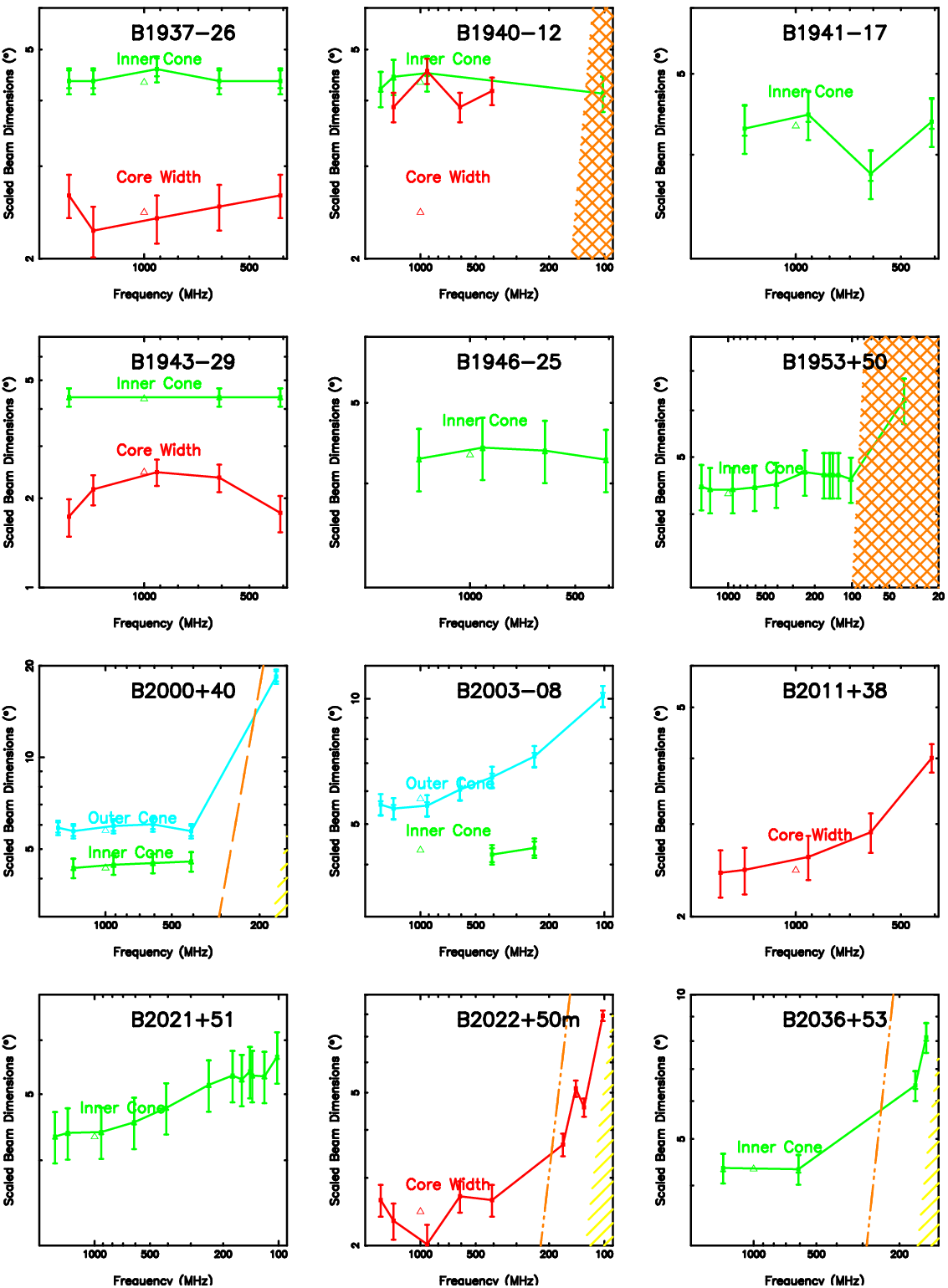}
\caption{Emission-beam geometry models as in Fig~\ref{figA1}.}
\label{figA13}
\end{center}
\end{figure*}

\begin{figure*}
\begin{center}
\includegraphics[width=180mm,height=225mm,angle=0.]{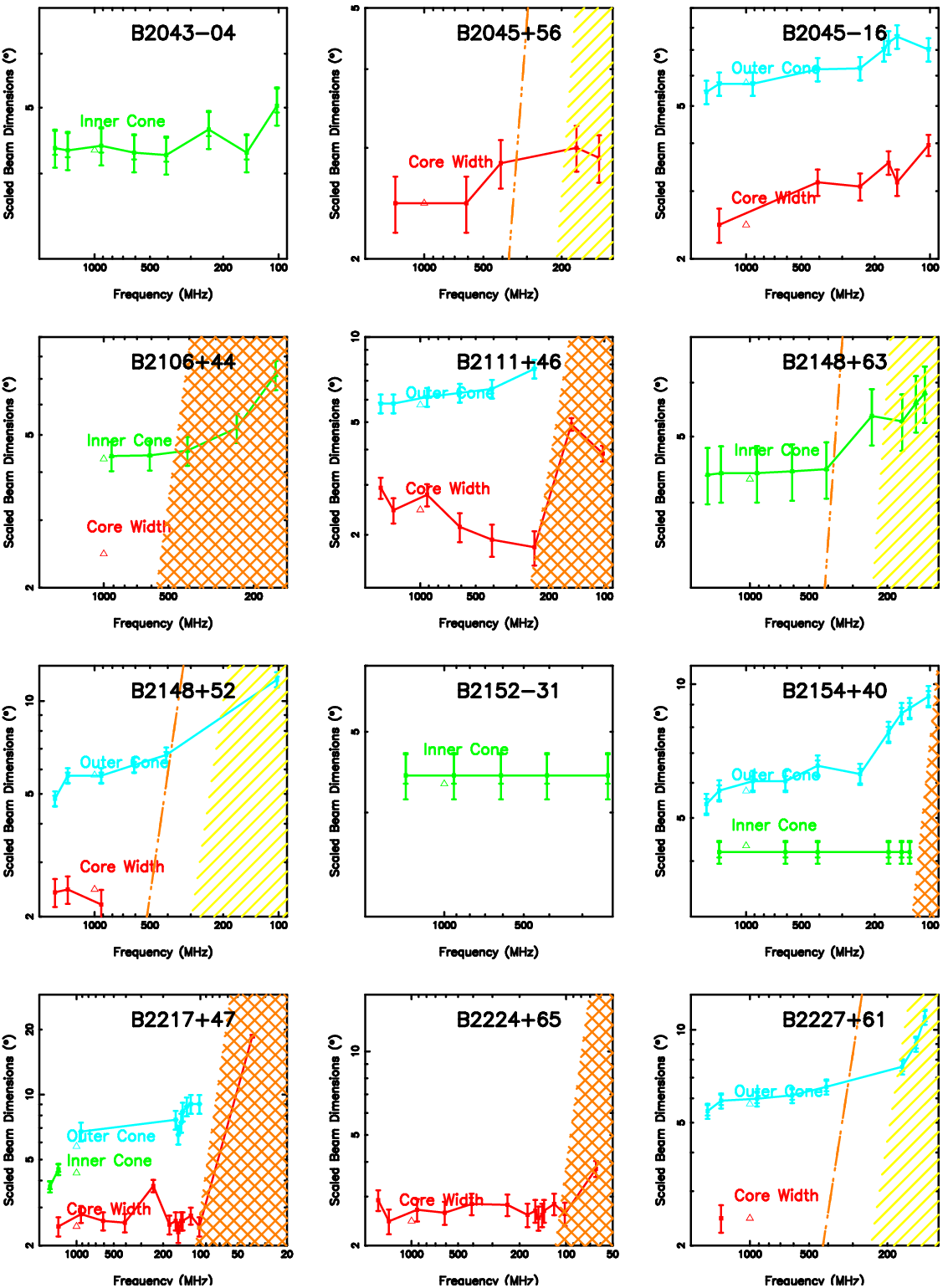}
\caption{Emission-beam geometry models as in Fig~\ref{figA1}.}
\label{figA14}
\end{center}
\end{figure*}

\begin{figure*}
\begin{center}
\includegraphics[width=180mm,height=225mm,angle=0.]{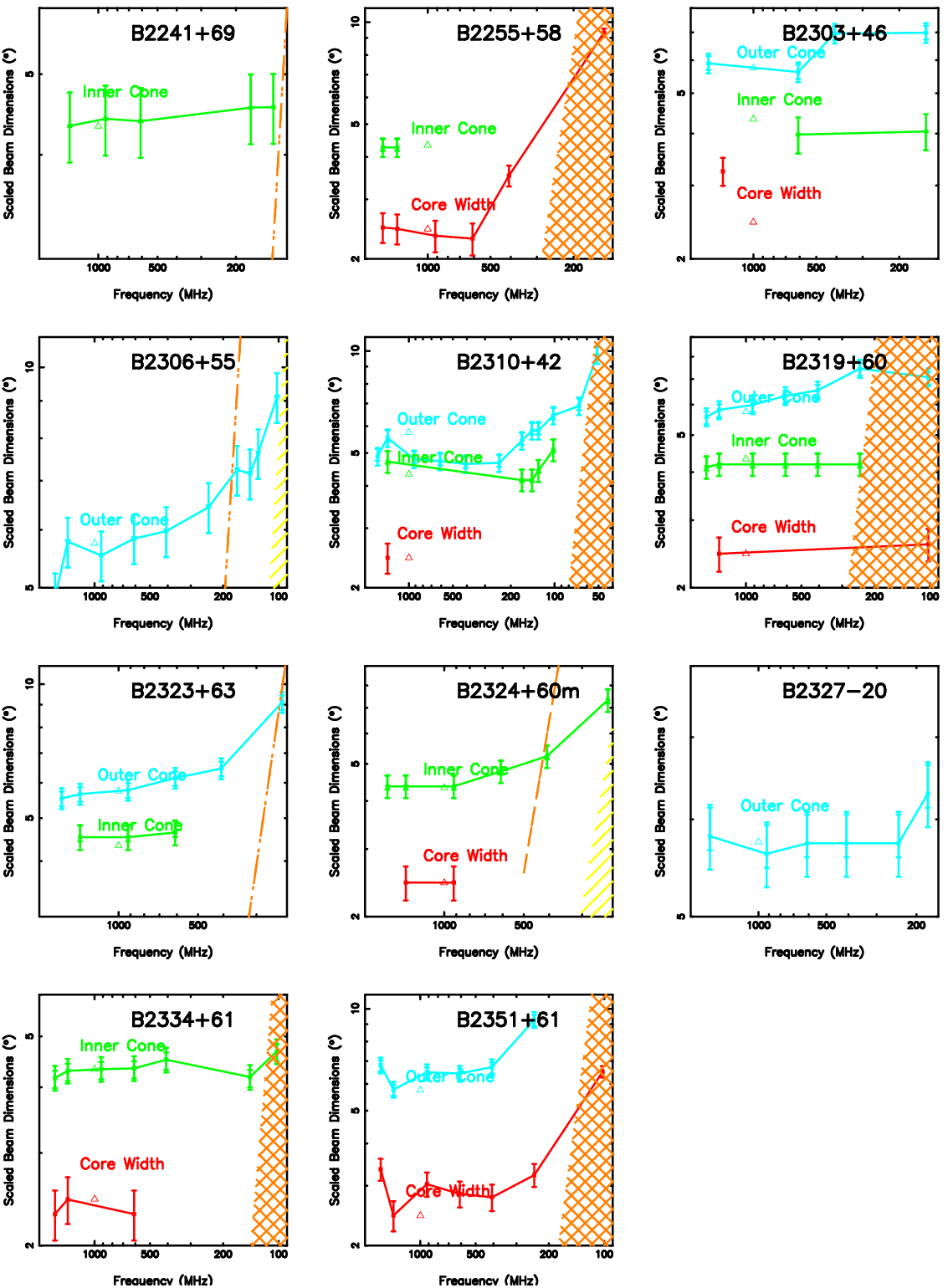}
\caption{Emission-beam geometry models as in Fig~\ref{figA1}.}
\label{figA15}
\end{center}
\end{figure*}

\begin{center}
\setlength{\tabcolsep}{3pt}
\begin{longtable}{lc|cccc|cccc|ccccc|ccc}
\caption{Gould \& Lyne Population Emission Beam Model Geometry} \label{tabA3}  \\
  \hline
      Pulsar &  Class & $W_{c}$ & $\alpha$ & $R$ & $\beta$ &  $W_i$ & $\rho_i$ & $W_o$  & $\rho_o$ & $W_{c}$ & $W_i$ & $\rho_i$    & $W_o$  & $\rho_o$ & $W_{c}$ & $W_{i,o}$  & $\rho_{i,o}$ \\
  &   & (\degr) & (\degr) & (\degr/\degr) & (\degr) & (\degr) & (\degr) & (\degr) & (\degr) & (\degr) & (\degr) & (\degr) & (\degr) & (\degr) & (\degr) & (\degr) & (\degr) \\
  \hline
  & & \multicolumn{4}{c|}{1-GHz Geometry} & \multicolumn{4}{c|}{1-GHz Cone Sizes} & \multicolumn{5}{c|}{100-MHz Cone Sizes} & \multicolumn{3}{c}{$<$100 MHz} \\
  \hline
  \hline
\endfirsthead
   \hline
      Pulsar &  Class & $W_{c}$ & $\alpha$ & $R$ & $\beta$ &  $W_i$ & $\rho_i$ & $W_o$  & $\rho_o$ & $W_{c}$ & $W_i$ & $\rho_i$    & $W_o$  & $\rho_o$ & $W_{c}$ & $W_{i,o}$  & $\rho_{i,o}$ \\
  &   & (\degr) & (\degr) & (\degr/\degr) & (\degr) & (\degr) & (\degr) & (\degr) & (\degr) & (\degr) & (\degr) & (\degr) & (\degr) & (\degr) & (\degr) & (\degr) & (\degr) \\
  \hline
  & & \multicolumn{4}{c|}{1-GHz Geometry} & \multicolumn{4}{c|}{1-GHz Cone Sizes} & \multicolumn{5}{c|}{100-MHz Cone Sizes} & \multicolumn{3}{c}{$<$100 MHz} \\
  \hline
  \hline
\endhead
  B0011+47 & cT? &  --- & 6.7 & -1.8 & +3.7 & 19.6 & 4.0 & 48 & 5.1 &  --- & 27 & 4.2 &  --- &  --- &  --- & 23 & 4.1 \\
B0031-07 & Sd &  --- & 6 & -1.0 & +6.0 &  --- &  --- & 17.0 & 6.1 &  --- &  --- &  --- & 34 & 6.5 &  --- & 39 & 6.6 \\
B0037+56 & Sd? &  --- & 90 & +18 & +3.2 & 3.7 & 3.7 &  --- &  --- &  --- & 19 & 10.0 &  --- &  --- &  --- &  --- &  --- \\
B0052+51 & D/T? &  --- & 50 & $\infty$ & 0.0 &  --- &  --- & 10.1 & 3.9 &  --- &  --- &  --- & 15.5 & 5.9 &  --- &  --- &  --- \\
B0053+47 & St? & $\sim$8 & {\bf 26} &  --- &  --- &  --- &  --- &  --- &  --- & 14.9 &  --- &  --- &  --- &  --- & 88 &  --- &  --- \\
\\[-3pt]
B0059+65 & T & $\approx$4? & {\bf 28} & -16 & +1.7 &  --- &  --- & 16.9 & 4.4 &  --- &  --- &  --- & 21.8 & 5.5 &  --- &  --- &  --- \\
B0105+65 & Sd &  --- & 18 & -5 & +3.5 & 7.5 & 3.8 &  --- &  --- &  --- & 7.7 & 3.8 &  --- &  --- &  --- &  --- &  --- \\
B0105+68 & T & $\sim$5.7? & {\bf 25} & $\infty$ & 0.0 &  --- &  --- & 26.7 & 5.5 & $\sim$6 &  --- &  --- & $\sim$33 & 6.8 &  --- &  --- &  --- \\
B0114+58 & St & 11.7 & {\bf 41} & +1.2 &  --- &  --- &  --- &  --- &  --- & 29 &  --- &  --- &  --- &  --- &  --- &  --- &  --- \\
B0136+57 & St & 7 & {\bf 42} & +5.3 & -7.3 & $\sim$10 & 7.9 &  --- &  --- &  --- & $\sim$18 & 9.2 &  --- &  --- &  --- &  --- &  --- \\
\\[-3pt]
B0138+59 & cQ? &  --- & 20 & -9 & +2.2 & $\sim$18? & 3.9 & 27 & 5.3 &  --- &  --- &  --- & 37.3 & 7.0 &  --- & 44 & 8.2 \\
B0144+59 & St? & $\approx$5? & 50 & $\approx$5 & +8.8 & $\sim$11 & 9.9 &  --- &  --- & 4.5 &  --- &  --- &  --- &  --- &  --- &  --- &  --- \\
B0148-06 & D &  --- & 14.5 & +7.4 & +1.9 &  --- &  --- & 32.0 & 4.7 &  --- &  --- &  --- & 91 & 12.0 &  --- &  --- &  --- \\
B0149-16 & T? & $\approx$3? & {\bf 84} & +30 & +1.9 & 8.2 & 4.5 &  --- &  --- &  --- & 23 & 11.6 &  --- &  --- &  --- &  --- &  --- \\
B0153+39 & Sd/D? &  --- & 16.5 & $\approx$-4 & +4.1 &  --- &  --- & $\sim$8 & 4.3 &  --- &  --- &  --- & 39 & 7.4 &  --- &  --- &  --- \\
\\[-3pt]
B0154+61 & St & 6.0 & {\bf 15} & -7? & +2.2 & $\sim$12? & 2.8 &  --- &  --- & 7.4 &  --- &  --- &  --- &  --- &  --- &  --- &  --- \\
B0226+70 & T & $\approx$3? & {\bf 40} & $\sim$9? & +4.1 &  --- &  --- & 7.1 & 4.7 & 2.1 &  --- &  --- & 9.1 & 5.1 &  --- & 16 & 6.7 \\
B0320+39 & Sd &  --- & 38 & +23 & +1.5 &  --- &  --- & 9.4 & 3.3 &  --- &  --- &  --- & 7.7 & 2.9 &  --- & 12 & 4.1 \\
B0329+54 & T/M? & 5.8 & {\bf 30} & -13.5 & +2.1 &  --- &  --- & 24.9 & 6.7 & 3.4 &  --- &  --- & 36.5 & 9.6 & 9.1 &  --- &  --- \\
B0331+45 & Sd &  --- & 33 & $\approx$+4 & +7.8 & 9.2 & 8.3 &  --- &  --- &  --- & 8.3 & 8.2 &  --- &  --- &  --- &  --- &  --- \\
\\[-3pt]
B0339+53 & Sd &  --- & 64 & -16 & +3.2 &  --- &  --- & 5.7 & 4.1 &  --- &  --- &  --- & 9.8 & 5.5 &  --- &  --- &  --- \\
B0353+52 & St & $\sim$9.4 & {\bf 36} &  --- &  --- &  --- &  --- &  --- &  --- & 12.4 &  --- &  --- &  --- &  --- &  --- &  --- &  --- \\
B0355+54 & St/T & $\sim$8.0 & {\bf 51} & -10 & +4.4 & 25 & 10.9 &  --- &  --- & 10 &  --- &  --- &  --- &  --- &  --- &  --- &  --- \\
B0402+61 & T/M? & $\approx$3? & {\bf 83} & +26 & +2.2 & $\sim$10? & 5.4 & 14 & 7.3 &  --- &  --- &  --- & 35 & 17.5 &  --- &  --- &  --- \\
B0410+69 & Sd &  --- & 64 & $\approx$-9 & +5.7 & 8.4 & 6.9 &  --- &  --- &  --- & 6.8 & 6.5 &  --- &  --- &  --- &  --- &  --- \\
\\[-3pt]
B0447-12 & T & $\sim$6.3? & {\bf 36} & $\infty$ & 0.0 & 22.5 & 6.6 &  --- &  --- & 7.8 & 27.3 & 8.0 &  --- &  --- &  --- &  --- &  --- \\
B0450-18 & T & $\sim$8 & {\bf 24} & +6 & +4.0 & 20 & 5.9 &  --- &  --- & $\sim$8 & 22 & 6.3 &  --- &  --- &  --- &  --- &  --- \\
B0450+55 & T & $\sim$9 & {\bf 28} & -9 & +3.0 & 29 & 7.7 &  --- &  --- & 9.9 & 33 & 8.6 &  --- &  --- & 9.8 & 38 & 9.7 \\
B0458+46 & T? & $\approx$6? & {\bf 31} & -7 & +4.2 & $\sim$12 & 5.3 &  --- &  --- &  --- &  --- &  --- &  --- &  --- &  --- &  --- &  --- \\
B0559-05 & T? & $\sim$10? & {\bf 23} & +3 & +7.5 &  --- &  --- & $\sim$22? & 8.9 & $\sim$10? &  --- &  --- & 22 & 8.9 &  --- &  --- &  --- \\
\\[-3pt]
B0621-04 & cQ &  --- & 31 & -60 & +0.5 & $\sim$16? & 4.2 & 21.4 & 5.6 &  --- &  --- &  --- & 28 & 7.3 &  --- &  --- &  --- \\
B0628-28 & Sd & --- & 14 & -4.2 & -3.3 & 18.9 & 3.9 &  --- &  --- &  --- & 20.5 & 4.0 &  --- &  --- &  --- & 30 & 4.6 \\
B0643+80 & cQ &  --- & 29 & +6 & +4.6 &  --- &  --- & 9.4 & 5.2 &  --- & 6.3 & 4.9 & 13.2 & 5.8 &  --- &  --- &  --- \\
B0655+64 & D &  --- & 90 & -7 & +8.2 & 10.6 & 9.8 &  --- &  --- &  --- & 10.2 & 9.7 &  --- &  --- &  --- & 8.6 & 9.3 \\
B0727-18 & T? & $\sim$4? & {\bf 70} & +12 & +4.5 &  --- &  --- & 13.9 & 8.0 &  --- &  --- &  --- & 20 & 10.5 &  --- &  --- &  --- \\
\\[-3pt]
B0740-28 & St & $\sim$10? & {\bf 37} & -3.5 & +9.9 & $\sim$12? & 10.6 &  --- &  --- & 16 &  --- &  --- &  --- &  --- &  --- &  --- &  --- \\
B0756-15 & Sd &  --- & 27 & $\sim$5 & +5.2 & 4.2 & 5.3 &  --- &  --- &  --- & 12.3 & 6.0 &  --- &  --- &  --- &  --- &  --- \\
B0809+74 & Sd &  --- & 9 & -2 & +4.5 &  --- &  --- & 21.5 & 4.9 &  --- &  --- &  --- & 17.9 & 4.8 &  --- & 57 & 7.0 \\
B0818-13 & Sd &  --- & 11.5 & +3 & +3.8 & 6.3 & 3.9 &  --- &  --- &  --- & 3.8 & 3.8 &  --- &  --- &  --- &  --- &  --- \\
B0826-34 & M? & 40 & {\bf 3} & +1.5 & +1.8 & $\sim$100? & 3.3 & $\sim$140 & 4.2 &  --- &  --- &  --- &  --- &  --- &  --- &  --- &  --- \\
\\[-3pt]
B0841+80 & cQ &  --- & 27.5 & $\infty$ & 0.0 &  --- &  --- & 19.5 &  --- &  --- &  --- &  --- & $\sim$20 & 4.6 &  --- &  --- &  --- \\
B0844-35 & St & $\sim$7? & {\bf 19} & -27 & +0.7 & $\sim$24 & 4.1 &  --- &  --- & 4.3 &  --- &  --- &  --- &  --- &  --- &  --- &  --- \\
B0853-33 & D & --- & 41 & +12 & +3.1 & 6.4 & 3.8 &  --- &  --- &  --- & 7 & 3.9 &  --- &  --- &  --- &  --- &  --- \\
B0906-17 & St/T? & $\sim$9 & {\bf 25} & -4.5 & +5.5 & 18.3 & 7.0 &  --- &  --- & 8.3 &  --- &  --- &  --- &  --- &  --- &  --- &  --- \\
B0917+63 & D &  --- & 13.5 & +3 & +4.5 &  --- &  --- & 8.1 & 4.6 &  --- &  --- &  --- & 12.2 & 4.8 &  --- & 15.2 & 4.9 \\
\\[-3pt]
B0942-13 & Sd &  --- & 55 & +9 & +5.2 & 4.9 & 5.6 &  --- &  --- &  --- & 8.8 & 6.4 &  --- &  --- &  --- &  --- &  --- \\
B1010-23 & Sd? & --- & 16 & $\approx$6? & +2.6 & $\sim$5 & 2.7 &  --- &  --- &  --- & 7.3 & 2.8 &  --- &  --- &  --- &  --- &  --- \\
B1016-16 & Sd &  --- & 28 & -11 & +2.4 & +8.3 & 3.2 &  --- &  --- &  --- & $\sim$10 & 3.5 &  --- &  --- &  --- &  --- &  --- \\
B1039-19 & M & $\sim$4? & {\bf 31} & -18 & +1.7 & $\sim$13? & 3.8 & 17 & 4.8 &  --- &  --- &  --- & 20 & 5.6 &  --- &  --- &  --- \\
B1112+50 & Sd &  --- & 34 & -11 & +2.9 & 5.9 & 3.4 &  --- &  --- &  --- & 6.2 & 3.4 &  --- &  --- &  --- & 39 & 11.6 \\
\\[-10pt]
B1254-10 & T & 3.9 & {\bf 53} & -15 & +3.1 &  --- &  --- & 16.2 & 7.3 & $\sim$15 &  --- &  --- &  --- &  --- &  --- &  --- &  --- \\
B1309-12 & Sd &  --- & 15 & -2.3 & +6.5 & 5.2 & 6.5 &  --- &  --- &  --- & $\sim$22 & 7.3 &  --- &  --- &  --- &  --- &  --- \\
B1322+83 & D? &  --- & 11 & +2.8 & +3.9 &  --- &  --- & $\sim$52 & 6.9 &  --- &  --- &  --- & $\sim$62 & 7.8 &  --- & 19.3 & 4.5 \\
B1508+55 & T & $\sim$4 & {\bf 45} & -15 & -2.7 & 11 & 4.7 &  --- &  --- & 5.6 & 15.8 & 6.1 &  --- &  --- & 8.7 &  --- &  --- \\
B1540-06 & Sd &  --- & 38 & -7 & +5.0 & 4.0 & 5.2 &  --- &  --- &  --- & 4.8 & 5.3 &  --- &  --- &  --- & 7.4 & 5.6 \\
\\[-1pt]
B1552-31 & cQ/M? & $\sim$4? & {\bf 52} & $\infty$ & $\sim$0 & 15.3 & 6.1 & 20.2 & 8.0 &  --- &  --- &  --- & 19.4 & 7.7 &  --- &  --- &  --- \\
B1552-23 & T & $\sim$8? & {\bf 25} & +5 & +4.8 & $\sim$16 & 6.0 &  --- &  --- & 8 & 17 & 6.2 &  --- &  --- &  --- &  --- &  --- \\
B1600-27 & T & $\sim$3.5 & {\bf 53} & -30 & +1.5 & $\sim$12 & 5.0 &  --- &  --- &  --- & 12 & 5.0 &  --- &  --- &  --- &  --- &  --- \\
B1607-13 & cT & cT/D? & 22 & +9 & +2.4 & $\sim$18 & 4.3 & $\sim$26 & 5.6 &  --- &  --- &  --- & $\approx$25? & 5.5 &  --- &  --- &  --- \\
B1612-29 & Sd? & --- & 20 & $\approx$6? & +3.3 &  --- &  --- & $\sim$8 & 3.6 &  --- & 8 & 3.6 &  --- &  --- &  --- &  --- &  --- \\
\\[-1pt]
B1620-09 & Sd &  --- & 65 & $\sim$15 & +3.5 & 3.4 & 3.8 &  --- &  --- &  --- & 4.2 & 4.0 &  --- &  --- &  --- &  --- &  --- \\
B1620-26 & St & 23.5 & {\bf 82} & -2 & +29.7 & $\sim$60 & 41.8 &  --- &  --- & 95 &  --- &  --- &  --- &  --- &  --- &  --- &  --- \\
B1642-03 & St & 4.2 & 70 & -50 & -1.1 & 14.4 & 6.8 &  --- &  --- & 5.3 &  --- &  --- &  --- &  --- &  --- &  --- &  --- \\
B1648-17 & D & --- & 34 & -11 & +2.9 & 11.5 & 4.4 &  --- &  --- &  --- & 11.5 & 4.4 &  --- &  --- &  --- &  --- &  --- \\
B1649-23 & cT/D? & --- & 60 & +100 & +0.5 & 7.6 & 3.3 &  --- &  --- &  --- & 7.5 & 3.3 &  --- &  --- &  --- &  --- &  --- \\
\\[-1pt]
B1657-13 & cT? & --- & 12 & -2.2 & +5.4 & $\sim$10 & 5.6 & $\sim$38 & 7.2 &  --- & $\sim$14 & 5.7 & $\sim$39 & 7.3 &  --- &  --- &  --- \\
B1700-32 & T & $\sim$3? & {\bf 50} & +24 & +1.8 &  --- &  --- & 12.3 & 5.1 &  --- &  --- &  --- & 14.7 & 6.0 &  --- &  --- &  --- \\
B1700-18 & Sd &  --- & 25 & -5.6 & +4.3 & 8.8 & 4.8 &  --- &  --- &  --- & $\sim$21 & 6.4 &  --- &  --- &  --- &  --- &  --- \\
B1702-19 & T & 4.5 & {\bf $\sim$90} & -14 & +4.1 & $\sim$14 & 8.1 &  --- &  --- &  --- & $\sim$16 & 9.0 & 0 & 0 &  --- &  --- &  --- \\
B1706-16 & T & $\sim$4? & {\bf 49} & -9 & -4.8 & $\sim$7? & 5.5 &  --- &  --- & 6.8 &  --- &  --- &  --- &  --- &  --- &  --- &  --- \\
\\[-1pt]
B1709-15 & Sd? & --- & 14? & $\approx$3? & +4.6 & 5.1 & 4.7 &  --- &  --- &  --- & $\sim$11 & 4.9 &  --- &  --- &  --- &  --- &  --- \\
B1714-34 & St?? & --- &  --- &  --- &  --- &  --- &  --- &  --- &  --- &  --- &  --- &  --- &  --- &  --- &  --- &  --- &  --- \\
B1717-16 & Sd & --- & 45 & $\sim$23 & +1.8 & 8.3 & 3.5 &  --- &  --- &  --- & $\sim$9 & 3.7 &  --- &  --- &  --- &  --- &  --- \\
B1717-29 & cQ & --- & 34.6 & -7 & +4.7 & $\sim$10 & 5.5 & $\sim$19 & 7.4 &  --- & 10 & 5.5 & 20 & 7.6 &  --- &  --- &  --- \\
B1718-02 & Sd/cT? & --- & 16 & -2.6 & +6.0 &  --- &  --- & $\sim$37 & 8.4 &  --- &  --- &  --- & $\sim$107 & 17.5 &  --- &  --- &  --- \\
\\[-1pt]
B1718-19 & Sd & --- & 60 & $\sim$12 & +4.1 & $\sim$3 & 4.3 &  --- &  --- &  --- & 7.5 & 5.3 &  --- &  --- &  --- &  --- &  --- \\
B1718-35 & St? & $\sim$9? & {\bf 31} & -6? & +4.9 &  --- &  --- &  --- &  --- &  --- &  --- &  --- &  --- &  --- &  --- &  --- &  --- \\
B1718-32 & T? & 5.6 & {\bf 39} & -15 & +2.4 & $\sim$18 & 6.3 &  --- &  --- &  --- & 16 & 5.7 &  --- &  --- &  --- &  --- &  --- \\
B1727-33 & St? & --- &  --- &  --- &  --- &  --- &  --- &  --- &  --- &  --- &  --- &  --- &  --- &  --- &  --- &  --- &  --- \\
B1730-22 & cT/cQ? & --- & 17 & $\approx$+4 & +4.4 &  --- &  --- & $\sim$26 & 6.1 &  --- &  --- &  --- & 27 & 6.2 &  --- &  --- &  --- \\
\\[-1pt]
B1732-02 & Sd & --- & 36 & $\sim$9 & +3.7 & $\sim$10 & 4.8 &  --- &  --- &  --- & $\sim$36 & 11.6 &  --- &  --- &  --- &  --- &  --- \\
B1732-07 & T? & 4.7 & {\bf 54} & $\infty$ & 0.0 & 17 & 6.8 &  --- &  --- & 6.5 &  --- &  --- &  --- &  --- &  --- &  --- &  --- \\
B1734-35 & St? & --- &  --- &  --- &  --- &  --- &  --- &  --- &  --- &  --- &  --- &  --- &  --- &  --- &  --- &  --- &  --- \\
B1735-32 & Sd? & --- & 22 & +4.5 & -4.8 & 6.6 & 5.0 &  --- &  --- &  --- & 8.9 & 5.1 &  --- &  --- &  --- &  --- &  --- \\
B1736-31 & St? & --- &  --- &  --- &  --- &  --- &  --- &  --- &  --- & 293.5 &  --- &  --- &  --- &  --- &  --- &  --- &  --- \\
\\[-1pt]
B1736-29 & St/T? & $\sim$4.3? & {\bf 90} & +8 & +7.0 & 6.6 & 7.7 &  --- &  --- & --- & --- & --- & --- & --- &  --- &  --- &  --- \\
B1737-30 & St & 3.7 & {\bf 58} & -4.5 & +10.9 &  --- &  --- &  --- &  --- & 24.4 &  --- &  --- &  --- &  --- &  --- &  --- &  --- \\
B1738-08 & cQ & --- & 26 & +15 & +1.7 & $\sim$11 & 3.0 & 16 & 4.0 &  --- & 11 & 3.0 & 17.2 & 4.2 &  --- &  --- &  --- \\
B1740-13 & T? & $\sim$8 & {\bf 27} & $\sim$7 & +3.8 & $\sim$22 & 6.5 &  --- &  --- &  --- & $\sim$15 & 5.2 &  --- &  --- &  --- &  --- &  --- \\
B1740-31 & Sd? & --- & 37 & +17 & +2.0 & 6.4 & 2.8 &  --- &  --- &  --- & 28 & 8.8 &  --- &  --- &  --- &  --- &  --- \\
\\[-1pt]
B1740-03 & T & $\sim$4 & {\bf 74} & +7.5 & +7.4 & 10.7 &  --- & 10.7 & 9.0 &  --- &  --- &  --- & $\sim$23 & 13.4 &  --- &  --- &  --- \\
B1742-30 & T/M? & $\sim$10? & {\bf 24} & -3.6 & 6.4 & $\sim$15 & 7.3 & 32 & 9.7 & $\sim$10 &  --- &  --- & 32 & 9.7 &  --- &  --- &  --- \\
B1745-12 & cQ? & $\sim$5? & 69 & -15 & +3.6 & 13.0 & 7.1 & 18 & 9.2 &  --- &  --- &  --- &  --- &  --- &  --- &  --- &  --- \\
B1746-30 & T/M? & $\sim$13 & {\bf 14} & -3.6 & +4.0 & $\sim$29 & 5.7 & $\sim$45 & 7.4 &  --- &  --- &  --- & 50 & 8.0 &  --- &  --- &  --- \\
B1747-31 & cT/T? & $\sim$7 & {\bf 22} & +20 & +1.1 &  --- &  --- & 31.5 & 6.0 &  --- &  --- &  --- & 35.7 & 6.8 &  --- &  --- &  --- \\
\\[-1pt]
B1749-28 & St & 4.9 & {\bf 42} & +13 & +2.9 & 13 & 5.3 &  --- &  --- & 6.3 &  --- &  --- &  --- &  --- &  --- &  --- &  --- \\
B1750-24 & St? & --- &  --- &  --- &  --- &  --- &  --- &  --- &  --- &  --- &  --- &  --- &  --- &  --- &  --- &  --- &  --- \\
B1753+52 & cQ? & --- & 19.5 & +9 & +2.1 &  --- &  --- & 16.9 & 3.6 &  --- &  --- &  --- & 25.3 & 4.9 &  --- &  --- &  --- \\
B1753-24 & cT & --- & 33 & +7 & +4.5 & $\sim$10 & 5.3 & $\sim$19 & 7.1 &  --- &  --- &  --- & 41.7 & 12.8 &  --- &  --- &  --- \\
B1754-24 & T & $\sim$7 & {\bf 46} & $\infty$ & 0.0 & 25.5 & 9.2 &  --- &  --- &  --- & 52.1 & 18.8 &  --- &  --- &  --- &  --- &  --- \\
\\[-10pt]
B1756-22 & St/T? & 3.6 & {\bf 90} & $\approx$-15? & +3.8 & $\sim$10 & 6.3 &  --- &  --- & 9.1 &  --- &  --- &  --- &  --- &  --- &  --- &  --- \\
B1757-24 & St? & 14.4 & {\bf 29} & -4.5 & +6.1 &  --- &  --- &  --- &  --- & 55.3 &  --- &  --- &  --- &  --- &  --- &  --- &  --- \\
B1758-23 & St? & $\sim$13 & {\bf 18} &  --- &  --- &  --- &  --- &  --- &  --- &  --- &  --- &  --- &  --- &  --- &  --- &  --- &  --- \\
B1758-03 & T & 4.2 & {\bf 37} & +12 & +2.9 & 11 & 4.5 &  --- &  --- & 4.4 &  --- &  --- &  --- &  --- &  --- &  --- &  --- \\
B1800-21 & T & $\sim$29? & {\bf 13} & -2 & +6.6 & 0 & 15.8 & $\sim$105 & 15.8 &  --- &  --- &  --- & 142 & 19.9 &  --- &  --- &  --- \\
\\[-1pt]
B1802-07 & St?? & 38.5 & {\bf 25} &  --- &  --- &  --- &  --- &  --- &  --- &  --- &  --- &  --- &  --- &  --- &  --- &  --- &  --- \\
B1804-27 & St? & $\sim$6 & {\bf 27} &  --- &  --- &  --- &  --- &  --- &  --- & 6.9 &  --- &  --- &  --- &  --- &  --- &  --- &  --- \\
B1804-08 & T/M? & $\sim$6.5 & {\bf 69} & -17 & +3.1 & $\sim$22 & 10.8 & 29.0 & 14.0 & $\sim$80 &  --- &  --- &  --- &  --- &  --- &  --- &  --- \\
B1805-20 & Sd/D? & --- & 13 & -3 & +4.3 & $\sim$12 & 4.6 &  --- &  --- &  --- & 79.9 & 11.0 &  --- &  --- &  --- &  --- &  --- \\
B1806-21 & St/Sd? & 5.5 & {\bf 32} &  --- &  --- &  --- &  --- &  --- &  --- & 15.9 &  --- &  --- &  --- &  --- &  --- &  --- &  --- \\
\\[-1pt]
B1809-173 & ?? & --- &  --- &  --- &  --- &  --- &  --- &  --- &  --- &  --- &  --- &  --- &  --- &  --- &  --- &  --- &  --- \\
B1809-176 & Sd/D? & --- &  --- &  --- &  --- &  --- &  --- &  --- &  --- &  --- &  --- &  --- &  --- &  --- &  --- &  --- &  --- \\
B1811+40 & cQ? & --- & 58 & +18 & +2.7 & 8.4 & 4.5 & 12.2 & 5.9 &  --- &  --- &  --- & 12.3 & 5.9 &  --- & 19.3 & 8.7 \\
B1813-17 & St? & 6.9 & {\bf 24} &  --- &  --- &  --- &  --- &  --- &  --- & 67.4 &  --- &  --- &  --- &  --- &  --- &  --- &  --- \\
B1813-26 & cT/cQ? & --- & 22 & -4 & +5.3 & $\sim$10 & 5.6 & $\sim$26 & 7.4 &  --- &  --- &  --- & 32.7 & 8.5 &  --- &  --- &  --- \\
\\[-1pt]
B1813-36 & T? & 6.1 & {\bf 40} & +9 & +4.1 & $\sim$17 & 7.0 &  --- &  --- & 6.1 & 17 & 7.0 &  --- &  --- &  --- &  --- &  --- \\
B1815-14 & St? & $\sim$9 & {\bf 30} &  --- &  --- &  --- &  --- &  --- &  --- & 175 &  --- &  --- &  --- &  --- &  --- &  --- &  --- \\
B1817-13 & St/Sd? & $\sim$5 & {\bf 31} &  --- &  --- &  --- &  --- &  --- &  --- & $\sim$245 &  --- &  --- &  --- &  --- &  --- &  --- &  --- \\
B1817-18 & Sd? & --- & 32 & $\sim$5? & +6.1 & $\sim$17 & 7.8 &  --- &  --- &  --- &  --- &  --- &  --- &  --- &  --- &  --- &  --- \\
B1818-04m & St/T & $\sim$5.5? & {\bf 35} & +9? & +3.7 & 14.4 & 5.7 &  --- &  --- & 15 &  --- &  --- &  --- &  --- &  --- &  --- &  --- \\
\\[-1pt]
B1819-22 & Sd & --- & 16 & +4 & +4.0 &  --- &  --- & $\sim$10 & 4.2 &  --- &  --- &  --- & 15.5 & 4.6 &  --- &  --- &  --- \\
B1820-14 & St? & 14.2 & {\bf 22} &  --- &  --- &  --- &  --- &  --- &  --- & $\sim$71 &  --- &  --- &  --- &  --- &  --- &  --- &  --- \\
B1820-11 & St/T? & $\sim$27? & {\bf 10} & -1.4 & +7.0 & $\sim$40 & 8.3 &  --- &  --- &  --- & 136 & 16.0 &  --- &  --- &  --- &  --- &  --- \\
B1820-30B & Sd? & --- & 37 & $\approx$-5? & +6.9 & $\sim$5 & 7.1 &  --- &  --- &  --- & 5.6 & 7.1 &  --- &  --- &  --- &  --- &  --- \\
B1820-31 & St? & 6.5 & {\bf 45} &  --- &  --- &  --- &  --- &  --- &  --- & 8.1 &  --- &  --- &  --- &  --- &  --- &  --- &  --- \\
\\[-1pt]
B1821-19 & St/T? & 5.1 & {\bf 90} & -25 & +2.3 & $\sim$19 & 10.0 &  --- &  --- & 47.6 &  --- &  --- &  --- &  --- &  --- &  --- &  --- \\
B1821-11 & St? & $\sim$12 & {\bf 18} &  --- &  --- &  --- &  --- &  --- &  --- & 99 &  --- &  --- &  --- &  --- &  --- &  --- &  --- \\
B1822-14 & St?? & $\sim$4.6? & {\bf 90} & $\infty$ & +0.0 &  --- &  --- &  --- &  --- &  --- &  --- &  --- &  --- &  --- &  --- &  --- &  --- \\
B1822-09m & St & $\sim$3 & {\bf $\sim$90} & $\infty$ & $\sim$0 & $\sim$10 & 4.9 &  --- &  --- & 5.7 &  --- &  --- &  --- &  --- & 11.3 &  --- &  --- \\
B1823-11 & D &  --- & 16.4 & $\sim$7 & +2.3 & 12.3 & 3.0 &  --- &  --- &  --- &  --- &  --- &  --- &  --- &  --- &  --- &  --- \\
\\[-1pt]
B1823-13 & D?? & --- & 18 & -6 & +3.0 &  --- &  --- & 112 & 18.2 &  --- &  --- &  --- &  --- &  --- &  --- &  --- &  --- \\
B1824-10 & St? & 36 & {\bf 8} &  --- &  --- &  --- &  --- &  --- &  --- & 143.4 &  --- &  --- &  --- &  --- &  --- &  --- &  --- \\
B1826-17 & T & $\sim$5? & {\bf 62} & $\infty$ & 0.0 & $\sim$18 & 8.0 &  --- &  --- & 59.3 &  --- &  --- &  --- &  --- &  --- &  --- &  --- \\
B1828-11 & St & $\sim$3.8 & {\bf 90} &  --- &  --- &  --- &  --- &  --- &  --- & 8.9 &  --- &  --- &  --- &  --- &  --- &  --- &  --- \\
B1829-08 & T? & 4.5 & {\bf 43} & -7.5 & -5.2 &  --- &  --- & $\sim$16 & 7.3 & 16.9 &  --- &  --- &  --- &  --- &  --- &  --- &  --- \\
\\[-1pt]
B1829-10 & St? & $\sim$9? & {\bf 28} & +7 & +3.9 &  --- &  --- &  --- &  --- & 186.9 &  --- &  --- &  --- &  --- &  --- &  --- &  --- \\
B1830-08 & St/T? & 22 & {\bf 22} & +1.8 & -12.2 & $\sim$70 & 15.2 &  --- &  --- & 99 &  --- &  --- &  --- &  --- &  --- &  --- &  --- \\
B1831-03 & T & 3.6 & {\bf 55} & -90 & +0.5 & 12.8 & 5.3 &  --- &  --- &  --- &  --- &  --- &  --- &  --- &  --- &  --- &  --- \\
B1831-04 & M & $\sim$26? & {\bf 10} & -5 & +2.0 & $\sim$85? & 8.2 & 113 & 10.6 & 32 &  --- &  --- &  --- &  --- &  --- &  --- &  --- \\
B1832-06 & St? & $\sim$11? & {\bf 24} & -9 & +2.6 & $\sim$36 & 8.0 &  --- &  --- &  --- &  --- &  --- &  --- &  --- &  --- &  --- &  --- \\
\\[-1pt]
B1834-04 & T?? & $\sim$4? & {\bf $\sim$90} & -20 & +2.9 & $\sim$13 & 7.1 &  --- &  --- &  --- &  --- &  --- &  --- &  --- &  --- &  --- &  --- \\
B1834-10 & St & 5.7 & {\bf 35} &  --- &  --- &  --- &  --- &  --- &  --- & 62.7 &  --- &  --- &  --- &  --- &  --- &  --- &  --- \\
B1834-06 & D/cQ? & --- & 16 & $\infty$ & 0.0 & 23.4 & 3.2 &  --- &  --- &  --- & 33.4 & 4.6 &  --- &  --- &  --- &  --- &  --- \\
B1839+56 & M & $\sim$3? & 40 & +12 & +3.0 &  --- &  --- & 9.9 & 4.4 & 2.6 & 9.1 & 4.3 &  --- &  --- & 2.8 &  --- &  --- \\
B1838-04 & St & $\sim$11? & {\bf 30} & -3 & +9.6 & $\sim$12 & 10.3 &  --- &  --- & $\sim$10 &  --- &  --- &  --- &  --- &  --- &  --- &  --- \\
\\[-1pt]
B1839-04 & D &  --- & 7.5 & +9 & +0.8 &  --- &  --- & $\sim$60 & 4.2 &  --- &  --- &  --- &  --- &  --- &  --- &  --- &  --- \\
B1841-05 & St & 11.4 & {\bf 25} &  --- &  --- &  --- &  --- &  --- &  --- &  --- &  --- &  --- &  --- &  --- &  --- &  --- &  --- \\
B1841-04 & Sd & --- & 23 & -4 & +5.7 &  --- &  --- & 4.9 & 5.8 &  --- &  --- &  --- & 11.7 & 6.3 &  --- &  --- &  --- \\
B1842-02 & ?? & --- &  --- &  --- &  --- &  --- &  --- &  --- &  --- &  --- &  --- &  --- &  --- &  --- &  --- &  --- &  --- \\
B1842-04 & T? & $\sim$5? & {\bf 50} & +14 & +3.1 & $\sim$14 & 6.3 &  --- &  --- &  --- & 25 & 10.2 &  --- &  --- &  --- &  --- &  --- \\
\\[-10pt]
B1844-04 & St/T? & $\sim$6 & {\bf 32} & 5.5 & +5.5 &  --- &  --- & $\sim$17 & 7.3 & $\sim$13 &  --- &  --- &  --- &  --- &  --- &  --- &  --- \\
B1845-19 & T? & $\sim$4 & {\bf 17} & -9 & +1.9 & $\sim$5 & 2.0 &  --- &  --- & $\approx$5? & 5.6 & 2.1 &  --- &  --- &  --- &  --- &  --- \\
B1845-01 & cT & --- & 39 & +8 & +4.5 & $\sim$9? & 5.4 & 16 & 6.9 &  --- & 24.9 & 9.4 &  --- &  --- &  --- &  --- &  --- \\
B1846-06 & St & 3.6 & {\bf 34} & $\infty$ & 0.0 & $\sim$13 & 3.7 &  --- &  --- & 14 &  --- &  --- &  --- &  --- &  --- &  --- &  --- \\
B1851-14 & Sd & --- & 34 & -7.8 & +4.1 &  --- &  --- & $\sim$12 & 5.4 &  --- &  --- &  --- & 12.7 & 5.6 &  --- &  --- &  --- \\
\\
B1857-26 & M & $\sim$7.5? & {\bf 25} & -11 & +2.2 & $\sim$24? & 5.6 & 33 & 7.5 & 8.1 &  --- &  --- &  --- &  --- &  --- &  --- &  --- \\
B1900-06 & T & 6.1 & {\bf 38} & +6 & -5.8 & 10.9 & 6.6 &  --- &  --- & $\sim$16 & 36 & 11.8 &  --- &  --- &  --- &  --- &  --- \\
B1905+39 & M & $\approx$4? & {\bf 33} & -15 & +2.1 & $\sim$12? & 4.0 & 16.4 & 5.1 &  --- &  --- &  --- & 22.5 & 6.7 &  --- &  --- &  --- \\
B1907-03 & T & 6.2 & {\bf 34} & +5.5 & +5.8 &  --- &  --- & 18.7 & 8.0 &  --- &  --- &  --- &  --- &  --- &  --- &  --- &  --- \\
B1911-04 & St & $\sim$3.0 & {\bf 64} & -27 & -1.9 & 9.8 & 4.8 &  --- &  --- & 5.6 &  --- &  --- &  --- &  --- &  --- &  --- &  --- \\
\\
B1937-26 & T? & 3.9 & {\bf 82} & -12 & +4.7 & 10 & 6.9 & 10.0 &  --- & 4.2 &  --- &  --- &  --- &  --- &  --- &  --- &  --- \\
B1940-12 & Sd & 4.1 & 74 & -14 & +3.9 & 4.5 & 4.5 &  --- &  --- &  --- & 2.9 & 4.2 &  --- &  --- &  --- &  --- &  --- \\
B1941-17 & Sd? & --- & 38 & +9 & +3.9 & $\sim$9 & 4.9 &  --- &  --- &  --- & 8.5 & 4.8 &  --- &  --- &  --- &  --- &  --- \\
B1943-29 & T & 4.8 & {\bf 31} & -8 & +3.7 & $\sim$9 & 4.5 &  --- &  --- & 3.5 &  --- &  --- &  --- &  --- &  --- &  --- &  --- \\
B1946-25 & Sd & --- & 41 & +9 & +4.2 & $\sim$5 & 4.5 &  --- &  --- &  --- & 3.7 & 4.4 &  --- &  --- &  --- &  --- &  --- \\
\\
B1953+50 & Sd & --- & 90 & -10 & +5.7 & 4.2 & 6.1 &  --- &  --- &  --- &  --- &  --- &  --- &  --- & 8.7 &  --- &  --- \\
B2000+40 & cT? & --- & 29 & +7.2 & +3.9 & $\sim$9.4? & 4.6 & 18.0 & 6.0 &  --- &  --- &  --- & 75 & 19.4 &  --- &  --- &  --- \\
B2003-08 & M & $\sim$14 & {\bf 13} & +4 & +3.3 & $\sim$35? & 5.5 & 53 & 7.5 &  --- &  --- &  --- & 104 & 13.3 &  --- &  --- &  --- \\
B2011+38 & St & $\sim$24 & {\bf 12} & -1.2 & +10.2 &  --- &  --- &  --- &  --- & 39.2 &  --- &  --- &  --- &  --- &  --- &  --- &  --- \\
B2021+51 & D/Sd &  --- & 23 & +4 & +5.6 & 10.3 & 6.0 &  --- &  --- &  --- & 25.1 & 7.8 &  --- &  --- &  --- &  --- &  --- \\
\\
B2022+50m & St & 3.8 & {\bf 90} &  --- &  --- &  --- &  --- &  --- &  --- & 13 &  --- &  --- &  --- &  --- &  --- &  --- &  --- \\
B2036+53 & Sd &  --- & 53 & $\sim$15? & +3.1 & 4.9 & 3.6 &  --- &  --- &  --- & 15 & 6.8 &  --- &  --- &  --- &  --- &  --- \\
B2043-04 & Sd & --- & 58 & +18 & +2.7 & 5.1 & 3.5 &  --- &  --- &  --- & $\sim$7 & 4.0 &  --- &  --- &  --- &  --- &  --- \\
B2045+56 & St & 8.9 & {\bf 23} &  --- &  --- &  --- &  --- &  --- &  --- & 10.5 &  --- &  --- &  --- &  --- &  --- &  --- &  --- \\
B2045-16 & T & $\sim$3? & {\bf 34} & -26 & -1.2 &  --- &  --- & 14 & 4.1 &  --- &  --- &  --- & $\sim$18 & 5.0 &  --- &  --- &  --- \\
\\
B2106+44 & D/cQ? & --- & 11.5 & -1.8 & +6.4 & $\sim$20 & 6.8 &  --- &  --- &  --- & 75 & 11.1 &  --- &  --- &  --- &  --- &  --- \\
B2111+46 & T & $\sim$15? & {\bf 9} & -6.7 & +1.4 &  --- &  --- & 65.4 & 5.8 & 23.8 &  --- &  --- &  --- &  --- &  --- &  --- &  --- \\
B2148+63 & Sd & --- & 10.5 & +1.5 & +7.0 & 13.9 & 7.2 &  --- &  --- &  --- & 54 & 9.4 &  --- &  --- &  --- &  --- &  --- \\
B2148+52 & T & 4.7 & {\bf 65} & -11 & +4.7 &  --- &  --- & $\sim$19 & 9.9 &  --- & 43 & 20.3 & $\sim$43 & 20.3 &  --- &  --- &  --- \\
B2152-31 & D? & --- & 32 & -9 & +3.4 & $\sim$10 & 4.4 &  --- &  --- &  --- & 10 & 4.4 &  --- &  --- &  --- &  --- &  --- \\
\\
B2154+40 & cT? & --- & 20 & +8Ê & +2.5 & $\sim$13 & 3.4 & $\sim$22 & 4.7 &  --- & $\sim$12 & 3.4 & $\sim$40 & 7.6 &  --- &  --- &  --- \\
B2217+47 & St & 5.0 & {\bf 42} & +8.5 & +4.5 & 12 & 6.1 &  --- &  --- & 5 &  --- &  --- &  --- &  --- & 38 &  --- &  --- \\
B2224+65 & St wP & 11 & {\bf 16} & -4.5 & +3.4 &  --- &  --- &  --- &  --- & 11.7 &  --- &  --- &  --- &  --- & $\sim$17 &  --- &  --- \\
B2227+61 & T? & $\sim$6.5? & {\bf 34} & 5.4 & +6.0 &  --- &  --- & 21 & 8.9 &  --- &  --- &  --- & 51 & 16.5 &  --- &  --- &  --- \\
B2241+69 & Sd  &  --- & 15 & -4.5 & +3.3 & 4.5 & 3.4 &  --- &  --- &  --- & 9 & 3.5 &  --- &  --- &  --- &  --- &  --- \\
\\
B2255+58 & St & 10.8 & {\bf 22} & -4 & +5.4 & $\sim$22 & 7.0 &  --- &  --- & 41 &  --- &  --- &  --- &  --- &  --- &  --- &  --- \\
B2303+46 & cT? &  --- & 34 & $\sim$9 & +3.6 &  --- &  --- & 14.9 & 5.6 &  --- & 5.6 & 3.9 & 19.7 & 6.8 &  --- &  --- &  --- \\
B2306+55 & D &  --- & 54 & +25 & +1.9 &  --- &  --- & $\sim$20 & 8.4 &  --- &  --- &  --- & 32 & 13.2 &  --- & $\sim$65 & 26.5 \\
B2310+42 & M? & $\sim$5? & {\bf 56} & +7 & +6.8 & 9.7 & 8.0 & $\sim$15 & 9.4 &  --- & 12.4 & 8.6 & 19.9 & 10.9 &  --- & $\sim$35 & 16.5 \\
B2319+60 & cQ? & $\approx$5.2? & {\bf 18} & -8 & +2.2 & $\sim$10 & 2.8 & 19 & 3.9 & 5.5 &  --- &  --- & 25 & 4.7 &  --- &  --- &  --- \\
\\
B2323+63 & cT/cQ? & --- & 14.5 & +5 & +2.9 & $\sim$17 & 3.7 & $\sim$28 & 4.8 &  --- &  --- &  --- & 51.9 & 7.6 &  --- &  --- &  --- \\
B2324+60m & T? & $\sim$10.5? & {\bf 29} & +3.8 & +7.3 & $\sim$20 & 9.0 &  --- &  --- &  --- & 50 & 15.2 &  --- &  --- &  --- &  --- &  --- \\
B2327-20 & T & 2.2 & {\bf 60} & +15 & +3.3 &  --- &  --- & 6.8 & 4.5 &  --- &  --- &  --- & 8.2 & 4.9 &  --- & $\sim$10 & 5.5 \\
B2334+61 & St & $\sim$6? & {\bf 33} & -9 & +3.5 & 17.7 & 6.1 &  --- &  --- &  --- & $\sim$20 & 6.7 &  --- &  --- &  --- &  --- &  --- \\
B2351+61 & T? & 3.8 & {\bf 42} & -11 & +3.5 &  --- &  --- & 14.1 & 5.9 & 10.1 & $\sim$40 & 14.1 &  --- &  --- &  --- &  --- &  --- \\

  \hline
\end{longtable}
\end{center}

\newpage
\setcounter{figure}{0}
\renewcommand{\thefigure}{B\arabic{figure}}
\renewcommand{\thetable}{B\arabic{table}}
\setcounter{table}{0}
\renewcommand{\thefootnote}{B\arabic{footnote}}
\setcounter{footnote}{0}

\section{Far South Population Tables, Notes and Model Plots}

\begin{center}
\begin{longtable}{llccc|l|}
\caption{Far South Multiband Population Observation Information.} \label{tabB1}  \\
   \hline
 Pulsar & J Name & P & DM & RM & References \\
    (B1950) & (J2000) & (s) & ($pc/cm^{3}$) & ($rad$-$m^{2}$)    \\
    \hline
    \hline
\endfirsthead
    \hline
 Pulsar & J Name & P & DM & RM  & References \\
    (B1950) & (J2000) & (s) & ($pc/cm^{3}$) & ($rad$-$m^{2}$)    \\
    \hline
    \hline
\endhead
B0203--40 & J0206-4028 & 0.63 & 12.9 & -4.0 & MHMb; Q95; JK18 \\
B0254--53 & J0255-5304 & 0.45 & 15.9 & 32.0 & MHMA; MHM; MHMb; W93; MHQ; ETIX; JK18 \\
B0403--76 & J0401-7608 & 0.55 & 21.7 & 19.0 & Q95; CMH; JK18 \\
B0529--66 & J0529-6652 & 0.98 & 103.2 & 4.0 & CMH \\
B0538--75 & J0536-7543 & 1.25 & 18.6 & 23.8 & Q95; MHQ; CMH; JK18 \\
\\[-6pt]
B0736--40 & J0738-4042 & 0.37 & 160.9 & 12.1 & MHM; MHMA; JKMG; JII; JK18 \\
B0743--53 & J0745-5353 & 0.21 & 121.4 & -71.0 & Q95; CMH; JK18 \\
B0808--47 & J0809-4753 & 0.55 & 228.3 & 105.0 & Q95; CMH; JK18 \\
B0818--41 & J0820-4114 & 0.55 & 113.4 & 57.7 & Q95; JK18 \\
B0833--45 & J0835-4510 & 0.09 & 68.0 & 31.4 & HMAK; MHMA; MHM; KJ06; JI; JK18 \\
\\[-6pt]
B0835--41 & J0837-4135 & 0.75 & 147.3 & 145.0 & HMAK; MHMA; MHM; MHMb; KJ06; JII; JK18 \\
B0839--53 & J0840-5332 & 0.72 & 156.5 & 81.0 & Q95; CMH; JK18 \\
B0855--61 & J0856-6137 & 0.96 & 95.0 & -70.0 & MHQ; JK18 \\
B0903--42 & J0904-4246 & 0.97 & 145.8 & 284.0 & Q95 \\
B0904--74 & J0904-7459 & 0.55 & 51.1 & 14.0 & Q95; JK18 \\
\\[-6pt]
B0905--51 & J0907-5157 & 0.25 & 103.7 & -23.3 & W93; Q95; JK18 \\
B0906--49 & J0908-4913 & 0.11 & 180.4 & 10.0 & W93; Q95 \\
B0909--71 & J0909-7212 & 1.36 & 54.3 & -18.0 & Q95; JK18 \\
B0922--52 & J0924-5302 & 0.75 & 152.9 & 150.0 & MHQ; JK18 \\
B0932--52 & J0934-5249 & 1.44 & 100.0 & 18.0 & MHQ; JK18 \\
\\[-6pt]
B0940--55 & J0942-5552 & 0.66 & 180.2 & -61.9 & MHMA; MHM; MHMb; KJ06; JK18 \\
B0941--56 & J0942-5657 & 0.81 & 159.7 & 135.0 & Q95; MHQ; JK18 \\
B0953--52 & J0955-5304 & 0.86 & 156.9 & -97.0 & Q95; MHQ \\
B0957--47 & J0959-4809 & 0.67 & 92.7 & 50.0 & JK18; BMM+ \\
B0959--54 & J1001-5507 & 1.44 & 130.3 & 297.0 & HMAK; MHMA; MHM; MHMb; MHQ; JK18 \\
\\[-6pt]
B1011--58 & J1012-5857 & 0.82 & 383.9 & 74.0 & Q95; JK18 \\
B1014-53 & J1016-5345 & 0.77 & 66.8 & -21.0 \\
B1036--58 & J1038-5831 & 0.66 & 72.7 & -15.0 & Q95; JK18 \\
B1039--55 & J1042-5521 & 1.17 & 306.5 & 155.0 & MHQ \\
B1046--58 & J1048-5832 & 0.12 & 128.7 & -155.0 & Q95; JK18 \\
B1054--62 & J1056-6258 & 0.42 & 320.3 & 4.0 & W93; CMH; JK18 \\
\\[-6pt]
B1055--52m & J1057-5226 & 0.20 & 29.7 & 47.2 & MHMA; MHM; MHMb; TvO; ETIX; WW09; JK18 \\
B1056--78 & J1057-7914 & 1.35 & 51.0 & -22.2 & CMH; JK18 \\
B1056--57 & J1059-5742 & 1.19 & 108.7 & -75.0 & W93; CMH \\
B1107--56 & J1110-5637 & 0.56 & 262.6 & 419.0 & Q95; JK18 \\
B1110--65 & J1112-6613 & 0.33 & 249.3 & -132.0 & CMH; JK18 \\
\\[-6pt]
B1112--60 & J1114-6100 & 0.88 & 677.0 & * & Q95; JK18 \\
B1114--41 & J1116-4122 & 0.94 & 40.5 & -37.0 & W93; Q95; MHQ; JK18; BMM+ \\
B1119-54 & J1121-5444 & 0.54 & 204.7 & 42.0 & JK18; D'Amico+98 \\
B1131--62 & J1133-6250 & 1.02 & 567.8 & 848.0 & Q95; JK18 \\
B1133--55 & J1136-5525 & 0.36 & 85.5 & 28.0 & W93; Q95; CMH; JK18 \\
\\[-6pt]
B1143--60 & J1146-6030 & 0.27 & 111.7 & -5.0 & CMH; JK18 \\
B1154--62 & J1157-6224 & 0.40 & 325.2 & 508.2 & MHMA; MHM; CMH; JK18 \\
B1159--58 & J1202-5820 & 0.45 & 145.4 & 139.0 & W93; MHQ; JK18 \\
B1221--63 & J1224-6407 & 0.22 & 97.7 & -3.6 & MHMA; MHM; MHMb; W93; ETIX; JK18 \\
B1222--63 & J1225-6408 & 0.42 & 415.1 & 337.0 & Q95; JK18 \\
\\[-6pt]
B1240--64 & J1243--6423 & 0.39 & 297.3 & 157.8 & MHMA; MHM; W93; ETIX; JK18 \\
B1259--63 & J1302--6350 & 0.05 & 146.7 & 21.1 & JNK98; JK18 \\
B1302--64 & J1305--6455 & 0.57 & 505.0 & -420.0 & CMH; JK18 \\
B1303--66 & J1306--6617 & 0.47 & 436.9 & 396.0 & Q95; JK18 \\
B1309--55 & J1312--5516 & 0.85 & 134.1 & 141.0 & Q95 \\
\\[-10pt]
B1316--60 & J1319--6056 & 0.28 & 400.9 & -280.6 & Q96; JK18 \\
B1317--53 & J1320--5359 & 0.28 & 97.1 & 141.0 & CMH; JK18 \\
B1323--58 & J1326--5859 & 0.48 & 287.3 & -579.6 & W93; CMH; JK18 \\
B1323--62 & J1327--6222 & 0.53 & 318.8 & -306.0 & MHMA; MHM;JK18 \\
B1325--43 & J1328--4357 & 0.53 & 42.0 & -22.9 & W93; MHQ; FDR; JK18; BMM+ \\
\\[-9pt]
B1325--49 & J1328--4921 & 1.48 & 118.0 & 170.0 & MHQ \\
B1334--61 & J1338--6204 & 1.24 & 640.3 & -459.0 & Q95; JK18 \\
B1338--62 & J1341--6220 & 0.19 & 719.7 & -921.0 & Q95; JK18 \\
B1353--62 & J1357--62 & 0.46 & 416.8 & -586.0 & MHM; KJ06; JK18 \\
B1356--60 & J1359--6038 & 0.13 & 293.7 & 33.0 & MHQ; W93; ETIX; JK18 \\
\\[-9pt]
B1358--63 & J1401--6357 & 0.84 & 98.0 & 62.0 & W93; Q95; JK18 \\
B1409--62 & J1413--6307 & 0.39 & 122.0 & 44.0 & Q95; JK18 \\
B1424--55 & J1428--5530 & 0.57 & 82.4 & 4.0 & W93; Q95; CMH; JK18 \\
B1426--66 & J1430--6623 & 0.79 & 65.3 & -19.2 & HMAK; MHMA; MHM; MHMb; ETIX; JI; JK18 \\
B1436--63 & J1440--6344 & 0.46 & 124.2 & 29.0 & MHQ \\
\\[-9pt]    
B1449--64 & J1453--6413 & 0.18 & 71.2 & -18.6 & HMAK; MHMA; MHM; MHMb; ETIX; JI; JK18 \\
B1451--68 & J1456--6843 & 0.26 & 8.6 & -4.0 & MHMA; MHM; MHMb; JI; JK18 \\
B1504--43 & J1507--4352 & 0.29 & 48.7 & -34.0 & JKMG; JK18 \\
B1518--58 & J1522--5829 & 0.40 & 199.9 & -24.2 & Q95; JK18 \\
B1523--55 & J1527--5552 & 1.05 & 362.7 & 34.0 & MHQ \\
\\[-9pt]
B1524--39 & J1527--3931 & 2.42 & 49.0 & 4.0 & MHQ; BMM+ \\
B1530--53 & J1534--5334 & 1.37 & 24.8 & -46.0 & HMAK; MHMA; MHM; MHMb; ETIX; JK18 \\
B1535--56 & J1539--5626 & 0.24 & 175.9 & -18.0 & Q95; JK18 \\
B1541--52 & J1544--5308 & 0.18 & 35.2 & -29.0 & MHQ; JK18 \\
B1555--55 & J1559--5545 & 0.96 & 212.9 & -150.0 & MHQ \\
\\[-9pt]
B1556--44 & J1559--4438 & 0.26 & 56.1 & -5.0 & MHMA; MHM; MHQ; W93; JII; ETIX; JKMG \\
B1557--50 & J1600--5044 & 0.19 & 262.8 & 119.0 & MHM; W93; JK18 \\
B1558--50 & J1602--5100 & 0.86 & 170.8 & 71.5 & MHM; MHMA; KJ06; JK18 \\
B1600--49 & J1604--4909 & 0.33 & 140.8 & 34.0 & W93; MHQ; JII; JK18; BMM+ \\
B1601--52 & J1605--5257 & 0.66 & 35.1 & 1.0 & Q95; JKMG; JK18; BMM+ \\
\\[-9pt]
B1610--50 & J1614--5048 & 0.23 & 582.4 & -451.0 & Q95; JK18 \\
B1620--42 & J1623--4256 & 0.36 & 295.0 & 109.6 & Q95; JK18 \\
B1629--50 & J1633--5015 & 0.35 & 398.4 & 406.1 & Q95; JK18 \\
B1641--45 & J1644--4559 & 0.46 & 478.8 & -626.9 & MHM; MHMA; ETIX; KJ06; JK18 \\
B1641--68 & J1646--6831 & 1.79 & 43.0 & 105.0 & W93; Q95; JK18 \\
\\[-9pt]
B1647--52 & J1651--5222 & 0.64 & 179.1 & -38.0 & W93; MHQ; JK18 \\
B1648--42 & J1651--4246 & 0.84 & 482.0 & -167.4 & W93; ETIX;  JK18 \\
B1657--45 & J1701--4533 & 0.32 & 526.0 & 4.0 & Q95; JK18 \\
B1659--60 & J1704--6016 & 0.31 & 54.0 & 50.0 & Q95 \\
B1703--40 & J1707--4053 & 0.58 & 360.0 & -179.7 & Q95; FDR; JK18 \\
\\[-9pt]
B1706--44 & J1709--4429 & 0.10 & 75.7 & 0.7 & Q95; JNK98; JI; JK18; BMM+ \\
B1719--37 & J1722--3712 & 0.24 & 99.5 & 104.0 & W93; MHQ; JK18 \\
B1727--47 & J1731--4744 & 0.83 & 123.1 & -429.1 & HMAK; MHMA; MHM; MHMb; JKMG; JK18 \\
B1737--39 & J1741--3927 & 0.51 & 158.5 & 204.0 & W93; MHQ; CMH; JK18; BMM+ \\
B1747--46 & J1751--4657 & 0.74 & 20.4 & 19.0 & HMAK;MHMA;MHM;MHMb;MHQ;JK18;BMM+ \\
\\[-9pt]
B1758--29 & J1801--2920 & 1.08 & 125.6 & -62.0 & Q95; JK18; BMM+ \\
B1800--27 & J1803--2712 & 0.33 & 165.5 & -165.0 & Q95: Q98; JK18 \\
B1806--53 & J1810--5338 & 0.26 & 45.0 & 58.0 & Q95 \\
B1851--79 & J1900--7951 & 1.28 & 39.0 & 18.6 & Q95 \\
B2048--72 & J2053--7200 & 0.34 & 17.3 & 17.0 & Q95; MHQ; JK18 \\
\\[-9pt]
B2123-67 & J2127--6648 & 0.33 & 35.0 & * &  JK18; D'Amico+98 \\
B2321--61 & J2324--6054 & 2.35 & 14.0 & 15.6 & Q95; TvO; JK18 \\
    \hline
\end{longtable}
Notes: CMH: \citet{CMH}; ETIX: \citet{ETIX}; ETV: \citet{rankin1993b}; HMAK: \citet{HMAK}; JKMG: \citet{JKMG2008};  JK18: \citet{jk18}; JI: \citet{JohnstonI}; JII: \citet{JohnstonII}; MHM: \citet{MHM}; MHMA: \citet{MHMA}; MHMb: \citet{MHMb}; MHQ: \citet{MHQ}; MM10: \citet{MM10}; Q95: \citet{QMLG95}; W93: \citet{WeWr09}
\end{center}

\begin{center}
\begin{longtable}{lccc|ccccccc}
\caption{Far South Multiband Pulsar Population Parameters} \label{tabB2}  \\
\hline
 Pulsar & L & B & Dist. & P & $\dot{P}$ & $\dot{E}$ & $\tau$ & $B_{surf}$ & $B_{12}/P^2$ & 1/Q \\ 
 (B1950) & (\degr) & (\degr) & (kpc) & (s) & ($10^{-15}$ s/s) & ($10^{32}$ ergs/s)  & (Myr) & ($10^{12}$ G) &   &   \\
\hline
\hline
\endfirsthead
    \hline
  Pulsar & L & B & Dist. & P & $\dot{P}$ & $\dot{E}$ & $\tau$ & $B_{surf}$ & $B_{12}/P^2$ & 1/Q \\ 
 (B1950) & (\degr) & (\degr) & (kpc) & (s) & ($10^{-15}$ s/s) & ($10^{32}$ ergs/s)  & (Myr) & ($10^{12}$ G) &   &   \\
    \hline
    \hline
\endhead B0203--40 & 258.60 & -69.63 & 1.26 & 0.631 & 1.20 & 1.90 & 8.4 & 0.88 & 2.2 & 0.9 \\
B0254--53 & 269.86 & -55.31 & 1.51 & 0.448 & 0.03 & 0.14 & 227 & 0.12 & 0.6 & 0.3 \\
B0403--76 & 290.31 & -35.91 & 1.01 & 0.545 & 1.54 & 3.80 & 5.6 & 0.93 & 3.1 & 1.2 \\
B0529--66 & 276.98 & -32.76 & 49.70 & 0.976 & 15.47 & 6.60 & 1.0 & 3.93 & 4.1 & 1.5 \\
B0538--75 & 287.16 & -30.82 & 0.14 & 1.246 & 0.58 & 0.12 & 34.2 & 0.86 & 0.6 & 0.3 \\
\\[-2pt]
B0736--40 & 254.19 & -9.19 & 1.60 & 0.375 & 1.38 & 10.0 & 4.3 & 0.73 & 5.2 & 1.7 \\
B0743--53 & 266.66 & -14.28 & 0.57 & 0.215 & 2.19 & 87.0 & 1.6 & 0.69 & 15.0 & 3.7 \\
B0808--47 & 263.30 & -7.96 & 6.49 & 0.547 & 3.08 & 7.40 & 2.8 & 1.31 & 4.4 & 1.5 \\
B0818--41 & 235.89 & 12.60 & 0.57 & 0.545 & 0.02 & 0.05 & 437 & 0.11 & 0.4 & 0.2 \\
B0833--45 & 263.55 & -2.79 & 0.28 & 0.089 & 125.01 & 69000 & 0.0 & 3.38 & 424 & 49 \\
\\[-2pt]
B0835--41 & 260.90 & -0.34 & 1.50 & 0.752 & 3.54 & 3.30 & 3.4 & 1.65 & 2.9 & 1.1 \\
B0839--53 & 270.77 & -7.14 & 0.57 & 0.721 & 1.64 & 1.70 & 7.0 & 1.10 & 2.1 & 0.9 \\
B0855--61 & 278.58 & -10.43 & 0.37 & 0.963 & 1.68 & 0.74 & 9.1 & 1.29 & 1.4 & 0.6 \\
B0903--42 & 265.07 & 2.86 & 0.68 & 0.965 & 1.88 & 0.82 & 8.2 & 1.36 & 1.5 & 0.7 \\
B0904--74 & 289.74 & -18.32 & 1.05 & 0.550 & 0.46 & 1.10 & 18.9 & 0.51 & 1.7 & 0.7 \\
\\[-2pt]
B0905--51 & 272.15 & -3.03 & 0.34 & 0.254 & 1.83 & 44.0 & 2.2 & 0.69 & 10.7 & 2.9 \\
B0906--49 & 270.27 & -1.02 & 1.00 & 0.107 & 15.10 & 4900 & 0.1 & 1.28 & 112 & 17 \\
B0909--71 & 287.73 & -16.26 & 0.75 & 1.363 & 0.33 & 0.05 & 66.0 & 0.68 & 0.4 & 0.2 \\
B0922--52 & 274.71 & -1.93 & 0.51 & 0.746 & 35.33 & 34.0 & 0.3 & 5.20 & 9.3 & 2.9 \\
B0932--52 & 275.69 & -0.70 & 0.35 & 1.445 & 4.65 & 0.61 & 4.9 & 2.62 & 1.3 & 0.6 \\
\\[-2pt]
B0940--55 & 278.57 & -2.23 & 0.30 & 0.664 & 22.68 & 31.0 & 0.5 & 3.93 & 8.9 & 2.7 \\
B0941--56 & 279.35 & -2.99 & 0.41 & 0.808 & 39.61 & 30.0 & 0.3 & 5.73 & 8.8 & 2.8 \\
B0953--52 & 278.26 & 1.16 & 0.40 & 0.862 & 3.53 & 2.20 & 3.9 & 1.76 & 2.4 & 1.0 \\
B0957--47 & 275.74 & 5.42 & 0.36 & 0.670 & 0.08 & 0.11 & 125 & 0.24 & 0.5 & 0.3 \\
B0959--54 & 280.23 & 0.08 & 0.30 & 1.437 & 51.58 & 6.90 & 0.4 & 8.71 & 4.2 & 1.6 \\
\\[-2pt]
B1011--58 & 283.71 & -2.15 & 3.19 & 0.820 & 17.80 & 13.0 & 0.7 & 3.87 & 5.8 & 2.0 \\
B1036--58 & 286.28 & -0.02 & 0.92 & 0.662 & 1.25 & 1.70 & 8.4 & 0.92 & 2.1 & 0.9 \\
B1039--55 & 285.19 & 3.00 & 2.79 & 1.171 & 6.72 & 1.70 & 2.8 & 2.84 & 2.1 & 0.9 \\
B1046--58 & 287.43 & 0.58 & 2.90 & 0.124 & 96.12 & 20000 & 0.0 & 3.49 & 228 & 31 \\
B1054--62 & 290.29 & -2.97 & 2.40 & 0.422 & 3.58 & 19.0 & 1.9 & 1.24 & 6.9 & 2.1 \\
\\[-2pt]
B1055--52m & 285.98 & 6.65 & 0.09 & 0.197 & 5.84 & 300. & 0.5 & 1.09 & 28.1 & 6.0 \\
B1056--78 & 297.57 & -17.57 & 1.44 & 1.347 & 1.33 & 0.21 & 16.1 & 1.35 & 0.7 & 0.4 \\
B1056--57 & 288.35 & 1.95 & 1.65 & 1.185 & 4.31 & 1.00 & 4.4 & 2.29 & 1.6 & 0.7 \\
B1107--56 & 289.28 & 3.53 & 2.45 & 0.558 & 2.06 & 4.70 & 4.3 & 1.09 & 3.5 & 1.3 \\
B1110--65 & 293.19 & -5.23 & 2.53 & 0.334 & 0.82 & 8.70 & 6.4 & 0.53 & 4.8 & 1.5 \\
\\[-2pt]
B1112--60 & 291.44 & -0.32 & 5.48 & 0.881 & 46.01 & 27.0 & 0.3 & 6.44 & 8.3 & 2.7 \\
B1114--41 & 284.45 & 18.07 & 0.28 & 0.943 & 7.95 & 3.70 & 1.9 & 2.77 & 3.1 & 1.2 \\
B1119--54 & 290.08 & 5.87 & 2.32 & 0.536 & 2.78 & 7.10 & 3.1 & 1.24 & 4.3 & 1.5 \\
B1131--62 & 294.21 & -1.30 & 7.45 & 1.023 & 0.45 & 0.17 & 35.9 & 0.69 & 0.7 & 0.4 \\
B1133--55 & 292.31 & 5.89 & 1.52 & 0.365 & 8.23 & 67.0 & 0.7 & 1.75 & 13.2 & 3.5 \\
\\[-2pt]
B1143--60 & 294.98 & 1.34 & 1.63 & 0.273 & 1.79 & 35.0 & 2.4 & 0.71 & 9.5 & 2.6 \\
B1154--62 & 296.71 & -0.20 & 4.00 & 0.401 & 3.93 & 24.0 & 1.6 & 1.27 & 7.9 & 2.4 \\
B1159--58 & 296.53 & 3.92 & 1.89 & 0.453 & 2.13 & 9.10 & 3.4 & 0.99 & 4.8 & 1.6 \\
B1221--63 & 299.98 & -1.42 & 4.00 & 0.216 & 4.95 & 190. & 0.7 & 1.05 & 22.4 & 5.1 \\
B1222--63 & 300.13 & -1.41 & 9.85 & 0.420 & 0.95 & 5.10 & 7.0 & 0.64 & 3.6 & 1.3 \\
\\[-2pt]
B1240--64 & 302.05 & -1.53 & 2.00 & 0.388 & 4.49 & 30.0 & 1.4 & 1.34 & 8.9 & 2.6 \\
B1259--63 & 304.18 & -0.99 & 2.63 & 0.048 & 2.28 & 8300 & 0.3 & 0.33 & 146 & 19.7 \\
B1302--64 & 304.41 & -2.09 & 11.90 & 0.572 & 4.03 & 8.50 & 2.3 & 1.54 & 4.7 & 1.6 \\
B1303--66 & 304.46 & -3.46 & 15.87 & 0.473 & 5.98 & 22.0 & 1.3 & 1.70 & 7.6 & 2.3 \\
B1309--55 & 306.01 & 7.46 & 4.19 & 0.849 & 5.70 & 3.70 & 2.4 & 2.23 & 3.1 & 1.2 \\
\\[-10pt]
B1316--60 & 306.31 & 1.74 & 11.85 & 0.284 & 1.53 & 26.0 & 3.0 & 0.67 & 8.2 & 2.4 \\
B1317--53 & 307.31 & 8.64 & 2.20 & 0.280 & 9.25 & 170. & 0.5 & 1.63 & 20.8 & 4.9 \\
B1323--58 & 307.50 & 3.57 & 3.00 & 0.478 & 3.24 & 12.0 & 2.3 & 1.26 & 5.5 & 1.8 \\
B1323--62 & 307.07 & 0.20 & 4.00 & 0.530 & 18.79 & 50.0 & 0.4 & 3.19 & 11.4 & 3.3 \\
B1325--43 & 309.87 & 18.42 & 1.42 & 0.533 & 3.01 & 7.90 & 2.8 & 1.28 & 4.5 & 1.6 \\
\\[-6pt]
B1325--49 & 309.12 & 13.07 & 8.40 & 1.479 & 0.61 & 0.07 & 38.4 & 0.96 & 0.4 & 0.3 \\
B1334--61 & 308.37 & 0.31 & 12.36 & 1.239 & 13.79 & 2.90 & 1.4 & 4.18 & 2.7 & 1.1 \\
B1338--62 & 308.73 & -0.04 & 12.60 & 0.193 & 253.11 & 14000 & 0.0 & 7.08 & 189 & 28 \\
B1353--62 & 310.47 & -0.57 & 6.48 & 0.456 & --- & --- & --- & --- & --- & --- \\
B1356--60 & 311.24 & 1.13 & 5.00 & 0.128 & 6.33 & 1200. & 0.3 & 0.91 & 55.9 & 10 \\
\\[-6pt]
B1358--63 & 310.57 & -2.14 & 1.80 & 0.843 & 16.83 & 11.0 & 0.8 & 3.81 & 5.4 & 1.9 \\
B1409--62 & 312.05 & -1.72 & 3.04 & 0.395 & 7.43 & 48.0 & 0.8 & 1.73 & 11.1 & 3.1 \\
B1424--55 & 316.43 & 4.80 & 1.90 & 0.570 & 2.09 & 4.40 & 4.3 & 1.10 & 3.4 & 1.2 \\
B1426--66 & 312.65 & -5.40 & 1.33 & 0.785 & 2.78 & 2.30 & 4.5 & 1.49 & 2.4 & 1.0 \\
B1436--63 & 314.65 & -3.38 & 3.45 & 0.460 & 1.12 & 4.60 & 6.5 & 0.73 & 3.4 & 1.2 \\
\\[-6pt]
B1449--64 & 315.73 & -4.43 & 2.80 & 0.179 & 2.74 & 190. & 1.0 & 0.71 & 22.0 & 5.0 \\
B1451--68 & 313.87 & -8.54 & 0.43 & 0.263 & 0.10 & 2.10 & 42.2 & 0.16 & 2.3 & 0.9 \\
B1504--43 & 327.34 & 12.46 & 1.39 & 0.287 & 1.58 & 27.0 & 2.9 & 0.68 & 8.3 & 2.4 \\
B1518--58 & 321.63 & -1.22 & 3.88 & 0.395 & 2.06 & 13.0 & 3.1 & 0.91 & 5.8 & 1.9 \\
B1523--55 & 323.64 & 0.59 & 5.32 & 1.049 & 11.27 & 3.90 & 1.5 & 3.48 & 3.2 & 1.3 \\
\\[-6pt]
B1524--39 & 333.05 & 14.02 & 1.71 & 2.418 & 19.06 & 0.53 & 2.0 & 6.87 & 1.2 & 0.6 \\
B1530--53 & 325.72 & 1.94 & 0.81 & 1.369 & 1.43 & 0.22 & 15.2 & 1.42 & 0.8 & 0.4 \\
B1535--56 & 324.62 & -0.81 & 3.54 & 0.243 & 4.85 & 130. & 0.8 & 1.10 & 18.6 & 4.4 \\
B1541--52 & 327.27 & 1.32 & 0.93 & 0.179 & 0.06 & 4.20 & 47.0 & 0.11 & 3.3 & 1.1 \\
B1555--55 & 327.24 & -2.03 & 4.13 & 0.957 & 19.92 & 9.0 & 0.8 & 4.42 & 4.8 & 1.7 \\
\\[-6pt]
B1556--44 & 334.54 & 6.37 & 2.30 & 0.257 & 1.02 & 24.0 & 4.0 & 0.52 & 7.8 & 2.2 \\
B1557--50 & 330.69 & 1.63 & 6.90 & 0.193 & 5.06 & 280. & 0.6 & 1.00 & 26.9 & 5.9 \\
B1558--50 & 330.69 & 1.29 & 8.00 & 0.864 & 69.41 & 42.0 & 0.2 & 7.84 & 10.5 & 3.2 \\
B1600--49 & 332.15 & 2.44 & 3.23 & 0.327 & 1.02 & 11.0 & 5.1 & 0.58 & 5.4 & 1.7 \\
B1601--52 & 329.73 & -0.48 & 0.93 & 0.658 & 0.26 & 0.35 & 40.7 & 0.42 & 1.0 & 0.5 \\
\\[-6pt]
B1610--50 & 332.21 & 0.17 & 5.15 & 0.232 & 494.94 & 16000 & 0.0 & 10.80 & 201 & 30 \\
B1620--42 & 338.89 & 4.62 & 21.56 & 0.365 & 1.01 & 8.20 & 5.8 & 0.61 & 4.6 & 1.5 \\
B1629--50 & 334.70 & -1.57 & 6.01 & 0.352 & 3.79 & 34.0 & 1.5 & 1.17 & 9.4 & 2.7 \\
B1641--45 & 339.19 & -0.20 & 4.50 & 0.455 & 20.09 & 84.0 & 0.4 & 3.06 & 14.8 & 3.9 \\
B1641--68 & 321.84 & -14.83 & 1.25 & 1.786 & 1.70 & 0.12 & 16.6 & 1.76 & 0.6 & 0.3 \\
\\[-6pt]
B1647--52 & 335.01 & -5.17 & 6.28 & 0.635 & 1.81 & 2.80 & 5.6 & 1.09 & 2.7 & 1.0 \\
B1648--42 & 342.46 & 0.92 & 5.20 & 0.844 & 4.75 & 3.10 & 2.8 & 2.03 & 2.8 & 1.1 \\
B1657--45 & 341.36 & -2.18 & 19.59 & 0.323 & 0.52 & 6.10 & 9.9 & 0.41 & 4.0 & 1.3 \\
B1659--60 & 329.76 & -11.37 & 1.59 & 0.306 & 0.91 & 12.0 & 5.3 & 0.53 & 5.7 & 1.8 \\
B1703--40 & 345.72 & -0.20 & 4.00 & 0.581 & 1.92 & 3.90 & 4.8 & 1.07 & 3.2 & 1.2 \\
\\[-6pt]
B1706--44 & 343.10 & -2.69 & 2.60 & 0.102 & 92.98 & 34000 & 0.0 & 3.12 & 297 & 38 \\
B1719--37 & 350.49 & -0.51 & 2.48 & 0.236 & 10.86 & 330. & 0.3 & 1.62 & 29.0 & 6.4 \\
B1727--47 & 342.57 & -7.67 & 0.70 & 0.830 & 163.63 & 110. & 0.1 & 11.80 & 17.1 & 4.7 \\
B1737--39 & 350.56 & -4.75 & 4.62 & 0.512 & 1.71 & 5.00 & 4.7 & 0.95 & 3.6 & 1.3 \\
B1747--46 & 345.00 & -10.18 & 0.74 & 0.742 & 1.30 & 1.30 & 9.1 & 0.99 & 1.8 & 0.8 \\
\\[-6pt]
B1758--29 & 1.44 & -3.25 & 3.01 & 1.082 & 3.29 & 1.00 & 5.2 & 1.91 & 1.6 & 0.7 \\
B1800--27 & 3.49 & -2.53 & 3.47 & 0.334 & 0.02 & 0.18 & 310 & 0.08 & 0.7 & 0.3 \\
B1806--53 & 340.29 & -15.90 & 1.65 & 0.261 & 0.39 & 8.60 & 10.7 & 0.32 & 4.7 & 1.5 \\
B1851--79 & 314.32 & -27.07 & 2.38 & 1.279 & 1.86 & 0.35 & 10.9 & 1.56 & 1.0 & 0.5 \\
B2048--72 & 321.87 & -35.00 & 1.03 & 0.341 & 0.20 & 2.00 & 27.4 & 0.26 & 2.3 & 0.9 \\
\\[-6pt]
B2123--67 & 326.39 & -39.78 & 5.99 & 0.326 & 0.23 & 2.60 & 22.8 & 0.28 & 2.6 & 0.9 \\
B2321--61 & 320.43 & -53.17 & 1.21 & 2.347 & 2.58 & 0.08 & 14.4 & 2.49 & 0.5 & 0.3 \\
\hline
\end{longtable}
\end{center}

\twocolumn
\noindent\textit{\textbf{B0203--40}}: The pulsar has a bright leading component and a weaker trailing one---with similar forms across the band of observations---as do many conal single profiles, and we so model it.  No scattering is apparent in the \cite{MHMb} 170-MHz profile, and no measurement is available.  
\vskip 0.09in
\noindent\textit{\textbf{B0254--53}}:  We depend on the usual set of far south profile observations, and a double profile of nearly constant width is seen in most.  Only at 270 MHz MHMb is the interior region is filled and the profile slightly wider.  The PPA traverse is chaotic at most frequencies, but the latter observation shows a --8.3\degr/\degr\ rate with a 90\degr\ ``jump'' on the trailing side.  We model the beam with an inner conal {\textbf D} configuration.  No fluctional spectral or scattering studies are available. 
\vskip 0.09in
\noindent\textit{\textbf{B0403--76}}: The profiles show a boxy structure with three parts.  The high quality \citet{jk18} observation is better resolved and shows what seems to be a broad core component marked by antisymmetric $V$.  We thus model its beams as a core-cone triple {\textbf T}.  A core width of nearly 9\degr\ is indicated for an inner cone (and nearly 7\degr\ for an outer one)---and as observations go only to 600 MHz these cannot be distinguished.  No scattering time measurement is available.  
\vskip 0.09in
\noindent\textit{\textbf{B0529--66}}:  The 600-MHz profile is single with a flat PPA traverse and a central depolarized interval that perhaps reflects a 180\degr\ rotation.  It is difficult to understand the geometry that would produce this situation.  We await better profiles over a larger frequency band.  
\vskip 0.09in
\noindent\textit{\textbf{B0538--75}}:  This far south pulsar has a filled trapazoidal profile that shows evidence of four or five conflated features over the entire band of observations.  The PPA traverse has a full ``S'' shape, so the PPA rate is very well determined.  We model it as a conal quadruple {c\textbf Q} only because the core is clearly apparent in no profile, though the central space is filled in all of them.  A core width near 9\degr\ is plausible.  No scattering is discernible at 400 MHz, nor is a measurement available.  
\vskip 0.09in
\noindent\textit{\textbf{B0736--40}}: We support the core-cone triple {\textbf T} classification of ET VI.  The pulsar lies too far south for GL98, but well measured profiles are available at 1612 (MHM), 1375 (JK18), 950 (TvO) and 631 MHz (MHMA).  Scattering is substantial, even at 600 MHz \citep{kmn+15}. 
\vskip 0.09in
\noindent\textit{\textbf{B0743--53}}: The available profiles are all broad and single, and all show a well defined PPA rate.  Probably this pulsar has a core-single {\textbf S$_t$} beam configuation, but no higher frequency profiles exist to check for conal ``outriders''.  The broadening and slight asymmetry at 600 MHz may indicate the onset of scattering in a manner compatible with the \citet{abs86} measurement.  
\vskip 0.09in
\noindent\textit{\textbf{B0808--47}}: The high quality \citet{jk18} profile shows a clear triple structure with a well defined PPA rate.  The lower frequency profiles are of poorer quality and hint of scattering at 600 MHz.  We model the geometry with  core/inner cone triple {\textbf S$_t$} beams.  \citet{kmn+15} give a measured scattering time.
\vskip 0.09in
\noindent\textit{\textbf{B0818--41}}: This very broad profile shows two components clearly, but its interior is filled over the entire band of observations, and the $V$ profile shows a clear antisymmetric signature.  We model it with an outer conal double {\textbf D} beam configuration, but more sensitive analysis may well show core and inner conal emission.  Also, the PPA rate is so shallow that it cannot be well determined, and the unknown sign of $\beta$ will affect the $\alpha$ value.
\vskip 0.09in
\noindent\textit{\textbf{B0833--45}}:  This important pulsar remains understudied as we know little about how its profiles at higher frequencies relate to those below 1 GHz.  We support the ET VI {\textbf S$_t$} geometry and model the 1612 (MHM), 631 (MHMA), and 400/338-MHz (HMAK) profiles as core components, while seeing clear scattering tails develop at the lower frequencies.  The question is how to account for the profile forms seen at very high frequencies.
\vskip 0.09in
\noindent\textit{\textbf{B0835--41}}:  The single component seen across all the observations supports the {\textbf S$_t$} geometry of ET VI.  Here we depend on the 1612 (MHM), 950 (TvO), 631 (MHMA), 338 (HMAK) and 270-MHz (M80) observations as the pulsar is too far south of GL98.  A PPA rate can be estimated from the first observation, and its broad wings suggest the emergence of conal outriders.  The lowest frequency profile is broader, but with no clear ``tail'' as might be expected from the substantial $t_{\rm scatt}$ by \cite{abs86}.
\vskip 0.09in
\noindent\textit{\textbf{B0839--53}}:  The profiles are single and show a well defined PPA rate at both 600 and 1400 MHz.  We model the geometry using a conal single {\textbf S$_d$} beam although a core single cannot be ruled out---especially if conal ``outriders'' were identified at a higher frequency.  A scattering time has been measured by \citet{mr01a}.  
\vskip 0.09in
\noindent\textit{\textbf{B0855--61}}: The three profiles (JK18, MHQ) all show the same 7\degr\ width and a shallow PPA rate.  This is most probably a conal single structure, and so we model it.  
\vskip 0.09in
\noindent\textit{\textbf{B0903--42}}: Only one 660-MHz profile \citep{QMLG95} is available, and it shows a conflated double profile with the leading feature stronger than the trailing.  The PPA rate can be estimated, so we model it with a usual inner conal {\textbf D} geometry.  A scattering measurement is reported by \citet{kmn+15}.
\vskip 0.09in
\noindent\textit{\textbf{B0904--74}}: The two profiles have single forms, a linear PPA traverse, and the edge depolarization characteristic of conal profiles---and so we model it as an inner conal single beam.  A curiosity is the feature seen clearly at 660 MHz trailing by about 35\degr.  It is cut off in the 1.4-GHz plot \citep{jk18}, but can be discerned in the full period EPN profile.  This may be another example of a postcursor.  No scattering measurement is available.   
\vskip 0.09in
\noindent\textit{\textbf{B0905--51}}: The profiles are triple with a weak separated leading component and a double trailing feature.  The PPA traverse is well defined but increasingly steep in the leading part of the profiles suggesting aberration/retardation.  No scattering is apparent at 660 MHz, but \citet{mr01a} provide a measured scattering time.  
\vskip 0.09in
\noindent\textit{\textbf{B0906--49}}: This energetic 107-ms interpulsar has a main pulse (MP) and interpulse (IP) whose main components are separated by very nearly 180\degr.  The sensitive \citet{jk18} 1.4-GHz profile shows structure in both---and that the main components of both have widths just over 4\degr, narrower than the 7.5\degr\ polar cap angular diameter---probably indicating a non-dipolar field structure at the emission height.  Apart then from arguing that both bright components are probably core emission, more detailed analysis is beyond the scope of this work.  \citet{kmn+15} give a scattering time scale. 
\vskip 0.09in
\noindent\textit{\textbf{B0909--71}}:  Both profiles show well resolved double forms.  Interestingly, the published \citet{jk18} profile is cut off and looks single; however the weak trailing feature is clear enough in the full period EPN version.  The PPA rate has to be estimated as a modestly steep value that accrues most of 180\degr\ between the two components. No scattering measurement is available.  
\vskip 0.09in
\noindent\textit{\textbf{B0922--52}}: The \citet{jk18} 1.4-GHz profile shows two distinct features of which the trailing seems to be a core component; however, it is barely broad enough to reflect the polar cap width, so we take $\alpha$ as 90\degr.  The PPA traverse is well defined and steep.  Perhaps the trailing skirt of the core hides a conal ``outrider''---which together with the leading component would be a pair.  If so, an outer conal double beam geometry is possible.  Observations at higher and lower frequencies are needed to clarify the geometry.  \citet{mr01a} measure a scattering time.  
\vskip 0.09in
\noindent\textit{\textbf{B0932--52}}: The quality \citet{jk18} profile seems to show three closely spaced features that may represent a conal triple configuration---while the two lower frequency profiles are less well resolved.  The PPA rate may be about -9\degr/\degr.  We then model the three features as an inner conal triple {c\textbf T} {\textbf S$_t$} beam system, though a {\textbf S$_t$} with a plausible core width of 2.0\degr\ is also possible geometrically.  All three profiles may show weak conal power on their extreme edges that would tilt toward an M system.  Further sensitive study is needed to resolve these questions.  A scattering time scale has been determined by \citet{mr01a}.
\vskip 0.09in
\noindent\textit{\textbf{B0940--55}}: Only the \citet{TvO} profile is new since the ET VI modeling, and it does reiterate a PPA rate around +30\degr/\degr.  We thus model the profile using an outer core-cone geometry, where the conal ``outriders'' seem to be discernible both at 1612 and 950 MHz.  The profile is substantially broader at 268 MHz (MHMb) and may be the effect of the scattering measured by \citet{mr01a}.
\vskip 0.09in
\noindent\textit{\textbf{B0941--56}}: The leading part of the bright feature in the profiles is very likely a core emission component, and the orderly PPA traverse suggests conal emission.  The high quality \citet{jk18} $L$ profile suggests that the core preceeds the trailing conal ``outrider'' and has a width close to 2\degr.  Thus we model the emission using an outer conal core-cone {\textbf T} geometry. No measured scattering time is available.  
\vskip 0.09in
\noindent\textit{\textbf{B0953--52}}: The 1440- and 658-MHz profiles are clearly triple with hints that the bright central component represents core emission.  Their widths are about 3\degr, and the conal width is about 13\degr\ in both cases.  The PPA rate is poorly determined but the steep central traverse at the lower frequency may be representative, and with this we model the profile with a core/outer conal beam geometry.  \citet{mr01a} measure a scattering time scale.  
\vskip 0.09in
\noindent\textit{\textbf{B0957--47}}: Only two profiles seem to have been published for this pulsar, a 1375- \citep{jk18} and a 408-MHz \citep{LM88}, both showing conal double components, but the former with the ``boxy'' structure often seen in {\textbf M} stars at high frequency.  \citet{basu2016} find both a rough 5-$P$ ``drift'' modulation and 100-$P$ amplitude fluctuations.  We model the pulsar as an outer cone-core triple {\textbf T}, estimating the core width, because no meaningful measurements can be made of the inner conal components.  No scattering timescale seems to have been measured.  
\vskip 0.09in
\noindent\textit{\textbf{B0959--54}}: We follow ET VI in modeling the pulsar as having a core-single {\textbf S$_t$} geometry---as beautifully exemplified by the JK18 observation.  Conal ``outriders'' may also be discernible in the \citet{MHM} and \citet{TvO} profiles, all compatible with an inner conal structure.  A scattering ``tail'' is visible on the 268-MHz \citep{MHMb} profile that may be compatible with the \citet{mr01a} $t_{\rm scatt}$ measurement. 
\vskip 0.09in
\noindent\textit{\textbf{B1011--58}}: The \citet{jk18} profile shows a symmetrical narrow component with breaks on its flanks [the \cite{QMLG95} profile is too poorly resolved to be useful].  We interpret this as a core component with unresolved low level broadening by a pair of conal outriders.  The PPA rate is poorly defined and requires a rough estimate.  We model this with a core/inner conal single geometry.  No lower frequency profiles are available, and most would probably show the effects of scattering given the large measured value by \citet{joh90}.  
\vskip 0.09in
\noindent\textit{\textbf{B1036--58}}: The 1.4-GHz profiles are double with a well defined PPA rate.  We model it with an inner conal double model as an outer cone is not possible.  The better \citet{jk18} profile shows a hint of core emission on the leading edge of the trailing component.  No scattering measurement is available.
\vskip 0.09in
\noindent\textit{\textbf{B1039--55}}: Only a 658-MHz \citep{MHQ} Gaussian shaped profile with a well defined PPA rate.  We model it using a conal single {\textbf S$_d$} geometry. \citet{kmn+15} have measured a scattering time scale.
\vskip 0.09in
\noindent\textit{\textbf{B1046--58}}: The \citet{jk18} 1.4-GHz profile shows a central bifurcated feature together with a clear pair of conal ``outriders'' [the \citet{QMLG95} is too poorly resolved to be useful].  The PPA trajectory is well defined and the $V$ only positive.  Interpreting the central dual feature as the core results in an $\alpha$ of 50\degr, and an inner conal beam model for the ``outriders''.  This seems to be a rare example of a bifurcated core component.  Interestingly, the \citet{JKW06} 8.4-GHz profile shows a similar structure apart from a much narrow core---as if the two parts had different spectra.  No lower frequency observation nor scattering measurement is available.  \vskip 0.09in
\noindent\textit{\textbf{B1054--62}}:   All the profiles have an asymmetric single form with a fast rise and slow trailing ``tail'', almost as if scattering was involved.  However, the \citet{abs86} timescale shows that scattering is important only at frequencies lower than 600 MHz \citep{CMH}.  Closer inspection of the best profiles \citep{JKW06,jk18} show a very shallow PPA traverse and possible 3-4 highly conflated features.  Such an energetic pulsar would usually have clear core emission, but this is not evident.  Even a conal quadruple model fails due to the very shallow PPA rate.  This is a rare example of a well studied pulsar for which no core/double cone model seems appropriate.
\vskip 0.09in 
\noindent\textit{\textbf{B1055--52}}:   This intriguing high energy interpulsar has attracted a number of studies beginning with \citet{MHAK}.  Its main pulse was modeled as a core-cone triple in ET VI, but the \citet{WeWr09} analysis and the references there cited have shown that this interpretation is incorrect.  The pulsar then provides a well studied example of a pulsar that seemingly cannot be fitted into the core/double-cone model.  
\vskip 0.09in 
\noindent\textit{\textbf{B1056--78}}: The two profiles have very similar single forms and dimensions, and only the 1.4-GHz \citep{jk18} gives even a hint of the PPA rate.  We model the profile as an inner conal single with a rough estimate of the PPA rate.  No scattering value is available.  
\vskip 0.09in 
\noindent\textit{\textbf{B1056--57}}: All the profiles have a primary single form, with little $L$ and a poorly defined PPA rate.  There is evidence for a pair of weak outriding components, especially in \citet{CMH}, but evidence for the bright central component being a core seems poor.  We thus model it with an inner conal single geometry.  Further study may show how to accommodate CMH's outriders.  \citet{mr01a} provide a scattering time measurement.  \vskip 0.09in 
\noindent\textit{\textbf{B1107--56}}:  The \citet{jk18} 1.4 GHz profile shows a filled conal double structure with a suggestion of core emission on the inside of the trailing component (the MHQ profile is too poorly resolved to be useful).  We can only model the geometry using an inner conal beam, but further study may reveal a core contribution to the profile.  No scattering measurement is available.  
\vskip 0.09in 
\noindent\textit{\textbf{B1110--65}}:  The well measured \citet{jk18} profile has a conflated double form, whereas the poorly resolved 600-MHz profile may have a triple form.  As both have similar outside conal widths, we model the beam system as a core-inner cone triple wherein the central component has a steeper relative spectrum.  No scattering time measurement is available.  
\vskip 0.09in \noindent\textit{\textbf{B1112--60}}: \citet{joh90} The \citet{jk18} 1.4-GHz profile is asymmetric and may already incur some effects of the large scattering \citep{joh90}.  Probably this pulsar has a core single beam geometry.  
\vskip 0.09in 
\noindent\textit{\textbf{B1114--41}}: All the profiles are Gaussian shaped and have only slightly increasing widths with wavelength---and this may reflect poorer resolution or scattering [as measured by \citet{abs86}]. 
\citet{basu2016} find a weak 20-$P$ amplitude modulation.  We model the beam with a core-single geometry.
\citet{abs86} measure a scattering time scale.
\vskip 0.09in
\noindent\textit{\textbf{B1119-54}}: The profile is very clearly triple with marked antisymmetric $V$ labeling the central component and a well define positive $R$. The core width seems to be about that of the polar-cap diameter, and the bare core in the 450-MHz profile \citep{D'Amico+98} shows that the beam geometry is that of a core-single {\textbf S$_t$}, not a {\textbf T}. Again, \citet{kmn+15} measure a significant level of scattering.
\vskip 0.09in	
\noindent\textit{\textbf{B1131--62}}: The 1.4-GHz profiles show a standard wide and widely separated conal double {\textbf D} profile.  No lower frequency profiles are available, but the \citet{joh90} measurement indicates substantial scattering.  
\vskip 0.09in
\noindent\textit{\textbf{B1133--55}}: The \citet{jk18} and \citet{WMLQ} profiles indicate a triple structure with the leading component stronger than the trailing one.  The PPA rate seems very shallow, arguing for a broad core width of perhaps 12\degr.  This in turn suggests that the core may not be complete, with a weak or absent trailing portion.  The 600-MHz profiles are poorly resolved but roughly compatible in overall width.  This core-cone triple model can only then be conjectural.    
\vskip 0.09in
\noindent\textit{\textbf{B1143--60}}: The \citet{jk18} 1.4-GHz profile shows both a clear triple structure and PPA traverse, whereas the 600-MHz profile \citep{CMH} is poorly resolved and conflates the features.  We thus model the profiles with a core-inner cone model.  No scattering time is available.  
\vskip 0.09in
\noindent\textit{\textbf{B1154--62}}: While the core-single classification goes back to ET VI/IX, the beautiful \citet{jk18} 1.4-GHz together with the \citet{kj06} 3.1-GHz profile confirm it absolutely.  The profiles permit a small change in the estimated $\alpha$ value.  No conal ``outriders'' are discernible in the older profiles, and the larger 631-MHz width probably signals the onset of the large scattering measured by \citet{abs86}.  
\vskip 0.09in
\noindent\textit{\textbf{B1159--58}}: The \citet{WMLQ} and \citet{MHQ} profiles are compatible and seem to be mainly core emission, but here the \citet{jk18} profiles seems aberrant. The PPA rate is unclear as is the nature of the conflated trailing feature.  No scattering time is available.  
\vskip 0.09in
\noindent\textit{\textbf{B1221--63}}:  Again the recent \citet{jk18} 1.4-GHz and \citet{sjd+21} 2.8-GHz profiles dramatically confirm the ET VI/IX findings of a triple {\textbf T} geometry.  Apart from the \citet{MHM} 1612-MHz profile, all the others permit only estimates of their dimensions---though there is a 268-MHz observation \citep{MHMb} that shows no obvious scattering effects.  No scattering timescale is available. 
\vskip 0.09in
\noindent\textit{\textbf{B1222--63}}: The 1.4-GHz profiles show a bright central feature with a well separated leading component and a weaker conflated trailing feature.  The PPA rate is complex, but a plausible estimate is about -12\degr/\degr.  These dimensions are compatible with an outer conal/core single or triple configuration.  No observations at other frequencies are available.  Nor is there a scattering time measurement.  
\vskip 0.09in
\noindent\textit{\textbf{B1240--64}}:  Little additional can be added to the ET VI and ET IX analyses apart from the very well measured profiles at 1.4 and 3.1 GHz by \citet{kj06} which seem to show the development of the conal ``outriders'', the leading of which may also be visible at 8.4 GHz \citep{JKW06}.  
\vskip 0.09in
\noindent\textit{\textbf{B1259--63}}: This highly energetic pulsar's broad pointed double profile with sharp inner edges has long been fascinating \citep[\eg][]{vonHoensbroech1997}.  High energy is usually associated with core radiation, but no core feature is seen.  The PPA rate is very well defined and very very shallow, with the result that the sign of $\beta$ is pertinent to modeling.  The profiles increase in width at lower frequencies, so we use an outer cone geometry.  Negative $\beta$ results in a smaller sightline circle that does not accommodate the increased profile width at 674 MHz \citep{JNK98}.  Positive $\beta$, however, gives a value of +16.5\degr, which would then pass the core beam by almost 3 times its radius.  This unusual geometry may explain the conal double profile of such an energetic pulsar.  No scattering measurement is available.   
\vskip 0.09in
\noindent\textit{\textbf{B1302--64}}: The \citet{jk18} profile shows a clear triple structure, while the 600-MHz \citep{CMH} observation is poorly resolved and conflated.  The PPA rates seem roughly compatible in the two observations.  We model the geometry with a core and outer conal beams.  \citet{abs86} provide a scattering time measurement.  
\vskip 0.09in
\noindent\textit{\textbf{B1303--66}}: The \citet{jk18} apparently double profile seems to have a weak conflated trailing feature.  We thus model it as a core-single or core-cone triple configuration, using the shallower PPA rate associated with the peak region.  A conal double interpretation is also possible, and only profiles at other frequencies can indicate which direction is correct.  \citet{joh90} measure a large scattering time scale.
\vskip 0.09in
\noindent\textit{\textbf{B1309--55}}: The 660-MHz profile \citep{QMLG95} shows a Gaussian shape with perhaps two conflated features as well as a well defined PPA rate.  The EPN site also shows a 1.4-GHz Stokes $I$ profile \citep{D'Amico+98} that has a similar shape and is perhaps slightly narrower.  Either a conal or core beam could generate this configuration, but in modeling it we tilt to the former.  No scattering timescale has been published.  
\vskip 0.09in
\noindent\textit{\textbf{B1316--60}}: The 1.4-GHz \citet{jk18} profile shows a tripartite structure with a break indicating a trailing feature.  We model it using a {\textbf S$_t$} or perhaps {\textbf T} geometry.  The PPA traverse provides no clear value but --10\degr/\degr is plausible.   \citet{joh90} give a scattering time estimate.  
\vskip 0.09in
\noindent\textit{\textbf{B1317--53}}: All the profiles \citep[and a 450-MHz $I$ \citep{D'Amico+98} in the EPN Database]{jk18,CMH} are Gaussian shaped, and the well measured 1.4 GHz one has a steep PPA rate on the trailing edge.  We thus model them with a core-single beam.  No scattering time scale is available.  
\vskip 0.09in
\noindent\textit{\textbf{B1323--58}}:  The \citet{jk18} and \citet{WMLQ} profiles suggest a triple structure, but the former gives only a hint of the trailing conal ``outrider''.  This is resolved by \citet{kj06} where it is clearly present at 3.1 GHz.  The PPA traverse is complex but shows a steep negative rate.  We thus model the above using an inner cone/core {\textbf T} geometry.  Scattering ``tails'' are seen on all the profiles below 1 GHz, and \citet{abs86} provide a $t_{\rm scatt}$ value.
\vskip 0.09in
\noindent\textit{\textbf{B1323--62}}:  The pulsar is too far south for GL98, so we depend on JK18, MHM, TvO, and MHMA.  As we are often finding the first profile is of higher quality and shows a tripartite structure with a well defined PPA rate.  Other profiles are available at high frequencies, and the 3.1-GHz \citet{kj06} shows the core-cone {\textbf T} structure clearly.  A scattering ``tail'' is apparent at 950 MHz and dominates the 631-MHz one, probably in keeping with the \citet{lkk15} measurement.
\vskip 0.09in
\noindent\textit{\textbf{B1325--43}}:  An number of profiles are available, but they show different forms that may indicate moding:  the \citet{Force2015} and \citet{WMLQ} observations do not show the structure seen in \citet{jk18} at high frequency, and the \citet{MHQ} and \citet{mbm+16} profiles are similarly different.  Nonetheless, the PPA rate is well determined, and if the sometimes discernible central component is a core with a plausible width of about 3.8\degr, a core/inner cone triple {\textbf T} model describes the geometry adequately.  No scattering measurement is available.  
\vskip 0.09in
\noindent\textit{\textbf{B1325--49}}: The lone \citet{MHQ} 434-MHz profile without PPA information leaves little to go on.  The emission is very likely conal dominated---as the strong and coherent 3.3-$P$ amplitude modulation \citep{basu2016} suggests---so we model it with a {\textbf D} geometry assuming a central sightline passage.  \citet{kmn+15} measure a very small and near average scattering level. 
\vskip 0.09in
\noindent\textit{\textbf{B1334--61}}: The 1.4-GHz \citet{jk18} profile shows perhaps three features distorted by a scattering ``tail'' as measured by \citet{joh90}.  (The Q95 profile is poor by comparison.) Nonetheless, the early part of the profile can be interpreted to argue a triple structure with a less prominent core component surrounded by conal features.  An inner conal/core model for the geometry then seems very workable.  
\vskip 0.09in
\noindent\textit{\textbf{B1338--62}}: The 1.4-GHz profiles show a long scattering ``tail'' probably compatible with the \citet{joh90} value, so no profile interpretation is possible.  However, the 8.4-GHz profile \citep{JKW06} shows what is probably a core-cone structure, but with no PPA information.  A toy model wherein $\alpha$ is 90\degr, $\beta$ 0\degr, and the conal width the 8.4-GHz value of 17\degr\ gives a slightly too small conal beam radius.  However, a 2\degr\ increase down to 1.4 GHz would correct it---strong evidence for this interpretation.
\vskip 0.09in
\noindent\textit{\textbf{B1353--62}}: The pulsar has a beautiful triple profile at both 1.4 (JK18) and 3.1 GHz \citep{kj06} with the expected weaker core at the higher frequency.  The PPA rate is estimated from the region just prior to the core.  No lower frequency profiles are available, so we model the profile using a core-single {\textbf S$_t$} configuration but a {\textbf T} could also be possible.  No spindown has been measured, so a number of parameters cannot be computed.  Neither has a scattering time been measured. 
\vskip 0.09in
\noindent\textit{\textbf{B1356--60}}:  \citet{kj06} shows a usual core-single evolution between 1.4 and 3.1 GHz, and \citet{JKW06} seems to confirm this at 8.4 GHz.  The \citet{WMLQ} profile is poorly resolved, but \citet{MHQ} and \citet{TvO} show the onset of scattering below 1 GHz.  The best profiles show a steepest-gradient PPA point on the trailing edge of the profile---a problematic circumstance---but we accept it at about +11\degr/\degr.  A core width can be measured from the 1.4- and 3.1-GHz profiles at about 7.2\degr, which implies that $\alpha$ is some 72\degr, and the conal width can be estimated both from 3.1- and 8.4-GHz profiles at about 11-12\degr\ (which we show at 1.4 and 1.6 GHz on the model plot).  These seemingly well determined values are interesting in that they imply a conal beam radius that is much smaller than the expected value for a pulsar of this rotation rate.  Several possible instances of ``more inner'' cones have been encountered previously, but this seems to be a potentially strong example.  \citet{lkk15} measure a large scattering time.  
\vskip 0.09in
\noindent\textit{\textbf{B1358--63}}:  The 1.4-GHz \citet{jk18} profile shows a triple structure with a conflated core and trailing conal ``outrider''.  The PPA rate is well determined and if the core width is plausibly about 4\degr, an inner cone/core {\textbf T} beam model works adequately.  No scattering measurement is available.  
\vskip 0.09in
\noindent\textit{\textbf{B1409--62}}: The \citet{jk18} 1.4-GHz observation shows a narrow profile with 3 breaks suggesting four features---and a difficult to interpret PPA traverse.  We take the outer two features to be conal and the inner two a bifurcated core, which combined have the right width to reflect the polar cap angular diameter, making $\alpha$ about 90/\degr.  An inner cone {\textbf S$_t$} geometry then requires a $\beta$ of about 5.7\degr\ or a PPA rate of about 10\degr/\degr---a value probably compatible with the sightline traverse, but not measurable from it.  
\vskip 0.09in
\noindent\textit{\textbf{B1424--55}}: The profiles are all Gaussian shaped with nearly constant widths, and the PPA rate is well defined by \citet{jk18}, so we model it as a classic core-single beam where no conal ``outriders'' are seen.  No scattering measurement.   
\vskip 0.09in
\noindent\textit{\textbf{B1426--66}}:  ET IX classified this pulsar as a core-inner cone triple; however, the broad leading component seems to have two conflated parts in some profiles.  Further, some profiles \citep{JohnstonI} show a weak trailing component.  The bright central component is often marked by antisymmetirc $V$, and we take it as a core component, though it is slightly narrower than the polar cap diameter.  Here we model the leading and trailing components as an outer cone, but this leaves the possibility that the inner cone is also active, perhaps with its trailing component conflated with the core---and this conjecture is supported by the quantitative geometry.  In addition to the above we depend on MHM, TvO, MHMA and JKW06 as well as unpublished Parkes 268- and 170-MHz profiles (MHMb) that show no obvious scattering effects.  
\vskip 0.09in
\noindent\textit{\textbf{B1436--63}}: Both the EPN 1.4-GHz and MHQ 658-MHz have the same form but only the latter is polarimetric and there is little clarity about the PPA traverse.  More information is need to interpret this pulsar's geometry.  
\vskip 0.09in
\noindent\textit{\textbf{B1449--64}}:  Sensitive recent observations \citep{jk18,sjd+21, kj06} support the ET VI core-inner cone triple {\textbf T} configuration.  Profiles have been obtained at 271 and 170 MHz (MHMb) and scattering may be visible on the extended trailing side of the latter, though no measurement is available.  
\vskip 0.09in
\noindent\textit{\textbf{B1451--68}}: ET VI found B1451--68 to be a well studied pulsar with a five-component {\textbf M} profile; however, the components are conflated at most frequencies.  We rely on the Gaussian decomposion and fitting of \citet{Wu1998}.  
\vskip 0.09in
\noindent\textit{\textbf{B1504--43}}: The \citet{JKMG2008} profiles show a single feature of roughly constant width, with a well defined PPA traverse everywhere.  Despite the absence of conal outriders at 3.1 GHz, we model it as having probable core-single {\textbf S$_t$} geometry.  
\vskip 0.09in
\noindent\textit{\textbf{B1518--58}}: Two 1.4-GHz profiles \citep{jk18,QMLG95} show an asymmetric triangular profile, but the former is much better resolved.  Within the profiles there seem to be three features, a leading conal ``outrider'', a bright core component, and a very weak trailing ``outrider''.  Only by virtue of the 8.4-GHz observation \citep{JKW06} can the latter two features be distinguished, and at 1.4 GHz they are probably conflated by the substantial reported scattering \citep{joh90}.  No lower frequency profiles seem to exist, perhaps because they are ``scattered out''.
\vskip 0.09in
\noindent\textit{\textbf{B1523--55}}: The lone 658-MHz profile \citep{MHQ} could be either core or conal; however, the antisymmetric $V$ tilts toward a core-single interpretation and model.  No effect of the modest  scattering level \citep{kmn+15} is seen in the profile.
\vskip 0.09in
\noindent\textit{\textbf{B1524--39}}: The EPN 1.4-GHz profile is single, whereas the \citet{MHQ} 658- and 434-MHz profiles are double---all with slightly increasing widths hardly more than 5\degr.  \citet{basu2016} find several long period phase-modulation features that deserve further study.  The PPA rate is well defined so we model it with a conal single {\textbf S$_d$} geometry.  No scattering measurement is available.  
\vskip 0.09in
\noindent\textit{\textbf{B1530--53}}: Here again we depend on MHM, MHMA, TvO and HMAK as well as \citet{MHMb} at 271 and 170 MHz.  At first glance all the profiles look scattered, but almost all have a total width of about 20\degr.  In most there is also a suggestion of a weak component following the strong leading one and in several a hint of what could be a trailing inner conal component.  This overall structure is often seen in conal quadruple profiles, and we so model it with a double conal {c\textbf Q} beam configuration.  The $L/I$ is low and the PPA rate poorly defined, but --30\degr/\degr\ is a reasonable estimate.  The inner conal dimensions are very rough estimates.  Only the 170-MHz profile has a larger overall width, perhaps due to scattering, though no measurement is available in the literature. 
\vskip 0.09in
\noindent\textit{\textbf{B1535--56}}: The \citet{jk18} 1.4-GHz profile shows a triple structure, especially in $L$ with an antisymmetric $V$ signature under the central component, while the \citet{QMLG95} profile is too poorly resolved to be useful.  This triple structure is even clearer in the \citet{JKW06} 8.4-GHz observation.  The PPA traverse is inconsistent with a +ve slope early, a set of ``kinks'' under the peak, and a trailing shallow region---all probably signaling an unresolved rotation, so we take $\beta$ to be zero.  We thus model the beam geometry using a core/inner cone configuration.  No lower frequency observation or scattering time value is available.  
\vskip 0.09in
\noindent\textit{\textbf{B1541--52}}: The \citet{jk18} sensitive 1.4-GHz profile shows an asymmetric tripartite structure that almost looks like scattering; however, the \citet{MHQ} 658-MHz profile has a very similar shape.  We therefore model it with a core/inner conal {\textbf S$_t$} geometry.  No scattering time measurement is available.   
\vskip 0.09in
\noindent\textit{\textbf{B1555--55}}: There is only the \citet{MHQ} 658-MHz profile that shows a triple form strongly suggesting an {\textbf S$_t$} or {\textbf T} geometry.  We so model it, guessing a PPA rate as the profile gives no information.  No scattering time is available.  
\vskip 0.09in
\noindent\textit{\textbf{B1556--44}}: We support the ET IX geometry with minor changes, including observations showing three components down to 600 MHz, so {\textbf S$_t$/T} gives a more accurate classification.  The profile structure at 3.1 GHz \citep{JohnstonII} is more complex and requires further study.
\vskip 0.09in
\noindent\textit{\textbf{B1557--50}}: Only the \citet{MHM} 1612-MHz profiles seems unscattered, reflecting a probable {\textbf S$_t$} geometry as in ET VI, perhaps with weak conal ``outriders''.  Both the 1368- and 950-MHz profiles \citep{jk18,TvO} show substantial ``tails'' compatible with the large scattering measured by \citet{ldk+13}.  
\vskip 0.09in
\noindent\textit{\textbf{B1558--50}}: Here we depend on MHM, MHMA, TvO and more recently on JK18---and only the latter is well measured enough to guide an attempt at interpretation.  This profile has at least four features and perhaps a weak central core component.  The outer components seems to represent an inner cone, and the small longitude width together with the steep PPA traverse render an outer conal geometry impossible.  So here is a case where the inner pair of conal components are may be a ``more inner'' cone as they do not seem to differ in linear polarization mode.  We thus interpret the geometry using a core-inner cone triple {\textbf T} beam geometry as in ET VI, leaving aside the interesting inner component pair.  The core width is only estimated, as it can be measured in none of the profiles; however, about 3\degr\ indicating a near orthogonal geometry is very plausible.  No scattering is apparent in the available profiles, though \citet{abs86} measure a value.  
\vskip 0.09in
\noindent\textit{\textbf{B1600--49}}: The triple profile given by \citet{jk18} is unusual only in that the central core component seems to have two conflated parts.  Nonetheless, its dimensions and PPA rate square well with the inner cone/core geometry we use to model its geometry---and both \citet{MHQ,mbm+16} provide 600-MHz profiles, but the latter is better resolved.  \citet{basu2016} find a long period ($\sim$50 $P$) amplitude modulation.  \citet{kmn+15} measure a scattering timescale for the pulsar.  
\vskip 0.09in
\noindent\textit{\textbf{B1601--52}}:  The complex filled double profiles at 1.4 GHz and 660 MHz \citep{QMLG95} and \citet{jk18} do not seem to be compatible with the 325-MHz of \citet{BMM2020}.  Perhaps the former profiles show some inner conal emission as well as the two framing outer conal components, and perhaps the profiles exhibit moding.  More detailed study is needed.  For the present we model the pulsar with an outer conal double geometry.  No scattering time scale is available.  
\vskip 0.09in
\noindent\textit{\textbf{B1610--50}}: The well measured 1.4-GHz \citet{jk18} profile shows the trailing ``tail'' and flat PPA characteristic of scattering, so no geometrical model is possible.  Nonetheless, this energetic pulsar is very likely core dominated with a probable {\textbf S$_t$} or {\textbf T} profile at higher frequencies.  \citet{joh90} give a values for the scattering time.  
\vskip 0.09in
\noindent\textit{\textbf{B1620--42}}: The \citet{jk18} profile shows a classic core-single {\textbf S$_t$} profile with a conal ``outrider'' pair, antisymmetric $V$, and a PPA traverse from which a rate can be estimated.  We model it with an inner cone.  A scattering timescale is available \citep{kmn+15}.
\vskip 0.09in
\noindent\textit{\textbf{B1629--50}}: The \citet{jk18} 1.4-GHz profile seems to have three parts with a bright central putative core component.  It may also be distorted by the substantial scattering \citep{joh90}---and the flat PPA traverse and slight ``tail'' may reflect this.  We thus model the geometry using a core/inner cone {\textbf S$_t$} model, taking the PPA rate at +6\degr/\degr, as it seems to be just after the early 90\degr\ ``jump''.  No higher frequency profile exists to check whether this interpretation is correct.
\vskip 0.09in 
\noindent\textit{\textbf{B1641--45}}: Conal ``outriders'' are very clear in the 8.4-GHz profile of \citet{JKW06} and perceptible in both the \citet{kj06} 3.1/1.4-GHz and \citep{jk18} profiles.  Even at 638-MHz in the MHMA profile, a leading conal component seems to survive the huge scattering distortion, leading us to revise the ET VI {\textbf S$_t$} classification.  A steep PPA traverse under the trailing edge of the core component is visible only at 3.1 GHz and appears to represent the principle difference between the latter two profiles.  Strong scattering in measured by \citet{ldk+13}.  
\vskip 0.09in
\noindent\textit{\textbf{B1641--68}}: The splendid 1.4-GHz \citet{jk18} profile shows a well resolved double cone structure with some $V$ signature in the center though a core component is not visible [clear structures not seen in the poorly resolved 1560-MHz \citep{WMLQ} profile].  A similar structure can be gleaned in the \citet{QMLG95} 660-MHz profile, though with less detail.  The inner conal dimensions are estimated as is the very steep PPA rate.  Given that no core component is visible, we model the profile dimensions using a conal quadruple {c\textbf Q} geometry.  
\vskip 0.09in
\noindent\textit{\textbf{B1647--52}}:  The 1368- and 658-MHz \citep{jk18,MHQ} profiles are similar and the PPA rate is well defined in both.  We thus model the geometry using a conal single {\textbf S$_d$} model.  No scattering time measurement is available.
\vskip 0.09in
\noindent\textit{\textbf{B1648--42}}: The very well measured 1.4-GHz profile of \citet{jk18} supports the ET IX analysis with a suggestion of a weak trailing conal feature. All the profiles \citep{jk18,TvO} show two broad components and a well defined PPA traverse---though the steepest point in on the trailing edge of the profiles.  The delayed PPA inflection suggests a core-cone triple or core-single geometry, and we so model it.  Observations only extend down to 950 MHz, so much could be learned with lower frequency observations and a fluctuation-spectral analysis.  No scattering measurement is available.  
\vskip 0.09in
\noindent\textit{\textbf{B1657--45}}: The only two profiles are at 1.4 GHz, and the \citet{jk18} shows two components and an ``S''-shaped PPA traverse along with what appears to be the beginnings of a scattering ``tail'' in accord with \citet{joh90}.  We model it with in inner conal double {\textbf D} geometry.
\vskip 0.09in
\noindent\textit{\textbf{B1659--60}}: \citet{QMLG95} provide the only polarimetry at 660 MHz, which shows a well resolved double profile and a short steep PPA traverse.  We model the geometry using a 1.4-GHz EPN profile \citep{D'Amico+98} and an old 408-MHz one \citep{LM88} as well.  No scattering measurement is available.  
\vskip 0.09in
\noindent\textit{\textbf{B1703--40}}: The three available profiles are all at 1.4 GHz, and all have long scattering ``tails''.  This is a heavily scattered pulsar as measured by \citet{joh90}.  The emission is probably a core beam, but a conal one cannot be ruled out.  Only observations at highter frequencies can assess the intrinsic beam geometry.  
\vskip 0.09in
\noindent\textit{\textbf{B1706--44}}: This highly energetic pulsar has a Gaussian shaped profile with nearly complete $L/I$ and an ``S''shaped PPA traverse.  No hint of conal ``outriders'' are seen at 8.4 GHz \citep{JKW06} or above. This seems to be a classic core single {\textbf S$_t$} beam traverse.  The EPN 450-MHz $I$ profile is much broader, but surprisingly no scattering analysis is available.
\vskip 0.09in
\noindent\textit{\textbf{B1719-37}}:  All the profiles are Gaussian shaped, and the \citet{jk18} shows the most detail as usual.  Although no conal ``outriders'' can be discerned, we model this energetic pulsar with a core-single {\textbf S$_t$} beam geometry.  The increased width at 333 MHz \citep{mbm+16} may owe to scattering as no measurement is available to assess it.  
\vskip 0.09in
\noindent\textit{\textbf{B1727-47}}:  This long studied southern pulsar was classified as a one-sided triple in Paper VI.  We now see that this was incorrect, as three components show clearly in most of the \cite{JKMG2008} profiles.  The core width cannot be accurately measured at any frequency, but about 3\degr\ is a plausible value and compatible with an inner conal geometry.  We also depend on the 1612 (MHM), 950 (TvO), 631 (MHMA), 400 (HMAK) and 270-MHz (MHMb) observations as the pulsar is too far south for GL98.  Evidence of a scattering ``tail'' is seen at 271 MHz, which seems compatible with the \cite{Krishnakumar2015} $t_{\rm scatt}$ value. 
\vskip 0.09in
\noindent\textit{\textbf{B1737--39}}: The \citet{jk18} profile shows the triple structure at 1.4 GHz very clearly.  However, it is the \citet{WMLQ} observation that seems to give the best guidance about the PPA rate.  With this a core/inner conal triple {\textbf S$_t$} geometry can be modeled successfully.  \citet{basu2016} see fluctuation power with a period around 50 $P$.  The lower frequency profiles are primarily core, and scattering sets in visibily, perhaps even at 950 MHz, probably in accordance with the \citet{kmn+15} time scale measurement.
\vskip 0.09in
\noindent\textit{\textbf{B1747--46}}:  The available profiles \citep{MHM,MHQ,TvO,MHMA, MHAK,MHMb} and now \citet{jk18} do show a perplexing evolution, where the two-component profile above 1 GHz is single at 600 MHz, double again at 400 MHz, single at 270 MHz and barely double at 170 MHz.  ET VI tried to see this as a core-cone evolution pattern, but here we believe it is basically a conal double {\textbf D} beam system in which either mode-switching or perhaps ``absorption'' (as in pulsar B0809+74) leads to the unusually and perplexingly complicated evolution.  Scattering may be seen in the 170-MHz profile according to the measurement by \cite{brg99}.   
\vskip 0.09in
\noindent\textit{\textbf{B1758--29}}:  The pulsar has a classic core/outer cone triple {\textbf T} geometry.  \citet{jk18} trace the PPA traverse at 1.4 GHz and \citet{mbm+16} at both 610 and 334 MHz.  \citet{Weltevrede2006} find a flat fluctuation spectrum, whereas \citet{basu2016} see fluctuations at around 2 $P$.  No scattering analysis is available.
\vskip 0.09in
\noindent\textit{\textbf{B1800--27}}: The three available profiles are all at 1.4 GHz and poor.  We model the beam using a conal single geometry, and while we use an inner one, it could be either.  
\vskip 0.09in
\noindent\textit{\textbf{B1806--53}}:  We have only a single 660-MHz profile from \citet{QMLG95}, and it shows a triple structure with a bright central component.  The PPA traverse is well defined, but may miss a steep central rotation.  We model the beams using an inner cone/core configuration, taking $\beta$ as zero.  No scattering time scale is available.  
\vskip 0.09in
\noindent\textit{\textbf{B1851--79}}: The 660-MHz profile shows two features and a long ``tail''---however, it cannot be due to scattering, as an EPN $I$ profile \citep{D'Amico+98} shows a similar form less well.  Several pulsars with conal quadruple c{\textbf Q} profiles have this form with the leading  pair much brighter than the trailing pair \citep[\eg][]{rankin2017}.  So we model it.  The PPA rate is probably reflects the latter several points, and of course the inner conal dimensions can only be estimated roughly.  No scattering time measurment is available.   
\vskip 0.09in
\noindent\textit{\textbf{B2048--72}}:  The various profiles show a filled double structure that changes little in overall width.  The PPA rate is well defined, and the central region shows a consistent $V$ signature.  We model the profile using an outer conal double {\textbf D} geometry, but a conflated core component with a width of some 6\degr\ may be active in the profile center---a structure possibly hinted at by the 450-MHz profile on EPN \citep{JNK98}---so a {\textbf T} geometry is also very possible.  No scattering time has been measured.
\vskip 0.09in
\noindent\textit{\textbf{B2123-67}}:  The broad asymmetric JK18 profile is very similar in form and width to that at 646 MHz, so does not appear to be distorted by the significant scattering measured by \citet{abs86}.  The PPA rate could be about +4\degr/\degr\ per the scant evidence. If this is a conal double profile with a weaker conflated second component, the overall width might be about 36\degr\ at both frequencies.  
\vskip 0.09in
\noindent\textit{\textbf{B2321--61}}: This object has an obvious conal double {\textbf D} geometry with an ``S''-shaped PPA traverse.  However, it also shows antisymmetic $V$ suggesting that some core radiation fills the center of the profile.  No scattering analysis is available. 
\vskip 0.09in

\onecolumn

\begin{center}
\begin{longtable}{lc|cccc|cccc|ccccc}
\caption{Far South Multiband Population Emission Beam Geometry} \label{tabB3}  \\
  \hline
      Pulsar &  Class & $W_{c}$ & $\alpha$ & $R$ & $\beta$ &  $W_i$ & $\rho_i$ & $W_o$  & $\rho_o$ & $W_{c}$ & $W_i$ & $\rho_i$    & $W_o$  & $\rho_o$ \\
  &   & (\degr) & (\degr) & (\degr/\degr) & (\degr) & (\degr) & (\degr) & (\degr) & (\degr) & (\degr) & (\degr) & (\degr) & (\degr) & (\degr) \\
  \hline
  & & \multicolumn{4}{c|}{1-GHz Geometry} & \multicolumn{4}{c|}{1-GHz Cone Sizes} & \multicolumn{5}{c|}{100-MHz Cone Sizes} \\
  \hline
  \hline
\endfirsthead
   \hline
      Pulsar &  Class & $W_{c}$ & $\alpha$ & $R$ & $\beta$ &  $W_i$ & $\rho_i$ & $W_o$  & $\rho_o$ & $W_{c}$ & $W_i$ & $\rho_i$    & $W_o$  & $\rho_o$ \\
  &   & (\degr) & (\degr) & (\degr/\degr) & (\degr) & (\degr) & (\degr) & (\degr) & (\degr) & (\degr) & (\degr) & (\degr) & (\degr) & (\degr) \\
  \hline
  & & \multicolumn{4}{c|}{1-GHz Geometry} & \multicolumn{4}{c|}{1-GHz Cone Sizes} & \multicolumn{5}{c|}{100-MHz Cone Sizes} \\
  \hline
  \hline
\endhead
  B0203--40 & Sd? & --- & 90 & +15 & +3.8 & $\sim$8 & 5.5 &  --- &  --- &  --- & 9 & 5.9 &  --- &  --- \\
B0254--53 & D & --- & 57 & -8.3 & +5.8 & 7.0 & 6.5 &  --- &  --- &  --- & 9 & 7.0 &  --- &  --- \\
B0403--76 & T & $\sim$9? & {\bf 22} & -4.3 & +5.1 & 14.5 & 5.9 &  --- &  --- &  --- & 16.5 & 6.2 &  --- &  --- \\
B0529--66 & ?? & --- &  --- &  --- &  --- &  --- &  --- &  --- &  --- &  --- &  --- &  --- &  --- &  --- \\
B0538--75 & cQ/M? & --- & 22 & -14 & +1.6 & $\sim$18 & 3.8 & 25 & 5.1 &  --- & 18 & 3.8 & $\sim$29 & 5.9 \\
\\[-3pt]
B0736--40 & T & $\sim$14 & {\bf 17} & +3.8 & +4.3 & 35 & 7.0 &  --- &  --- &  --- & 46 & 8.5 &  --- &  --- \\
B0743--53 & St? & 18 & {\bf 17} & -2.8 & +6.0 &  --- &  --- &  --- &  --- & 23 &  --- &  --- &  --- &  --- \\
B0808--47 & T & 5.5 & {\bf 37} & -11 & +3.1 & $\sim$16 & 5.9 &  --- &  --- & 12 &  --- &  --- &  --- &  --- \\
B0818--41 & D/M? & --- & 4 & -0.7 & +5.7 &  --- &  --- & $\sim$100 & 7.8 &  --- &  --- &  --- & $\sim$120 & 8.5 \\
B0833--45 & St & 7.8 & {\bf 90} & -4.8 & +12.0 & $\sim$14 & 13.9 &  --- &  --- & 18 &  --- &  --- &  --- &  --- \\
\\[-2pt]
B0835--41 & St & 3.7 & {\bf 50} & +8 & +5.5 &  --- &  --- & $\sim$10 & 6.8 & 5.2 &  --- &  --- &  --- &  --- \\
B0839--53 & Sd? & --- & 31 & +7.5 & +3.9 & $\sim$12 & 5.1 &  --- &  --- &  --- & 12 & 5.1 &  --- &  --- \\
B0855--61 & Sd? & --- & 21 & +5 & +4.1 & $\sim$7 & 4.3 &  --- &  --- &  --- & $\sim$8 & 4.4 &  --- &  --- \\
B0903--42 & D? & --- & 39 & -12 & +3.0 & $\sim$10 & 4.4 &  --- &  --- &  --- & 10 & 4.4 &  --- &  --- \\
B0904--74 & Sd & --- & 36 & +7 & +4.8 & 11 & 5.9 &  --- &  --- &  --- & 12 & 6.1 &  --- &  --- \\
\\[-3pt]
B0905--51 & T & 18 & {\bf 16} & -14 & +1.1 & $\sim$60 & 8.4 &  --- &  --- &  --- & 62 & 8.6 &  --- &  --- \\
B0906--49 & ?? & --- & 0 & 0 & 0 & 0 & 0.0 & 0 & 0.0 &  --- &  --- &  --- &  --- &  --- \\
B0909--71 & D & --- & 25 & $\approx$-15? & -1.6 &  --- &  --- & 23 & 4.9 &  --- &  --- &  --- & $\sim$36 & 7.4 \\
B0922--52 & T? & $\sim$2.8? & {\bf 90} & -45 & +1.3 & 9.5 & 4.9 &  --- &  --- & 2.8 & 10 & 5.2 &  --- &  --- \\
B0932--52 & cT/St? & $\sim$2? & 30 & -9 & 3.2 & $\sim$7 & 3.7 &  --- &  --- &  --- & 7 & 3.7 &  --- &  --- \\

\\[-2pt]
B0940--55 & St & 5.3 & {\bf 35} & +30 & +1.1 &  --- &  --- & $\sim$25 & 7.1 & $\sim$10 &  --- &  --- &  --- &  --- \\
B0941--56 & T? & 2.7 & {\bf 90} & +10 & +5.7 &  --- &  --- & $\sim$5 & 6.3 &  --- &  --- &  --- & $\sim$6 & 6.5 \\
B0953--52 & T & 3 & {\bf 62} & +25 & +2.0 & $\sim$13 &  --- & $\sim$13 & 6.1 & $\sim$3 &  --- &  --- & 13 & 6.1 \\
B0957--47 & T/M? & $\sim$14? & {\bf 12} & +4 & +3.0 &  --- &  --- & $\sim$54 & 7.0 &  --- &  --- &  --- & $\sim$58 & 7.4 \\
B0959--54 & St & 3.7 & {\bf 34} & +12 & +2.6 & $\sim$9 & 3.7 &  --- &  --- & 7.5 &  --- &  --- &  --- &  --- \\
\\[-3pt]
B1011--58 & St? & 2.7 & {\bf 90} & -18 & +3.2 & $\sim$7 & 4.7 &  --- &  --- &  --- &  --- &  --- &  --- &  --- \\
B1036--58 & D & --- & 65 & +15 & +3.5 & 9.0 & 5.4 &  --- &  --- &  --- & 6 & 4.4 &  --- &  --- \\
B1039--55 & Sd & --- & 32 & +9 & +3.4 & $\sim$8 & 4.0 &  --- &  --- &  --- &  --- &  --- &  --- &  --- \\
B1046--58 & St & 9.1 & {\bf 50} & +5 & +8.8 & 22 & 12.5 &  --- &  --- &  --- &  --- &  --- &  --- &  --- \\
B1054--62 & ?? & --- &  --- &  --- &  --- &  --- &  --- &  --- &  --- &  --- &  --- &  --- &  --- &  --- \\
\\[-2pt]
B1055--52m & T & $\sim$5.5? & {\bf 90} &  --- &  --- &  --- &  --- &  --- &  --- &  --- &  --- &  --- &  --- &  --- \\
B1056--78 & Sd? & --- & 18 & +5 & +3.5 & $\sim$6.5 & 3.7 &  --- &  --- &  --- & 6.3 & 3.7 &  --- &  --- \\
B1056--57 & Sd? & --- & 55 & -15 & +3.1 & 5.5 & 3.9 &  --- &  --- &  --- & 5.5 & 3.9 &  --- &  --- \\
B1107--56 & D/T? & --- & 34 & +8 & +4.0 & 14.5 & 5.8 &  --- &  --- &  --- &  --- &  --- &  --- &  --- \\
B1110--65 & T? & $\sim$5? & {\bf 58} & -20 & +2.4 & $\sim$16 & 7.3 &  --- &  --- & $\sim$5 & 16 & 7.3 &  --- &  --- \\
\\[-3pt]
B1112--60 & St? & 12 & {\bf 13} &  --- &  --- &  --- &  --- &  --- &  --- &  --- &  --- &  --- &  --- &  --- \\
B1114--41 & St & 4.0 & {\bf 39} & -12 & +3.0 &  --- &  --- &  --- &  --- & 5.0 &  --- &  --- &  --- &  --- \\
B1119--54 & St & $\sim$3 & {\bf $\sim$90} & +30 & +1.9 &  --- &  --- & $\sim$15 & 7.7 & $\sim$8 &  --- &  --- &  --- &  --- \\
B1131--62 & D & --- & 5 & -1.8 & +2.8 &  --- &  --- & $\sim$93 & 5.6 &  --- &  --- &  --- &  --- &  --- \\
B1133--55 & T? & $\approx$12? & {\bf 20} & +2.3 & +8.5 & $\sim$22 &  --- & $\sim$22 & 9.5 &  --- &  --- &  --- & 22 & 9.5 \\
\\[-2pt]
B1143--60 & T & $\sim$5.4? & {\bf 60} & -10 & +5.0 & 14.5 & 8.1 &  --- &  --- &  --- & 17 & 9.0 &  --- &  --- \\
B1154--62 & St & 11 & {\bf 21} &  --- &  --- &  --- &  --- &  --- &  --- & 17 &  --- &  --- &  --- &  --- \\
B1159--58 & St? & $\sim$6 & {\bf 37} &  --- &  --- &  --- &  --- &  --- &  --- &  --- &  --- &  --- &  --- &  --- \\
B1221--63 & T & $\sim$6? & {\bf 65} & +7 & +7.5 & 12.3 & 9.4 &  --- &  --- &  --- & 16.5 & 10.7 &  --- &  --- \\
B1222--63 & St/T? & 4.0 & {\bf 71} & -15 & +3.6 &  --- &  --- & $\sim$17 & 8.9 &  --- &  --- &  --- &  --- &  --- \\
\\[-3pt]
B1240--64 & St & 4.5 & {\bf 61} & +14 & +3.6 & $\sim$13 & 6.9 &  --- &  --- & 9 &  --- &  --- &  --- &  --- \\
B1259--63 & D & --- & 9 & -0.6 & +16.5 &  --- &  --- & $\sim$173 & 26.4 &  --- &  --- &  --- & $\sim$205 & 28.8 \\
B1302--64 & T & $\sim$10 & {\bf 19} & -2.8 & -6.6 &  --- &  --- & $\sim$27 & 7.5 &  --- &  --- &  --- & $\sim$25 & 7.4 \\
B1303--66 & St/T? & 15 & {\bf 14} & -2.4 & +5.7 & $\sim$18 & 6.2 &  --- &  --- &  --- &  --- &  --- &  --- &  --- \\
B1309--55 & Sd/St? & --- & 26 & +6 & +4.2 & 9 & 4.7 &  --- &  --- &  --- & 10 & 4.8 &  --- &  --- \\
\\[-10pt]
B1316--60 & St/T? & 5 & {\bf 67} & -10 & +5.3 & 13.5 & 8.2 &  --- &  --- &  --- &  --- &  --- &  --- &  --- \\
B1317--53 & St? & 10 & {\bf 28} & -7.5 & +3.5 &  --- &  --- &  --- &  --- & 10.1 &  --- &  --- &  --- &  --- \\
B1323--58 & T & 4.8 & {\bf 48} & -25 & +1.7 & 16 & 6.2 &  --- &  --- & $\sim$24 &  --- &  --- &  --- &  --- \\
B1323--62 & T & $\sim$3.4? & {\bf 85} & -13 & +4.4 & $\sim$8 & 5.9 &  --- &  --- &  --- & 25 & 13.2 &  --- &  --- \\
B1325--43 & T & $\sim$3.8? & {\bf 62} & +13 & +3.9 & 9.5 & 5.8 &  --- &  --- &  --- & 9.5 & 5.8 &  --- &  --- \\
\\[-6pt]
B1325--49 & D? & --- & 23 & $\infty$ & 0.0 & $\sim$18 & 3.5 &  --- &  --- &  --- &  --- &  --- &  --- &  --- \\
B1334--61 & St/T? & 9 & {\bf 14} & -10 & +1.4 & $\sim$28 & 3.8 &  --- &  --- &  --- &  --- &  --- &  --- &  --- \\
B1338--62 & St? & 5.6 & 90 & $\infty$ & 0.0 & $\sim$19 & 9.5 &  --- &  --- &  --- &  --- &  --- &  --- &  --- \\
B1353--62 & St/T? & 7.4 & {\bf 60} & -20 & +2.5 & $\sim$22 & 7.7 &  --- &  --- &  --- &  --- &  --- &  --- &  --- \\
B1356--60 & St & 7.2 & {\bf 72} & +11 & +5.0 & $\sim$12 & 7.6 &  --- &  --- & $\sim$16 &  --- &  --- &  --- &  --- \\
\\[-6pt]
B1358--63 & St & $\sim$4 & {\bf 42} & +12 & +3.2 & $\sim$10 & 4.7 &  --- &  --- & $\sim$5.0 &  --- &  --- &  --- &  --- \\
B1409--62 & St/T? & 3.9 & {\bf 88} & +10 & +5.7 & 7.5 & 6.9 &  --- &  --- &  --- &  --- &  --- &  --- &  --- \\
B1424--55 & St & 8.5 & {\bf 22} & +5 & +4.4 &  --- &  --- &  --- &  --- & $\sim$9 &  --- &  --- &  --- &  --- \\
B1426--66 & T/M? & 2.2 & {\bf 90} & 150 & -0.4 & $\sim$10 & 5.0 & $\sim$13 & 6.5 & 4.5 &  --- &  --- & 13 & 6.5 \\
B1436--63 & ?? & --- &  --- &  --- &  --- &  --- &  --- &  --- &  --- &  --- &  --- &  --- &  --- &  --- \\
\\[-6pt]
B1449--64 & St & 7 & {\bf 56} & +7 & +6.8 & 18.3 & 10.4 &  --- &  --- & 8.9 &  --- &  --- &  --- &  --- \\
B1451--68 & M & $\sim$7.6 & {\bf 39} & +5.7 & -6.3 & 19.7 & 8.5 & 33.1 & 11.5 & 9.0 &  --- &  --- & $\sim$52 & 16.3 \\
B1504--43 & St? & 5.6 & {\bf 55} & -18 & +2.6 &  --- &  --- &  --- &  --- & 6.2 &  --- &  --- &  --- &  --- \\
B1518--58 & St/T? & $\approx$9? & {\bf 26} & -4 & -6.2 & $\sim$15 & 6.8 &  --- &  --- &  --- &  --- &  --- &  --- &  --- \\
B1523--55 & St? & 5 & {\bf 29} &  --- &  --- &  --- &  --- &  --- &  --- &  --- &  --- &  --- &  --- &  --- \\
\\[-6pt]
B1524--39 & D & --- & 52 & -23 & +2.0 & $\sim$5 & 2.8 &  --- &  --- &  --- & 5 & 2.8 &  --- &  --- \\
B1530--53 & cQ & --- & 28 & -30 & +0.9 & $\sim$15 & 3.7 & 20 & 4.8 &  --- &  --- &  --- & 25 & 6.0 \\
B1535--56 & T & $\sim$6 & {\bf 56} & $\infty$ & 0.0 & $\sim$22 & 9.1 &  --- &  --- &  --- &  --- &  --- &  --- &  --- \\
B1541--52 & T & 6.2 & {\bf 69} & -23 & +2.3 & $\sim$21 & 10.2 &  --- &  --- & 6.1 & 22 & 10.6 &  --- &  --- \\
B1555--55 & St? & 3.5 & {\bf 46} & +25 & +1.6 & 11.5 & 4.5 &  --- &  --- &  --- & 11.5 & 4.5 &  --- &  --- \\
\\[-6pt]
B1556--44 & St/T? & 8.7 & {\bf 34} & -13 & -2.4 & $\sim$30 & 8.4 &  --- &  --- & 9.0 &  --- &  --- &  --- &  --- \\
B1557--50 & St & $\sim$8? & {\bf 46} & +4.5 & 9.3 & $\sim$10 & 10.1 &  --- &  --- & 15 &  --- &  --- &  --- &  --- \\
B1558--50 & T? & $\sim$3? & {\bf 70} & +80 & +0.7 & $\sim$10 & 4.8 &  --- &  --- &  --- &  --- &  --- &  --- &  --- \\
B1600--49 & T & 4.9 & {\bf 61} & -23 & +2.2 & 16.5 & 7.6 &  --- &  --- & 4.8 & 19.5 & 8.9 &  --- &  --- \\
B1601--52 & D/cQ? & --- & 18 & -4.5 & +3.9 &  --- &  --- & $\sim$36 & 7.3 &  --- &  --- &  --- & $\sim$40 & 7.8 \\
\\[-6pt]
B1610--50 & St? & --- &  --- &  --- &  --- &  --- &  --- &  --- &  --- &  --- &  --- &  --- &  --- &  --- \\
B1620--42 & St & 9 & {\bf 27} & -6 & +4.3 & 23 & 7.0 &  --- &  --- &  --- &  --- &  --- &  --- &  --- \\
B1629--50 & St/T? & $\sim$7 & {\bf 36} & +6 & +5.6 & $\sim$15 & 7.3 &  --- &  --- &  --- &  --- &  --- &  --- &  --- \\
B1641--45 & T & 6.5 & {\bf 34} & +50 & +0.6 & $\sim$23 & 6.5 &  --- &  --- & $\sim$40 &  --- &  --- &  --- &  --- \\
B1641--68 & cQ/M? & 3.4 & {\bf 33} & -60 & +0.5 & $\sim$12 & 3.3 & 15.5 & 4.3 & 3.35 & 12 & 3.3 & 17 & 4.7 \\
\\[-6pt]
B1647--52 & Sd & --- & 43 & -9 & +4.3 & $\sim$9 & 5.4 &  --- &  --- &  --- & 9 & 5.4 &  --- &  --- \\
B1648--42 & St/T? & $\sim$17 & {\bf 9} & -7.5 & +1.2 & $\sim$55 & 4.7 &  --- &  --- &  --- &  --- &  --- &  --- &  --- \\
B1657--45 & D & --- & 34 & -7 & +4.6 & 21 & 7.7 &  --- &  --- &  --- &  --- &  --- &  --- &  --- \\
B1659--60 & D & --- & 11 & -30 & +0.4 &  --- &  --- & $\sim$110 & 10.3 &  --- &  --- &  --- & $\sim$110 & 10.3 \\
B1703--40 & ?? & --- &  --- &  --- &  --- &  --- &  --- &  --- &  --- &  --- &  --- &  --- &  --- &  --- \\
\\[-6pt]
B1706--44 & St & 21 & {\bf 21} & +2.3 & +9.1 &  --- &  --- &  --- &  --- & 44 &  --- &  --- &  --- &  --- \\
B1719--37 & St & 5.8 & {\bf 60} & +13 & +3.8 &  --- &  --- &  --- &  --- & $\sim$9 &  --- &  --- &  --- &  --- \\
B1727--47 & T & $\sim$3? & 0 & 0 & 0 & 9.8 & 5.0 &  --- &  --- &  --- & 11.5 & 5.6 &  --- &  --- \\
B1737--39 & St & 5.5 & {\bf 38} & +30 & +1.2 & $\sim$19 & 6.1 &  --- &  --- & $\sim$7 &  --- &  --- &  --- &  --- \\
B1747--46 & D & --- & 90 & -20 & +2.9 & 8.1 & 5.0 &  --- &  --- &  --- & 8.8 & 5.2 &  --- &  --- \\
\\[-6pt]
B1758--29 & T & 3.8 & {\bf 38} & +45 & +0.8 &  --- &  --- & 17.5 & 5.5 &  --- &  --- &  --- & 19 & 6.0 \\
B1800--27 & Sd & --- & 14 & +2 & +6.9 & $\sim$22 & 7.7 &  --- &  --- &  --- &  --- &  --- &  --- &  --- \\
B1806--53 & T & $\sim$8 & {\bf 37} & $\infty$ & 0.0 & $\sim$29 & 8.7 &  --- &  --- &  --- &  --- &  --- &  --- &  --- \\
B1851--79 & cQ? & --- & 17 & -5 & -3.4 & $\sim$13 & 3.8 & $\sim$29 & 5.1 &  --- & 13 & 3.8 & $\sim$29 & 5.1 \\
B2048--72 & D/T? & $\sim$6.2? & {\bf 43} & +18 & +2.2 &  --- &  --- & $\sim$28 & 9.9 &  --- &  --- &  --- & 29 & 10.2 \\
\\[-6pt]
B2123--67 & D?? & --- & 18 & +4 & +4.4 & $\approx$36? & 7.6 &  --- &  --- &  --- & $\approx$36? & 7.6 &  --- &  --- \\
B2321--61 & D/T? & --- & 49 & -18 & +2.4 &  --- &  --- & 7.5 & 3.8 &  --- &  --- &  --- & $\sim$9 & 4.2 \\

  \hline
\end{longtable}
\end{center}

\begin{figure*}
\begin{center}
\includegraphics[width=180mm,height=218mm,angle=0.]{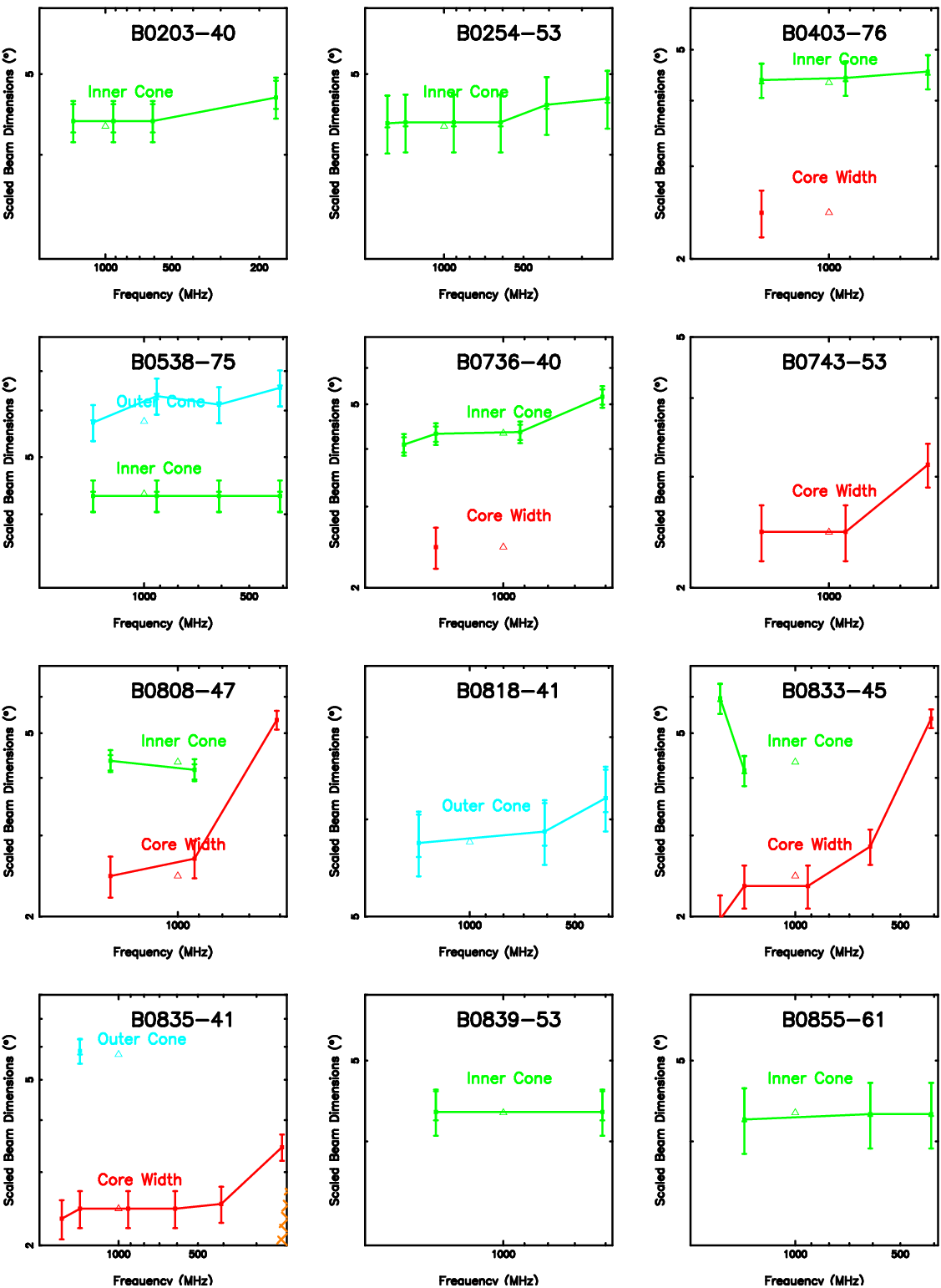}
\caption{Emission-beam geometry models for the Far South population.  The inner and outer conal beam radii and core widths are plotted as a function of radio frequency, scaled to a 1-second orthogonal rotator configuration (see text).  The error bars reflect 10\% uncertainties in measuring both the widths and the PPA sweep rate $R$.  The triangles at 1 GHz show the nominal beam dimensions.}
\label{figB1}
\end{center}
\end{figure*}

\begin{figure*}
\begin{center}
\includegraphics[width=180mm,height=225mm,angle=0.]{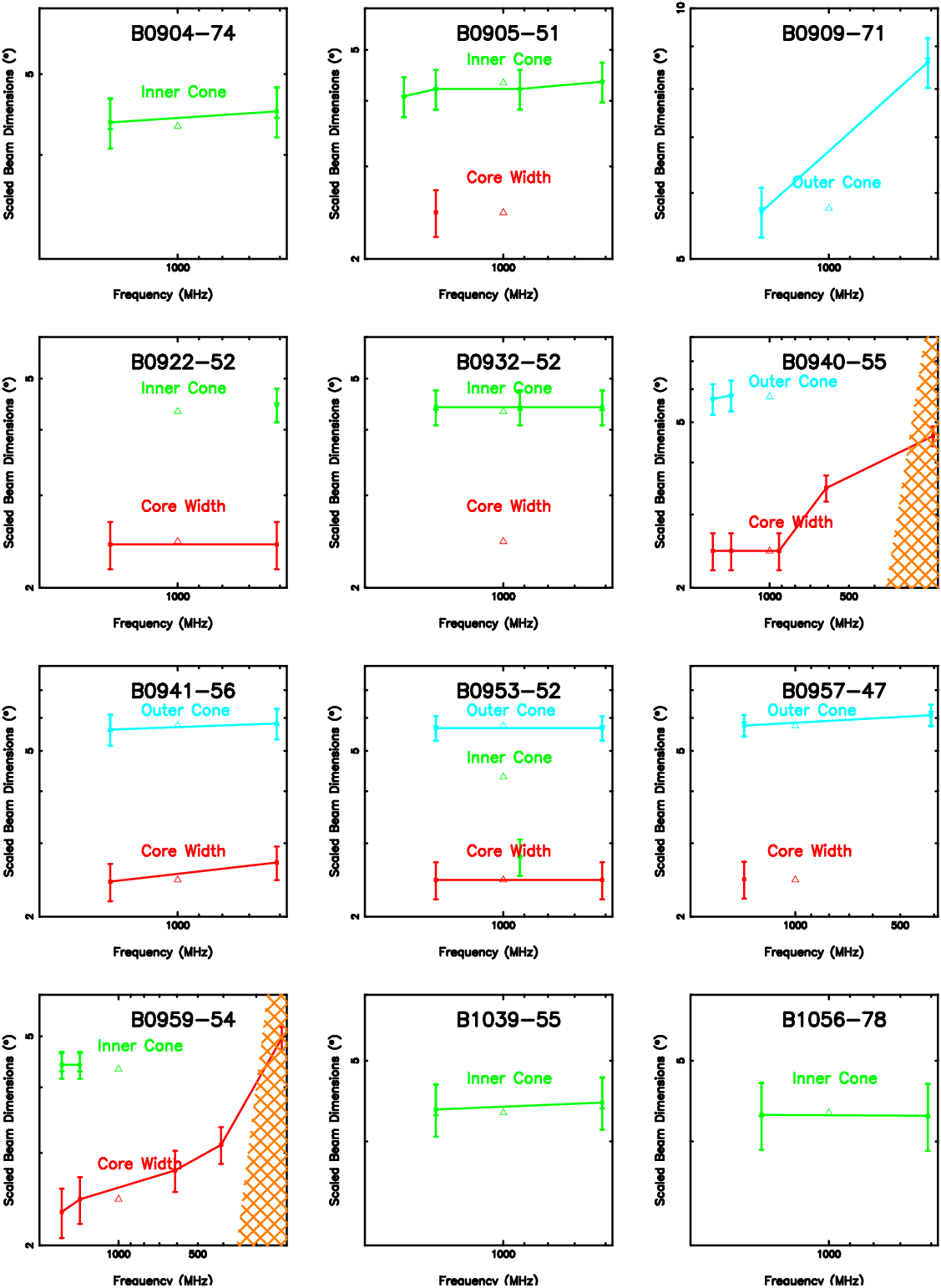}
\caption{Emission-beam geometry models as in Fig~\ref{figB1}.}
\label{figB2}
\end{center}
\end{figure*}

\begin{figure*}
\begin{center}
\includegraphics[width=180mm,height=225mm,angle=0.]{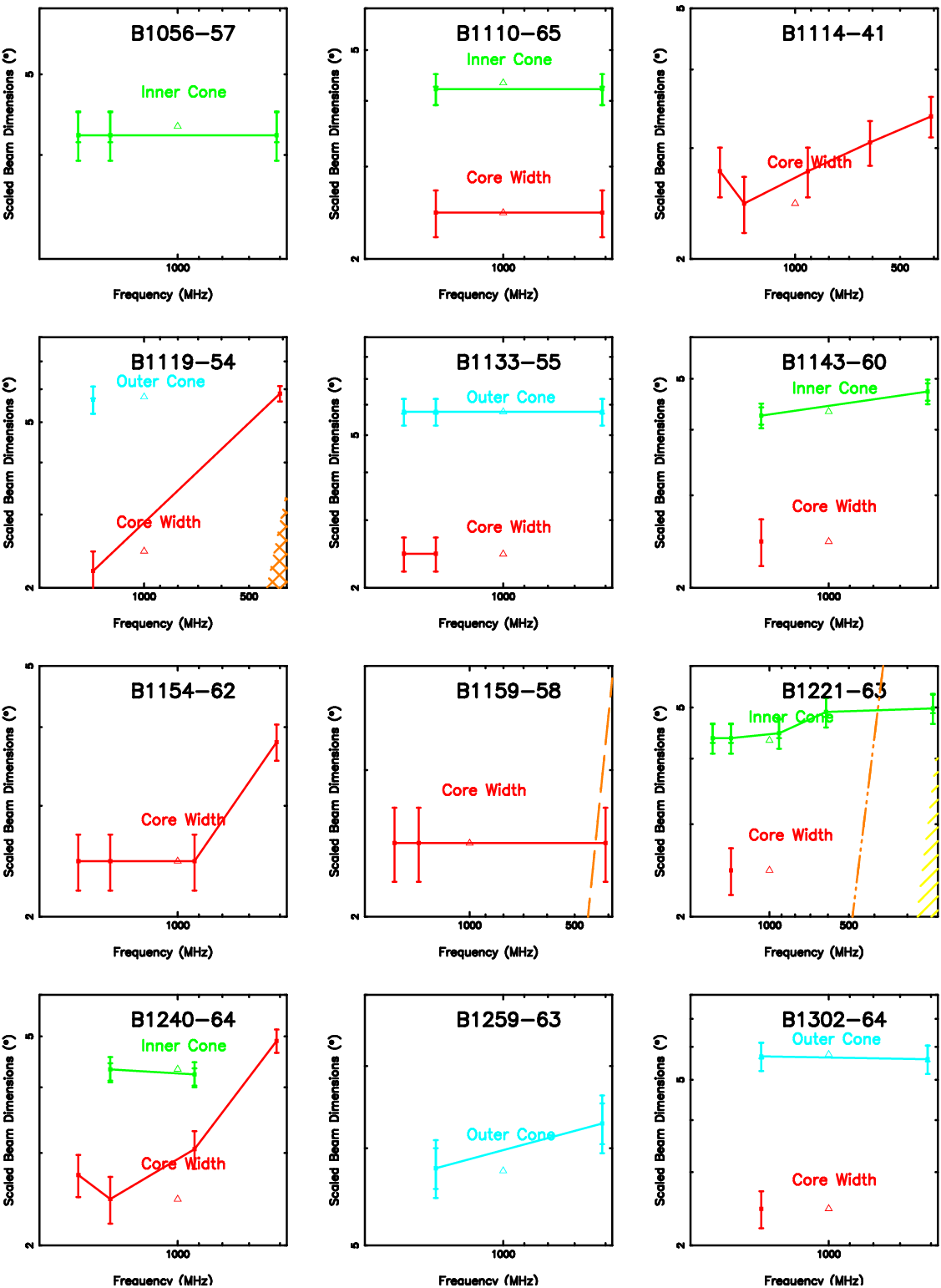}
\caption{Emission-beam geometry models as in Fig~\ref{figB1}.}
\label{figB3}
\end{center}
\end{figure*}

\begin{figure*}
\begin{center}
\includegraphics[width=180mm,height=225mm,angle=0.]{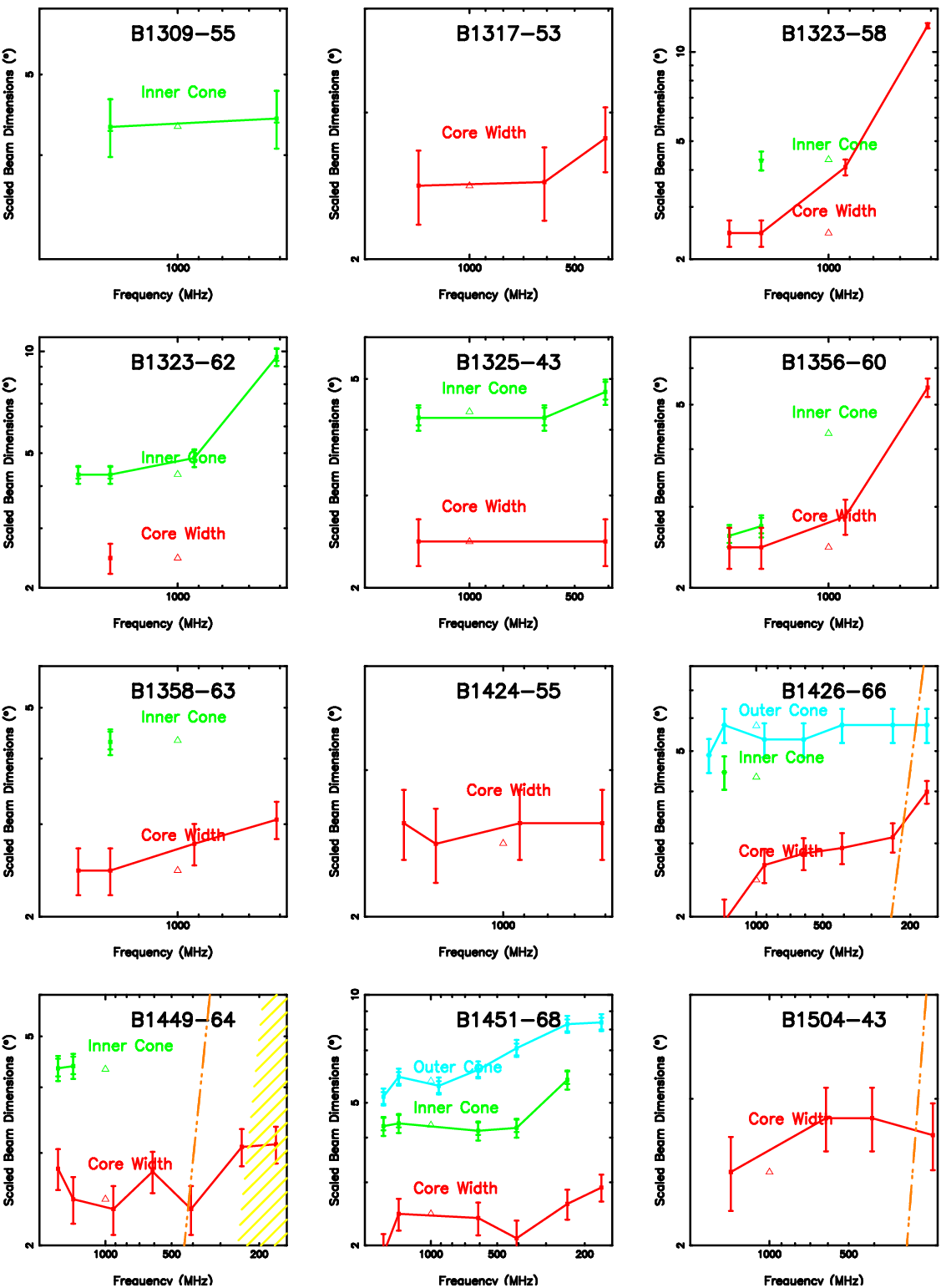}
\caption{Emission-beam geometry models as in Fig~\ref{figB1}.}
\label{figB4}
\end{center}
\end{figure*}

\begin{figure*}
\begin{center}
\includegraphics[width=180mm,height=225mm,angle=0.]{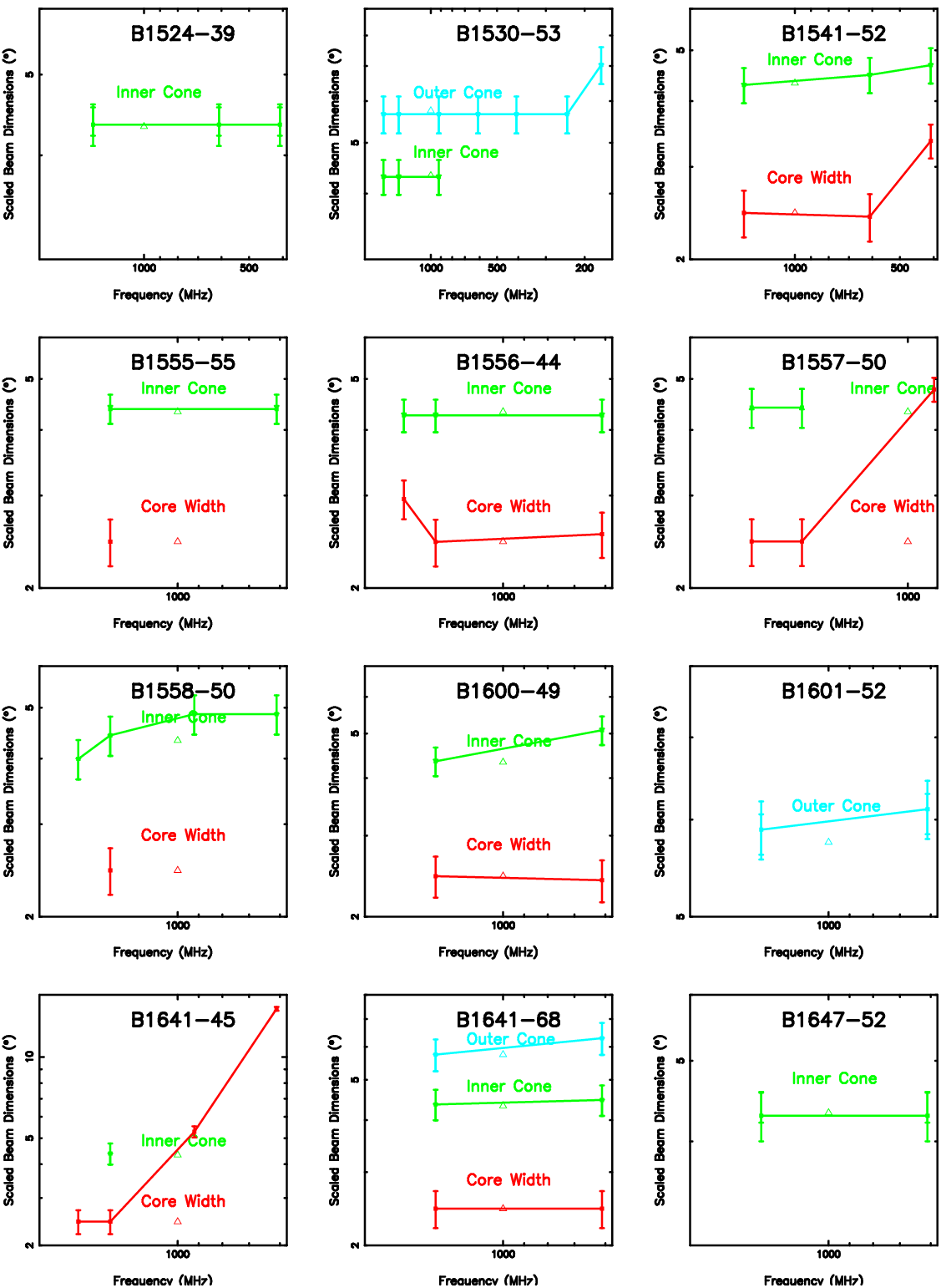}
\caption{Emission-beam geometry models as in Fig~\ref{figB1}.}
\label{figB5}
\end{center}
\end{figure*}

\begin{figure*}
\begin{center}
\includegraphics[width=180mm,height=225mm,angle=0.]{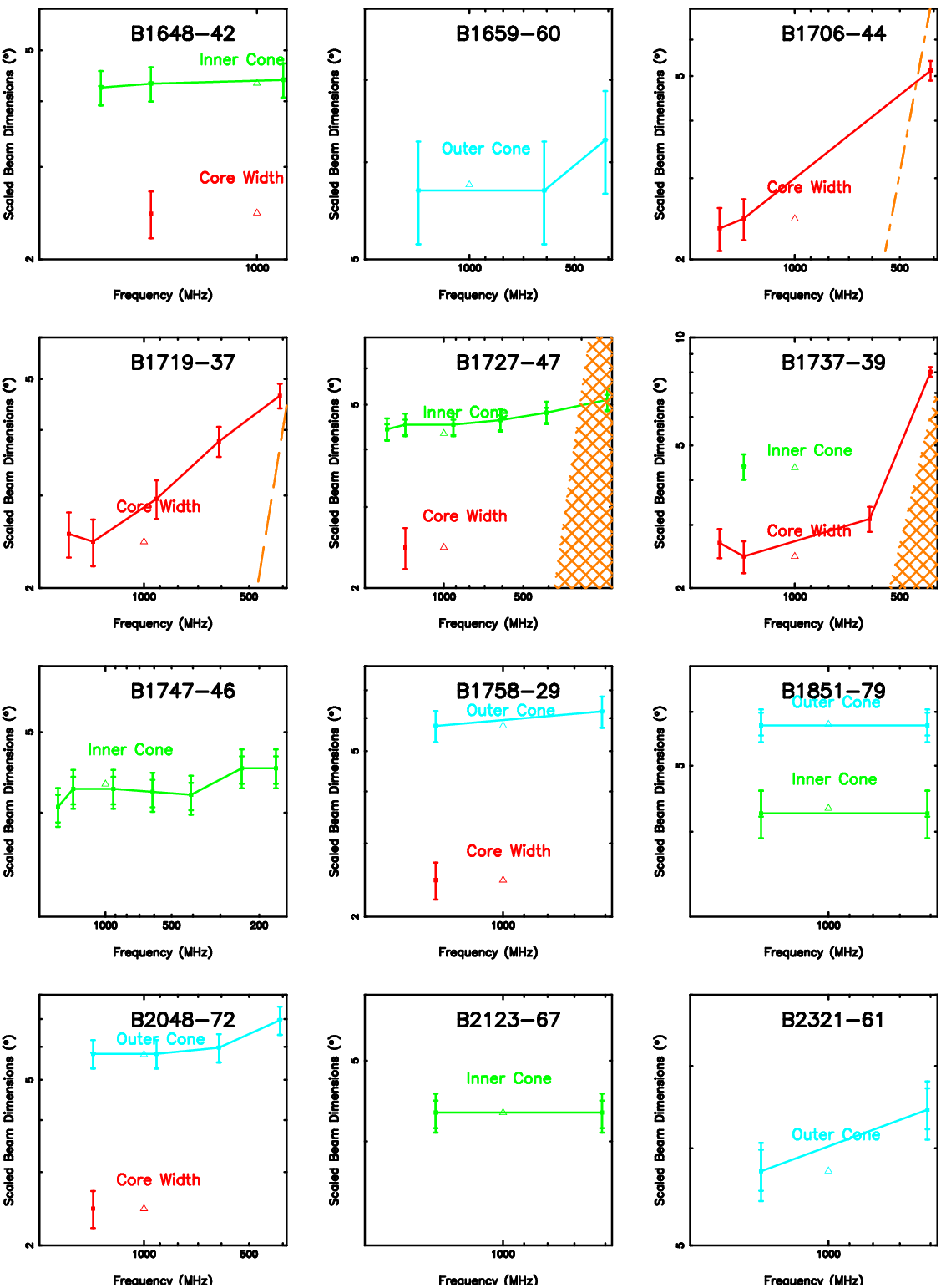}
\caption{Emission-beam geometry models as in Fig~\ref{figB1}.}
\label{figB6}
\end{center}
\end{figure*}



\begin{table*}
\caption{Far South 1.4-GHz Population Parameters}
\begin{tabular}{lcc|ccc|ccccccc}
\toprule
 Pulsar & DM & RM & L & B & Dist. & P & $\dot{P}$ & $\dot{E}$ & $\tau$ & $B_{surf}$ & $B_{12}/P^2$ & 1/Q \\ 
 (B1950) & ($pc/cm^{3}$) & ($rad$-$m^{2}$) & (\degr) & (\degr) & (kpc) & (s) & ($10^{-15}$ s/s) & ($10^{32}$ ergs/s)  & (Myr) & ($10^{12}$ G) &   &   \\
\midrule
\midrule
B0840--48 & 196.85 & 145.00 & 267.18 & -4.10 & 3.10 & 0.644 & 9.56 & 14.0 & 1.1 & 2.51 & 6.0 & 2.0 \\
B0901--63 & 72.72 & -59.20 & 280.39 & -11.08 & 0.19 & 0.660 & 0.11 & 0.15 & 97.8 & 0.27 & 0.6 & 0.3 \\
B0923--58 & 57.40 & -45.00 & 278.39 & -5.60 & 0.11 & 0.740 & 4.92 & 4.80 & 2.4 & 1.93 & 3.5 & 1.3 \\
B0950--38 & 162.88 & 331.70 & 268.70 & 12.03 & 0.52 & 1.374 & 0.58 & 0.09 & 37.5 & 0.90 & 0.5 & 0.3 \\
B1001--47 & 98.49 & 18.00 & 276.04 & 6.12 & 0.37 & 0.307 & 2.07 & 28.0 & 2.4 & 0.81 & 8.6 & 2.5 \\
\\ 
B1014--53 & 66.80 & -21.00 & 281.20 & 2.45 & 0.12 & 0.770 & 1.93 & 1.70 & 6.3 & 1.23 & 2.1 & 0.9 \\
B1015--56 & 438.70 & 332.80 & 282.73 & 0.34 & 3.51 & 0.503 & 3.14 & 9.70 & 2.5 & 1.27 & 5.0 & 1.7 \\
B1030--58 & 418.20 & 100.00 & 285.92 & -1.01 & 3.01 & 0.464 & 1.80 & 7.10 & 4.1 & 0.92 & 4.3 & 1.5 \\
B1044--57 & 240.20 & 133.00 & 287.07 & 0.73 & 2.34 & 0.369 & 1.15 & 9.00 & 5.1 & 0.66 & 4.8 & 1.6 \\
B1105--59 & 158.40 & -31.00 & 290.25 & 0.52 & 1.92 & 1.517 & 0.34 & 0.04 & 70.7 & 0.73 & 0.3 & 0.2 \\
\\ 
B1110--69 & 148.40 & -36.00 & 294.42 & -8.22 & 2.08 & 0.820 & 2.82 & 2.00 & 4.6 & 1.54 & 2.3 & 0.9 \\
B1118--79 & 27.40 & -11.00 & 298.71 & -17.50 & 0.81 & 2.281 & 3.67 & 0.12 & 9.9 & 2.93 & 0.6 & 0.3 \\
B1124--60 & 280.27 & -41.00 & 292.83 & 0.29 & 2.43 & 0.203 & 0.28 & 13.0 & 11.4 & 0.24 & 5.9 & 1.7 \\
B1236--68 & 94.30 & -49.00 & 301.88 & -5.69 & 1.58 & 1.302 & 11.88 & 2.10 & 1.7 & 3.98 & 2.3 & 1.0 \\
B1237--41 & 44.10 & 17.00 & 300.69 & 21.41 & 1.68 & 0.512 & 1.74 & 5.10 & 4.7 & 0.96 & 3.6 & 1.3 \\
\\ 
B1256--67 & 94.70 & -55.70 & 303.69 & -4.83 & 1.57 & 0.663 & 0.86 & 1.20 & 12.3 & 0.76 & 1.7 & 0.7 \\
B1322--66 & 209.60 & -47.00 & 306.31 & -4.37 & 6.73 & 0.543 & 5.33 & 13.00 & 1.6 & 1.72 & 5.8 & 1.9 \\
B1323--627 & 294.91 & 87.00 & 306.97 & -0.43 & 6.22 & 0.196 & 1.53 & 80.0 & 2.0 & 0.56 & 14.4 & 3.5 \\
B1323--63 & 502.70 & 226.40 & 306.75 & -1.53 & 11.99 & 0.793 & 3.09 & 2.40 & 4.1 & 1.58 & 2.5 & 1.0 \\
B1336--64 & 76.99 & -2.00 & 308.05 & -2.56 & 1.40 & 0.379 & 5.05 & 37.0 & 1.2 & 1.40 & 9.8 & 2.8 \\
\\ 
B1503--66 & 129.80 & -42.70 & 315.86 & -7.30 & 4.79 & 0.356 & 1.15 & 10.0 & 4.9 & 0.65 & 5.1 & 1.6 \\
B1508--57 & 627.47 & 510.00 & 320.77 & -0.11 & 6.84 & 0.129 & 6.85 & 1300 & 0.3 & 0.95 & 57.4 & 10 \\
B1509--58 & 252.50 & 216.00 & 320.32 & -1.16 & 4.40 & 0.152 & 1529 & 1.7E5 & 0.0 & 15.4 & 670 & 75 \\
B1510--48 & 51.50 & 18.00 & 325.87 & 7.84 & 1.24 & 0.455 & 0.93 & 3.90 & 7.8 & 0.66 & 3.2 & 1.2 \\
B1530--539 & 190.82 & -86.80 & 325.46 & 1.48 & 3.75 & 0.290 & 1.54 & 25.0 & 3.0 & 0.68 & 8.1 & 2.3 \\
\\ 
B1550--54 & 210.00 & 113.00 & 327.19 & -0.90 & 3.79 & 1.081 & 15.72 & 4.90 & 1.1 & 4.17 & 3.6 & 1.4 \\
B1556--57 & 176.55 & -131.00 & 325.97 & -3.70 & 4.18 & 0.194 & 2.13 & 110.0 & 1.5 & 0.65 & 17.2 & 4.1 \\
B1607--52 & 127.35 & -79.00 & 330.92 & -0.48 & 2.95 & 0.182 & 5.17 & 340.0 & 0.6 & 0.98 & 29.5 & 6.3 \\
B1609--47 & 161.20 & -138.00 & 334.57 & 2.84 & 3.52 & 0.382 & 0.63 & 4.50 & 9.6 & 0.50 & 3.4 & 1.2 \\
B1611--55 & 124.48 & 10.00 & 329.04 & -3.46 & 3.19 & 0.792 & 2.00 & 1.60 & 6.3 & 1.27 & 2.0 & 0.9 \\
\\ 
B1626--47 & 498.00 & -348.00 & 336.40 & 0.56 & 5.00 & 0.576 & 22.29 & 46.0 & 0.4 & 3.63 & 10.9 & 3.2 \\
B1630--44 & 474.10 & 159.00 & 338.73 & 1.98 & 14.94 & 0.437 & 6.20 & 29.0 & 1.1 & 1.67 & 8.8 & 2.6 \\
B1630--59 & 134.90 & 110.30 & 327.75 & -8.31 & 7.21 & 0.529 & 1.37 & 3.60 & 6.1 & 0.86 & 3.1 & 1.1 \\
B1634--45m & 193.23 & 10.00 & 338.48 & 0.76 & 3.44 & 0.119 & 3.19 & 750.0 & 0.6 & 0.62 & 44.2 & 8.3 \\
B1635--45 & 258.91 & -28.00 & 338.50 & 0.46 & 3.79 & 0.265 & 2.89 & 62.0 & 1.5 & 0.89 & 12.6 & 3.3 \\
\\ 
B1636--47 & 586.32 & -411.00 & 337.71 & -0.44 & 4.91 & 0.517 & 42.11 & 120.0 & 0.2 & 4.72 & 17.6 & 4.6 \\
B1643--43 & 490.40 & -62.00 & 341.11 & 0.97 & 6.23 & 0.232 & 112.8 & 3600 & 0.0 & 5.17 & 96.4 & 17 \\
B1647--528 & 164.00 & 42.10 & 334.59 & -5.52 & 5.81 & 0.891 & 2.12 & 1.20 & 6.7 & 1.39 & 1.8 & 0.8 \\
B1650--38 & 207.20 & -82.00 & 345.88 & 3.27 & 5.44 & 0.305 & 2.79 & 39.0 & 1.7 & 0.93 & 10.0 & 2.8 \\
B1658--37 & 303.40 & -605.90 & 347.76 & 2.83 & 12.95 & 2.455 & 11.12 & 0.30 & 3.5 & 5.29 & 0.9 & 0.5 \\
\\ 
B1713--40 & 306.90 & -809.00 & 346.76 & -1.89 & 7.28 & 0.888 & 3.70 & 2.10 & 3.8 & 1.83 & 2.3 & 1.0 \\
B1715--40 & 386.60 & -218.00 & 347.65 & -1.53 & 9.83 & 0.189 & 1.67 & 97.0 & 1.8 & 0.57 & 15.9 & 3.8 \\
B1718--36 & 416.20 & -307.00 & 350.93 & 0.00 & 3.99 & 0.399 & 4.46 & 28.0 & 1.4 & 1.35 & 8.5 & 2.5 \\
B1729--41 & 195.30 & -198.00 & 347.98 & -4.46 & 7.24 & 0.628 & 12.84 & 20.0 & 0.8 & 2.87 & 7.3 & 2.3 \\
B1730--37 & 153.18 & -335.00 & 351.58 & -2.28 & 3.15 & 0.338 & 15.04 & 150.0 & 0.4 & 2.28 & 20.0 & 4.9 \\
\\ 
B1804--12 & 122.41 & 255.90 & 17.14 & 4.42 & 3.01 & 0.523 & 1.41 & 3.90 & 5.9 & 0.87 & 3.2 & 1.2 \\

\bottomrule
\end{tabular}
\label{tabC12} 
\end{table*}

\twocolumn

\noindent\textit{\textbf{B0840--48m}}: This interpulsar has narrow components and a nearly 180\degr\ spacing, so it may well be emitting from both poles.  Both MP and IP are similar with angular widths close to that of the polar cap which strengthens the above case.  The PPA rate seems to be large suggesting a central traverse---and the MP seems to have some structure---but neither can be measured with any confidence.  So we suggest that the MP represents a core-single geometry.  A scattering time scale has been measured \citep{kmn+15}.   
\vskip 0.07in
\noindent\textit{\textbf{B0901--63}}:  The single profile with a shallow PPA rate and edge depolarization suggests conal emission, though it is unusual to see the prominent antisymmetric $V$ in such a profile.  Nonetheless, we tilt toward modeling it with a conal single {\textbf S$_d$} beam model.  
\vskip 0.07in
\noindent\textit{\textbf{B0923--58}}: This symmetrical single profile could reflect either core or conal emission.  Owing to the large $L/I$ and edge depolarization we tilt toward a conal model, using an inner cone but an outer one is also possible.  
\vskip 0.07in
\noindent\textit{\textbf{B0950--38}}: Little can be done with this marginally detected profile.  A small scattering value is available \citep{kmn+15}.
\vskip 0.07in
\noindent\textit{\textbf{B1001--47}}: The 1.4-GHz profile shows a triple structure with a steep PPA rate and a hint of antisymmetric $V$.  The bright central feature is surely core emission, so the beam seems to have a core/inner core 
{\textbf T} or {\textbf S$_t$} structure. No scattering time measurement is available.  
\vskip 0.07in
\noindent\textit{\textbf{B1014--53}}: The narrow single depolarized profile hints of being conal in origin, but no beam model is possible.  No scattering time 
has been published.  
\vskip 0.07in
\noindent\textit{\textbf{B1015--56}}: The profile shows three components, the bright central one with substantial RH $V$ and a well defined PPA rate.  The core width is a little smaller than the polar cap diameter, suggesting that it is incomplete.  Otherwise, this seems to have a classic core/inner core {\textbf T} or {\textbf S$_t$} beam structure.  \citet{kmn+15} measure a small but significant scattering level.
\vskip 0.07in
\noindent\textit{\textbf{B1030--58}}: The poorly defined profile with no discernible PPA track makes modeling impossible.  \citet{abs86} measure a $t_{\rm scatt}$ value.  
\vskip 0.07in
\noindent\textit{\textbf{B1044--57}}:  The profile seems to have a pair of components around the central one, the leading relatively bright and the trailing much weaker.  The central component has a width of close to 4\degr, which is just the polar cap size, probably confirming it as the core.  The PPA rate cannot be measured, the suggestion of flatness suggests a central sightline traverse, and with this model in turn is clearly inner conal.  We cannot know from a single observation whether the profile is core-single {\textbf S$_t$} or triple {\textbf T}.  \citet{kmn+15} have measured a scattering time scale.  
\vskip 0.07in
\noindent\textit{\textbf{B1105--59}}: The poor single depolarized profile hints of being conal in origin, but no beam model is possible.  No scattering time has been published.   
\vskip 0.07in	
\noindent\textit{\textbf{B1110--69}}: The two closely separated components appear to indicate a conal double structure.  One has to trust that the poorly defined PPA traverse can be interpreted accurately, but if so an inner conal 
double beam model is appropriate.  
\vskip 0.07in	
\noindent\textit{\textbf{B1118--79}}: The pulsar shows a well separated two-component profile, and we model it with a usual outer conal double geometry---though no lower frequencies are available to check for width increase at lower frequencies.  \citet{kmn+15} measure a significant scattering level.
\vskip 0.07in	
\noindent\textit{\textbf{B1124--60}}: The marginal quality of this profile's polarization make it a stretch to interpret, but if its interpulse reflects emission from the second pole, $\alpha$ must be hear 90\degr.  Then, if $R$ is about --12\degr/\degr\ 
and the conal width about 17\degr, an inner cone {\textbf S$_t$} beam geometry is indicated.  No scattering information is available.  
\vskip 0.07in	
\noindent\textit{\textbf{B1236--68}}: This profile with a clear strong leading and weak trailing component could be entirely conal, but the antisymmetric $V$ under what seems to be a conflated core feature tilts toward it having a core-cone triple {\textbf S$_t$} configuration---and if the PPA rate is actually about --20\degr/\degr, then an inner cone is indicated with a core width close to 3\degr.  No scattering time is available.  
\vskip 0.07in	
\noindent\textit{\textbf{B1237--41}}: Little can be said about this pulsar.  The profile is not measured accurately enough to assess its structure. 
\vskip 0.07in	
\noindent\textit{\textbf{B1256--67}}: Very possibly a usual conal double profile.  If the PPA traverse is about --16\degr/\degr\ as the interior values seem to suggest, an outer conal geometry is indicated.  No scattering data has been published.  
\vskip 0.07in	
\noindent\textit{\textbf{B1322--66}}:  The profile seems to reflect a core-cone triple beam system.  The PPA rate and conal dimensions are well determined, but the core width can only estimated.  A value near 9\degr\ seems to be plausible and would require and inner conal geometry.  
\vskip 0.07in	
\noindent\textit{\textbf{B1323--627}}: This profile exemplifies what we do not understand about the emission of fast, energetic pulsars:  There seems to be no core emission.  (Interpreting the bright component as a core feature gives a bizarre geometry.)  So perhaps this is a conal double profile with two 90\degr\ ``jumps''.  The profile is wide and the PPA rate is shallow.  Maybe the first ``jump'' is associated with edge depolarization of the leading component.  A conal double model is possible where $\alpha$ is some 12\degr\ and $\beta$ 8\degr.  Interestingly, the polar-cap diameter is some 5.5\degr, so the sightline would miss the core by three core beam radii.  Is this enough for the core to be invisible?
\vskip 0.07in	
\noindent\textit{\textbf{B1323--63}}: The profile is a classic core-cone triple with antisymmetric $V$ and a probable positive PPA rate.  We model it using a core/inner cone geometry.  \citet{abs86} measure significant scattering at low frequency.  
\vskip 0.07in	
\noindent\textit{\textbf{B1336--64}}: The profile seems to have three conflated features with a poorly defined PPA traverse.  If the PPA rate is near +9\degr/\degr\ and the core width is plausibly some 4\degr, an outer conal configuration is indicated.  No scattering time has been reported.  
\vskip 0.07in	
\noindent\textit{\textbf{B1503--66}}:  The narrow profile has a bright component that is hardly 4\degr\ wide, just the size here of the polar cap, strongly suggesting that this is core emission.  The PPA rate can only be guessed at, but may be some +9\degr/\degr.  The trailing component may be a conal ``outrider'' and a second weak one may be just visible on the leading edge.  If all this proves out, this is an inner conal core single profile and so we model it.  
\vskip 0.07in	
\noindent\textit{\textbf{B1508--57}}:  \citet{joh90} measure a large scattering time scale for this pulsar, so the profile asymmetry is probably due to this effect with some consequent flattening of the PPA traverse.  Maybe the intrinsic width measured on the leading edge is as low as 10\degr, and the PPA rate stepper than some +5\degr/\degr.  We can then model the profile provisionally as a core-single.  No structure is seen to indicate conal outriders.
\vskip 0.07in	
\noindent\textit{\textbf{B1509--58}}: The highly linearly polarized single profile of this high $\dot E$ pulsar is probably comprised of core radiation.  The PPA rate is well defined but very shallow, and no conal components are discernible.  No $t_{\rm scatt}$ measurement is 
available.
\vskip 0.07in	
\noindent\textit{\textbf{B1510--48}}:  Little definite can be said about this profile because no PPA traverse or rate can be discerned.  The two bright components are probably conal, but weak core emission could fill the center of the profile.  
\vskip 0.07in	
\noindent\textit{\textbf{B1530--539}}: The profile seems to be triple, but its form makes accurate measurement difficult, and the fragmented PPA traverse leaves no option but assuming a central sightline traverse.  An inner conal triple {\textbf T} configuration then results.  No $t_{\rm scatt}$ value has been published. 
\vskip 0.07in	
\noindent\textit{\textbf{B1550--54}}: The 1.4-GHz profile seems to have the usual core/inner cone {\textbf S$_t$} geometry, though the PPA track is inconsistent but seemingly steep, and the core width value is poorly determined.  No $t_{\rm scatt}$ has been measured.  
\vskip 0.07in	
\noindent\textit{\textbf{B1556--57}}: The profile is triple with a brighter, narrower trailing conal outrider, so the PPA rate cannot be determined and a central traverse is assumed.  The dimensions then give an outer conal triple configuration.  No $t_{\rm scatt}$ measurement is available.
\vskip 0.07in	
\noindent\textit{\textbf{B1607--52}}: We can make no sensible interpretation of this energetic pulsar's narrow profile without more information.  It seems to have a 3-4-part structure, but the bright feature is too narrow to be a core component.  Even if one conjectures a two-part core, the remaining structure is too narrow to interpret as a conal component pair.  
\vskip 0.07in	
\noindent\textit{\textbf{B1609--47}}:  Very possibly a conal double structure if the shallow poorly determined $R$ can be relied upon.  \citet{kmn+15} find a small scattering value that does not significantly distort the profile we have.  
\vskip 0.07in	
\noindent\textit{\textbf{B1611--55}}: Likely a usual closely spaced conal double structure, but the nearly complete depolarization frustrates any possibility of a suitable model.  No scattering time is available.  
\vskip 0.07in	
\noindent\textit{\textbf{B1626--47}}:  The 1.4-GHz profile is too scattered to be useful, but the \citet{JKW06} 8.4-GHz observation suggests a core-cone triple configuration with roughly compatible dimensions for an inner cone. A large $t_{\rm scatt}$ value has been measured by \citet{joh90}.  
\vskip 0.07in	
\noindent\textit{\textbf{B1630--44}}:  The profile shows a clear scattering tail and no other is available, so no beam model is possible.  \citet{joh90} measure a very large scattering time. 
\vskip 0.07in	
\noindent\textit{\textbf{B1630--59}}:  The noisy profile shows three components in what appears to be a core-cone triple configuration.  The PPA rate seems to be positive and about +8\degr/\degr.  With a plausible core width of about 5\degr, an outer conal geometry is indicated.  No $t_{\rm scatt}$ value has been measured.
\vskip 0.07in	
\noindent\textit{\textbf{B1634--45m}}: The fast interpulsar probably has a core single main pulse.  Its width is a little larger than the polar cap size, and no structure is discernible, but some broadening due to scattering is possible---especially given the flat PPA---though no measurement is available.  The IP seems to be a little broader.  Given the accurate 180\degr\ separation, we presume that $\alpha$ 
is about 90\degr.  
\vskip 0.07in	
\noindent\textit{\textbf{B1635--45}}:  Little can be said about this noisy profile with no clear PPA rated apart than that \citet{kmn+15} have measured a scattering time scale.  
\vskip 0.07in	
\noindent\textit{\textbf{B1636--47}}: The single profile seems only a little distorted by the very large scattering time measured by \citet{joh90}, but the flat PPA track is also indicative.  Probably the emission is core-dominated but no beam model is possible.  
\vskip 0.07in	
\noindent\textit{\textbf{B1643--43}}: The asymmetric single profile shows what seems to be a scattering ``tail'', and \citet{joh90} measure a very large scattering time.  While probably a core-dominated beam, no model is then possible.   
\vskip 0.07in	
\noindent\textit{\textbf{B1647--528}}:  The pulsar's three part profile seems to have a core single or triple geometry, despite its somewhat small $\dot E$.  The PPA rate is well defined, and an inner conal model seems to work well.  
\vskip 0.07in	
\noindent\textit{\textbf{B1650--38}}:  One can only guess at a quantiative model for this pulsar, but it seems likely that the bright component is mostly core radiation.  The leading component does not have the flat PPA seen in many precursors, so may be a leading conal ``outrider''---though no trailing one is clearly apparent.  If the PPA rate is about +14\degr/\degr as might be compatible with the entire traverse and the profile width down to the far trailing edge about 14\degr, an inner conal geometry would be indicated. \citet{joh90} measure a scattering time.  
\vskip 0.07in	
\noindent\textit{\textbf{B1658--37}}: This well measured profile appears to have a conal quadruple {c\textbf Q} beam system where the inner conal width can be estimated.  The large $V$ is unusual if the emission is entirely conal, so some core power may be conflated in the profile center.  The double cone model needs a PPA rate of about --12\degr/\degr, and this may be accommodated if one takes the PPA rate as a sort of average over the traverse ignoring the steeper central portion.  \citet{kmn+15} give a scattering timescale. 
\vskip 0.07in	
\noindent\textit{\textbf{B1713--40}}: The bright single profile shows a clear scattering ``tail'' compatible with the very large scattering measured by \citet{khs+14}.  No beam model is then possible.  
\vskip 0.07in	
\noindent\textit{\textbf{B1715--40}}: This somewhat noisy profile suggests three components and gives a hint of the PPA rate.  We model it as an inner conal core single successfully.  \citet{joh90} measures a large scattering time.
\vskip 0.07in	
\noindent\textit{\textbf{B1718--36}}:  This is another example of an energetic pulsar with a difficult profile to interpret---that is, where core emission is expected and not obviously present.  An outer  conal double model is possible where the sightline would miss the core by about three core-beam radii, so perhaps any core emission is conflated with the conal in the profile center.  Observations at other frequencies may resolve what is the full geometry here.  A scattering timescale has been measured 
\citep{kmn+15}.
\vskip 0.07in	
\noindent\textit{\textbf{B1729--41}}:  The profile is double with what seems to be a conflated narrow core feature in the center.  The PPA traverse is very poorly defined, but may be steep on the poor basis of two points near the $L$ maximum.  Taking the core width as about that of the polar cap, an inner cone/core {\textbf S$_t$} model results.  \citet{kmn+15} find a small scattering value. 
\vskip 0.07in	
\noindent\textit{\textbf{B1730--37}}:  This energetic pulsar seems to have a precursor some 40\degr/ ahead of a core-single feature.  The contrasting flat PPA for the former and clear negative traverse for the latter show this clearly.  No scattering time has been published.  
\vskip 0.07in	
\noindent\textit{\textbf{B1804--12}}:  The 1.4-GHz profile shows a well resolved double form but with substantial filling and some hints of an inner feature.  The PPA rate is steep and very well defined.  We model it with an inner conal double {\textbf D} geometry, leaving open the possibility of some barely resolved core emission in the center, and the needed width of about 6\degr\ is plausible.  \citet{kmn+15} measure a scattering time scale. 
\vskip 0.07in

\begin{table}[hbt!]
\caption{Far South Multiband Population Emission Beam Geometry}
\setlength{\tabcolsep}{2pt}
\begin{tabular}{lc|cccc|cccc|}
  \hline
      Pulsar &  Class & $W_{c}$ & $\alpha$ & $R$ & $\beta$ &  $W_i$ & $\rho_i$ & $W_o$  & $\rho_o$ \\
  &   & (\degr) & (\degr) & (\degr/\degr) & (\degr) & (\degr) & (\degr) & (\degr) & (\degr)  \\
  \hline
  & & \multicolumn{4}{c|}{1-GHz Geometry} & \multicolumn{4}{c|}{1-GHz Cone Sizes} \\
  \midrule
  \midrule
  B0840-48m & St? & $\sim$3 & {\bf 90} &  --- &  --- &  --- &  --- &  --- &  --- \\
B0901-63 & Sd? & 0 & 16 & +3 & +5.3 & $\sim$10 & 5.5 &  --- &  --- \\
B0923-58 & Sd/St? & 0 & 19 & -6 & +3.0 & $\sim$24 & 5.2 &  --- &  --- \\
B0950-38 & ?? & 0 &  --- &  --- &  --- &  --- &  --- &  --- &  --- \\
B1001-47 & St/T? & 7.5 & {\bf 36} & +25 & +1.4 & $\sim$25 & 7.6 &  --- &  --- \\
\\ 
B1014-53 & Sd? & 0 &  --- &  --- &  --- &  --- &  --- &  --- &  --- \\
B1015-56 & St/T? & $\sim$3 & {\bf 90} & +30 & +1.9 & $\sim$10 & 5.4 &  --- &  --- \\
B1030-58 & ?? & 0 &  --- &  --- &  --- &  --- &  --- &  --- &  --- \\
B1044-57 & St/T? & $\sim$4 & {\bf 90} & $\infty$ & 0.0 & $\sim$14 & 7.0 &  --- &  --- \\
B1105-59 & Sd? & 0 &  --- &  --- &  --- &  --- &  --- &  --- &  --- \\
\\
B1110-69 & D? & 0 & 55 & +20 & +2.3 & $\sim$10 & 4.8 &  --- &  --- \\
B1118-79 & D? & 0 & 16 & +30 & +0.5 &  --- &  --- & $\sim$27 & 3.8 \\
B1124-60 & St? & $\sim$5 & {\bf $\sim$90} & -12 & +4.8 & $\sim$17 & 9.7 &  --- &  --- \\
B1236-68 & St/T? & $\sim$3 & {\bf 46} & -20 & -2.1 & 9.5 & 3.9 &  --- &  --- \\
B1237-41 & ?? & 0 &  --- &  --- &  --- &  --- &  --- &  --- &  --- \\
\\
B1256-67 & D? & 0 & 70 & -16 & +3.4 &  --- &  --- & $\sim$13 & 7.0 \\
B1322-66 & T & $\approx$9? & {\bf 21} & +5.5 & +3.8 & $\sim$24 & 6.0 &  --- &  --- \\
B1323-627 & D?? & 0 & 12 & -1.5 & +8.0 & $\sim$45 & 10.0 &  --- &  --- \\
B1323-63 & St/T? & $\sim$4 & {\bf 43} & +12 & +3.3 & $\sim$10 & 4.8 &  --- &  --- \\
B1336-64 & St/T? & $\sim$4 & {\bf 65} & +9 & +5.8 &  --- &  --- & $\sim$16 & 9.4 \\
\\
B1503-66 & St? & $\sim$4 & {\bf $\sim$90} & +9 & +6.7 & $\sim$6 & 7.3 &  --- &  --- \\
B1508-57 & St? & $\sim$10 & {\bf 43} & +5 & +7.9 &  --- &  --- &  --- &  --- \\
B1509-58 & St? & $\sim$27 & {\bf 13} & +12 & +1.1 &  --- &  --- &  --- &  --- \\
B1510-48 & D/T? & 0 &  --- &  --- &  --- &  --- &  --- &  --- &  --- \\
B1530-539 & T? & $\sim$6 & {\bf 47} & $\infty$ & +0.0 & $\sim$22 & 8.1 &  --- &  --- \\
\\
B1550-54 & St/T? & $\sim$4 & {\bf 40} & +45 & +0.8 & $\sim$13 & 4.3 &  --- &  --- \\
B1556-57 & T? & $\sim$8 & {\bf 43} & $\infty$ & +0.0 &  --- &  --- & $\sim$39 & 13.2 \\
B1607-52 & ?? & 0 &  --- &  --- &  --- &  --- &  --- &  --- &  --- \\
B1609-47 & D? & 0 & 28 & -4 & -6.7 & $\sim$9 & 7.0 &  --- &  --- \\
B1611-55 & D? & 0 &  --- &  --- &  --- &  --- &  --- &  --- &  --- \\
\\
B1626-47 & St? & $\sim$10 & {\bf 19} & $\infty$ & +0.0 & $\sim$35 & 5.6 &  --- &  --- \\
B1630-44 & ?? & 0 &  --- &  --- &  --- &  --- &  --- &  --- &  --- \\
B1630-59 & St/T? & $\sim$5 & {\bf 42} & +8 & +4.8 &  --- &  --- & $\sim$18 & 8.0 \\
B1634-45m & St? & $\sim$10 & {\bf $\sim$90?} &  --- &  --- &  --- &  --- &  --- &  --- \\
B1635-45 & ?? & 0 &  --- &  --- &  --- &  --- &  --- &  --- &  --- \\
\\
B1636-47 & ?? & 0 &  --- &  --- &  --- &  --- &  --- &  --- &  --- \\
B1643-43 & ?? & 0 &  --- &  --- &  --- &  --- &  --- &  --- &  --- \\
B1647-528 & St/T? & $\approx$5? & {\bf 31} & +19 & 1.5 & $\sim$17 & 4.7 &  --- &  --- \\
B1650-38 & St? & $\sim$4 & {\bf $\sim$90} & +14 & +4.1 & $\sim$14 & 8.1 &  --- &  --- \\
B1658-37 & cQ & 0 & 26 & -12 & +2.1 & $\sim$8 & 2.8 & $\sim$13 & 3.6 \\
\\
B1713-40 & ?? & 0 &  --- &  --- &  --- &  --- &  --- &  --- &  --- \\
B1715-40 & St? & $\sim$6 & {\bf $\sim$90} & -8 & +7.2 & $\sim$14 & 10.0 &  --- &  --- \\
B1718-36 & D/T? & 0 & 28 & -4.1 & +6.6 &  --- &  --- & $\sim$24 & 9.0 \\
B1729-41 & St/T? & $\sim$3 & {\bf $\sim$90} & +18 & +3.2 & $\sim$9 & 5.5 &  --- &  --- \\
B1730-37 & PC/St? & $\sim$14 & {\bf 18} & -1.25 & -13.9 &  --- &  --- &  --- &  --- \\
\\
B1804-12 & D/T? & $\approx$6? & {\bf 33} & -30 & 1.0 & 21.8 & 6.0 &  --- &  --- \\
  \bottomrule
\end{tabular}
 \label{tabC3}
\end{table}


\end{document}